\documentclass[nofootinbib,showpacs,preprintnumbers,amsmath,amssymb,floatfix,superscriptaddress,twocolumn]{revtex4-1}
\usepackage{graphicx,graphics,wrapfig,rotating}         
\usepackage{dcolumn}                            
\usepackage{bm,fancybox}                       
\usepackage{times,euscript,eufrak,oldgerm}              %
\usepackage[english]{babel}          
\usepackage{mathrsfs}
\usepackage{bbm,bm,fancybox}
\usepackage{natbib}
\usepackage{times}
\usepackage[usenames,dvipsnames]{color}
\usepackage[normalem]{ulem}
\usepackage{hyperref}

\newcommand{\rom}[1]{\mathrm{#1}}

\newcommand{\ket}[1]{\ensuremath{|#1\rangle}}
\newcommand{\bra}[1]{\ensuremath{\langle#1|}}

\numberwithin{equation}{section}

\newcommand{\sys}{\mathcal{S}}
\newcommand{\res}{\mathcal{R}}
\newcommand{\beq}{\begin{equation}}
\newcommand{\eeq}{\end{equation}}
\newcommand{\bea}[1]{\begin{equation}\begin{array}{#1}}
\newcommand{\eea}{\end{array}\end{equation}}
\newcommand{\beqn}{\begin{eqnarray}}
\newcommand{\eeqn}{\end{eqnarray}}
\DeclareMathOperator{\tr}{Tr}
\providecommand{\openone}{\mathbbm{1}}

\newcommand{\hil}{\mathcal{H}}
\newcommand{\proj}[1]{\ensuremath{|#1\rangle \langle #1|}}

\newtheorem{theorem}{Theorem}

\newtheorem{definition}[theorem]{Definition}

\newtheorem{criterion}[theorem]{Criterion}

\begin{document}
\title{Open-System Dynamics of Entanglement}
\author{Leandro Aolita}
\affiliation{Dahlem Center for Complex Quantum Systems, Freie Universit\"{a}t Berlin, Berlin, Germany}
\author{Fernando de Melo}
\affiliation{Centro Brasileiro de Pesquisas F\'{\i}sicas, Rua Dr. Xavier Sigaud 150, Rio de Janeiro, RJ 22290-180, Brazil}
\author{Luiz Davidovich}
\affiliation{Instituto de F\'{\i}sica, Universidade Federal do Rio de
Janeiro, Caixa Postal 68528, Rio de Janeiro, RJ 21941-972, Brazil}

\begin{abstract}

One of the greatest challenges in the fields of quantum information processing and quantum technologies is the detailed coherent control over each and all of the constituents of quantum systems with an ever increasing number of particles. Within this endeavor, the harnessing  of many-body entanglement against the detrimental effects of the environment is a major and pressing issue. Besides being an important concept from a fundamental standpoint, entanglement has been recognised as a crucial resource for quantum speed-ups or performance enhancements over classical methods. Understanding and controlling many-body entanglement in open systems may have strong implications in quantum computing, quantum simulations of many-body systems, secure quantum communication or cryptography, quantum metrology, our understanding of the quantum-to-classical transition, and other important questions of quantum foundations.

In this paper we present an overview of recent theoretical and experimental efforts to underpin the dynamics of entanglement under the influence of noise. Entanglement is thus taken as a dynamic quantity on its own, and we survey how it evolves due to the unavoidable interaction of the entangled system with its surroundings.  We analyse several scenarios, corresponding to different families of states and environments, which render a very rich diversity of dynamical behaviours. 

In contrast to single-particle quantities, like populations and coherences, which typically vanish only asymptotically in time, entanglement may disappear at a finite time. In addition, important classes of entanglement display an exponential decay with the number of particles when subject to local noise,  which poses yet another threat to the already-challenging scaling of quantum technologies. Other classes, however, turn out to be extremely robust against local noise. Theoretical results and recent experiments regarding the difference between local and global decoherence are summarized. Control and robustness-enhancement techniques, scaling laws, statistical and geometrical aspects of multipartite-entanglement decay are also reviewed; all in order to give a broad picture of entanglement dynamics in open quantum systems addressed to both theorists and experimentalist inside and outside the field of quantum information.

\end{abstract}


\maketitle
%
\pagenumbering{roman}
%


%
%

%

\tableofcontents 


\pagenumbering{arabic}

\section{Introduction}
\label{1}

Since the seminal paper by Albert Einstein, Boris Podolski, and Nathan Rosen \cite{epr35} in 1935, and the famous series of papers published by Erwin Schr\"odinger in the years 1935 and 1936 \cite{schrodinger35,schrodinger2,schrodinger3}, entanglement has occupied a central position in quantum physics. This peculiar phenomenon has posed formidable challenges to several generations of physicists. In fact, it took about 30 years since the 1935 papers for this mathematical property to gain a physical  consequence, as was demonstrated by John S. Bell  \cite{bellepr64,bell04}; and nearly 30 further years for  it to be identified as a resource for quantum information processing and transmission \cite{ekert91, bennett92a, bennett93, shor97, steane98,bennetnature,nielsenchuang,bouwmeester00}. 

Schr\"odinger summarized, in a way that in modern terms would be based on the notion of information, the main ingredient of this phenomenon. In the first paper of the series of three published in Naturwissenchaften in 1935 \cite{schrodinger35}, he states that ``this is the reason that knowledge of the individual systems can decline to the scantiest, even to zero, while knowledge of the combined system remains continually maximal. Best possible knowledge of a whole does not include best possible knowledge of its parts - and that is what keeps coming back to haunt us.'' 

This is the case for a singlet state of two spin one-half particles. Even though the two-party state is completely known (pure state, corresponding to a total spin equal to zero), each part is described by a statistical mixture with a 50-50 chance for each particle to have spin up or down. On the other hand, measurement of the spin of one of the particles determines the spin of the other, even if the two parties are far apart. This was referred to by Einstein, in a letter to Born in 1947 \cite{einsteinborn}, as a ``spooky action at a distance.''

Evolving from a daunting concept to a useful resource, entanglement is nowadays known to be at the heart of many potential applications, such as the efficient transmission of information through dense coding \cite{bennett92a,mattle}  or teleportation \cite{bennett93,davidovich0,zeilinger97,boschi,Riebe04}, the security of transmitted data through entanglement-based quantum cryptography \cite{ekert91,gisin02}, including the recent development of device-independent quantum cryptography \cite{Barrett05a,Acin2006a,Acin07}, as well as both device-independent randomness generation \cite{Colbeck07,Pironio10} or amplification \cite{Colbeck12,Gallego13,Brandao13}, quantum metrology \cite{Braunstein1996,leibfried04,escher,GLM2011:222}, the efficient solution of the factorization problem \cite{shor97}, the efficient quantum simulation of many-body physical problems that may be classically intractable \cite{feynman82,lloyd96,Jaksch98, Greiner02,Bloch08,cirac2012,bloch2012,blatt2012,guzik2012,houck2012}, or of sampling problems proven (modulo widely accepted complexity-theoretical assumptions) classically hard \cite{AA11}, and universal quantum computing in general \cite{nielsenchuang}.

Motivated by these potential applications, and also by the fundamental role played by entanglement in quantum mechanics, important experimental results have been obtained in the last few years, concerning the generation of multiparty entangled states, the transfer of entanglement between two systems, macroscopic signatures of entanglement, and the dynamics of entangled states under the influence of the environment. These results were made possible by the development of  experimental methods that allowed measuring and manipulating individual quantum systems, pioneered by David Wineland and Serge Haroche, awarded the Nobel Prize in Physics in 2012. Examples are the step-by-step generation of multiparticle entanglement among atoms and photons in a microwave cavity \cite{arno}, the demonstration of entanglement between a single neutral  atom -- or charged ion -- and its spontaneously-emitted single photon  with the assistance of an optical cavity \cite{Kuhn02,McKeever04,keller04}, between two photons sequentially emitted by the same single atom in the cavity \cite{Wilk08}, and even between a charged ion and its emitted photon without the assistance of any cavity \cite{blinov}, the mapping of photonic entanglement into and out of an atomic-ensemble quantum memory \cite{Julsgaard04, choi08,Clausen11}, the generation of multiparticle entanglement of trapped ions \cite{sackett00,leibfried05,haeffner05,monz2011},  of multiphoton entangled states \cite{pan00,zhao,Walther05,lu07,Prevedel07, Pino08, Prevedel09, Ceccarelli09, tenqubitpan,pan2012,Yao2012,Huang2012}, of entanglement among separate atomic samples \cite{julsgaard01,Choi10}, of artificial-atom \cite{Majer07} and photonic \cite{Matthews09} entanglement in on-chip integrated circuits, and the demonstration that the magnetic susceptibility at low temperatures yields information on the ground-state entanglement of magnetic materials \cite{susan}. 

And yet many fundamental problems remain unsolved. Among them, the characterization of entanglement for multiparticle systems or bipartite systems of large dimensions in general (mixed) states, and the dynamics of entanglement for a system in contact with its environment. This last problem is the main focus of this paper. It is directly related to important practical questions: the robustness of quantum communication schemes, quantum simulators and quantum computers, and the ultimate precision in the estimation of parameters, subject at the core of quantum metrology. It also concerns a fundamental problem in modern physics: the subtle relation between the classical and the quantum world. 

This very question is present in one of the first papers published by Schr\"odinger in 1926  \cite{schrodinger26a}, where, considering the behavior of the eigenfunctions of the harmonic oscillator, he remarks that ``at first sight it  appears very strange to try to describe a process, which we previously regarded as belonging to particle mechanics, by a system of such proper vibrations.'' In order to demonstrate ``{\it in concreto} the transition to macroscopic mechanics,'' he then remarks that "a group of proper vibrations'' of high-order quantum number $n$  and of relatively small-order quantum number differences may represent a particle executing the motion expected from usual mechanics, i. e. oscillating with a constant frequency. This ``group of proper vibrations'' was actually a coherent state, later studied by Glauber \cite{glauber1,glauber2} in great detail.

Schr\"odinger realized however that this argument was not enough to guarantee that the new quantum physics would correctly describe the classical world. In Section 5 of his three-part essay on ``The Present Situation in Quantum Mechanics,'' published in 1935 \cite{schrodinger35},  he notes that ``an uncertainty originally restricted to the atomic domain has become transformed into a macroscopic uncertainty, which can be resolved through direct observation.'' This remark was prompted by his famous Schr\"odinger-cat example, in which a decaying atom leads to a coherent superposition of two macroscopically distinct states, corresponding respectively to a cat that is either dead or alive. He adds that ``this inhibits us from accepting in a naive way a `blurred model' as an image of reality... There is a difference between a shaky or not sharply focused photograph and a photograph of clouds and fogbanks.'' This problem is also mentioned by Einstein in a letter to Max Born in 1954 \cite{einsteinborn}, where he considers a fundamental problem of quantum mechanics ``the inexistence at the classical level of the majority of states allowed by quantum mechanics,'' namely coherent superpositions of two or more macroscopically localized states.

These comments are very relevant to quantum measurement theory, as pointed out by Von Neumann \cite{vonneumann32,wheeler83}. Indeed, let us assume for instance that a microscopic
two-level system (states $|+\rangle$ and $|-\rangle$) interacts with a
macroscopic measuring apparatus in such a way that the pointer of the
apparatus points to a different (and classically distinguishable!)
position for each of the states $|+\rangle$ and $|-\rangle$. That is, we assume that the the joint atom-apparatus initial state transforms into
\begin{eqnarray}\label{vn0}
|+\rangle|\uparrow\rangle&\rightarrow&|+\rangle^\prime|\nearrow\,\,\rangle\,,\nonumber
\\
|-\rangle|\uparrow\rangle&\rightarrow&|-\rangle^\prime|\nwarrow\,\,\rangle\,,
\end{eqnarray}
where we allow for a change in the state of the system due to the interaction. The linearity of quantum mechanics implies that, if the system
is prepared in say the coherent superposition 
$|\Psi\rangle=(|+\rangle+|-\rangle)/\sqrt{2}$, the final state of the joint system should be a coherent superposition of two product states, each of which corresponds to a different position of the
pointer:
\begin{eqnarray}\label{vn}
&{}&(1/\sqrt{2})(|+\rangle+|-\rangle)|\uparrow\rangle\nonumber\\
&\rightarrow&(1/\sqrt{2})(|+\rangle^\prime|\nearrow\,\,\rangle+|-\rangle^\prime|\nwarrow\,\,\rangle)\nonumber\\
&\doteq&(1/\sqrt{2})(|\nearrow\,\,\rangle^\prime+|\nwarrow\,\,\rangle^\prime)\,.
\end{eqnarray}
In the last step, it is assumed that the two-level system is 
incorporated into the measurement apparatus after their interaction
(for  instance, an atom that gets stuck to the detector). One gets,
therefore, as a result of the interaction between the microscopic and the
macroscopic system, a coherent superposition of two classically distinct states of
the macroscopic apparatus. 
This would imply that one should be able
in principle to get interference between the two states of the pointer:
it is precisely the lack of evidence of such phenomena in the macroscopic
world that motivated Einstein's concern.

One knows nowadays that decoherence plays a fundamental role in the emergence of the classical world from quantum physics \cite{caldeira,joos,paz01,zurek:715,  schlosshauer07}. Theoretical \cite{caldeira,zurek:715,paz01, joos,davidovich1,davidovich2} and experimental \cite{enscat,myatt2,Deleglise08} research have demonstrated that a coherent superposition of two macroscopically distinguishable states (a ``Schr\"odinger-cat-like'' state) decays to a mixture of the same states with a characteristic time that is inversely proportional to some macroscopicity parameter. The decay law is, within a very good approximation, exponential. 

An important question remains, however, about the ultimate limits of applicability of quantum mechanics for macroscopic systems \cite{leggett02}. Recent experiments, involving entanglement between  macroscopic objects \cite{julsgaard01,lee11}, or micro-macro entanglement between a single photon and a macro system involving up to a hundred million photons \cite{demartini,lvovsky2013,bruno}, have pushed these limits further. Micro-macro entanglement is precisely the one involved in Eq.~\eqref{vn0}. Pushing quantum superpositions or entangled states to ever increasing macroscopic scales submits quantum physics to stringent tests, involving for instance probing the effect of the gravitational field, when massive objects like micro-mirrors are involved \cite{gigan06,arcizet,kleckner}. Controlling decoherence in this case is of utmost importance. 

For multiparty entangled states, the environment may affect local properties, like the excitation and the coherences of each part, and also global properties, like the entanglement of the state. The above-mentioned studies on decoherence lead to natural questions regarding the dynamics of entanglement: What is the decay law? Is it possible to introduce a decay rate, in this case? How does the decay of entanglement scale with the number of entangled parts? How robust is the entanglement of different classes of entangled states, and are there efficient ways to improve such robustnesses? Under which conditions does entanglement grow due to the interaction with the environment? 

Recent theoretical \cite{rajagopal,karol0,duan00,simon02,diosi03,jamroz03,dodd04,duer04,ficek1,carvalho04,yu04,serafini04,lidar04,hein05,fine05,mintert05b,aravind05,yu06,yu062,santos06,liu06,benatti06,yonac061,ficek2,eberly07,yu:459,yonac:s45,zubairy07,liu07,seligman07,seligman2007,sabrina07,terra01,ficek08,lopez-2008,marek08,guo08,james08,concentration,hu08,lai08,Ferraro08,Paz:220401,paz08,yu08b,aolita08,xu09,Paz:032102,cavalcanti09,hor09,terra02,yu09,mazzola09,zell09,sumanta,papp09,viviescas10,cavalcanti10,dur11,aolita2011} and experimental \cite{almeida07,Laurat,alejo,nussenzveig09,barbosa10,Barreiro10,monz2011,farias2012,osvaldo2012} work, involving both continuous and discrete variables, has given partial answers to these questions. It is now known that the dynamics of entanglement can be quite different from that of a single particle interacting with the environment. The pioneer contributions of Rajagopal and Rendell \cite{rajagopal}, who analyzed the dynamics of entanglement for two initially entangled harmonic oscillators, under the action of local environments, and {\.{Z}}yczkowski {\it et al.} \cite{karol0}, who considered the dynamics of entanglement for two  two-level systems under the action of local stochastic environments, represented the first studies specifically focussed on entanglement dynamics of which we have record. They established that entanglement may disappear before coherence decays, and also showed that revivals of entanglement may occur.  Different models have been studied since then, involving particles interacting with individual and independent environments \cite{simon02,diosi03,jamroz03,dodd04,duer04,carvalho04,yu04,serafini04,lidar04,hein05,fine05,mintert05b,aravind05,yu06,yu062,santos06,liu06,benatti06,yonac061,ficek2,eberly07,yu:459,yonac:s45,zubairy07,liu07,seligman07,seligman2007,sabrina07,terra01,alejo,ficek08,lopez-2008,marek08,guo08,james08,concentration,hu08,lai08,Paz:220401,paz08,yu08b,xu09,Paz:032102,cavalcanti09,terra02,yu09,mazzola09,zell09,sumanta,papp09,barbosa10,Barreiro10,viviescas10,dur11}, or with the same environment \cite{ficek1,ficek2,ficek08,hor09,adriana10} or yet  combinations of both situations \cite{ficek2,monz2011}.

The preliminary conclusions in \cite{rajagopal,karol0} turned out to be quite general. Entanglement decay with time does not follow an exponential law, even in the Markovian regime, and may vanish at finite times, much before coherence disappears. Initially entangled states may decay under the action of independent local environments, while particles may become entangled when interacting with the same environment. Revivals of entanglement may also occur \cite{rajagopal,karol0,ficek2}. Finite-time disentanglement, sometimes referred to as ``entanglement sudden death'' \cite{yu062,yu09}, has been experimentally demonstrated \cite{almeida07,Laurat,nussenzveig09,barbosa10}. Moreover, the entanglement of important classes of multipartite states  exhibit, for a fixed time, an exponential decay with the number of parties \cite{carvalho04,aolita08,aolitapra09}, which contributes to the concerns regarding the viability of large-scale quantum information processing.  For the case of collective decoherence, however, it is possible to construct {\it decoherence-free subspaces} of entangled states immune to the noise \cite{kwiat00,haeffner05b}. Furthermore, it is possible to produce and protect quantum states by engineering artificial reservoirs \cite{poyatos,andre01,pielawa,pielawa2}; and, remarkably, through similar techniques, even to implement dissipation-induced universal quantum computation \cite{diehl08,kraus08,verstraete09}. Feedback control has also been proposed for the purpose of stabilizing entanglement \cite{andre07,andre08}. The stabilization of entanglement through engineered dissipation has been demonstrated in recent experiments \cite{devoret,lin}. 

Stabilization techniques may help increase the robustness of quantum communication and information processing tasks, and may also be applied to quantum metrology, where the presence of decoherence tends to drive the precision in the estimation of parameters from the ultimate quantum limit (sometimes called the ``Heisenberg limit'') \cite{helstrom,Braunstein1996,lloyd04}, to the classical standard limit \cite{escher,escherbjp,escher12}. The use of entangled states in quantum metrology has been advocated by several authors, especially for frequency estimation in ion traps \cite{bollinger,leibfried04,blatt08} or Ramsey spectroscopy \cite{Huelga1997}, and phase estimation in optical interferometers \cite{dowling02,Kacprowicz2010,bryn}. The proposed states are however highly sensitive to decoherence \cite{Kacprowicz2010}. Knowledge of techniques to sustain entanglement is crucial for further developments of this field. 

The aim of this review article is to specifically address the dynamics of the entanglement in quantum open systems. We have tried to make this review self-contained and pedagogic enough so that it is accessible to both theorists and experimentalists within and outside the subfield of quantum information. However, we have refrained from an encyclopaedic treatment of the subject. Excellent reviews, previously published, cover in detail the mathematics and physics of entanglement \cite{mintert05b,KarolBook,amico,plenio07,horodecki09,guehne09,arealaw} and decoherence \cite{zurek:715, paz01, joos, breuer, schlosshauer07}. We direct the reader to these references for further details. Here, in contrast, we focus on the {\it effects of decoherence on entanglement}.

In Section II, we discuss the concept of entanglement, its quantification and measurement. In Section III, we consider open-system dynamics, as well as the different families of noise channels. Section IV reviews the theory of entanglement dynamics of bipartite systems, while Section V addresses the theory of multipartite entanglement decay. Experimental results are reviewed in Section VI.  Finally, in Section VII, we present some perspectives and open problems, and summarize the conclusions of the paper.

\section{The concept of entanglement}
\label{2}
In this section we introduce the basic concepts about entanglement, as well as some of the existing criteria to detect it, and the main methods to quantify it in its different classes. As mentioned in the introduction, the goal of this report is not to focus on entanglement itself but on its dynamic features under decoherence, therefore the brief revision about the formalism of entanglement presented in this section cannot -- and must not -- be considered exhaustive. For excellent and in-depth reviews on the formalism of entanglement  we refer the reader to Refs. \cite{mintert05b,plenio07,amico,horodecki09,guehne09,arealaw}.

\subsection{Definition}
\label{2.1}
Let us consider a multipartite system ${\mathcal S}$ of $N$ parties. The corresponding space of  states is a Hilbert space ${\mathcal H}_{\sys}$ resulting from the tensor product of the $N$ individual Hilbert spaces of the subsystems: ${\mathcal H}_{\sys}\equiv{\mathcal H}_{1}\otimes\ ... \ \otimes{\mathcal H}_{N}$, where ${\mathcal H}_{i}$,
with $1\le i \le N$, is the $d_i$-dimensional Hilbert space associated to the $i$-th subsystem. The dimension $d_{\sys}$ of the total space ${\mathcal H}_{\sys}$ is $d_{\sys}\equiv\prod_{i=1}^N d_i$. One should note that for many systems of interest -- like for instance harmonic oscillators -- the dimension $d_i$ may be infinite. Due to the vector nature of the total Hilbert space (stemming from the quantum superposition principle),  not all its elements are necessarily products of some others. In other words,  calling  $\ket{j_i}$, with $0\leq j_i\leq d_i-1$, the $j$-th element of some convenient basis of  ${\mathcal H}_{i}$, the superposition principle allows to write the most general $N$-partite quantum state as:
\begin{equation}
\label{Psi}
\ket{\Psi}=\sum_{j_{1}\ ... \  j_{N}}\Psi_{j_{1}\ ... \  j_{N}}\ket{j_1}\otimes\ ... \ \otimes\ket{j_N}.
\end{equation}
The product basis $\ket{j_1}\otimes\ ... \ \otimes\ket{j_N}$, for which we use the short-hand notation $\ket{j_1}\ ... \ \ket{j_N}$, or simply $\ket{j_1\ ... \  j_N}$, depending on convenience,  is called from now on the {\it computational basis} of ${\mathcal H}_{\sys}$. State~\eqref{Psi}  cannot in general be written as a product of the individual states of the subsystems. In other words, it is in general not possible to attribute a state vector to each individual subsystem, which is precisely the formal statement of the phenomenon of entanglement: 

\begin{definition}[Separable pure states] \label{definition:seppure}
A pure state $\ket{\Psi}\in{\mathcal H}_{\sys}$ is {\it separable} if it is a product state. That is, if it can be expressed as
\begin{equation}
\label{Psi2}
\ket{\Psi}=\ket{\Psi_1}\otimes\ ... \ \otimes\ket{\Psi_N},
\end{equation}
 for some $\ket{\Psi_1}\in{\mathcal H}_{1}$, $ ... $ and $\ket{\Psi_N}\in{\mathcal H}_{N}$.
\end{definition}

\begin{definition}[Entangled pure states] \label{definition:entpure}
A pure state $\ket{\Psi}\in{\mathcal H}_{\sys}$ is {\it entangled} if it is not separable.
\end{definition}

\noindent  For any pure state $\ket{\Psi}$ of a bipartite system,  there always exists a product basis $\ket{\phi^1_{j}\phi^2_{j}}$ in terms of which one can write $\ket{\Psi}=\sum_{j=0}^{r-1}\varsigma_j\ket{\phi^1_{j}\phi^2_{j}}$, with integer $r\leq d\doteq\min\{d_1,d_2\}$ (the dimension of the smallest subsystem) and  $\varsigma_j>0$ for all $j$. This is the well-known  Schmidt decomposition \cite{schmidt07}, and $r$ and $\varsigma_j$ are called respectively the Schmidt rank and Schmidt coefficients of $\ket{\Psi}$. A pure state is entangled if and only if  $r>1$. For finite-dimensional systems, the  maximally entangled states are all the pure states whose Schmidt decomposition is given by
\begin{equation}
\label{maximally}
\ket{\Phi^+_{d}}\equiv\frac{1}{\sqrt{d}}\sum_{j=0}^{d-1}\ket{\phi^1_{j}\phi^2_{j}}.
\end{equation}
Infinite-dimensional maximally entangled states will be discussed in Sec. \ref{2.2.1a}. Since all product bases are connected through local unitary transformations, the maximally entangled states are the ones local-unitarily related to $\ket{\Phi^+_{d}}$. Arguably the most popular example  of maximally entangled states is given for the case of two qubits ($d_1=2=d_2$) by the four Bell states, expressed in the computational basis as:
\begin{equation}
\label{Bellstate}
\ket{\Psi^{\pm}}\equiv\frac{1}{\sqrt{2}}(\ket{01}\pm\ket{10})\ \text{and}\ \ket{\Phi^{\pm}}\equiv\frac{1}{\sqrt{2}}(\ket{00}\pm\ket{11}),
\end{equation}
which constitute a maximally entangled basis of ${\mathcal H}_{1}\otimes{\mathcal H}_{2}$. 

\par Maximally entangled states  possess a remarkable property: the reduced density matrix of the smallest subsystem is given, for finite-dimensional systems, by the maximally mixed state $\frac{\openone}{d}$, with $\openone$ the identity operator. This contains no information at all (maximal entropy) and therefore any measurements on it  yield completely random outcomes. Still,  the available information about the whole two-qubit system is maximal, because the state is pure (zero entropy). This is the formal statement of Schr\"odinger's quotation already mentioned in the introduction:  ``{\it The best possible knowledge of a whole does not include best possible knowledge of its parts}".  Furthermore, there are correlations between local measurements on both subsystems that cannot be described by models based on local hidden-variables, which would determine the values of the local observables at each run of the experiment.  As we will see in the following sections, these peculiarities constitute the strongest manifestations of how much the notion of quantum entanglement defies classical intuition.

\par All in all, mixed states are much more abundant than pure ones (in fact,  they describe realistic laboratory situations, where decoherence and imperfect  operations lead to incomplete information about the state vector describing the system). They are represented by trace-1 normalised density matrices $\varrho$ belonging to the space ${\mathcal D}({\mathcal H}_{\sys})$ of positive-semidefinite operators acting on a Hilbert space ${\mathcal H}_{\sys}$. The definition of entanglement for mixed states is more subtle then the one for pure states, and is given by \cite{werner89}:

\begin{definition}[Separable states] \label{definition:separable}
A state $\varrho\in{\mathcal D}({\mathcal H}_{\sys})$ is {\it separable} if it can be expressed as a convex combination of pure product states, {\it i.e.} if
\begin{equation}
\label{Rho}
\varrho=\sum_{\mu} p_{\mu} \ket{{\Psi_{\mu}}_1}\bra{{\Psi_{\mu}}_1}\otimes\ ... \ \otimes\ket{{\Psi_{\mu}}_N}\bra{{\Psi_{\mu}}_N},
\end{equation}
for some $\ket{{\Psi_{\mu}}_{i}}\in{\mathcal H}_i$, and $p_{\mu}\geq0$ such that $\sum_{\mu} p_{\mu}\equiv 1$.  
\end{definition}
Expression \eqref{Rho} can be thought of as a pure-state ensemble decomposition of $\varrho$. In that case, $\ket{{\Psi_{\mu}}_{i}}$ is the state of particle $i$ in the $\mu$-th member of the ensemble in question.

\begin{definition}[Entangled states] \label{definition:net}
A state $\varrho\in{\mathcal D}({\mathcal H}_{\sys})$ is {\it entangled} if it is not separable.
\end{definition}

\par The multipartite scenario is much richer, as a variety of separability subclasses arises there. We introduce the corresponding  sub-classifications beginning with the following definitions.
 
\begin{definition}[$k$-separability with respect to a partitioning] \label{definition:kseppart}
For any $2\leq k\leq N$, a state $\varrho\in{\mathcal D}({\mathcal H}_{\sys})$ is {\it $k$-separable} with respect to a particular $k$-partitioning of the $N$ parts if it can be expressed as a convex combination of pure states all $k$-factorable in the partition. That is, if
\begin{equation}
\label{RhoKseppart}
\varrho=\sum_{\mu} p_{\mu} \ket{{\Psi'_{\mu}}_{1}}\bra{{\Psi'_{\mu}}_{1}}\otimes\ ... \ \otimes\ket{{\Psi'_{\mu}}_{k}}\bra{{\Psi'_{\mu}}_{k}},
\end{equation}
for some $\ket{{\Psi'_{\mu}}_{i}}\in{\mathcal H}'_{i}$, where subindex $1\leq i\leq k$ labels now each of the $k$ subsets, ${\mathcal H}'_{i}$ is the composite Hilbert space of the parts in the $i$-th subset, and $p_{\mu}\geq0$ such that $\sum_{\mu} p_{\mu}\equiv 1$. 
\end{definition}

\noindent For example, consider three qubits shared among parts $A$, $B$ and $C$, conventionally called Alice, Bob, and Charlie, respectively, in the composite state $\ket{\Psi_{ABC}}\doteq\ket{\Phi^{+}_{AB}}\otimes\ket{0_{C}}\in{\mathcal H}_{\sys}={\mathcal H}_{A}\otimes{\mathcal H}_{B}\otimes{\mathcal H}_{C}$. In the notation of \eqref{RhoKseppart}, this corresponds to a single term $\mu$, ${\mathcal H}'_{1}\doteq{\mathcal H}_{A}\otimes{\mathcal H}_{B}$, and ${\mathcal H}'_{2}\doteq{\mathcal H}_{C}$. That is,  Alice and Bob share the Bell state $\ket{{\Psi'}_{1}}\doteq\ket{\Phi^{+}_{AB}}\in{\mathcal H}_{A}\otimes{\mathcal H}_{B}$ and Charlie has the pure state  $\ket{{\Psi'}_{2}}\doteq\ket{0_{C}}\in{\mathcal H}_{C}$.  The composite state is not separable because it possesses entanglement, but it clearly factorizes with respect to the splitting ``Alice and Bob versus Charlie". It is therefore $k$-separable for $k=2$, commonly referred to as {\it biseparable}, in the split  $AB|C$.

\begin{definition}[$k$-separable states] \label{definition:ksep}
For any $2\leq k\leq N$, a state $\varrho\in{\mathcal D}({\mathcal H}_{\sys})$ is {\it $k$-separable}  if it can be expressed as a convex combination of states each $k$-separable with respect to at least one of the $k$-partitions. That is, if
\begin{equation}
\label{RhoKsep}
\varrho=\sum_{\mu} p_{\mu} \ket{{\Psi'_{\mu}}_{1}}\bra{{\Psi'_{\mu}}_{1}}\otimes\ ... \ \otimes\ket{{\Psi'_{\mu}}_{k}}\bra{{\Psi'_{\mu}}_{k}},
\end{equation}
for some $\ket{{\Psi'_{\mu}}_{i}}\in{{\mathcal H}'_{\mu}}_{i}$, where subindex $1\leq i\leq k$ labels again each of the $k$ subsets, but with the subsets now in general varying with $\mu$, so that ${{\mathcal H}'_{\mu}}_{i}$ is now the composite Hilbert space of the parts in the $i$-th subset of the  $\mu$-th $k$-splitting (that of the $\mu$-th member of the decomposition), and $p_{\mu}\geq0$ such that $\sum_{\mu} p_{\mu}\equiv 1$ as usual.  
\end{definition}

\noindent In this terminology, the separable states of Definition \ref{definition:separable} are called $N$-separable, or simply {\it fully separable}. 

Analogously, entanglement also admits sub-classifications in terms of the number of parts actually taking place in the correlations:

\begin{definition}[Blockwise $M$-party entanglement] \label{definition:blockent} 
Given any $2< M\leq N$, and a particular $M$-partitioning of the $N$ parts, a state $\varrho\in{\mathcal D}({\mathcal H}_{\sys})$ is blockwise $M$-partite entangled with respect to the $M$-partition, if it cannot be expressed as a convex combination of states each biseparable with respect to some bipartition of the $M$ blocks.
\end{definition}

\noindent  When $M=N$, this definition reduces in turn to the following crucial case.

\begin{definition}[Genuine multipartite entanglement] \label{definition:genmultent} 
A state $\varrho\in{\mathcal D}({\mathcal H}_{\sys})$ is {\it  $N$-partite entangled}, or  {\it genuinely multipartite entangled}, if it is not biseparable. 
\end{definition}

\par
\begin{figure}[t]
\begin{center}
\includegraphics[width=1\linewidth]{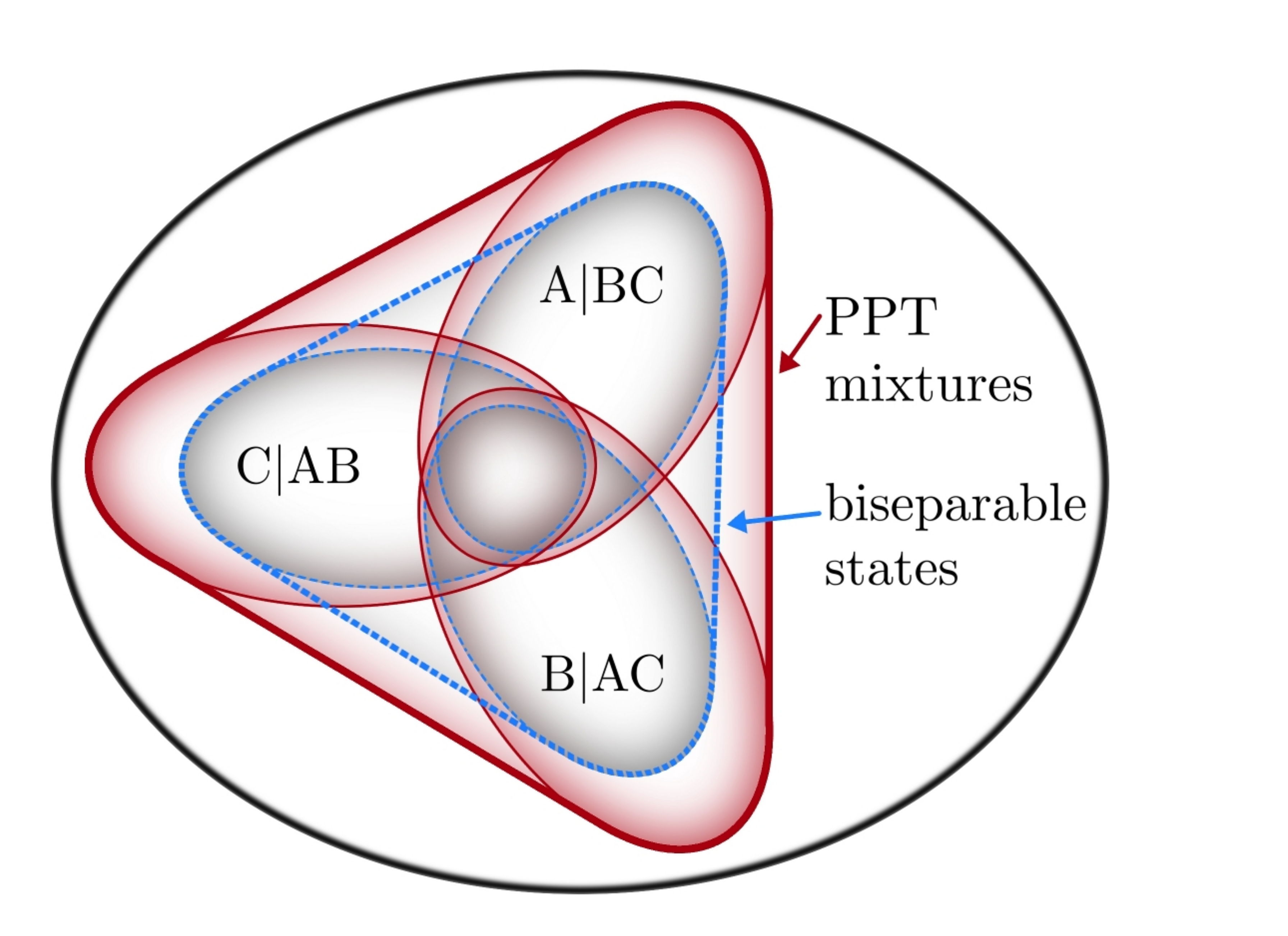}
\caption{
\label{guehne}
Schematic representation of three-qubit states. There are three convex sets (in thin dashed blue) of states separable with respect to each bipartition, $A|BC$, $B|AC$, and $C|AB$. The minimum convex set containing these three sets, i. e. their  convex hull (in thick dashed blue), is the set of biseparable states. Each biseparable set is in turn contained by the set of states that are PPT with respect to the corresponding bipartition (in thin solid red). Their convex hull forms the set of PPT mixtures (in thick solid red). Reprinted figure with permission from B. Jungnitsch, T. Moroder, and O. G\"uhne, \href{http://link.aps.org/doi/10.1103/PhysRevLett.106.190502}{Phys. Rev. Lett. {\bf 106}, 190502 (2011)}. Copyright (2011) by the American Physical Society.}
\end{center}
\end{figure}

 Once again, three qubits are enough for a very illustrative example of how the above sub-classifications apply. Consider the mixed state \cite{guehne09}
\begin{eqnarray}
\label{ABCexample}
\nonumber
\varrho_{ABC}&=&\frac{1}{3}\big(\ket{\Phi^{+}_{AB}}\bra{\Phi^{+}_{AB}}\otimes\ket{0_{C}}\bra{0_{C}}\\
\nonumber
&+&\ket{\Phi^{+}_{BC}}\bra{\Phi^{+}_{BC}}\otimes\ket{0_{A}}\bra{0_{A}}\\
&+&\ket{\Phi^{+}_{CA}}\bra{\Phi^{+}_{CA}}\otimes\ket{0_{B}}\bra{0_{B}}\big),
\end{eqnarray}
It is immediate to verify (for instance, with the PPT criterion discussed in Sec. \ref{2.2.1}) that this state is entangled in all its three bisplittings. However, since it is a convex combination of biseparable states with respect to the three splits, it is by definition biseparable and therefore not genuinely multipartite entangled. This simple example teaches us a very important lesson: The presence of entanglement in all the bipartitions does not imply genuine multipartite entanglement. The situation is pictorially represented in Fig. \ref{guehne} for the three-qubit scenario. 

\par On the other hand, two archetypical examples of genuinely multipartite entangled states are the GHZ states
\begin{equation}
\label{GHZdef0}
\ket{{\text{GHZ}}_N} \doteq \frac{1}{\sqrt{2}} \big( \ket{000\ldots 0} + \ket{111\ldots 1} \big)
\,,
\end{equation}
named after Greenberger, Horne, and Zeilinger, who were the first to introduced this state in its  three-qubit version \cite{ghz89,ghz90,ghz93}; and the W states
\begin{eqnarray}
\label{WdefN0}
\nonumber
\ket{{\text{W}}_N}&\doteq&\frac{1}{\sqrt{N}}(\ket{00\ ...\  01}+\ket{00\ ...\  10}+\\
\ &...&\ +\ket{01\ ...\  00}+\ket{10\ ...\  00}),
\end{eqnarray}
originally introduced, also in its three-qubit version, by D\"{u}r, Vidal, and Cirac in Ref. \cite{duer00b}.

Interestingly, the two states (\ref{GHZdef0}) and (\ref{WdefN0}) cannot be obtained from each other through stochastic local operations and classical communication, that is, probabilistic operations carried out locally on each part and eventually coordinated (correlated) among all parts by means of classical communication. In this sense, they represent two inequivalent classes of genuine multipartite entanglement \cite{duer00b}. The concept of local operations and classical communication will be developed in Sec. \ref{2.2.2}. Other families of genuine multipartite entangled states will be encountered in Chapter \ref{V}.

\subsection{PPT-ness, entanglement witnesses, biseparability criteria, PPT mixtures, free and bound entanglement}
\label{2.2}
The zoology of criteria \cite{horodecki09} establishing sufficient conditions for entanglement (or, equivalently, necessary ones for separability) is tremendously vast. Nevertheless, there exists yet no criterium that allows one to unambiguously or -- specially -- efficiently\footnote{Distinguishing between separable and  entangled 
mixed states is indeed known to be NP-Hard problem  \cite{Gurvits03}.} guarantee if a generic state, in the case of more than two particles, or two particles of arbitrary dimensions, is or not entangled. In what follows, we briefly describe just the best-known criteria. 
\subsubsection{The positive-partial-transpose criterion}
\label{2.2.1}
The criterion of the positive partial-transpose (PPT), first discovered by Peres \cite{peres96}, establishes a necessary condition for separability in the general bipartite case. It involves a simple algebraic calculation without any optimization and is capable of detecting a large family of entangled states, called the negative partial-transpose (NPT) states.   For an arbitrary state $\varrho\in{\mathcal D}({\mathcal H}_{\sys})$, and any splitting of $\sys$ into two subgroups of particle, $A$ and $B$, with respective Hilbert spaces $\mathcal{H}_{A}$ and $\mathcal{H}_{B}$, such that  $\mathcal{H}_{A}\otimes\mathcal{H}_{B}\doteq\mathcal{H}_{AB}={\mathcal H}_{\sys}$, it is stated as  follows.

\begin{criterion}[Positive-Partial-Transpose] \label{criterion:PPT}If $\varrho\in\mathcal{D}(\mathcal{H}_{\sys})$ is separable in the split $A:B$, then its partially transposed matrix with respect to $B$, $\varrho^{T_{B}}$, of matrix elements
\begin{eqnarray}
\label{TP}
\nonumber
\varrho^{T_{B}}_{j_Aj_Bj'_Aj'_B}&\equiv&\bra{j_A}\bra{j_B}\varrho^{T_{B}}\ket{j'_A}\ket{j'_B}\\
\nonumber
&\doteq&\bra{j_A}\bra{j'_B}\varrho\ket{j'_A}\ket{j_B}\\
&\equiv&\varrho_{j_Aj'_Bj'_Aj_B} ,
\end{eqnarray}
for $\{\ket{j_A}\}$ and $\{\ket{j_B}\}$ any orthonormal bases of $\mathcal{H}_{A}$ and $\mathcal{H}_{B}$, respectively, is also in $\mathcal{D}(\mathcal{H}_{\sys})$.  
\end{criterion}

\noindent That is, it asserts that $\varrho^{T_{B}}$ is also a bounded, positive-semidefinite, trace-1 normalized operator acting on $\mathcal{H}_{AB}$. The operation  $T_{B}$, called partial transposition with respect to subsystem $B$, corresponds to the transposition of the matrix indices associated only to $\mathcal{H}_B$. Any state satisfying the criterion is called a PPT state (with respect to the bipartition in question). The criterion automatically implies that if the partially transposed matrix of a state is negative (possesses at least one negative eigenvalue), then the state must necessarily be entangled. These are precisely the NPT states mentioned above.

\par The simplicity of the criterion makes it arguably the most popular separability criterion of all. Indeed, it is simple to understand how it works.  Consider then an arbitrary state separable in the bipartite cut $A:B$: $\varrho_{AB}\equiv\sum_{\mu} p_{\mu} \varrho_{A_{\mu}}\otimes\varrho_{B_{\mu}}$. Next, partially transpose it to obtain $\varrho_{AB}^{T_{B}}\equiv\sum_{\mu} p_{\mu} \varrho_{A_{\mu}}\otimes\varrho_{B_{\mu}}^{T}$. Since the transposition is a positive operation, the transposed $\varrho_{B_{\mu}}^{T}$ of any density operator $\varrho_{B_{\mu}}\in{\mathcal D}({\mathcal H}_{B})$ is also in ${\mathcal D}({\mathcal H}_{B})$. Therefore, the partially transposed composite operator $\varrho_{AB}^{T_{B}}$ constitutes a valid element of ${\mathcal D}({\mathcal H}_{A})\otimes{\mathcal D}({\mathcal H}_{B})\equiv{\mathcal D}({\mathcal H}_{AB})={\mathcal D}({\mathcal H}_{\sys})$. Thus, at the heart of the efficacy of the criterion is the fact that the  transposition is positive but the partial transposition is not. Technically, this means that the transposition does not belong to the more general family of {\it completely-positive} operations, which will be discussed in Sec. \ref{2.2.2}.

\par With the PPT criterion, Peres established a necessary condition for separability. Soon afterwards, the Horodecki family complemented it \cite{horodecki96} with the fundamental discovery that, for the particular cases of arbitrary-dimensional bipartite systems in pure states, or systems of dimensions $2\times 2$ or $2\times3$ in arbitrary states, it actually provides both necessary and sufficient conditions for separability. 

\par The continuous-variable-system version of the PPT criterion, discussed in the following, sets also both necessary and sufficient conditions for entanglement in the particular case of Gaussian states \cite{duan00,Simon00}. In  these cases, the criterion provides a complete characterization of the state's entanglement. Beyond these particular cases though, mixed entangled states are known that are PPT\footnote{Examples of criteria capable of detecting some PPT entangled states are the  range criterion \cite{Horodecki97b} and the  computable cross norm, or  realignment, criterion \cite{Chen03b,Rudolph05}.}. 

\subsubsection{Entanglement, PPT-ness, and separability in continuous-variable systems}
\label{2.2.1a}

 A thorough discussion of entanglement in continuous-variable (CV) systems may be found for instance in Refs. \cite{Ferraro05, adesso2007}. Here we limit ourselves to a very short introduction to this subject.  
 
 A maximally entangled CV state, corresponding to two quantum modes, or {\it qumodes}, with position quadrature operators $q_1$ and $q_2$, and momentum quadrature operators $p_1$ and $p_2$, is a common eigenstate of the operators $q_1+q_2$ and $p_1-p_2$, with $[q_j,p_{k}]=i\delta_{jk}$, for $j,k\in\{1,2\}$, where $\delta$ is the Kronecker delta. This implies that the sum of the variances of these two operators (total variance) should be zero. However, this state, often called EPR state, after Einstein, Podolski, and Rosen, who introduced it in their famous 1935 paper \cite{epr35}, is not physical, since it involves infinite energies. It is rather used as an abstract limit which physical states can approach. Physical, non-maximally entangled, approximations of it correspond to two-mode squeezed states \cite{scully97, gardiner}, for which the total variance is different from zero but approaches it as the degree of squeezing increases. This suggests that the total variance could lead to a criterium for separability. Indeed, this is the approach taken by Duan {\it et al.} in Ref. \cite{duan00}, who lower-bounded the total variances of separable states through the following criterion.
\begin{criterion}[Separability of generic two-qumode states] 
\label{criterion:Duan} 
For any separable two-mode state $\varrho$, and EPR-like operators $u$ and $v$ defined by
\begin{eqnarray}
 u&=&aq_1+\frac{1}{a}q_2\,,\nonumber\\
v&=& ap_1-\frac{1}{a}p_2\,,
\end{eqnarray}
with $a$ any positive real, the total-variance bound
\begin{equation}\label{duan}
\langle(\Delta u)^2\rangle_{\varrho}+\langle(\Delta v)^2\rangle_{\varrho}\ge a^2+{1\over a^2}\,
\end{equation}
holds, where $\Delta u\doteq u-\langle u\rangle_{\varrho}$ and $\Delta v\doteq v-\langle v\rangle_{\varrho}$.
\end{criterion}

An alternative approach was followed by Simon \cite{Simon00}, who formulated the Peres-Horodecki criterium in the CV setting. To this end, he considered the Wigner function, which offers a phase-space representation of states  equivalent to the density operator representation. For the particular case two qumodes, for instance, it is defined in terms of $\varrho$ as $W(\textbf{q},\textbf{p})={\pi}^{-2}\int d^2\textbf{q}^\prime\langle \textbf{q}-\textbf{q}^\prime|\varrho|\textbf{q}+\textbf{q}^\prime\rangle\exp(2i\textbf{q}^\prime\cdot \textbf{p})$, where $\textbf{q}=(\rm{q}_1,\rm{q}_2)\in\mathbb{R}^2$ and $\textbf{p}=(\rm{p}_1,\rm{p}_2)\in\mathbb{R}^2$ are respectively the real and imaginary parts of the coordinates of points in the associated two-dimensional complex phase space. He showed that, for CV mode  states, the transposition operation is equivalent to the mirror reflection in phase space of the momentum coordinate, or, which is the same, to time reversal of the Schr\"odinger equation. That is, for a two-mode state $\varrho$ with  by a Wigner description $W(\textbf{q},\textbf{p})$, the partial transposition of the corresponding density matrix with respect to the second mode is equivalent to the Wigner-distribution transformation $W(\rm{q}_1,\rm{p}_1,\rm{q}_2,\rm{p}_2)\rightarrow W(\rm{q}_1,\rm{p}_1,\rm{q}_2,-\rm{p}_2)$. Thus, Criterion \eqref{criterion:PPT} translates to the CV case as the necessary condition that the mirror-reflected function $W(\rm{q}_1,\rm{p}_1,\rm{q}_2,-\rm{p}_2)$ also be a valid Wigner distribution, for any separable $\varrho$. That is,  $W(\rm{q}_1,\rm{p}_1,\rm{q}_2,-\rm{p}_2)$ must describe a trace-one positive-semidefinite operator. 

A necessary condition for this, in turn, is that the phase-space distribution renders the correct uncertainty relations. This is convenient because these can be expressed in a concise way in terms of just the second moments of the distribution, as
\begin{equation}
\label{simoncov}
\gamma+i\Omega\ge0\,,
\end{equation}
where $\gamma$ is a 4$\times$4 real symmetric matrix, called the {\it covariance matrix} of $\varrho$, with matrix entries 
\begin{equation}
\gamma_{ij}=\langle\Delta\epsilon_i\Delta\epsilon_j+\Delta\epsilon_j\Delta\epsilon_i\rangle\,.
\end{equation}
Here, operator $\epsilon_i$, for $1\leq i\leq 4$, is the $i$-th component of the vector  $\epsilon=(q_1,p_1, q_2,p_2)$, $\Delta\epsilon_i=\epsilon_i-\langle\epsilon_k\rangle_{\varrho}$. The expectation value $\langle O \rangle_{\varrho}\doteq\text{Tr}[\varrho O]$ of a generic operator $O$ can be evaluated explicitly in the Wigner representation as the convolution of $W(\textbf{q},\textbf{p})$ with the Wigner function of $O$ \cite{Ferraro05, adesso2007}. $\Omega$, in turn, is the 4$\times$4 antisymmetric matrix
 \begin{equation}
\Omega\doteq\omega_1\oplus \omega_2\text{, where }\omega_i\doteq
\begin{pmatrix}
0&1\\
-1&0
 \end{pmatrix}\,
 \end{equation}
 is the {\it symplectic matrix} of mode $i$, for $i=1$, 2. When no particular mode is specified, one typically refers to $\Omega$ simply as the symplectic matrix. 
 
 Transformation $p_2\to-p_2$ corresponds to $\omega_2\to-\omega_2$ in \eqref{simoncov}, which leads us finally to the best-known form of the PPT criterion for CV systems:
\begin{criterion}[Positive-Partial-Transpose for two qumodes] \label{criterion:PPTCV}
If $\varrho\in\mathcal{D}(\mathcal{H}_{\sys})$ is a separable two-mode state with covariance matrix $\gamma$, then 
\begin{equation}
\label{PopularppT}
\gamma+i(\omega_1\oplus -\omega_2)\ge0\,.
\end{equation}

\end{criterion}
The operation of mirror reflection of the Wigner distribution, and therefore also Criterion \ref{criterion:PPTCV}, is straightforwardly generalized to any bipartition $N_A|N_B$ of a system with $N_A+N_B$ qumodes.  

Criteria \ref{criterion:Duan} and \ref{criterion:PPTCV} provide necessary conditions for separability, but, remarkably, for Gaussian states these conditions become also sufficient \cite{duan00,Simon00}. Gaussian states play a crucial role in quantum information and quantum optics \cite{Ferraro05, adesso2007}. They are defined as those whose Wigner representation is a Gaussian function. For these states, the covariance matrix captures all the correlations and  completely determines the state up to local unitary displacements. Covariance matrices transform according to {\it symplectic transformations}, which describe the transformations of phase-space coordinates associated to the physical transformations of quantum states. This are characterised by the group of all 4$\times$4 real matrices $F$ such that $F\Omega F^{T}=\Omega$, with $F^{T}$  the transposed of $F$. Williamson's Theorem guarantees that any covariance matrix can be diagonalised by a symplectic transformation \cite{Ferraro05}. This is called symplectic diagonalization, and the (non-negative) eigenvalues obtained, $e_i$, are the symplectic eigenvalues of $\gamma$. In terms of these, condition \eqref{PopularppT} expresses as ``$e_i\geq1$ for all $i$", which constitutes an equivalent formulation of the PPT criterion.

For arbitrary Gaussian states of $1+N$ qumodes, Werner and Wolf showed that PPTness of a bipartition $1|N$  is a necessary and sufficient condition for biseparability in the bipartition \cite{werner01}. Furthermore, if symmetries are present, the PPT criterium can be shown to be equivalent to biseparability for more general partitions. This is the case for bisymmetric $N_A+N_B$-mode Gaussian states, which are invariant under internal permutations of the qumodes in each subset $A$ or $B$. Then, it can be shown that PPTness is a necessary and sufficient condition for separability in the  split $N_A|N_B$ \cite{serafini05}. This implies that, for a fully symmetrical mixed Gaussian state, of an arbitrary number of qumodes, PPTness is equivalent to biseparability with respect to all bipartitions of the modes.

To end up with, Shchukin and Vogel \cite{shchukin} derived a general hierarchy of necessary and sufficient  conditions for separability of  two-qumode states. This can be expressed in terms of higher-order momenta of the two modes involved and is applicable to non-Gaussian states. Indeed, it has been used to test the separability of non-Gaussian states in optical experiments \cite{gomes}.

\subsubsection{Entanglement witnesses}
\label{2.2.2}
\par Entanglement witnesses \cite{horodecki96,terhal00,lewenstein00,bruss02b, guehne09} constitute a  very useful tool for the detection of entangled states in both the bipartite and multipartite cases. They give sufficient conditions for states to be entangled and possess a remarkable property: they can be directly obtained in the laboratory as the expectation value of physical observables, as is discussed in Sec. \ref{ExpEntWit}. That is, for every non-$k$-separable state  $\varrho\in{\mathcal D( {\mathcal H}_{\sys})}$, with $2\leq k\leq N$, there exists a Hermitean operator $W_k$ acting on ${\mathcal H}_{\sys}$ such that \cite{horodecki96} 
\begin{equation}
\label{sigmaW}
\text{Tr}[W_k\varrho_{sep}]\geq0,
\end{equation}
for every  $k$-separable $\varrho_{sep}\in{\mathcal D( \mathcal{H}_{\sys})}$, and
\begin{equation}
\label{Wdefinition}
\text{Tr}[W_k\varrho]<0.
\end{equation}
One says then that the non-$k$-separability of $\varrho$ is ``witnessed"  (detected) by the negative expectation value of the witness.  If on the other hand the expectation value of some particular witness is positive nothing can be concluded about the state being or not entangled.

\par Unfortunately, every witness succeeds to detect only a restricted portion of states. However, it is precisely this property what can make entanglement witnesses able to detect not only if a given state is entangled, but also if it belongs specifically to some particular entanglement family of interest. For example, if, for $k=2$, some  $W_2$ detects  $\varrho$ (that is, $\text{Tr}(W_k.\varrho)<0$), then it is known not only that  $\varrho$ is entangled, but also that it is genuinely $N$-partite entangled. More generally, if $\varrho$ is detected by some  $W_k$, then it is revealed to be genuinely $l$-partite entangled, with $l$ such that $N/(k-1)\leq l$, for $N$ an integer multiple of $k-1$, and $\lceil N/(k-1)\rceil\leq l$ otherwise, being $\lceil x\rceil$ the ceiling of $x$ (the smallest integer greater than, or equal to, $x$).  Furthermore, witnesses can even   be tailored  so as to also discriminate  inequivalent classes of genuine multipartite entanglement \cite{Acin01}, as we mention below. 
\begin{figure}[t!]
\begin{center}
\includegraphics[width=1\linewidth]{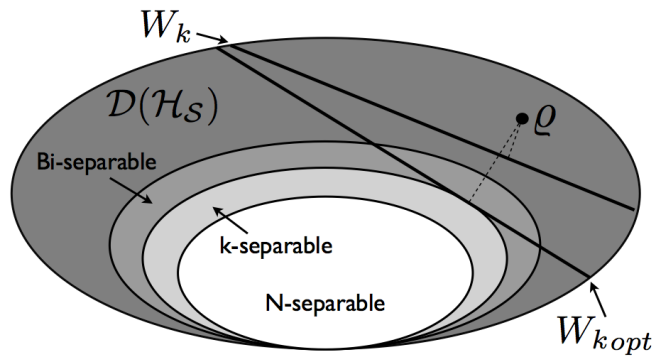}
\caption{
\label{Witnesses}
Schematic representation of the internal geometry of the set of density matrices  ${\mathcal D}({\mathcal H}_{\sys})$. Since the set of  $k$-separability is convex, there always exists a hyper-plane such that the set lies entirely at one side of it. Associated to the hyper-plane, there is an entanglement witness $W_k$ that detects a non-$k$-separable state $\varrho$. Adjacent to the set of $k$-separability in turn there is a hyper-plane that maximizes the distance to $\varrho$. This corresponds always  to some  optimal witness ${W_k}_{opt}$.}
\end{center}
\end{figure}

 \par The trace of the product of two Hermitean operators acting on  ${\mathcal H}_{\sys}$ defines their Hilbert-Schmidt inner product. Therefore the expression $\text{Tr}[W_k.\varrho]=0$ can be interpreted  as the defining equation of a hyperplane in ${\mathcal D}({\mathcal H}_{\sys})$, where $W_k$ plays the role of the vector orthogonal to the hyperplane and $\text{Tr}[W_k.\varrho]$ of the component of  $\varrho$ orthogonal to the hyperplane (times the norm of $W_k$). Thus, $W_k$ splits ${\mathcal D}({\mathcal H}_{\sys})$ into two semi-spaces: one corresponding to the states that it witnesses  [Eq.~\eqref{Wdefinition}], and the other to those it does not [Eq.~\eqref{sigmaW}].  In Fig.~\ref{Witnesses} we can see a schematic representation of the internal geometry of  ${\mathcal D}({\mathcal H}_{\sys})$ according to the sets of  $k$-separability, and including the division defined by the hyperplane corresponding to $W_k$. The dashed segment that goes from state $\varrho$ perpendicularly to this hyperplane pictorially represents $|\text{Tr}[W_k.\varrho]|$, which can also be taken as a distance between the hyper-plane and  $\varrho$. In the figure we can also see that there exists a hyperplane, adjacent to the set of  $k$-separability, that maximizes such distance. This hyper-plane is defined by some {\it optimal witness}  ${W_k}_{opt}$. 
 
 \par More precisely, a witness is said to be optimal if there exists no other witness {\it finer} than it \cite{lewenstein00},  meaning that no other witness  detects all entangled states detected by the former plus some other(s). This implies that a witness is optimal iff it is impossible to subtract from it any positive operator in a way such that the resulting observable is still a witness\footnote{There exists an alternative definition \cite{terhal00} that addresses optimality of witnesses relative to a given state. According to this, the optimal witness ${W_k}_{opt}$ for an entangled state  $\varrho$ is the one that maximizes $-\text{Tr}[W_k.\varrho]$ over  some compact subset ${\mathcal T_k({\mathcal H}_{\sys})}$ of witnesses on ${\mathcal H}_{\sys}$ (so that the maximization's convergence is guaranteed). Typical choices can be  ${\mathcal T_k({\mathcal H}_{\sys})}\doteq\{W_k\text{ s. t. Tr}[W_k]=K\}$ or ${\mathcal T_k({\mathcal H}_{\sys})}\doteq\{W_k\text{ s. t. }W_k\leq K\openone\}$, with $K$ some positive constant. Note that every optimal witness according to this definition is also 
 optimal accordingly to the deÞnition above, whereas the converse may not be true.} From this it 
can be seen that $W_k$ is optimal if all the $k$-factorable vectors $\ket{\Phi}$ in the Kernel of  $W_k$, $\bra{\Phi}W_k\ket{\Phi}=0$, possess enough linear independence so as to span the whole of ${\mathcal H}_{\sys}$. Systematic recipes for the optimization of witnesses were introduced in Ref. \cite{lewenstein00}.

\par  Here we just discuss two simple examples of how to construct witnesses. Before that, we introduce the notion of {\it decomposable witnesses}. A bipartite witness $W_{AB}$ is decomposable if it can be expressed as $W_{AB}= P+ Q^{T_B}$, where $P\geq 0$ and $Q\geq 0$ are positive semidefinite operators acting on  ${\mathcal H}_{AB}$, and $T_B$ represents the partial transposition with respect to B \cite{lewenstein00}. Observables with this structure automatically satisfy \eqref{sigmaW} for all PPT (and therefore all separable) states. This comes about because, for any PPT state $\varrho_{AB}$, one has $\text{Tr}[(P+ Q^{T_B})\varrho_{AB}]\equiv\text{Tr}[P\varrho_{AB}]+ \text{Tr}[Q^{T_B}\varrho_{AB}]\equiv\text{Tr}[P\varrho_{AB}]+ \text{Tr}[Q\varrho^{T_B}_{AB}]\geq0$, where the identity $\text{Tr}[Q\varrho^{T_B}_{AB}]\equiv\text{Tr}[Q^{T_B}\varrho_{AB}]$ was used. Indeed, a bipartite witness can detect entangled PPT states if, and only if, it is {\it non-decomposable} \cite{lewenstein00}.

The first example concerns witnesses for NPT states. Any NPT state $\varrho_{AB}$ is detected by a bipartite witness of the form \cite{lewenstein00}
\begin{equation}
\label{recipenpt}
{W_{AB}}\equiv\ket{\eta^-}\bra{\eta^-}^{T_{B}},
\end{equation}
where $\ket{\eta^-}$ is an eigenstate of $\varrho_{AB}^{T_{B}}$ with negative eigenvalue. This follows again from the identity $\text{Tr}[\ket{\eta^-}\bra{\eta^-}^{T_{B}}\varrho_{AB}]\equiv\text{Tr}[\ket{\eta^-}\bra{\eta^-}\varrho^{T_B}_{AB}]$. Witness \eqref{recipenpt} is by construction optimal and decomposable (it cannot detect any PPT entangled state).
 
\par The second one comes from the intuition that a state sufficiently close to an entangled state should also be entangled. Given any pure entangled state $\ket{\Psi}$, the observable
\begin{equation}
\label{recipeopt}
{W_k}\equiv{\alpha_k}\openone-\ket{\Psi}\bra{\Psi},
\end{equation}
with ${\alpha_k}\doteq\max_{\varrho \ k\text{-separable}}\text{Tr}[\varrho\ket{\Psi}\bra{\Psi}]\equiv\max_{\ket{\Phi}\ k\text{-factorable}}|\langle\Phi\ket{\Psi}|^2$, defines a valid witness. The previous equivalence is due to the fact that the maximum of a linear function over a convex set (mixed states) is always attained at its extremal points (pure states). If the fidelity $\text{Tr}[\ket{\Psi}\bra{\Psi}\varrho]=\bra{\Psi}\varrho\ket{\Psi}$ of a state $\varrho$ with $\ket{\Psi}$ goes beyond the critical value ${\alpha_k}$ then $\varrho$ is detected as non-$k$-separable. For the  bipartite case, the maximization of ${\alpha_2}$ is known \cite{Bourennane04} to be given always by the squared maximal Schmidt  coefficient of $\ket{\Psi}$. The maximization of ${\alpha_k}$ in the general multipartite domain is not a simple task, but yet some analytical results are known \cite{Acin01}. For instance, if $\ket{\Psi}$ is a  genuinely three-qubit entangled state as $\ket{\text{W}}$ or $\ket{\text{GHZ}}$ (defined in Sec. \ref{prelimi}), then ${\alpha_2}$ is $2/3$ or $1/2$, respectively. Furthermore, for  $\ket{\Psi}=\ket{\text{GHZ}}$ and ${\alpha_2}=3/4$, then not only does the resulting witness detect genuine three-partite entanglement but it also identifies it as GHZ-type entanglement. It can be shown that witnesses of the form \eqref{recipeopt}  can also only detect NPT entanglement \cite{guehne09}.

\par As we will see in Sec. \ref{ExpEntWit}, entanglement witnesses constitute one of the most versatile and useful toolboxes for the experimental detection of entanglement.

\subsubsection{Biseparability criteria}
\label{BisepCrit}
\par In Refs. \cite{GuehneSeevinck, Huber10}, very powerful biseparability criteria were introduced. These can be tailored to target at different genuinely multipartite entangled states. For instance, for states in the vicinity of GHZ or W states, they take very simple expressions, which we next present in the form introduced in \cite{GuehneSeevinck}:

\begin{criterion}[Biseparability of $3$ qubits (GHZ)] 
\label{criterion:BisepGHZ}
Any $3$-qubit biseparable state $\varrho$ necessarily fulfills 
\begin{equation}
\label{BisepcritGHZ}
\mathfrak{D}^{\ket{\text{GHZ}_3}}(\varrho)\leq\sqrt{\varrho_{001}\varrho_{110}}+\sqrt{\varrho_{010}\varrho_{101}}+\sqrt{\varrho_{011}\varrho_{100}},
\end{equation}
where $\mathfrak{D}^{\ket{\text{GHZ}_3}}(\varrho)\doteq|\bra{000}\varrho\ket{111}|$, and $\varrho_k\doteq\bra{k}\varrho\ket{k}$, for $001\leq k\leq 110$.  
\end{criterion}
In turn, the $W$-state version of the criterion is
\begin{criterion}[Biseparability of $3$ qubits (W)] 
\label{criterion:BisepW}
Any $3$-qubit biseparable state $\varrho$ necessarily fulfills 
\begin{eqnarray}
\label{BisepcritW}
\nonumber
\mathfrak{D}^{\ket{\text{W}_3}}(\varrho)&\leq&\sqrt{\varrho_{000}\varrho_{011}}+\sqrt{\varrho_{000}\varrho_{101}}+\sqrt{\varrho_{000}\varrho_{110}}\\
&+&\frac{1}{2}(\varrho_{001}+\varrho_{010}+\varrho_{100}),
\end{eqnarray}
where $\mathfrak{D}^{\ket{\text{W}_3}}(\varrho)\doteq|\bra{001}\varrho\ket{010}|+|\bra{001}\varrho\ket{100}|+|\bra{010}\varrho\ket{100}|$, and $\varrho_k\doteq\bra{k}\varrho\ket{k}$, for $000\leq k\leq 110$.  
\end{criterion}

In both criteria, the abbreviation $\mathfrak{D}^{\ket{\Psi}}(\varrho)$ stands  for the sum of the absolute values of the off-diagonal elements in the
upper triangle of density matrix $\varrho$, for which the corresponding entries $\ket{\Psi}\bra{\Psi}$ are not null. The violation of either of the criterions implies, of course, genuine $3$-qubit entanglement. It is important to emphasize that both criteria are valid for all states. The presence of the target states $\ket{\text{GHZ}_3}$ and $\ket{\text{W}_3}$ in \eqref{BisepcritGHZ} and \eqref{BisepcritGHZ}, respectively, just makes reference to the fact that Criterion 10 is especially good at detecting genuine multipartite-entangled states in the vicinity of $\ket{\text{GHZ}_3}$ and Criterion 11 in the vicinity of $\ket{\text{W}_3}$.

In spite of their simplicity, these criteria are stronger than all known entanglement witnesses. They can detect some bound entangled states (separable with respect to all partitions but not fully separable) \cite{GuehneSeevinck}. Also, \eqref{BisepcritW} is violated by 3-qubit W states mixed with white noise, that is $\rho^{W}=(1-p)|W\rangle\langle W|+p\openone/8$, for $p<8/17\approx0.471$  \cite{GuehneSeevinck}; whereas the corresponding best known entanglement witness \cite{guehne09}, ${\cal W}=(2/3)(\openone-|111\rangle\langle 111|)-|W\rangle\langle W|$ detects these states only for $p<8/19\approx0.421$. Furthermore, both Criterion 10 and Criterion 11 were observed in \cite{osvaldo2012} to perform significantly better than the GHZ and W fidelity-based witnesses given by \eqref{recipeopt} in identifying genuine tripartite entanglement of experimentally obtained  mixed three-qubit states.  

In Sec. \ref{prelimi}, we present a generalization of Criterion \ref{criterion:BisepGHZ} to $N$-qubit systems, as well as an extension of Criterion \ref{criterion:BisepW} to four-qubit W and Dicke states \cite{GuehneSeevinck}. We also notice that similar criteria have been derived in Ref. \cite{guehne11b} for four-qubit cluster-diagonal states, defined in Sec. \ref{Graphprelim}.

\subsubsection{Multipartite PPT mixtures}
\label{2.2.2}
As in the case of entanglement, also for NPT-ness does the multipartite scenario display some curious features. In analogy with biseparable states, one defines {\it PPT mixtures} as all possible convex combinations of PPT states. These are contained in the convex hull of the states PPT with respect to some bipartition, pictorially represented in Fig.~\ref{guehne}, together with the biseparability convex hull, for an exemplary three-partite system. There, we can see that there exist states that despite being NPT with respect to all bipartitions still lie inside the PPT-mixture region. In fact, as already anticipated, the example  \eqref{ABCexample} studied above is NPT with respect to all splits but is biseparable and therefore also a PPT mixture. 

In Ref. \cite{guehne2011}, Jungnitsch, Moroder, and  G\"uhne proposed a powerful  approach to characterise PPT mixtures through entanglement witnesses. The idea is to find a witness $W$ such that ({\it i}) $\text{Tr}[W\varrho_{PPT}]\geq0$ for all PPT mixtures $\varrho_{PPT}$ and ({\it ii}) $\text{Tr}[W\varrho_{NPT}]<0$ for some state $\varrho_{NPT}$ outside the PPT-mixture convex hull, of a given multipartite system. A natural way to automatically satisfy ({\it i}) is to demand that $W$ is decomposable, as defined in Sec. \ref{2.2.2}, with respect to all the bipartitions, i.e. that there exists operators $P_{\lambda}\geq 0$ and $Q_{\lambda}\geq 0$ such that $W= P_{\lambda}+ Q_{\lambda}^{T_{\lambda}}$ for all bipartitions $\lambda$, where $T_{\lambda}$ denotes the partial transposition operation with respect to subpart $\lambda$. When $W$ allows for such decompositions, the authors call it a {\it fully decomposable} witness. Conversely, any non-PPT-mixture state $\varrho_{NPT}$ can be detected by a fully decomposable witness  \cite{guehne2011}.

This problem defines a convex optimization that, in contrast to the characterization of biseparability, can be formulated as a linear semidefinite program: Given a multipartite state $\varrho$, find
\begin{eqnarray}
\label{SDP} 
&&\min \text{Tr}[W\varrho]\\
\nonumber 
&&\text{s.t. Tr}[W]=1,\\
\nonumber
&&\text{with } W=P_{\lambda}+ Q_{\lambda}^{T_{\lambda}}, P_{\lambda}\geq 0, Q_{\lambda}\geq 0\ \text{for all}\ \lambda.
\end{eqnarray}
Semidefinite programming is rather efficient and has the advantage that global optimality of the found solution $\text{Tr}[W_{\text{min}}\varrho]$ can be guaranteed. Notice also that decomposability with respect to a bipartition $\lambda$ automatically implies decomposability also with respect to the complementary bipartition $\overline{\lambda}$, so that only half the bipartitions need actually be considered. For system sizes of up to seven qubits, optimization \eqref{SDP} can be easily handled \cite{guehne11b}.

This approach thus renders a necessary condition for biseparability of multipartite systems, analogous to the PPT Criterion \ref{criterion:PPT} for separability of bipartite ones: 

\begin{criterion}[PPT mixtures] \label{criterion:PPTmixtures}If a multipartite state $\varrho\in\mathcal{D}(\mathcal{H}_{\sys})$ is biseparable, then
\begin{eqnarray}
\label{PPTmix}
\rm{Tr}[W_{\text{min}}\varrho]\geq 0,
\end{eqnarray}
where $\rm{Tr}[W_{\text{min}}\varrho]$ is the solution of \eqref{SDP}.  
\end{criterion}
For some subfamilies of states, as for instance the four-qubit cluster-diagonal states, defined in Sec. \ref{Graphprelim}, the condition is also sufficient \cite{guehne11b}. In general, as we describe in Sec. \ref{MultNeg}, the violation of \eqref{PPTmix} can be used to quantify genuinely multipartite entanglement.

\subsubsection{Local operations assisted by classical communication}
\label{2.2.2}
\par General physical processes will be discussed in Sec. \ref{Krausop} in the context of open-system dynamics. At this point, however, we introduce, for the sake of characterizing entanglement, a prominent subclass of physical processes, described by the celebrated {\it local operations and classical communication} (LOCCs) \cite{bennett96b},  operations carried out locally by the users but with the help of classical communication among them. The idea of LOCCs is that distant users, each one in possession of one out of $N$ parts of the system, apply arbitrary operations locally but in such a way that different local operations by each user are coordinated (correlated) among all parts by means of classical communication. A remarkable example of an LOCC protocol is the {\it quantum teleportation}, discovered by Bennett {\it et. al.} \cite{bennett93}. There, two distant users -- canonically called Alice and Bob -- share two qudits (quantum particles of $d$ levels each) in a maximally entangled state as \eqref{maximally}. These two qudits constitute the  teleportation channel. Alice wishes to teleport towards Bob's location another qudit, in an unknown, arbitrary state. First Alice locally measures her part of the channel together with this extra qudit in a basis of maximally-entangled two-qudit states. Next she communicates the classical outcome of her measurement to Bob. Finally, he applies a local operation to his qudit conditioned to the outcome communicated by Alice. As a result, the qudit in Bob's possession ends up precisely in the state of the initial qudit Alice wanted to teleport, achieving thus the desired goal. 

\par The mathematical form of the family of LOCC maps is rather involved \cite{donald02}. There exists though a more general family, called the {\it separable maps}, which include all LOCCs and possesses a much simpler mathematical characterization:
\begin{equation}
\label{LOCC} 
\mathcal{E}_{sep}(\varrho)\equiv\sum_{\mu} {K_1}_{\mu}\otimes\ ... \ \otimes {K_N}_{\mu}\varrho {K_1}_{\mu}^{\dagger}\otimes\ ... \ \otimes {K_N}_{\mu}^{\dagger},
\end{equation}
with  $\sum_{\mu}{K_1}_{\mu}^{\dagger}{K_1}_{\mu}\otimes\ ... \ \otimes {K_N}_{\mu}^{\dagger}{K_N}_{\mu}=\openone$, so that $\text{Tr}[\mathcal{E}_{sep}(\varrho)]\equiv 1\ \forall\ \varrho\in{\mathcal D}({\mathcal H}_{\sys})$, and where each operator\footnote{${K_1}_{\mu}\otimes\ ... \ \otimes {K_N}_{\mu}$ are in turn  called {\it Kraus operators}, which will be touched upon in detail in Sec. \ref{Krausop}. } ${K_i}_{\mu}$ acts on ${\mathcal H}_i$, with $1\leq i\leq N$. Thus, instead of dealing directly with LOCCs, one usually deals with separable operations and concludes properties about the former by inclusion. Since they involve classical communication, LOCC (as well as generic separable) operations can increase classical correlations. On the other hand, no separable map can increase entanglement when acting on pure states \cite{Gheorghiu}, or on general separable states, of course. More generally, for arbitrary states, we will see in Sec. \ref{2.3.2} that entanglement never increases under LOCCs.

\par Finally,  LOCC operations are the deterministic case of a wider family of operations, the {\it stochastic local operations and classical communication} (SLOCCs) \cite{duer00b,bennett00}, which describe also LOCC processes, but happening with a probability not necessarily equal to one. The map $\mathcal{E}_{SLOCC}$ describing this type of processes is $\varrho\rightarrow\mathcal{E}_{SLOCC}(\varrho)/\text{Tr}[\mathcal{E}_{SLOCC}(\varrho)]$, with $\mathcal{E}_{SLOCC}$ an LOCC map but with the normalization not necessarily equal one, that is $\text{Tr}[\mathcal{E}_{SLOCC}(\varrho)]\leq1$.
\subsubsection{Distillable, bound, and PPT entanglement}
\label{2.2.3} 
Most protocols for quantum information processing and quantum communication exploit maximally entangled pure states. However, in practice, due to non-perfect operations and noise, only mixed states are at hand. The problem of how to extract pure-state entanglement from mixed entangled states was considered in the seminal work by Bennett {\it et al.} \cite{bennett96c}. They established there the paradigm of  {\it entanglement distillation},  also  sometimes called {\it entanglement purification} or {\it concentration}. Once again, let us consider two distant users $A$ and $B$ who  share now $n$ identical copies of the state $\varrho_{AB}$ containing some noisy entanglement. They can apply an LOCC protocol $\mathcal{E}_{\text{Dist}}$  acting collectively on all $n$ copies of $\varrho_{AB}$ so as to obtain a smaller number $m(n)>0$ of copies of a state closer to a pure maximally entangled state than the original state $\varrho_{AB}$. Errors are allowed, but they must vanish in the asymptotic limit $n\rightarrow\infty$, and the obtained state must tend to the target maximally entangled state. When this is possible, $\mathcal{E}_{\text{Dist}}$ is an  entanglement distillation protocol for $\varrho_{AB}$ with efficiency 
\begin{equation}
\label{eta_D} 
\eta_{D}\doteq\lim_{n\rightarrow\infty}\frac{m(n)}{n}.
\end{equation}
The optimal protocol is the one that maximizes $\eta_{D}$. This optimal efficiency defines in turn the {\it distillable}, or {\it free},  {\it entanglement} $E_D(\varrho_{AB})$ of  $\varrho_{AB}$:
\begin{equation}
\label{E_D} 
E_{D}(\varrho_{AB})\doteq\sup_{\mathcal{E}_{\text{Dist}}\in \text{LOCC}}\eta_{D}.
\end{equation}
Accordingly, $\varrho_{AB}$ is said to be {\it distillable}, and possesses $E_{D}$ {\it ebits} (entanglement bits) of distillable entanglement. 

\par The inverse process, sometimes called {\it entanglement dilution}, is also possible. Starting from $m$ pure maximally entangled pairs, $A$ and $B$ apply an LOCC map $\mathcal{E}_{\text{Dil}}$ to obtain a larger number $n(m)$ of identical copies of  $\varrho_{AB}$. Then $\mathcal{E}_{\text{Dil}}$ is an entanglement dilution protocol for $\varrho_{AB}$, with a cost, in ebits per copy, given by the conversion rate 
\begin{equation}
\label{eta_C} 
\eta_{C}\doteq\frac{m(n)}{n}.  
\end{equation}
The optimal protocol is now the one that minimizes $\eta_{C}$, and the optimal cost defines the {\it entanglement cost} $E_{C}(\varrho_{AB})$ of $\varrho_{AB}$:
\begin{equation}
\label{E_C} 
E_{C}(\varrho_{AB})\equiv\inf_{\mathcal{E}_{\text{Dil}}\in\text{LOCC}}\lim_{m\rightarrow\infty}\eta_{C}.
\end{equation}

\par The natural question that arises is whether entanglement distillation and dilution are reversible processes or not.  Surprisingly, the answer is no. There are mixed entangled states from which it is not possible to distill any entanglement at all.  It turns out that the Peres-Horodecki PPT criterion is intimately related to this curious phenomenon. In fact, it was again the Horodecki family who soon after the discovery of the criterion found out \cite{horodecki98} that {\it every PPT state is non-distillable}. This means, in view of the fact that for mixed systems larger than $2\times 3$ there exist PPT entangled states, that there are entangled states that are non-distillable. These are precisely the celebrated {\it bound entangled states}. Clearly, we   have then in general $E_D\leq E_C$, the equality holding necessarily only for pure states, or for systems of  $d\leq2\times 3$ in arbitrary states.

\begin{figure}[t!]
\begin{center}
\includegraphics[width=1\linewidth]{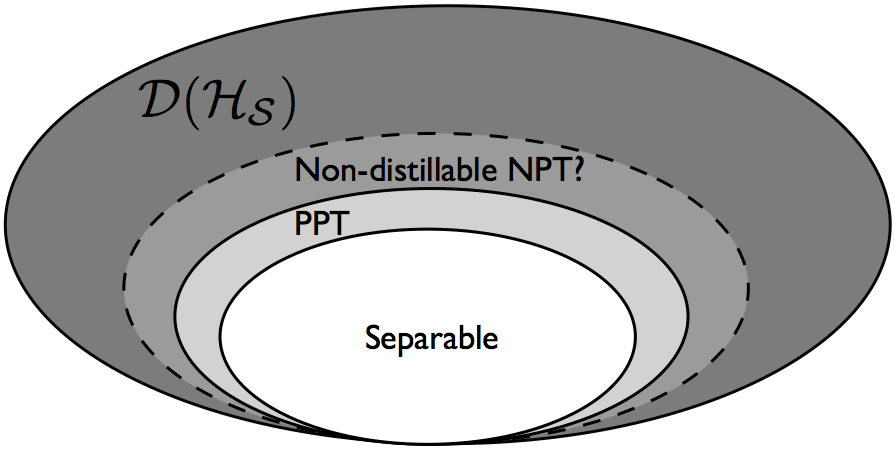}
\caption{
\label{Geometry}
Schematic representation of the inner geometry of the set of the density matrices ${\mathcal D}({\mathcal H}_{\sys})$, with the borders between the sets of the separable states, the PPT ones, the conjectured NPT non-distillable ones, in dashed line, and finally the rest of ${\mathcal D}({\mathcal H}_{\sys})$: the distillable states.} 
\end{center}
\end{figure}
\par On the other hand, it is not known whether there exist bound entangled states other than the PPT ones. This is a question that remains open since the very discovery of bound entanglement and has been called ``the problem of the NPT bound entanglement''. Every PPT state is non-distillable, but the conjecture \cite{DiVincenzo00,Duer00d} is that there could be a gap between the set of all distillable states and that of the PPT ones.  The situation is described  in Fig. \ref{Geometry}, which schematically represents the internal  geometry of ${\mathcal D}({\mathcal H}_{\sys})$ with the borders between the sets of the separable states, the PPT ones, the hypothetic NPT non-distillable ones (in dashed), and finally the rest of   ${\mathcal D}({\mathcal H}_{\sys})$: the distillable states. Proving, or disproving, the conjecture is one of the big fundamental open questions in entanglement theory. See \cite{clarisse06} for a review on the problem.
\subsubsection{Multipartite distillability}
\label{multidistil}
It is also possible to define the notion of distillability for multipartite systems. We consider then $N$ users that apply an LOCC map
on $n$ copies of an arbitrary $N$-particle state $\varrho$, with the aim of obtaining $m(n)\leq n$ copies of some pure entangled  target state. The LOCC map acts collectively on all $n$ copies of $\varrho$, but the ``L" in LOCC  stands for local with respect to each of the $N$ parts individually, who can correlate their local operations only through classical communication, as in the bipartite case. When, in the asymptotic limit of $n\to\infty$, $m(n)>0$, $\varrho$ is said to be {\it distillable}. If the distilled target pure state is genuinely $N$-partite entangled, then $\varrho$ is  in addition {\it genuinely multipartite  distillable}, or {\it $N$-party distillable}.  

\par Interestingly, genuinely multipartite entangled states that belong to inequivalent entanglement classes at the single-copy level can be mapped into one another by LOCCs when sufficiently many copies of them are available. In particular, this is the case for instance of the GHZ and W states for $N$-qubit systems (see for instance \cite{Yu13} and references therein). In general, a necessary and sufficient condition for genuine $N$-party distillability is the distillation of a pure maximally entangled state between every pair of  parties \cite{Dur00}. The reason behind this fact is that, from sufficiently many copies of these pairs, any pure genuinely multipartite entangled state can be obtained with LOCCs, and vice versa. An example of a bipartite-distillation based protocol for the distillation of genuine multipartite graph-state entanglement  is explicitly described in Sec. \ref{DuerGraph}. Still, protocols for the direct distillation of  multipartite entangled states exist, as we briefly mention in Sec. \ref{particudist} (see Ref. \cite{Duer&Briegel07} for a  review). A(n only) necessary (but much simpler to check) condition for genuine $N$-partite distillability is given by the following criterion \cite{duer00c,Dur00}:
\begin{criterion}[Multiparite distillability (necessary)] \label{criterion:genNdist}
If an arbitrary $N$-qubit state $\varrho\in{\mathcal D}({\mathcal H}_{\sys})$ is $N$-party distillable, then each and all of its bipartite splits are NPT. 
\end{criterion}
A simple way to convince oneself of the validity of the assertion is because, if any split is PPT, then it is certainly not possible to distill genuine $N$-party entanglement. On the other hand, the converse assertion cannot hold if NPT bound entanglement exists, as discussed in the previous subsection.

\par An important distinction must be made at this point. Genuinely mutipartite distillability and genuinely multipartite entanglement turn out to be inequivalent notions. On the one hand, mixed states that are PPT with respect to any choice of bipartite cuts (and therefore not $N$-party distillable) but at the same time display genuine $N$-partite  entanglement can be constructed \cite{Piani&Mora07}. On the other hand, there are biseparable states that are $N$-party distillable. That is,  one can distill genuine-multipartite entanglement from states with no genuine-multipartite entanglement at all. The reason behind this is that the distillation of pure maximally entangled states between every pair, which is sufficient for genuine $N$-party distillation, is certainly not enough to guarantee genuine-multipartite entanglement.  This is the case precisely of the 3-qubit state \eqref{ABCexample} studied above. The state is biseparable, but a pure maximally entangled state between Alice and Bob, and another between Alice and Charlie, can be distilled. With them, Alice can teleport one qubit of a locally created GHZ state to Bob and another to Charlie, thereby obtaining a GHZ state among all three users. This example  also shows that the set of non-genuinely mutipartite distillable states is not convex, because a convex combination of three non-genuinely mutipartite distillable states renders a genuinely mutipartite distillable one. 

\par Both the definition of $N$-party distillability as well as Criterion \ref{criterion:genNdist} can of course be directly extended to blockwise $M$-party distillability. There, the $N$ parts are grouped into $M$ blocks, each of which is treated as a single subpart of larger dimension, and the aim is to distill a blockwise $M$-partite pure entangled  state with respect to the $M$-partition.  Multi-party distillable states are in general easier to experimentally prepare, or detect, than multipartite entangled ones \cite{guehne09}. For the same reasons, they can  also be much more robust against noise. These points are elaborated in Sec. \ref{V}, where in particular we discuss other criteria for $N$-party and blockwise $M$-party distillability, for GHZ entanglement.

\subsubsection{Multipartite bound and unlockable entanglement}
\label{boundmulti}
An $N$-partite state is bound entangled if it is entangled and not $N$-party distillable. One of the first examples was obtained by Bennett {\it et al.}~in Ref.~\cite{bennett99}, where an entangled three-qubit state was found to be separable with respect to all its bipartitions. This state is clearly not  distillable because not even a singlet between any of the qubits can be extracted with $N$-party LOCCs. Another popular example is the Smolin state \cite{smolin01}:
\begin{equation}
\label{Smolin} 
\varrho^{\rm Smolin}_{ABCD}\equiv\frac{1}{4}\sum_{\mu=1}^{4}\ket{\Psi^{\mu}_{AB}}\bra{\Psi^{\mu}_{AB}}\otimes\ket{\Psi^{\mu}_{CD}}\bra{\Psi^{\mu}_{CD}},
\end{equation}
for four qubits $A$, $B$, $C$, and $D$, and where $\ket{\Psi^{\mu}}$ are the four maximally-entangled Bell states \eqref{Bellstate}. By construction, the split $AB:CD$ is separable. In addition, the state is invariant under the permutation of any of its parts. This can be seen by noticing that $\varrho^{\rm Smolin}_{ABCD}\equiv\frac{1}{4}(\openone^{\otimes 4}+\sum_{\mu=1}^{3}{{\sigma^{\mu}}}^{\otimes 4})$, with {$\sigma^{\mu}$} the three Pauli matrices. Therefore, it is separable with respect to any two-versus-two cut. This implies, again, that no  entanglement between any pair of qubits can be extracted with $N$-party LOCCs. So the state is non-distillable. However, it is immediate to check that all its one-versus-three  qubit bipartitions are NPT. So the state is entangled.

\par The Smolin state features in addition another exotic property. From \eqref{Smolin}, one sees that the pairs $AB$ and $CD$ have an equal probability of being in a Bell state, but without knowing which one. If, say, parts $A$ and $B$ are brought together, they can {\it unlock} the entanglement between $C$ and $D$ by performing a joint Bell-state measurement on their qubits  and then communicating their outcome via a classical channel to $C$ and $D$. The outcome works as a {\it flag} for  $C$ and $D$, because it marks onto which Bell state their subsystem  has been projected. As a result, $C$ and $D$ are left with a pure maximally-entangled state (and they know which one!). This process is known as {\it entanglement unlocking}, and the bound entanglement of $\varrho^{Smolin}_{ABCD}$ is said to be {\it unlockable}. As discussed in Sec. \ref{65}, state \eqref{Smolin} has already been the target of experimental investigations \cite{Amselem09,Lavoie10,Barreiro10}.  

\par Finally, we will see in Sec.~\ref{NatMultiBound} that multipartite bound entanglement is,  exotic as it may seem though, actually a rather common phenomenon, which can appear due to natural physical processes. Namely, we will see that multipartite bound entangled states  can arise in multipartite-entangled systems due to the  interaction with the environment.

\subsection{Entanglement measures}
\label{2.3}
\par The quantification of entanglement of general states happens to be a formidably difficult task, not less complex than the separable-versus-entangled problem. It typically involves optimizations whose required  computational effort grows so fast with the system size that, already for a handful of particles, calculations for arbitrary states become in practice impossible. Given the complexity and importance of the problem, there exists a wide variety of proposed quantifiers. Some are based on efficiencies of quantum-information protocols, some on geometrical aspects, on axiomatic approaches, etc, and each of them is more advantageous than others in some particular sense. In this subsection, we present barely some of the most popular proposals. We refer the interested reader to Refs. \cite{horodecki09,plenio07} for detailed reviews on the subject.
\subsubsection{Operational measures}
\label{Operational}
This type of quantifiers is based on the premise that entanglement is a resource for physical tasks.
Accordingly, the entanglement in a given state $\varrho_{AB}$ is given by its efficacy as a resource for a particular task. A simple example of this is the {\it maximal teleportation fidelity} $f_{max}$, defined as the fidelity of teleportation of a qudit attainable when $\varrho_{AB}$ is used as the teleportation channel, averaged over all possible input states and maximized over all possible teleportation strategies. If $\varrho_{AB}$ is maximally entangled,  the teleportation is faithful and $f_{max}(\varrho_{AB})=1$. Whereas if $\varrho_{AB}$ is separable or bound entangled,  the fidelity takes the maximum value attainable by classical means: $f_{max}(\varrho_{AB})=\frac{2}{d+1}$ \cite{Bruss99}. 

\par The two most important operational quantifiers are the {\it distillable entanglement} $E_{D}$ and the {\it entanglement cost} $E_{C}$, both defined in Sec. \ref{2.2.3}. $E_{D}$ is more powerful than $f_{max}$ at detecting entanglement because, whereas the latter restricts to the  single-copy regime, the former addresses the usefulness as a physical resource of asymptotically many copies of $\varrho_{AB}$. On the other hand, $E_{C}$  quantifies the ebits necessary for the LOCC production, in the asymptotic sense, of $\varrho_{AB}$. Since LOCCs can only map separable states into separable states, every entangled state costs a non-null number of ebits. That is,  $E_{C}>0$ for {\it every} entangled state, including the bound ones.  In this sense, $E_{C}$ is more powerful than both $E_{D}$ and $f_{max}$, which are  sensitive only to free entanglement. However, all these measures involve optimizations that make their numeric evaluation in practice a very hard problem. 

\subsubsection{Axiomatic measures}
\label{2.3.2}

Vedral and collaborators introduced in  \cite{vedral97} the idea of an axiomatic definition of an entanglement measure, such that any function that satisfies some reasonable postulates can be considered an entanglement quantifier.

The most important postulate, already proposed in \cite{bennett96b}, and on which there is absolute consensus within the community, is that of
\begin{itemize}
\item {\bf Monotonicity under LOCCs}:  Entanglement cannot increase due to local operations assisted by classical communication.
\end{itemize}
Mathematically, if $\varrho$ is any arbitrary state and $E(\varrho)$ is a measure of its entanglement, this axiom demands that
\begin{equation}
\label{monotony} 
E(\varrho)\geq E[\mathcal{E}_{LOCC}(\varrho)],
\end{equation}
for all LOCC maps $\mathcal{E}_{LOCC}$. 

\par There exists also another monotonicity condition that --- in spite of being more restrictive than \eqref{monotony} --- is satisfied by all known entanglement measures and used to be considered as the fundamental requirement. This condition is called  
\begin{itemize}
\item {\bf Strong monotonicity under LOCCs}:  Entanglement cannot increase {\it on average} due to local operations and classical communication.
\end{itemize}
Mathematically,
\begin{equation}
\label{strongmonotony} 
E(\varrho)\geq\sum_{\mu}p_{\mu}E(\sigma_{\mu}),
\end{equation}
where $\{p_{\mu},\sigma_{\mu}\}$ is the  {\it ensemble} obtained from $\varrho$ via the LOCC in question. The idea behind formulation \eqref{strongmonotony} of monotonicity is that the system undergoes a local process where information is gained about which member of the ensemble is actually realized. These processes can always be thought of as general local measurements, which lead to a {\it flagging information}, so some mixedness is always removed. The available entanglement is then that of the average over the resulting ensemble, which  is typically higher than, or equal to, the bare one of the full mixture\footnote{Strictly speaking, this is necessarily true only  when $E$ is a convex function, which --as we discuss below-- is another very general property satisfied by most entanglement measures.} considered in \eqref{monotony}.

\par  Vidal originally suggested \cite{vidal00} to consider strong LOCC monotonicity  as the only fundamental postulate required by entanglement measures, all other properties either being derived from this basic axiom or being optional. However, nowadays there is common agreement \cite{horodecki09} that condition  (\ref{monotony}) is the only fundamentally necessary requirement, as it concerns directly the amount of entanglement after any LOCC, even those by which no flagging information is acquired.  On the other hand, strong monotonicity is frequently easier to prove than simple monotonicity and, when the measure $E$ is  convex, the former automatically implies the latter. Therefore, condition  \eqref{strongmonotony} is still the most commonly used formulation and, usually, every function satisfying this condition is called  an {\it entanglement monotone}. We  denote entanglement monotones by ${\mathcal M}$\footnote{An important distinction between LOCC and separable maps turns up here:
Pathologic examples have been found of mixed entangled states for which the value of some particular entanglement monotone can 
increase under separable maps \cite{Chitambar09}.}. 

\par A fundamental property imposed by either monotonicity conditions, \eqref{monotony} or  \eqref{strongmonotony}, and whose importance on its own deserves an explicit comment, is the following: 
\begin{itemize}
\item  Entanglement is invariant under local unitary transformations: 
\end{itemize}
\begin{equation}  
\label{localunitinv} 
E(\varrho)\equiv E(U_1\otimes \ ...\otimes U_N\varrho U^{\dagger}_1\otimes  ...\otimes U^{\dagger}_N), 
\end{equation}
for all local unitary operators $U_1$, $U_2$,  $ ...\ U_N$  acting respectively on $\mathcal{H}_1$,  $\mathcal{H}_2$, $\ .... \mathcal{H}_N$.
To see it, notice that  when $\mathcal{E}_{LOCC}$ is  a local unitary operation it is invertible, its inverse $\mathcal{E}_{LOCC}^{-1}$ being simply the inverse local unitary operation. Then the only way for $E$ to be monotonous under both transformations $\varrho\rightarrow\mathcal{E}_{LOCC}(\varrho)$ and $\mathcal{E}_{LOCC}(\varrho)\rightarrow\mathcal{E}_{LOCC}^{-1}(\mathcal{E}_{LOCC}(\varrho))\equiv\varrho$ is  necessarily  to remain invariant in such processes. This is a very sensible characteristic of an entanglement quantifier, since local unitary transformations are nothing but  local basis changes.

Since every separable state can be transformed into any other separable state via  LOCCs \cite{vidal00}, $E$ must be constant over the set of separable states. In addition, this constant must set the minimal entanglement, because every separable state can be obtained from any other state via LOCCs. It is then convenient to set this constant as zero, so that the entanglement of a separable state is null. That is, if $\varrho$ is separable, then $E(\varrho)\equiv 0$. Note that, except for the arbitrariness in the exact value of the constant, this property is fully derived from monotonicity under LOCCs.

\par {\bf Other possible axioms}: There exist other properties that, while not necessarily required for every entanglement measure, can be convenient and natural in certain contexts. They are essentially \cite{horodecki09,mintert05b,plenio07,bruss02} the following ones:
\begin{itemize}
\item {\it Convexity}. The entanglement is a convex function in  ${\mathcal D}({\mathcal H}_{\sys})$:
\begin{equation}
\label{convexidade} 
E(p\varrho+ (1-p)\varrho')\leq pE(\varrho)+(1-p)E(\varrho'),
\end{equation}
for $0\leq p\leq 1$. Up to recently,  convexity used to be considered a necessary ingredient for monotonicity. These days, it is just a convenient mathematical property satisfied by most measures. Indeed, almost all entanglement measures mentioned in this review -- except perhaps for the distillable entanglement, whose convexity is still an open question related to the existence of NPT bound entanglement \cite{shor01, shor03} -- are convex.
\item {\it Continuity}. The entanglement is a continuous function:
\begin{equation}
\label{continuity} 
||\varrho-\sigma||\rightarrow 0\Rightarrow |E(\varrho)-E(\sigma)|\rightarrow 0, \ \forall\ \sigma, \varrho\in{\mathcal D}({\mathcal H}_\sys),
\end{equation}
where ``$||\ ||$" stands for the trace norm.
\item {\it Additivity}. The entanglement contained in $k$ copies of  $\varrho$  is equal to $k$ times the entanglement of $\varrho$:
\begin{equation}
\label{aditivity} 
E(\varrho^{\otimes k})\equiv kE(\varrho).
\end{equation}
This axiom is sometimes called {\it weak additivity}, the term {\it additivity} being  reserved for the more restrictive condition $E(\varrho\otimes\sigma)\equiv E(\varrho)+E(\sigma)$, with $\varrho$ and $\sigma$ any two states of arbitrary systems.
\item {\it Subadditivity}. For any two systems in arbitrary states, $\varrho$ and $\sigma$, the total entanglement is not greater than the sum of both individual entanglements:
\begin{equation}
\label{subaditivity} 
E(\varrho\otimes\sigma)\leq E(\varrho)+E(\sigma).
\end{equation}
\end{itemize}

\par {\bf Monotonicity for pure states}: For pure states strong monotonicity reduces to
\begin{equation}
\label{strongpure} 
E(\Psi)\geq\sum_{\mu}p_{\mu}E(\Psi_{\mu}),
\end{equation}
where $\{p_{\mu},\ \ket{\Psi_{\mu}}\}$ is the {\it ensemble} obtained from $\ket{\Psi}$ via any LOCC. Indeed, for pure bipartite states there exists a recipe for the construction of entanglement monotones \cite{vidal00}. Let $\ket{\Psi_{AB}}$ be a pure bipartite state, and $\varrho_{R}\in{\mathcal D}({\mathcal H}_{R})$ the reduced density matrix of subsystem $R=A$ or $B$. Then any function $f(\varrho_R):{\mathcal D}({\mathcal H}_{R})\rightarrow\Re$ that is
\begin{itemize}
\item {\it unitarily invariant}, meaning that $f(\varrho_{R})\equiv f(U\varrho_{R}U^{\dagger})$, for all unitary operator $U$ acting on ${\mathcal H}_{R}$, and 
\item {\it concave in} ${\mathcal D}({\mathcal H}_{R})$, meaning that $f(\varrho_{R})\geq pf(\sigma_{R})+(1-p)f(\sigma'_{R})$, for any $\sigma_{R}$ and $\sigma'_{R}\in{\mathcal D}({\mathcal H}_{R})$ such that $\varrho_{R}=p\sigma_{R}+(1-p)\sigma'_{R}$, and with $0\leq p\leq 1$,
\end{itemize}
yields an entanglement monotone. Notice that given the invariance of  $f$ under unitary operations, it can only be a function of unitary invariants, {\it i.e.}, of the spectrum of $\varrho_{R}$. Consequently, here  it is not necessary to distinguish between $\varrho_{A}$ and $\varrho_{B}$, because they both have the same non-null eigenvalues. For this reason we can simply use subindex $R$ in a generic way or, alternatively, refer directly to $\ket{\Psi}$: ${\mathcal M}(\Psi)\equiv f(\varrho_{R})$.

\par The most prominent choice satisfying the above conditions is the von Neumann entropy $S$, defined as
\begin{equation}
\label{vonneumannentropy} 
S(\varrho_{R})\equiv -\text{Tr}[\varrho_{R}\ln(\varrho_{R})].
\end{equation}
This entropy is the formal quantifier of the uncertainty, or lack of information, in $\varrho_{R}$. In the context of entanglement theory, it is frequently called {\it entanglement entropy}, or simply {\it entanglement}, of $\ket{\Psi}$, $E_E(\Psi)$.
\par Another important choice is the linear entropy $S_L$:
\begin{equation}
\label{linearentropy} 
S_L(\varrho_{R})\equiv 1 -\text{Tr}[\varrho_{R}^2].
\end{equation}
The trace of the squared  reduced matrix in  (\ref{linearentropy}) measures the {\it purity} of the reduced state. For this reason the linear entropy is frequently called also as the {\it mixedness} (or {\it degree of mixedness}{).} 

\par The idea of quantifying the entanglement of pure states via the impurity of their reduced subsystems is another direct consequence of  Schr\"odinger's seminal observation  that entangled pure states give us more information about the system as a whole than about any of its parts (see Sec.~\ref{2.1}). Thus, for pure states, lack of information of the subsystems can only be due to entanglement of the composite system.

\par  {\bf Monotonicity of mixed states: the ``convex roof''}: Whereas for pure states there is a relatively simple recipe for the construction of entanglement monotones,  for mixed states it is much more difficult to discriminate between classical and quantum correlations. Vidal showed  \cite{vidal00} that
an entanglement monotone  for an arbitrary mixed state $\varrho$ can be defined by taking any valid pure-state monotone and extending it to mixed states by means of the so-called {\it convex roof} (or {\it convex hull}) construction   \cite{bennett96b,uhlmann00}:
\begin{equation}
\label{ConvexRoof} 
{\mathcal M}(\varrho)\equiv\inf_{\{p_{\mu},\Psi_{\mu}\}}\sum_{\mu}p_{\mu}{\mathcal M}(\Psi_{\mu}),
\end{equation}
where the infimum is over all possible pure-state decompositions $\{p_{\mu},\Psi_{\mu}\}$ of the state, $\varrho=\sum_{\mu}p_{\mu}\ket{\Psi_{\mu}}\bra{\Psi_{\mu}}$. It yields the {\it infimum average entanglement}\footnote{Notice that if $\{p_{\mu},\Psi_{\mu}\}$ is the {\it ensemble} resulting from an LOCC on some pure state, expression~\eqref{ConvexRoof} is equivalent to the minimization of the right-hand side of~\eqref{strongmonotony}.}. The search of such infimum is typically a high-dimensional optimization problem, and this is the root of the difficulty in numerically evaluating these measures.

\par {\bf Entanglement of formation and concurrence}: The most popular entanglement quantifier is the {\it entanglement of formation} $E_F$ \cite{bennett96b}. It can be defined via Vidal's recipe for monotones described above with  the von Neumann entropy  (\ref{vonneumannentropy}) as ${\mathcal M}$:
\begin{equation}
\label{Entofform} 
E_F(\varrho_{AB})\equiv\inf_{\{p_{\mu},\Psi_{\mu}\}}\sum_{\mu}p_{\mu}S({\varrho_{R}}_{\mu})\equiv\inf_{\{p_{\mu},\Psi_{\mu}\}}\sum_{\mu}p_{\mu}E_E(\Psi_{\mu}),
\end{equation}
with ${\varrho_{R}}_{\mu}$ the reduced state corresponding to subsystems $A$ or $B$ of $\ket{\Psi_{\mu}}$. That is, the entanglement of formation is the infimum average entropy of entanglement over all possible pure-state decompositions of the state.  

\par For pure states the entanglement of formation coincides with the entanglement cost \cite{bennett96b}. Therefore, $E_F(\varrho_{AB})$ quantifies the cost in ebits of the formation of $\varrho_{AB}$ via LOCCs in a restricted scenario where each member $\ket{\Psi_{\mu}}$ of the  ensemble composing $\varrho_{AB}$ is formed independently (and then later on all members mixed with probabilities $p_{\mu}$). This is where the term ``of formation" historically came from. Furthermore, the asymptotically regularized version of $E_F$ has long been known to coincide with $E_C$ \cite{hayden01}:
\begin{equation}
\label{formandcost} 
\lim_{k\rightarrow\infty}\frac{E_F(\varrho_{AB}^{\otimes k})}{k}\equiv E_C(\varrho_{AB});
\end{equation}
and over the years it was believed \cite{Shor04, Matsumoto} that $E_F$ should be additive, {\it i. e.} that both sides of \eqref{formandcost} should actually be identically equal to $E_F(\varrho_{AB})$.

Very recently, Hastings solved \cite{Hastings09} this long-standing problem, known as the ``additivity conjecture".  He was able to show formally that counterexamples to this conjecture exist and that $E_C$ can actually be strictly less than $E_F$. This, in simple words, means that it takes less ebits to form many copies of $\varrho_{AB}$ simultaneously than one by one. The above-mentioned strategy with each pure-state member of the ensemble being formed independently is non-optimal. Today, we know that in general the inequalities 
\begin{equation}
E_D\leq E_C\leq E_F
\end{equation}
 hold, with the equalities necessarily holding only in the case of pure states. 

\par Unfortunately, $E_F$ is no exception in terms of the computational difficulty in its calculation, except for the two-qubit case. For two qubits there exists a closed analytical expression for $E_F$ in terms of an auxiliary quantity of immediate algebraic evaluation, the {\it concurrence}  $C$. It was first introduced in Ref. \cite{hills97} for the case of matrices of rank 2, and later generalized in Ref. \cite{wooters98} to any two-qubit state. It is defined as
\begin{subequations}
\label{concu1} 
\begin{align}
&C(\varrho_{AB})\equiv\max\{0, \Lambda \}\, ,\\
&\Lambda\equiv \xi_1-\xi_2-\xi_3-\xi_4\,,
\end{align}
\end{subequations}
where 
 $\xi_1$, $\xi_2$, $\xi_3$, and $\xi_4$ are the square roots, in decreasing order, of the eigenvalues of the matrix $\varrho_{AB}.\tilde{\varrho}_{AB}$, being $\tilde{\varrho}_{AB}\equiv Y\otimes Y\varrho_{AB}^{*}Y\otimes Y$,
with $Y$ the second Pauli matrix, and ``$^{*}$"\ the  complex conjugation in the computational basis. It is clear that concurrence coincides with $\Lambda$ when $\Lambda\ge 0$ and is equal to zero when $\Lambda<0$. Once obtained the concurrence of the state in question, an analytical formula for $E_F$ exists \cite{wooters98}:
\begin{equation}
\label{EntofformConcu} 
E_F(\varrho_{AB})\equiv H_2\Big(\frac{1}{2}+\frac{1}{2}\sqrt{1-C^{2}(\varrho_{AB})}\Big),
\end{equation}
where $H_2$ is the dyadic Shannon entropy function, defined as
\begin{displaymath}
H_2(x)=\left\{
\begin{array}{ll}
x\log x-(1- x)\log(1-x), & \forall\ x\in (0,1],\ \text{e}\\
0, & \text{for}\ x=0.
\end{array}\right.
\end{displaymath}
\par Owing to the simplicity of its algebraic evaluation, concurrence \eqref{concu1}  constituted a big step forward in the quantification of entanglement and, though originally motivated by the calculation of $E_F$, it is an alternative monotone that has gained the status of entanglement measure in its own right \cite{wooters98}. Its monotonicity  stems from the fact that,  for pure states $\varrho_{AB}=\ket{\Psi_{AB}}\bra{\Psi_{AB}}$, concurrence \eqref{concu1} 
becomes \cite{rungta01}:
\begin{eqnarray}
\label{concupure} 
C(\varrho_{AB})&&\equiv C(\Psi_{AB})\equiv\sqrt{1 -Tr[\varrho_{A}^2]+1 -Tr[\varrho_{B}^2]}\nonumber\\
&&\equiv\sqrt{2(1 -Tr[\varrho_{R}^2])}\equiv\sqrt{2S_L(\varrho_{R})},
\end{eqnarray}
with $S_L$ the linear entropy (\ref{linearentropy}). As a matter of fact, expression (\ref{concupure}) can also be considered an alternative definition of $C$ and, as such, can be generalized to the arbitrary-dimensional bipartite \cite{rungta01, uhlmann00} or multipartite \cite{meyer02,brennen03, carvalho04} case. 

\par We next present one such generalization \cite{carvalho04}. Notice that the square root in the bipartite definition  \eqref{concupure} contains the sum of the linear entropies of both reduced subsystems. Analogously, for an arbitrary $N$-partite system in a pure, normalized state  $\ket{\Psi}$,  the $N$-partite concurrence $C_N$ can be defined as \cite{carvalho04}
\begin{equation}
\label{ConcAndre}
C_N(\ket{\Psi})\doteq2^{1-N/2}\sqrt{(2^{N}-2)-\sum_{I}\mbox{Tr}\varrho_{R_{I}}^{2}},
\end{equation}
where subindex  $I$ labels the  $2^{N}-2$  possible non-trivial subpartitions  of the system, and $\varrho_{R_{I}}$ is the reduced density operator of the $I$-th subset for state $\ket{\Psi}$. The radicand in \eqref{ConcAndre} is the  average linear entropy of all reduced subsets of the $N$-partite system. Therefore, the multipartite concurrence $C_N$ is said to quantify the average  mixedness upon partial trace over all subsystems. The extension  to  mixed states in turn is achieved via the convex roof construction. Finally, apart from its simplicity, a particularly advantageous feature of generalization \eqref{ConcAndre} is that, as will be seen in Sec. \ref{copias}, it allows for direct experimental evaluations through projective measurements when two copies of the state are simultaneously available.
\subsubsection{Geometric measures}

This class of quantifiers is based on geometrical aspects of  ${\mathcal D}({\mathcal H}_{\sys})$,  involving the  notion of proximity among states. They rely on the  very intuitive idea that the further away a state  is from the separable states, the more entangled it should be. They apply to possibly-mixed, arbitrary-dimensional states $\rho$ with any number $N$ of constituent subsystems.  We concentrate here on the  {\it relative entropy of entanglement} \cite{plenio97,plenio}, which relies on how distinguishable state $\rho$ is from its closest separable. This approach was originally formulated in terms of separable states, but here we present it in a more general way, regarding  $k$-separable ones.

\par The notion of proximity is realized by the use of a formal distance $\Delta$ in  ${\mathcal D}({\mathcal H}_{\sys})$. The distance between $\varrho$ and the set of $k$-separability is defined as the minimum distance $\min_{\zeta} \Delta(\varrho ||\zeta)$ between $\varrho$ and any $k$-separable state $\zeta$. A pictorial representation is given in Fig.~\ref{Witnesses} as the dashed segment joining $\varrho$ with its closest state $\zeta$ on the border of the $k$-separability set. Such distance leads, upon a proper choice of metric, to a measure of the entanglement for $\varrho$.  A prominent choice for this metric is the von Neumman relative entropy $S_R(\varrho ||\zeta)\equiv\text{Tr}[\varrho(\log(\varrho)-\log(\zeta))]$, which quantifies how {\it distinguishable} $\varrho$ and  $\zeta$ are \cite{vedral97}.  This yields the {\it relative entropy of entanglement}  \cite{plenio97,plenio}:
\begin{equation}
\label{relentro}
E^R_k(\varrho)\doteq\min_{\zeta\ k\text{-separable}}S_R(\varrho ||\zeta).
\end{equation}

\par Besides $S_R$, other choices of mathematical distances are also possible \cite{plenio97,plenio}. In particular, if one restricts to pure states, the notion of distance between states becomes equivalent to that of the angle between vectors. This allows one to quantify entanglement  in terms of the overlap between $\varrho\equiv\ket{\Psi}\bra{\Psi}$ and its closest state $\zeta\equiv\ket{\Phi}\bra{\Phi}$, with $\ket{\Phi}$ a  $k$-factorable vector\footnote{Notice that this overlap is nothing but ${\alpha_k}$ of Eq. \eqref{recipeopt}.}, leading to a pure-state entanglement monotone known as  the {\it geometric measure of entanglement} \cite{Wei03}: 
\begin{equation}
\label{geometricmeasure}
E^G_k(\varrho)\equiv E^G_k(\Psi)\equiv1-\max_{\ket{\Phi}\ k\text{-factorable}}|\langle\Psi\ket{\Phi}|^2,
\end{equation}
generalizable to the mixed-state case through the convex-roof extension \eqref{ConvexRoof}.

\par Geometric quantifiers feature by definition the desirable property of detecting all non $k$-separable states. However, their numeric evaluations require not only optimizations but also being able to establish if a state is $k$-separable, for which, as we know, there is no general efficient criterion.

\subsubsection{Negativity}
\label{Outros}
There are some quantifiers that do not fit straightforwardly into  any of the two categories (axiomatic and geometric quantifiers) described so far. An important one is the  {\it negativity} \cite{vidal02},  originally introduced in  \cite{ziczkowski98}. It is the known monotone of simplest algebraic evaluation, as even for mixed states it does not involve any optimization. Given state $\varrho_{AB}$, it is defined as  \cite{vidal02}
\begin{equation}
\label{Negativity} 
Neg(\varrho_{AB})\doteq\frac{||\varrho_{AB}^{T_{B}}|| -1}{2},
\end{equation}
 where $||\varrho_{AB}^{T_{B}}||$ is the trace norm -- the sum of the absolute value of each eigenvalue -- of $\varrho_{AB}^{T_{B}}$. Negativity is based on Criterion \ref{criterion:PPT}. It can be recast as the absolute value of the sum of the negative eigenvalues of  $\varrho_{AB}^{T_{B}}$. That is, $Neg(\varrho_{AB})$ quantifies how much  $\varrho_{AB}$ fails to satisfy the PPT criterion, and as such it is only sensitive to NPT entanglement. This implies that, for arbitrary-dimensional bipartite systems in pure states, or systems of $2\times 2$ or $2\times 3$ levels in general states, $Neg$ is able to detect all the bipartite entanglement content; but for mixed states of dimension $d_\sys>6$ it fails to detect all PPT entangled states. 

As already mentioned, the list of entanglement measures proposed  in the literature is simply huge. In addition, the interplay between different members of this zoo features very curious peculiarities. For example, two different states can be assigned two different orderings depending on which entanglement measure orders them \cite{Eisert99,Virmani00,Verstraete01,Zykowski02,Miranowicz04}. This can happen even for the simple case of 2-qubit mixed states \cite{Eisert99,Miranowicz04}, or for pure states of larger dimensionality \cite{Zykowski02}. Negativity, for instance, coincides exactly \cite{vidal99} with concurrence, studied in Sec. \ref{2.3.2}, for two-qubit pure states, but gives some pairs of 2-qubit mixed states opposite orderings to the latter \cite{Miranowicz04}.  General conditions for  equality between negativity and concurrence have been established in Refs.~\cite{audenaert,Verstraete01}.  For detailed treatments of the connections among different measures we refer the interested reader once again to Refs. \cite{horodecki09,plenio07}.
\subsubsection{Multipartite negativity}
\label{MultNeg}
A generalization of the negativity to the genuinely multipartite case was proposed by  Jungnitsch, Moroder, and G\"uhne in Ref. \cite{guehne2011}. For any multipartite state $\varrho$, it can be defined as 
\begin{equation}
\label{MultiNegativity} 
Neg_{\text{Multi}}(\varrho)\doteq\min\{-\text{Tr}[W_{\text{min}}\varrho],0\},
\end{equation}
where $\text{Tr}[W_{\text{min}}\varrho]$ is given by \eqref{SDP}. Despite not possessing a closed analytical expression,  $\text{Tr}[W_{\text{min}}\varrho]$ can be found rather efficiently through semidefinite programming, as described in Sec. \ref{2.2.2}. $Neg_{\text{Multi}}(\varrho)$ quantifies how much $\varrho$ fails to satisfy Criterion \ref{criterion:PPTmixtures}, necessary for membership of the PPT mixtures, and therefore also for biseparability, so it is analogous to the negativity. As a matter of fact, it reduces (up to normalisation) to negativity \eqref{Negativity} for the bipartite case. Additionally, it is shown to be non-increasing under LOCCs, i.e. an entanglement monotone \cite{guehne2011}. For these reasons, it can be taken as a valid measure of genuinely multipartite entanglement. For instance, GHZ states have maximal multipartite negativity (equal to $1/2$). Other states with $Neg_{\text{Multi}}=1/2$ are mentioned in Sec. \ref{Ali}.

\subsection{Experimental detection of entanglement}
\label{ExpDetec}

\par In this subsection we mention the main techniques for the experimental verification of entanglement. Again, since this is not the central topic of this review, we treat it very briefly. For an excellent review on the subject, we refer the interested reader to Ref.  \cite{guehne09}.
\subsubsection{Bell inequalities}
\label{nonlocality}
\par  As  mentioned in the introduction, in 1964 Bell  was able to formalize the EPR argument \cite{epr35}, and to disprove it. He derived  simple inequalities satisfied by any local hidden variable (LHV) model that is to reproduce the perfect correlations of the singlet; and showed that quantum mechanics violates these inequalities \cite{bellepr64}. This is nowadays known  as {\it Bell's Theorem} \cite{bell04}. Even though it was enough to formally rule out any attempt of completion of quantum theory with subjacent LHVs, the  experimental violation of these inequalities would involve the observation of perfect correlations.  It was thus five years later when -- inspired by Bell --  Clauser, Horne, Shimony and Holt (CHSH) came up  \cite{CHSH69,CHSH70} with a remarkable family of new inequalities satisfied by {\it any} LHV model, without any assumption of consistency with quantum correlations. Again, the new inequalities were violated by quantum mechanics, but in addition their violation did not require perfect correlations, providing thus the first experimentally-checkable criteria to conclusively rule out interpretations based on any sort of LHV models. These inequalities belong to what is today known as {\it Bell inequalities}, or {\it non-locality tests}. As we see below, entanglement is a necessary condition for quantum correlations to violate any of them. Therefore, non-locality tests constitute in fact the oldest criteria for entanglement detection. 

\par Let us briefly formalize these notions. Suppose that Alice and Bob, in space-like separated laboratories, perform local measurements on their systems.  Alice measures one of two arbitrary dichotomic observables $A_0$ and $A_1$, each one with outcomes $a=1$ or $a=-1$, and  Bob  $ B_0$ or $ B_1$, with outcomes $b=1$ or $b=-1$. The  correlations between both systems are encapsulated in the joint probability $P(a,b|x,y)$ of Alice obtaining $a$ and Bob  $b$, given that she measured $ A_x$ and he  $B_y$, for $x$, $y=0$ or 1. The most general description of $P(a,b|x,y)$ by {\it any} LHV model --from now on referred to simply as {\it local model}--  is given by
\begin{equation}
\label{LHV}
P(a,b|x,y)\equiv\sum_{\lambda} P(\mathcal{E}) P(a|x,\lambda) P(b|y,\lambda),
\end{equation}
where $\lambda$ is a short-hand notation for all LHVs that may characterize the composite state. Expression \eqref{LHV} manifests the constraint that the measurement outcomes should at most be classically correlated through $\mathcal{E}$, whose values might have resulted from some local interaction in the past (in some common region in the past of both measurement's light cones). In turn, for any local correlations of the form \eqref{LHV}, it is immediate to show that the  following statistical inequality must hold:
\begin{equation}
\label{CHSH}
\langle A_0  B_0\rangle +\langle A_1  B_0\rangle +\langle A_0  B_1\rangle - \langle A_1  B_1\rangle \leq 2,
\end{equation}
where $\langle A_x  B_y\rangle\equiv\sum_{a,b=-1,1}a\times b\times P(a,b|x,y)$. This  is the CHSH inequality mentioned above \cite{CHSH69,CHSH70}, the best-known and simplest non-locality test. 

\par  Let us next bring quantum physics back into scene. In the quantum formalism the left hand side of  \eqref{CHSH} is expressed as the expectation value $\langle\beta_{\text{CHSH}}\rangle\doteq\bra{\Psi}\beta_{\text{CHSH}}\ket{\Psi}$, with respect to some quantum state $\ket{\Psi}$,  of the ``Bell operator" $\beta_{\text{CHSH}}\doteq A_0\otimes B_0+A_1\otimes B_0+A_0\otimes B_1-A_1\otimes B_1$,
where $A_x$ and $B_y$ are now  dichotomic Hermitian quantum observables. It is immediate to check that if for instance $A_0=Z$,  $A_1=X$,  $B_0=\frac{X+Z}{\sqrt{2}}$,  $B_0=\frac{X-Z}{\sqrt{2}}$, and $\ket{\Psi}=\ket{\Psi^-}$, then $\langle\beta_{\text{CHSH}}\rangle=2\sqrt{2}$.  This value is in fact the maximum attainable  by any quantum state and is known as the Tsirelson bound \cite{Cirelson80}. Thus, quantum mechanics violates the CHSH inequality by a factor of $\sqrt{2}$. 

\par As  mentioned, Bell inequalities are based exclusively on the assumption of locality. Therefore, their violation only tells us  that the observed correlations are non-local, {\it i. e.} cannot be written as \eqref{LHV}. However, if these correlations come from a quantum state, then in addition the state must  be entangled. To see this consider again the definition of a separable state,  $\varrho_{AB}=\sum_{\mu} p_{\mu} \varrho_{A_{\mu}}\otimes\varrho_{B_{\mu}}$. For it, one has $P(a,b|x,y)\doteq\langle a_x|\otimes\langle b_y|\varrho_{AB}|a_x\rangle\otimes|b_y\rangle=\sum_{\mu} p_{\mu} \langle a_x|\varrho_{A_{\mu}} |a_x\rangle\langle b_y|\varrho_{B_{\mu}}|b_y\rangle$, with $|a_x\rangle$ and $|b_y\rangle$  respectively the eigenstates of  $A_x$ and $B_y$ of eigenvalues $a$ and $b$. Next, identify $\mu$ with $\lambda$, $p_{\mu}$ with $P(\lambda)$, $\langle a_x|\varrho_{A_{\mu}} |a_x\rangle$ with $P(a|A_x,\lambda)$, and $\langle b_y|\varrho_{B_{\mu}} |b_y\rangle$  with $ P(b|B_y,\lambda)$. With this, one readily  recognizes $P(a,b|x,y)$  given explicitly in the form \eqref{LHV}. This simple consideration shows that all separable states exhibit only local correlations, which in turn implies that non-local quantum states cannot be separable. The same argument extends of course to greater numbers of measurement-settings, outcomes, users, and to general non-orthogonal measurements. 

\par For the pure-state case the converse is also true: every pure entangled state violates a Bell inequality \cite{Gisin91,Gisin92,Popescu92}. This is today known as {\it Gisin's Theorem}, after Gisin formalized the result in Refs. \cite{Gisin91,Gisin92}, but the fact is known for the bipartite case since long before (seee for example \cite{Carpasso73}). The extension to the multipartite case is due to Popescu and Rohrlich \cite{Popescu92}. In the general mixed-state case,  in contrast, there are mixed entangled states whose correlations are local \cite{werner89,barret02,acin06,Cavalcanti11}. Consider, for instance, the two-qubit Werner state  \cite{werner89}
\begin{equation}
\label{Wernerstate}
\varrho_{\text{Werner}}(p)\equiv p\ket{\Psi^-}\bra{\Psi^-}+(1-p)\frac{\openone}{4},
\end{equation}
where $p$ is some probability. This state is NPT (and therefore entangled)  for any $p$ greater than $\frac{1}{3}$, but it is known to violate the CHSH inequality\footnote{The optimal measurement settings of the CHSH inequality for any two-qubit state were characterized  in Ref. \cite{Horodecki95}.} only for $p>\frac{1}{\sqrt{2}}\approx0.71$ and another two-outcome Bell inequality involving more settings only for $p>0.7056$ \cite{Vertesi08}. Furthermore, all its correlations revealed by two-outcome measurements can be explicitly accounted for by Werner's original local model \cite{werner89} for $p\leq \frac{1}{2}$, for von Neumann projective measurements, and by more recent local models for $p\leq0.6595$, for both von Neumann  \cite{acin06} and general measurements \cite{Cavalcanti11}\footnote{For the fully general case of multi-outcome non-projective measurements a local model is known that simulates the correlations in $\varrho_{\text{Werner}}(p)$ for $p\leq \frac{5}{12}$ \cite{barret02}.}. That is, for  $\frac{1}{3}< p\leq0.6595$ $\varrho_{\text{Werner}}(p)$ is entangled but local under dichotomic measurements. In the gap $0.6595<p<0.7056$ in turn, nothing of its non-local nature is known. It is both disappointing and at the same time exciting that such fundamental problems still remain open even for the simplest composite system of all, two qubits. 

\par The  CHSH  inequality was successfully violated in a first experiment \cite{Freedman72} by Freedman and Clauser and, later on, in the  conclusive works by Aspect and collaborators \cite{aspect81,aspect82}. Since then, several remarkable experiments around the world  have repeatedly confirmed the violation of  \eqref{CHSH} \cite{Tapster94,Weihs98, tittel99,Rowe01,Baas08,Matsukevich08} and of other important bipartite Bell tests with different numbers of settings or outcomes \cite{Cinelli05,Yang05,Pomarico11,Aolita11}. On the multipartite side in turn several violations of locality have been reported \cite{raschenbeutel00,Bowmeester99, pan00,zhao04,Chen06b,Lavoie09}, including the violation of Mermin's inequality \cite{Mermin90} by 39 standard deviations \cite{Chen06b}, and of Svetlichny's inequality  \cite{Svetlichny87}, which accounts for genuine three-partite non-locality, by 3.6 standard deviations \cite{Lavoie09}.
All these experiments have shown not only that nature does not admit local descriptions but also that  it turns out instead to follow  quantum mechanics. It seems therefore that, on the grounds of physical fact, one is forced to abandon the ``comfort of LHVs" and accept the counter-intuitiveness of quantum non-locality. Nevertheless, it is important to mention that open loopholes exist, which in principle allow nature  to ``confabulate" against us, in such way  that all reported experiments are still describable by LHV models. The most important two are the  locality loophole and the detection, or fair-sampling, loophole. The detection loophole has been closed in experiments with matter qubits: with ions \cite{Rowe01}, superconducting circuits \cite{Ansmann09} and atoms \cite{Hofmann12}, where highly efficient detection is possible. Separately, the locality loophole has been closed in photonic experiments \cite{Weihs98,tittel99,Baas08}, which naturally allow for greater distances between the qubit. Up to date a fully loophole-free experimental Bell violation is still an open challenge. However, more recently, a particular type of Bell inequality, due to Eberhard \cite{eberhard93}, that is more resistant to lower detection efficiencies allowed for experimental violations with entangled photons that closed the detection loophole \cite{zeilinger2013,kwiat2013}. These may be considered an important step towards eventually closing, in a same experiment, both main loopholes together. See Ref. \cite{Brunner13} for an excellent recent review on non-locality.
\subsubsection{Entanglement witnesses and other criteria}
\label{ExpEntWit}
As first pointed out by Terhal \cite{terhal00}, there is an intimate connection between Bell inequalities and entanglement witnesses. As a matter of fact, Bell inequalities are, for a given choice of measurement settings, non-optimal entanglement witnesses. For example, the two-qubit observable
\begin{equation}
\label{WCHSH}
W_{\text{CHSH}}=2\openone-\beta_{\text{CHSH}}
,
\end{equation}
where $\beta_{\text{CHSH}}=Z\otimes \frac{X+Z}{\sqrt{2}}-X\otimes \frac{X+Z}{\sqrt{2}}-Z\otimes \frac{X-Z}{\sqrt{2}}+X\otimes \frac{X-Z}{\sqrt{2}}$ is the CHSH Bell operator defined in the previous subsection, constitutes an entanglement witness. This witness detects all entangled states whose correlations upon measurements along $A_0=Z$ or  $A_1=X$ for Alice and   $B_0=\frac{X+Z}{\sqrt{2}}$ or  $B_1=\frac{X-Z}{\sqrt{2}}$ for Bob violate inequality \eqref{CHSH}, {\it i. e.} those sufficiently close to the singlet $\ket{\Psi^-}$. It is of course a non-optimal witness though, as is clear from the discussion about entangled local states \eqref{Wernerstate}. 

\par Something readily manifest  in expression \eqref{WCHSH} is that it explicitly provides the local bases in which each user must measure so as to reconstruct the expectation value of the witness. Indeed, in experimental scenarios, entanglement witnesses are most-usefully taken advantage of when decomposed into local, fully factorable observables (otherwise one runs into the impractical situation of having to measure in entangled bases, in order to detect entanglement). In general, any complete basis of  ${\mathcal D}({\mathcal H}_{\sys})$ composed exclusively of products of single-particle observables yields one such decomposition. For example, in the case of $N$-qubit systems, this could be the set of all products of Pauli or identity operators of each particle. However, for large $N$ this is also impractical, because the number of local measurement settings required grows exponentially with $N$. The general problem of finding the 
optimal decomposition of witnesses, with the minimal amount of local measurement settings, has been studied in Refs. 
\cite{Guehne02,Guehne03}. There, the authors obtained analytical solutions for some small-sized  
bipartite \cite{Guehne02,Guehne03} and multipartite \cite{Guehne03} systems, including some examples capable of witnessing bound entangled states.

\par The first experimental demonstration of an entanglement witness was reported in Ref. \cite{Barbieri03} for two-qubit photonic systems, where the entanglement of states close to the Werner state was characterized with only three local measuring settings.  
Three and four-photon genuine multipartite entanglement was detected and classified in  Ref. \cite{Bourennane04}, again using witnesses decomposed into few local measurements. Up-to-six-qubit photonic Dicke \cite{kiesel07,Prevedel09,Wieczorek09}, GHZ \cite{lu07,Radmark09} and graph \cite{lu07} states  have also been detected with the help of witnesses. Using non-linear entanglement witnesses, nonlinear properties  as the Renyi entropy were extracted from  two-photon entangled states in Ref. \cite{bovino05}. In atomic systems,  three- \cite{roos04a}, four- \cite{sackett00} and six-ion \cite{leibfried05} GHZ states, up to eight-ion W-states \cite{haeffner05}, and fourteen-ion \cite{monz2011} GHZ states have been detected with the help of entanglement witnesses, as discussed in Sec. \ref{65}. 
Graph and Dicke states are genuine multiqubit-entangled states. They, as well as the W and GHZ states of more than three qubits, are discussed in detail Sec. \ref{prelimi}.

\par 
Finally, apart from entanglement witnesses, other entanglement criteria have been very helpful in the experimental verification of entanglement, with different criteria often used in a same experiment. For instance, GHZ entanglement in the experimental fourteen-ion states of \cite{monz2011} was corroborated with three different methods: using the GHZ fidelity-based witnesses  \eqref{recipeopt}; with biseparability Criterion \ref{criterion:BisepGHZN} of Sec. \ref{GHZprelim}, which is the $N$-qubit generalisation of Criterion \ref{criterion:BisepGHZ} of Sec. \ref{BisepCrit}; and with $N$-distillability Criterion \ref{criterion:DistMDist}, also in Sec. \ref{GHZprelim}. The common advantage of the three criteria is that only partial information about the experimental state is necessary for their evaluation, which is in striking contrast with the entanglement criteria whose evaluation require the full reconstruction of the state's density matrix, as discussed in the next subsection. However, it is important to keep in mind that, since every entanglement witness or criteria is sensitive only to a restricted subset of entangled states, in practice some knowledge of the state to measure is always required in advance. This is the main drawback of these technique. Still, when there is indeed some prior knowledge available, witnesses and these other efficiently evaluable criteria constitute an extremely useful tool for state characterization with economic detection resources, as proven by the experiments mentioned above.

\subsubsection{State tomography}

The full experimental reconstruction of any density matrix can always be done via quantum state tomography \cite{Leonhardt95, white99,roos04b}. The way this is typically accomplished is by measuring a complete set of local orthonormal observables, from which all the inputs of the density matrix can be derived. For example, any density matrix describing a  two-qubit system can be decomposed as:
\beq
\varrho = \frac{1}{4} \sum_{i,j=0}^3 \varrho_{i,j} \, \sigma^i\otimes\sigma^j;
\eeq
where $\sigma_0 = \openone$, and $\sigma^i$ with $i=1,2,3$ are respectively the Pauli matrices $X$, $Y$, and $Z$. These matrices satisfy the orthogonality condition $\tr [\sigma^i.\sigma^j] = 2 \delta_{i,j}$, where $\delta_{i,j}$ is the Kronecker delta. The task is to determine the real coefficients $\varrho_{i,j}$ that define $\varrho$, subject to $\varrho_{0,0}=1$ for normalization and $\sum_{i,j=0}^3 \varrho_{i,j}^2\leq4$ for positiveness.  This can be done by measuring all the correlation functions $\langle\sigma^i\otimes\sigma^j\rangle$ on the state, i.e. the expectation value of every observable $\sigma^i\otimes\sigma^j$: $\tr[\varrho \sigma_k\otimes\sigma_l]=\frac{1}{4} \sum_{i,j=0}^3\varrho_{i,j}\tr[\sigma_i\sigma_k\otimes\sigma_j\sigma_l]=\sum_{i,j=0}^3\varrho_{i,j}\delta_{i,k} \delta_{j,l}= \varrho_{k,l}$.

\begin{figure}
\begin{center}
\includegraphics[width=.45\linewidth]{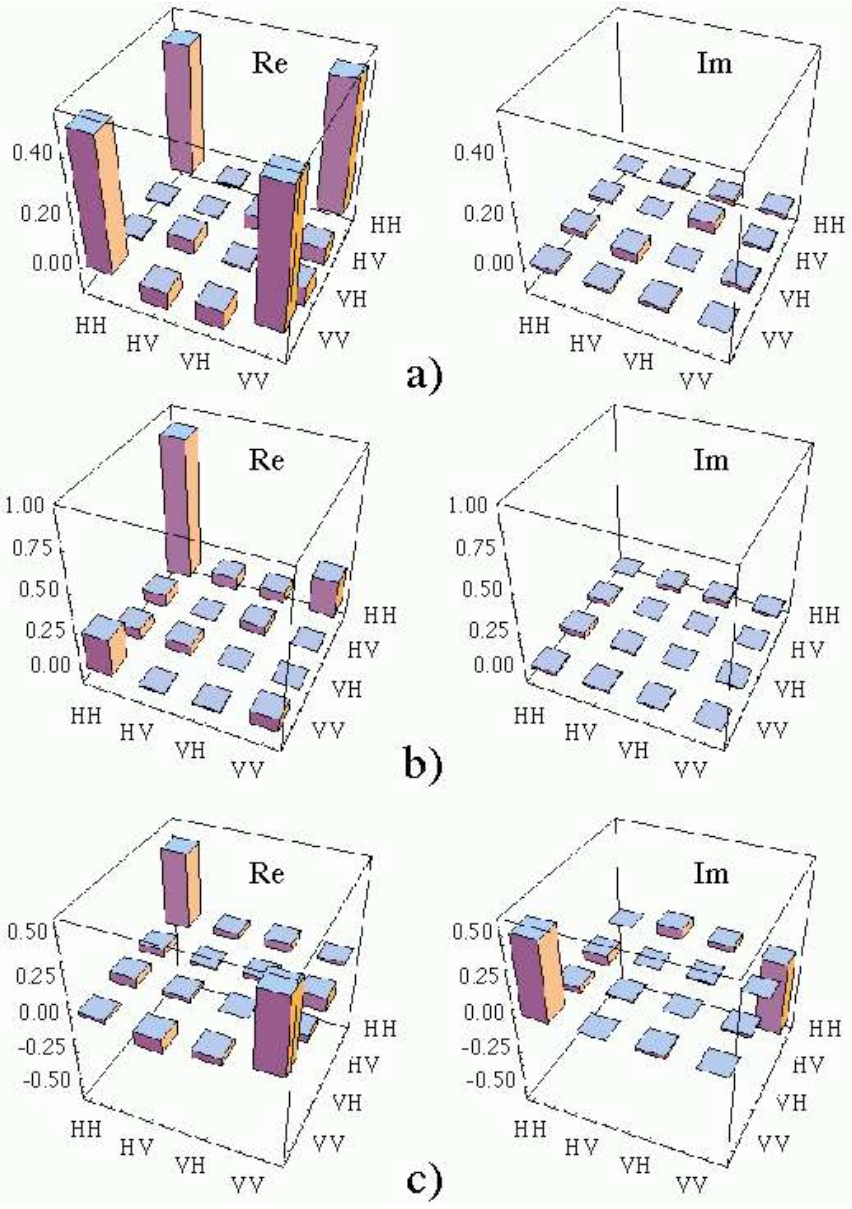}
\caption{
\label{tomo}
Real and imaginary parts of the elements of  topographically reconstructed density matrices corresponding to the experimentally created entangled states a) $\ket{H}\ket{H}+\ket{V}\ket{V}$, b) $\ket{H}\ket{H}+0.3\ket{V}\ket{V}$ and c) $\ket{H}\ket{H}-i\ket{V}\ket{V}$ (normalization omitted). Reprinted figure with permission from A.~White \emph{et al.}, \href{http://link.aps.org/doi/10.1103/PhysRevLett.83.3103}{Phys.~Rev.~Lett {\bf 83}, 3103 (1990)}. Copyright (1990) by the American Physical Society.} 
\end{center}
\end{figure}

\par In Fig. \ref{tomo} we can see the three measured density matrices of photonic-polarization two-qubit systems done in the experiment \cite{white99}. The density matrices are expressed in the local basis $\{\ket{H},\ket{V}\}$, with $H$ and $V$ corresponding respectively to horizontal and vertical polarizations of a photon. Once the complete density matrix has been reconstructed one can apply any valid entanglement criterion, or calculate the value of any valid entanglement quantifier, to see for its entanglement. In fact, many experiments where the presence of entanglement is verified using witnesses or other criteria, as for example some of the ones mentioned above  \cite{roos04a,bovino05,haeffner05,kiesel07,Lavoie09}, do not directly measure the witnesses or quantities involved in the criteria but rather perform state tomography and apply the witnesses or criteria to the tomographically-reconstructed states.

\par Notice that all  measurements involved are local and that the method does not require any prior  knowledge at all of the state in question.  In addition,  quantum state tomography extends of course to higher dimensions, including continuous variables \cite{Smithey93,Leibfried96,Lutterbach97,deMelo06}, and to multipartite systems. However, since the number of measurement settings grows exponentially with the number of system components, the technique has  disadvantageous scaling properties. This was clearly evidenced in the eight-ion experiment of Ref. \cite{haeffner05}. There, ten hours of data aquisition -- implementing 
measurements in $3^8=6561$  detection bases, each one involving a different
laser-pulse configuration --, followed by computationally expensive data processing, 
were necessary to reconstruct the experimentally prepared eight-ion 
state.  This bare approach is therefore not scalable in practice to more than a few particles. Recently, tomographic methods exploiting $t$-designs \cite{Bendersky08,Schmiegelow10} and compressed sensing \cite{Gross10} have been investigated. The experimental or computational resources required by these methods still scale (at least) exponentially with the system-size, but the methods are significantly more efficient than conventional tomography.

\subsubsection{Direct detection using copies of the state}
\label{copias}
The last approach we briefly describe is that in which the entanglement of an unknown state is directly assessed  through projective measurements  when two copies of the state are simultaneously at hand. The basic idea behind this technique comes from the fact that any polynomial function of the elements of a density matrix can be directly accessed via projective measurements  on as many copies of the state as the degree of the polynomial \cite{brun04}.   Among all  entanglement quantifiers, concurrence presented in Sec. \ref{2.3} plays a unique roll in this context, since for pure states its square is given by a quadratic function of the density matrix inputs, as is clear from Eqs.  \eqref{concupure} and \eqref{ConcAndre}. This implies that the squared concurrence of any pure state  can be directly determined via projective measurements on only two copies of the state.
\par
\begin{figure}
\begin{center}
\includegraphics[width=1\linewidth]{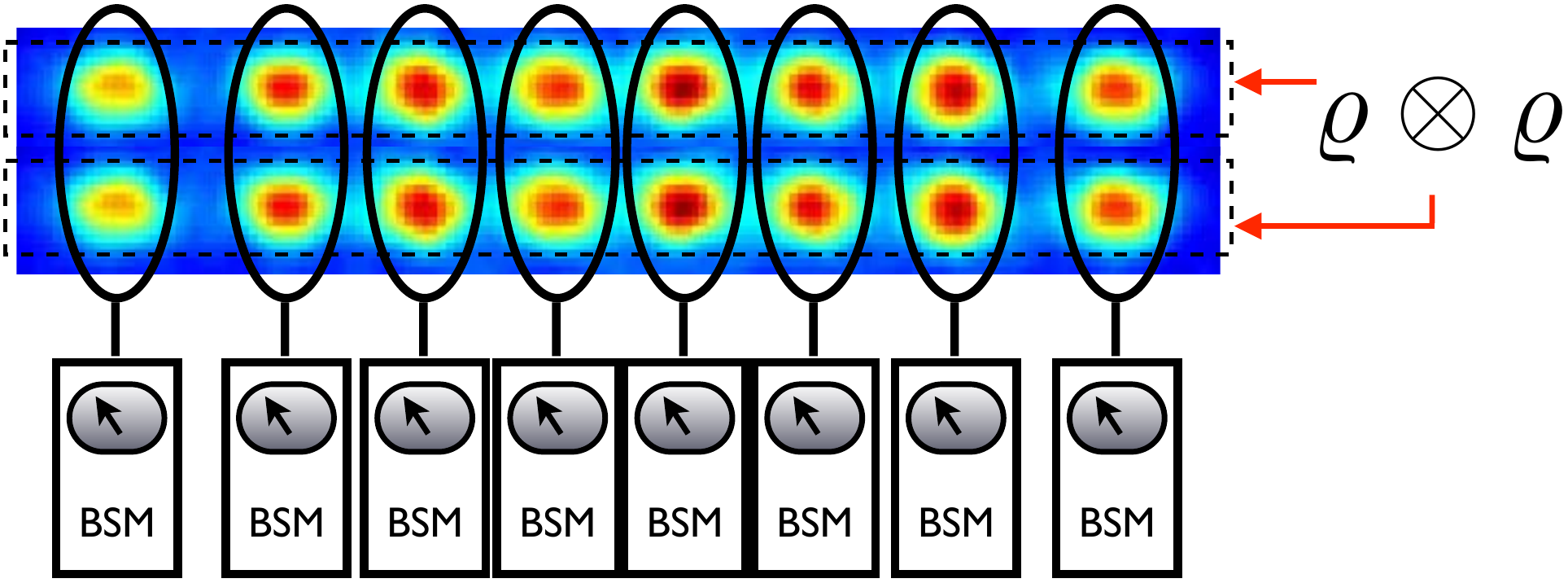}
\caption{
\label{8ion}
Direct experimental quantification of entanglement in an exemplary trapped-ion scenario. An eight-qubit pure state $\varrho$, together with its copy, is encoded  for example in strings of two-level ions, and is subject to Bell-state measurements (BSMs) on each pair. The concurrence of $\varrho$ depends exclusively on the joint probability $p^{8}_{+}$ of simultaneous appearance of 8 triplets. Reprinted figure with permission from L.~Aolita, F.~Mintert, and A.~Buchleitner, \href{http://link.aps.org/doi/10.1103/PhysRevA.78.022308}{Phys.~Rev.~A {\bf 78}, 022308 (2008)}. Copyright (2008) by the American Physical Society.} 
\end{center}
\end{figure}

  These projective measurements, as was shown by Mintert and collaborators  \cite{mintert05a,mintert05b}, turn out to be local measurements of the parity of each constituent part of the system together with its counterpart in the copy. Furthermore, Aolita and Mintert showed \cite{aolita06concu} that multipartite concurrence can  be directly quantified by the expectation value of {\it a single fully factorizable observable}. They showed that, for an arbitrary-dimensional $N$-partite  pure state $\varrho\equiv\ket{\Psi}\bra{\Psi}$, concurrence \eqref{ConcAndre} can be expressed as
\begin{equation}
\label{Product}
C_N(\varrho)=2\sqrt{\bra{\Psi}\otimes\bra{\Psi}\big(\openone-\bigotimes _{j=1}^N P^{j}_{+}\big)\ket{\Psi}\otimes\ket{\Psi}},
\end{equation}
where $P^{j}_{+}$, with $1\leq j\leq N$, is the projector onto the symmetric subspace ${\cal H}_{j}\odot{\cal H}_{j}$  -- corresponding to all states invariant under the exchange of both copies of $j$ -- of the Hilbert space ${\cal H}_{j}\otimes{\cal H}_{j}$ of two copies of the $j$-th subsystem, and $\openone$ is the identity operator in ${\cal H}\otimes{\cal H}$.Expression \eqref{Product} implies that a single local-measurement setting is required throughout the entire detection process. Indeed, pure-state concurrence depends only on a unique probability $p^{N}_{+}$, of finding each and all of the $N$ particles in a symmetric state with their respective  copies: $C_N(\ket{\Psi}\bra{\Psi})=2\sqrt{1-p^{N}_{+}}$. For example, for $N$-qubit systems each local symmetric subspace is spanned by the triplet states $\ket{\Psi^{+}}$, $\ket{\Phi^{+}}$ and $\ket{\Phi^{-}}$, defined in Eqs. \eqref{Bellstate}, and the detection protocol reduces thus to Bell-state measurements on each particle-copy subsystem, as is schematically sketched in Fig. \ref{8ion}. This is to be compared for example with the exponential number of settings required for state tomography. 

\par This technique bears of course the built-in drawback of requiring two simultaneous copies of $\varrho$, which in view of the impossibility of unknown-state perfect cloning \cite{wootters82,Dieks82} appears as a fundamental obstacle. Nevertheless, since several copies of the state must be created anyway, to build up the measurement statistics, one can keep a copy of the state until the next copy is available, and then collectively measure both copies, instead of  measuring individually in a sequential way. Another possibility is to create two copies of the state at the same time. The latter was demonstrated by Walborn {\it et al.} in Ref. \cite{walborn06},  constituting the first direct experimental quantification of entanglement. There, two copies of a two-qubit almost-pure state were simultaneously encoded respectively  in the polarization and spatial degrees of freedom of two photons, and Bell-state detection was carried out between these two degrees of freedom of the same photon.  

\par  Even though the detection with copies described above is not able to yield the exact value of concurrence when the state is mixed, the technique can be used to obtain considerably tight lower \cite{mintert07b,aolita07concu} and upper \cite{Sun07,Zhang08} bounds,  some of which have already been useful in experiments \cite{Sun07}. The general theory of measurements with two copies of a quantum state has been studied in Ref. \cite{Bendersky09}.

\par The generalization of this procedure to mixed states of two quits was presented in \cite{horodecki03}. There, collective measurements of up to eight copies of the state are necessary, in order to obtain the concurrence through direct measurements.

\par All in all, direct detection with copies of the state appears as a versatile method that complements the other  approaches previously mentioned, specially when scaling properties are a matter of concern.

\section{Open system dynamics}
\label{Open}

In this section we review conceptual and formal aspects of the evolution of quantum systems in contact with the environment. Our treatment is a non-exhaustive one. For a more complete and detailed description we refer the interested reader to Refs.~\cite{zurek:715,breuer,schlosshauer:1267,joos,gardiner} and references therein.
\subsection{Completely-positive maps as the most general physical evolution}
\subsubsection{Open-system dynamics and entanglement}
\label{Superop0}

We consider first a simple example that illustrates the relation between open-system dynamics and entanglement. Let a system $\mathcal{S}$, associated to Hilbert space $\mathcal{H}_{\sys}$, be initially in a coherent superposition of two orthonormal states, $\ket{\chi_1}$ and $\ket{\chi_2}$, and another  system $\res$, associated to Hilbert space $\mathcal{H}_{\res}$, be in some generic state denoted by $\ket{0}_{\res}$. The total state of the composite system is then given by the  product
\beq
\label{initial}
\ket{\Psi_{(0)}} = (\alpha\ket{\chi_1} + \beta\ket{\chi_2})\otimes\ket{0}_{\res},
\eeq
with $|\alpha|^2+|\beta|^2=1$. This implies that, initally, the two systems are uncorrelated. Now, suppose that $\mathcal{S}$ and $\res$ interact during a time $t$, undergoing a unitary evolution such that  
\begin{eqnarray}
\ket{\chi_1}\ket{0}_{\res}&\rightarrow& \ket{\chi_1}\ket{\phi_1}\,, \nonumber\\
\ket{\chi_2}\ket{0}_{\res}&\rightarrow& \ket{\chi_1}\ket{\phi_2}\,, 
\end{eqnarray}
where $\ket{\phi_i}$, with $i=1$ or $2$, represent two possible evolved states for $\ket{0}_{\res}$.   These equations are a special case of \eqref{vn0}: we assume here for simplicity that the states $\ket{\chi_1}$ and $\ket{\chi_2}$ of the system do not change. The coherent superposition of these two state does change however, as in \eqref{vn}:  
\beq
\ket{\Psi_{(t)}}= \alpha\ket{\chi_1}\ket{\phi_1}+\beta\ket{\chi_2}\ket{\phi_2}.
\label{statetil}
\eeq
 Any observable acting non-trivially only on $\mathcal{S}$ can be measured without resort to $\res$. Its expectation value depends only on the reduced state $\varrho_{(t)}$ of $\sys$, obtained by tracing  $\res$ out, and given by 
\beq
\varrho_{(t)}=\tr_{\res}\big[\ket{\Psi_{(t)}}\bra{\Psi_{(t)}}\big]=\left(\begin{array}{cc}
								|\alpha|^2&\alpha\beta^* \bra{\phi_2}\phi_1\rangle\\
								\alpha^*\beta \bra{\phi_1}\phi_2\rangle&|\beta|^2
								\end{array}\right),
\label{evolvedAD1q}
\eeq
where the matrix representation on the right-hand side is in the basis $\{\ket{\chi_1}, \ket{\chi_2}\}$. We observe that the coherences in the off-diagonal elements are now proportional to the scalar product of the state vectors of $\res$. In particular, if both evolved states for $\res$ coincide, $|\bra{\phi_1}\phi_2\rangle|=1$, we have a product state as in the initial situation. However, if both states are different, $|\bra{\phi_1}\phi_2\rangle| < 1$, $\sys$ and $\res$ have have become entangled. In this case the coherences have decreased and the system state is no longer pure. In the extreme case when $|\bra{\phi_1}\phi_2\rangle| = 0$ the coherences vanish and $\varrho_{(t)}$ becomes equivalent to a classical probability distribution. 

\par
This simple example conveys an important conceptual message that will appear repeatedly throughout this review: The generation of entanglement between two systems under a unitary evolution (so that the composite system is closed) implies that the evolution of either of them individually is not unitary, because the composite evolution does not preserve the purity of each subsystem. If one has access to both parties the composite evolution can be reversed by applying the inverse unitary transformation that maps state \eqref{statetil} onto \eqref{initial}, therefore disentangling $\sys$ and $\res$.  In contrast, when system $\res$ is actually an environment that surrounds $\sys$, typically with very many degrees of freedom and a complex internal dynamics, one does not have control of it. In this case only the subsystem under scrutiny is at one's disposal and its dynamics is irreversible, which characterizes an open system. From now on, subsystem $\res$ will denote the reservoir and, unless explicitly specified, the term system will be reserved for the subsystem of interest $\sys$. 

\par In the example above only the coherences of the system state are affected. The effects due to an arbitrary interaction with the environment may be more intricate though, as in the examples described in Sec. \ref{Noise_models}, but they are still often lumped under the term decoherence. In general, the  system is described by the reduced state
\beq
\label{generic}
\varrho_{(t)} = \tr_\res \left[\varrho_{\sys \res(t)}\right],
\eeq
where $\varrho_{\sys \res(t)}$ is a general composite state of $\sys$ and $\res$ at time $t$. However, since one has no access to the composite-system state, one cannot explicitly describe the system dynamics from this expression. Therefore, a formulation that accounts for the interaction with the environment but involves the states of the system alone is required. {\it Completely positive channels} provide such a formulation.

\subsubsection{Superoperators, complete positivity, the Choi-Jamio\l kowsky isomorphism, and the Kraus representation}
\label{Krausop}
\par
\begin{figure}[t!]
\begin{center}
\includegraphics[width=\linewidth]{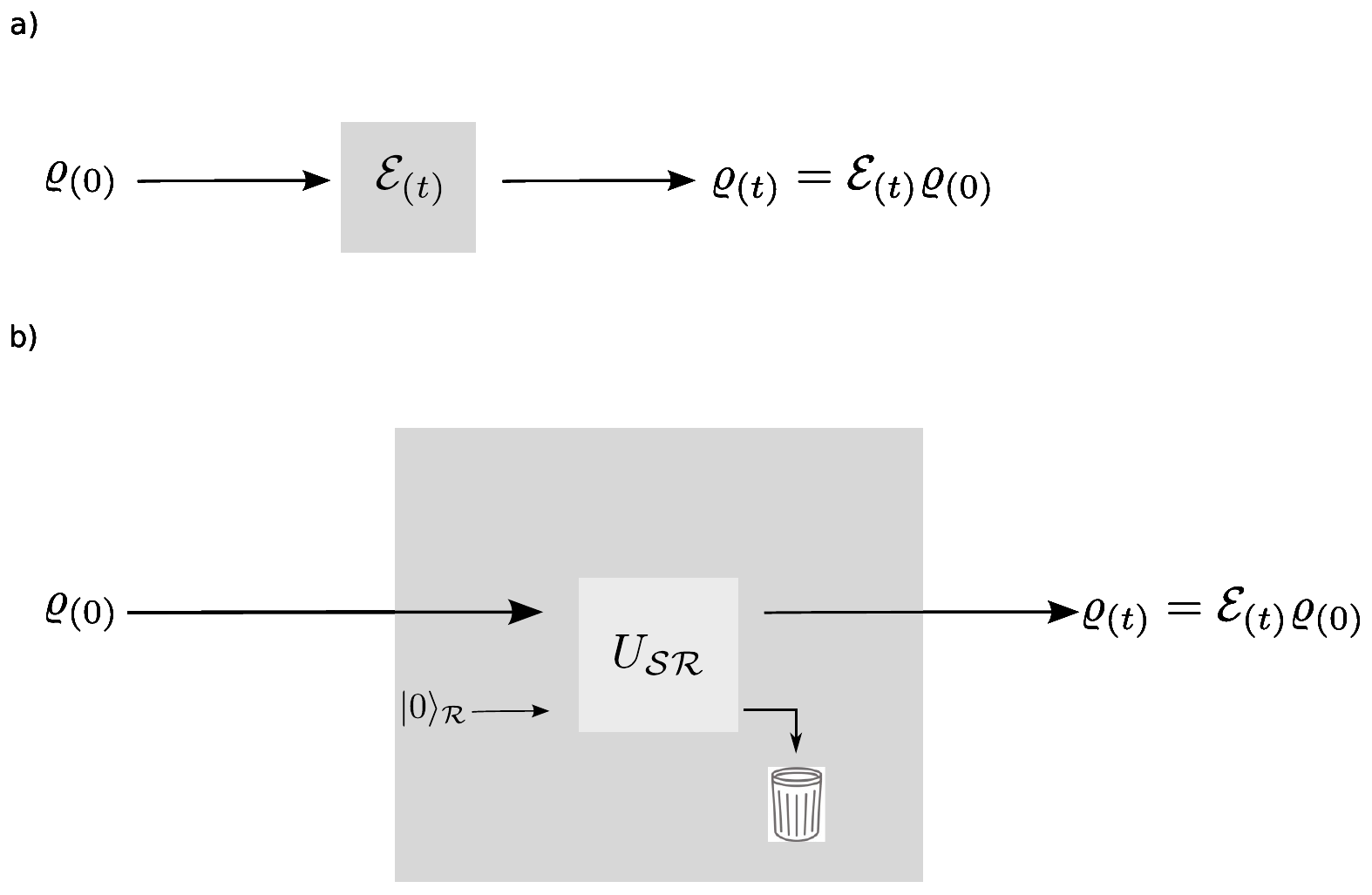}
\caption{Schematic description of superoperators. a) An input state $\varrho_{(0)}$ undergoes the action of a dynamical map $\mathcal{E}_{(t)}$, which  describes for instance a time evolution over some time period $t$. The output state is $\varrho_{(t)}\doteq\mathcal{E}_{(t)}\varrho_{(0)}$. b) The most general physical evolution for the open system $\sys$ can be thought of as an arbitrary unitary evolution in a larger Hilbert space that includes an auxiliary system $\res$ in some state $\ket{0}_\res$,  playing the role of an effective environment. The trash can represents the partial trace over $\res$.} 
\label{qChannel}
\end{center}
\end{figure}

 Quantum channels are transformations $\mathcal{E}_{(t)}$ acting on ${\cal D}({\cal H}_\sys)$ that satisfy three fundamental properties, described below: {\it convex linearity},   {\it trace preservation}, and {\it complete positivity}. They constitute a linear input-output theory for the transmission/evolution of quantum states under generic open-system conditions, and are often called {\it quantum operations}, {\it dynamical maps,} {\it quantum channels}, or simply {\it superoperators},  explicitly referring to the fact that they operate on density operators. The general picture is schematically shown in Fig.~\ref{qChannel}. As in the example discussed in the beginning of this section, superoperators give rise to non-unitary evolutions of the system, due to the generation of correlations between the system and its environment. A useful expression for superoperators can be obtained by assuming again that $\sys$ and $\res$ are initially uncorrelated, with $\sys$ in a generic initial state $\varrho_{(0)}$ and $\res$ in some generic pure state $\ket{0}_\res\bra{0}$. The composite evolution during a time period $t$ can be expressed in terms of a unitary operator $U_{\sys \res}$ as $\varrho_{\sys \res(t)}=U_{\sys \res}\left(\varrho_{(0)} \otimes \ket{0}_\res\bra{0}\right)U_{\sys \res}^{\dagger}$. From this and Eq. \eqref{generic},
one obtains an explicit characterization of the reduced dynamics for $\sys$ alone:
\begin{eqnarray}
\nonumber
\varrho_{(t)}&=& \tr_\res  \left [U_{\sys \res} \left( \varrho_{(0)} \otimes \ket{0}_\res\bra{0}\right) U_{\sys \res}^\dagger\right]\\
\nonumber
&\equiv&\sum_\mu \;\bra{\mu} U_{\sys \res} \ket{0}_\res \;\varrho_{(0)} \;\bra{0} U_{\sys \res}^\dagger\ket{\mu}_\res\\
&\doteq&\sum_\mu K_\mu \varrho_{(0)} K^\dagger_\mu\doteq \mathcal{E}_{(t)}\varrho_{(0)}\,,
\label{krausForm}
\end{eqnarray}
which defines the corresponding linear dynamical map $\mathcal{E}_{(t)}$. Here, the trace over $\res$ has been taken in an orthonormal basis $\{\ket{\mu}_\res\}$ of $\mathcal{H}_{\res}$ and the operators $K_\mu= \bra{\mu} U_{\sys \res} \ket{0}_\res$ have been introduced. These operators, which act solely on $\mathcal{H}_{\sys}$, are called the {\it Kraus operators}, while the {\it Kraus representation} \cite{kraus83} of $\mathcal{E}_{(t)}$, also referred to as  {\it operator-sum} representation~\cite{preskill98,nielsenchuang}, is defined by Eq. \eqref{krausForm}.  We previously mentioned it in Sec. \ref{2.2.2} in the context of LOCC operations -- see \eqref{LOCC}. Since this representation depends on the particular basis of $\mathcal{H}_{\res}$ chosen to take the trace, it is non-unique. 

As anticipated, every superoperator satisfies three fundamental properties. The Kraus form \eqref{krausForm} allows one to see it immediately: 
\begin{itemize}
\item{\bf Convex linearity}: This means that $\mathcal{E}_{(t)}[\lambda\varrho_1+(1-\lambda\varrho_2]=\lambda\mathcal{E}_{(t)}\varrho_1+(1-\lambda)\mathcal{E}_{(t)}\varrho_2$, for all $0\le\lambda\le1$.
\end{itemize}
\begin{itemize}
\item {\bf Trace preservation}:  $\mathcal{E}_{(t)}$ preserves the trace norm of all states, i.e. $\tr [\mathcal{E}_{(t)}\varrho]=\tr [\varrho],\ \forall\ \varrho\in\mathcal{D}(\mathcal{H}_{\sys})$. 
\end{itemize}
To see this, notice first that
\begin{eqnarray}\label{completenessKraus}
\sum_\mu K^\dagger_\mu K_\mu &=& \sum_\mu \bra{0} U_{\sys \res}\ket{\mu}_\res \bra{\mu} U^\dagger_{\sys \res}\ket{0}_\res\nonumber\\
&=&\bra{0} U_{\sys \res} U^\dagger_{\sys \res}\ket{0}_\res=\openone_\sys.
\end{eqnarray}
Here we have used that  $\{\ket{\mu}_\res\}$ is a complete basis and that $U_{\sys \res}$ is unitary. Then 
\bea{rcl}
\tr[\mathcal{E}_{(t)}\varrho]&\equiv&\tr \left[ \sum_\mu K_\mu \varrho K^\dagger_\mu\right]= \tr \left[ \left(\sum_\mu  K^\dagger_\mu K_\mu\right) \varrho \right]\nonumber\\
&=& \tr \left[\openone_\sys\varrho \right]= \tr \left[\varrho \right],
\eea
where we have used the linearity and invariance under cyclic permutations  of the trace. $\square$
\begin{itemize}
\item {\bf Complete-positivity}:  The trivial extension of $\mathcal{E}_{(t)}$ to any auxiliary system $\mathcal{A}$ of Hilbert space $\hil_\mathcal{A}$ preserves the positivity of all states, i.e. $\mathcal{E}_{(t)}\otimes\openone_{\mathcal{A}}\rho \ge 0,\ \forall\ \varrho\in\mathcal{D}(\mathcal{H}_{\sys}\otimes\hil_\mathcal{A})$. 
\end{itemize}
Complete positivity was previously mentioned in Sec. \ref{2.2.1} in the context of partial transposition. It expresses the fundamental requirement that if a map represents a physically valid transformation then its trivial extension to any auxiliary system, where the main system undergoes the transformation and the auxiliary system is not affected, should also render a physically valid transformation. By ``physically valid" we refer to transformations that map positive-semidefinite operators into positive-semidefinite operators. 

A powerful tool to, among other things, check if a given map is completely positive is the {\it Choi-Jamio\l kowsky isomorphism}~\cite{jamiol}. It states that a channel $\mathcal{E}_{(t)}$ acting on $\mathcal{D}(\mathcal{H}_{\sys})$ is completely-positive iff its trivial extension on $\mathcal{D}(\mathcal{H}_{\sys}\otimes\hil_\mathcal{A})$, with $\hil_\mathcal{A}$ of the same dimension $d_\sys$ as $\mathcal{H}_{\sys}$, applied to the maximally entangled state $\ket{\Phi^+_{d_\sys}}
\in\mathcal{H}_{\sys}\otimes\hil_\mathcal{A}$, renders a positive-semidefinite operator $\varrho_{\mathcal{E}_{(t)}}\in\mathcal{D}(\mathcal{H}_{\sys}\otimes\hil_\mathcal{A})$, \emph{i.e.} iff 
\beq
\label{CJDualism}
\varrho_{\mathcal{E}_{(t)}} \doteq \mathcal{E}_{(t)}\otimes\openone \proj{\Phi^+_{d_\sys}}\geq0. 
\eeq
This state-channel dualism holds even if $\mathcal{E}_{(t)}$ is not trace-preserving (in which case $\varrho_{\mathcal{E}_{(t)}}$ is not normalised). When $\mathcal{E}_{(t)}$ is trace-preserving then $\varrho_{\mathcal{E}_{(t)}}$ is necessarily restricted to fulfil $\text{Tr}_\sys[\varrho_{\mathcal{E}_{(t)}}]=\openone/d_\sys$. Conversely, the dualism also guarantees that for every positive-semidefinite operator $\varrho\in\mathcal{D}(\mathcal{H}_{\sys}\otimes\hil_\mathcal{A})$ there exists a unique completely-positive channel $\mathcal{E_\varrho}$, acting on $\mathcal{D}(\mathcal{H}_{\sys})$, such that $\varrho = \mathcal{E_\varrho}\otimes \openone\proj{\Phi^+_{d_\sys}}$. Thus, a bipartite mixed state and the corresponding single-partite channel contain equivalent information. One says that $\varrho$ is the dual state of channel $\mathcal{E_\varrho}$, and vice versa.

We show next that every superoperator in the Kraus form is completely positive. For all $\ket{\psi} \in \hil_\sys\otimes\hil_\mathcal{A}$, one has
\bea{rcl}
\bra{\psi}\mathcal{E}_{(t)}\otimes\openone_{\mathcal{A}}\rho\ket{\psi} &\equiv& \sum_\mu (\bra{\psi} K_\mu\otimes\openone_{\mathcal{A}}) \varrho (K^\dagger_\mu\otimes\openone_{\mathcal{A}} \ket{\psi}) \nonumber\\
&\doteq&  \sum_\mu \bra{\psi^\prime_\mu} \varrho  \ket{\psi^\prime_\mu} \ge 0,
\eea
where  we have used that every $\ket{\psi^\prime_\mu}\doteq K^\dagger_\mu\otimes\openone_{\mathcal{A}} \ket{\psi}$ is a (non-normalized) pure state in $\hil_\sys\otimes\hil_\mathcal{A}$. 
Conversely, in turn, every completely positive map has a Kraus representation. The implication in both directions is known as the {\it Kraus representation theorem} \cite{kraus83}: A convex linear map $\mathcal{E}_{(t)}$ is completely-positive if, and only if, it can be expressed in the form \eqref{krausForm}.

With the Kraus representation one can for instance find purifications of $\mathcal{E}_{(t)}$ in the sense of Fig. \ref{qChannel} b), i.e.  unitary operators $U^\prime_{\sys\res^\prime}$ and auxiliary environmental systems $\res^\prime$ such that the reduced system dynamics is given by $\mathcal{E}_{(t)}$. To this end, one first chooses an orthonormal basis $\{\ket{\phi_i}_\sys\}$ of $\mathcal{H}_{\sys}$ and an orthonormal basis $\{\ket{\mu}_{\res^\prime}$, with $0\leq \mu\leq d_{\res^\prime}\}$ of an arbitrary Hilbert space $\mathcal{H}_{\res^\prime}$. The dimension $d_{\res'}$ of $\mathcal{H}_{\res^\prime}$, associated to $\res^\prime$, is given by the number of Kraus operators $K_{\mu}$ in the particular Kraus decomposition. One then defines an operator $U^\prime_{\sys\res^\prime}$ such that 
\beq
U^\prime_{\sys\res^\prime} \ket{\phi_i}_\sys \ket{0}_{\res^\prime} = \sum_{\mu} K_{\mu} \ket{\phi_i}_\sys \ket{\mu}_{\res^\prime}.
\label{unitaryKraus}
\eeq
It is easy to check, by using \eqref{completenessKraus}, that $U^\prime_{\sys\res^\prime}$ is inner-product preserving in $\sys$. This implies that it can be extended to a unitary operator in $\mathcal{H}_{\sys}\otimes \mathcal{H}_\res$ \cite{preskill98,nielsenchuang}. Furthermore, one can trivially verify that, upon taking the partial trace over $\res^\prime$ in the basis $\{\ket{\mu}_{\res^\prime}\}$, this evolution yields exactly the Kraus form defined by the Kraus operators $K_{\mu}$. In contrast, if the partial trace is taken in another basis, a different Kraus decomposition to the starting one is obtained. However, since the trace operation is independent of the particular basis, the new Kraus form still corresponds to the same process $\mathcal{E}_{(t)}$. It is important to emphasize that the effective reservoir dimension $d_{\res'}$ in a given purification of $\mathcal{E}_{(t)}$ may in general be smaller than the (possibly infinite) dimension of the real reservoir $\res$ originally giving rise to $\mathcal{E}_{(t)}$ in the derivation above. Indeed, the maximum number of Kraus operators required to represent any superoperator $\mathcal{E}$ acting on a system $\sys$ of dimension $d_\sys$ is equal to $d^2_\sys$. This comes from the facts that the number of linearly independent operators in $\mathcal{D}(\mathcal{H}_{\sys})$  (its dimension) is $d^2_\sys$ and that $\mathcal{E}_{(t)}$ is a linear map \cite{preskill98,nielsenchuang}.

 Superoperators satisfy a final crucial property:
\begin{itemize}
\item {\bf Semigroup property}:  The set of  completely positive trace-preserving maps  on $\mathcal{D}(\mathcal{H}_{\sys})$ forms a semigroup with respect to the map composition. 
\end{itemize}
A given set, together with an operation, is said to form a semigroup if: ({\it i}) the set is closed under the operation  (closure), ({\it ii})  the operation is associative over the set  (associativity), ({\it iii})  there is an identity element in the set with respect to the operation (identity). Completely positive maps, together with their multiplication (composition), define a semigroup. Indeed, ({\it i}) the composition of any two completely positive maps yields  a completely positive map; ({\it ii}) the composition of two completely positive maps composed with a third one is equivalent to the composition of the first one with the composition of the second with the third one; and ({\it iii}) the identity map is a completely positive map. The missing condition for a semigroup to be a group is invertibility, i.e. that every element has an inverse within the set, with respect to the operation. A general completely positive trace-preserving map cannot be inverted (by another completely positive trace-preserving map). The only completely positive trace-preserving maps with an inverse within the set are the subset of unitary maps.

The fact that generic completely positive maps cannot be inverted can be intuitively understood as follows. Since a superoperator is defined by the partial trace over a unitary operation on an extended space, there is loss of information that is in general irreversible. This is pictorially represented in Fig.~\ref{qChannel} b) with the trash can. This means that, in general, due to the interaction with an uncontrollable environment, an arrow of time naturally appears: A system can loose coherence due to its interaction with the environment, but, in the limit of infinite environmental degrees of freedom, the environment never restores this coherence. The exception is of course given by the particular case of evolutions for which system and environment decouple. There, the partial trace is redundant and the reduced system dynamics is left unitary, as already mentioned. For example, in Sec. \ref{Examples} we discuss a simple case in which two qumodes subsequently entangle and disentangle in a periodic way. Analogously, control schemes on messoscopic environments can restore coherence by disentangling system and environment, as shown in \cite{yao}. However, this requires full control over all parts involved, which is excluded in our description of an environment as a very large system whose internal dynamics is out of our control. 

\subsubsection{Entanglement-breaking channels}
\label{EntBreaking}
An exemplary  family of completely positive maps that is important for the study of entanglement dynamics is that of {\it entanglement-breaking channels}. A  map $\mathcal{E}$  on $\mathcal{D}(\mathcal{H}_{\sys})$ is called entanglement-breaking (EB) if its trivial extension to $\mathcal{D}(\mathcal{H}_{\sys})\otimes\mathcal{D}(\mathcal{H}_{\sys})$,  $\mathcal{E}\otimes\openone$, outputs only separable states. That is, if for every state $\varrho\in\mathcal{D}(\mathcal{H}_{\sys})$ the output state $\mathcal{E}\otimes\openone(\varrho)\in\mathcal{D}(\mathcal{H}_{\sys})\otimes\mathcal{D}(\mathcal{H}_{\sys})$ is separable. 

These channels are fully understood for any system dimension $d_{\sys}$~\cite{ebc}. In particular, in order to know if an arbitrary map  $\mathcal{E}$ is EB it is not necessary to inspect the output of $\openone\otimes\mathcal{E}$ for all possible inputs. Using the Choi-Jamio\l kowsky isomorphism \eqref{CJDualism}, it is possible to show that $\mathcal{E}$ is EB iff $\mathcal{E}\otimes\openone \proj{\Phi^+_{d_\sys}}$ is separable.  This is used in Sec. \ref{IsoDPD} to calculate the time at which paradigmatic noise models as the depolarising or phase-damping channels become EB, for any $d_\sys$.

Also, an explicit decomposition of EB channels is known. A map $\mathcal{E}$ is EB iff it can be written as 
\beq
\label{EBHolevoform}
\mathcal{E}\varrho=\sum_{k} \bra{\phi_k}\varrho\ket{\phi_k}\ket{\psi_k}\bra{\psi_k},
\eeq
where $\ket{\phi_k}$ and $\ket{\psi_k}$ are arbitrary normalized pure states, with $\sum_{k}\ket{\phi_k}\bra{\phi_k}\leq\openone$. In the particular case 
$\sum_{k}\ket{\phi_k}\bra{\phi_k}=\openone$, the EB map is in addition trace-preserving and $\{ \ket{\phi_k}\}$ defines a positive-operator valued measure (POVM). Expression \eqref{EBHolevoform} is called the {\it Holevo form}, since it was originally introduced by Holevo \cite{Holevo99}. It  describes the situation where one first applies the measurement defined by $\{ \ket{\phi_k}\}$ and then, for each measurement outcome $k$, one prepares the state  $\ket{\psi_k}$.

\subsection{Quantum Markov processes}
\subsubsection{Quantum dynamical semigroups}
\label{one_parameter}
The semigroup property in the general form  presented above is satisfied by all completely positive trace-preserving maps. However,  dynamical maps very often fulfill a yet more restrictive condition: the additive-composition dynamical semigroup property. More precisely, we have seen that the complete time evolution of $\sys$ over any time period $t\geq0$ is given by a one-parameter family $\mathcal{E}\doteq\{\mathcal{E}_{(t)}|0\leq t\}$ of dynamical maps. The term ``one-parameter" is due to the fact that each member of the semigroup is specified only by the time period $t$. The additive composition rule means that each element of  $\mathcal{E}$ satisfies 
\beq
\label{dynamicalsemigroup}
\mathcal{E}_{(t+\tilde{t})}=\mathcal{E}_{(\tilde{t})}\circ\mathcal{E}_{(t)},\ \forall\ t,\ \tilde{t}\ \geq 0.
\eeq
If this is the case, $\mathcal{E}$ is typically referred to as a {\it quantum dynamical semigroup}. Trivial examples thereof are all unitary transformations generated by time-independent Hamiltonians. 

Notice that property \eqref{dynamicalsemigroup} automatically implies that the evolution is local in time. That is, for any arbitrarily small $\tilde{t}$, the system state $\varrho_{(t+\tilde{t})}=\mathcal{E}_{(t+\tilde{t})}\varrho_{(0)}$ at time $t+\tilde{t}$ can be directly obtained (through $\mathcal{E}_{(\tilde{t})}$) from the state $\varrho_{(t)}=\mathcal{E}_{(t)}\varrho_{(0)}$ at time $t$, without any regard of the previous history of the system. This is the quantum analogue of the classical local-in-time probabilistic {\it Markov processes}. The essential idea behind quantum Markov processes \cite{Davies76} is that the time  $\tau_{\res}$ over which all reservoir correlation functions decay is much shorter than any relevant time of the system's dynamics, characterized by the characteristic time $\tau_{\sys}$. Therefore, in a coarse-grained time regime with temporal resolution $\tau_{\text{cg}}$, with $\tau_{\res}<<\tau_{\text{cg}}<<\tau_{\sys}$, reservoir-memory effects on the evolution of $\sys$ can be neglected. This approximation is known as the {\it Markovian approximation}. The precise physical conditions underlying the Markovian approximation can be stated more rigorously through microscopic derivations of the {\it Markovian quantum master equation}. The latter is a linear, first-order, differential equation for the time evolution of $\sys$, which generates quantum dynamical semigroups in the same way as the Schr\"odinger equation generates unitary evolutions. Microscopic derivations are in turn those where the semigroup generator is obtained from first principles explicitly from the system-environment interaction Hamiltonian. Such derivations involve some approximations, apart from the Markovian one, as well as certain assumptions (for detailed discussions see for instance Refs. \cite{cohenAP, carmichael,breuer}). In this review, we adopt an axiomatic approach, taking condition \eqref{dynamicalsemigroup} as the definition of quantum Markov processes, and deriving the Markovian quantum master equation from it. The physics behind the main approximations required for microscopic derivations are however discussed in Sec. \ref{Examples} with the help of some examples.

Let us first  show that any dynamical semigroup must be generated by a linear, first-order differential equation. The explicit form of the generator is saved for Sec.~\ref{lindblad}. We first derive both sides of Eq.~\eqref{dynamicalsemigroup} with respect to $\tilde{t}$. Since the left-hand side depends exclusively on the sum  $t+\tilde{t}$, its derivative with respect to $\tilde{t}$ is identically equal to its derivative with respect to $t$. This gives $\dot{\mathcal{E}}_{t+\tilde{t}}\equiv\mathcal{E}'_{\tilde{t}}\circ\mathcal{E}_{(t)}$, where $\dot{\mathcal{E}}_{t+\tilde{t}}\doteq \frac{d\mathcal{E}_{(t+\tilde{t})}}{dt}$ and $\mathcal{E}_{(\tilde{t})}'\doteq \frac{d\mathcal{E}_{(\tilde{t})}}{d\tilde{t}}$. Since this holds for any $\tilde{t}\geq0$, we can take the limit $\tilde{t}\to0$ and obtain 
\beq
\label{first_order}
\dot{\mathcal{E}}_{(t)}=\dot{\mathcal{E}}_{(0)}\circ\mathcal{E}_{(t)},\ \forall\ t\geq0,
\eeq
where we have used that $\mathcal{E}'_{(0)}=\dot{\mathcal{E}}_{(0)}$. There are two possibilities: Either $\dot{\mathcal{E}}_{(0)}=0$ or $\dot{\mathcal{E}}_{(0)}\neq0$, where  $0$  stands for the null map. If the former is true, then Eq.~\eqref{first_order} yields $\dot{\mathcal{E}}_{(t)}=0,\ \forall\ t\geq0\Rightarrow\mathcal{E}_{(t)}\equiv\openone,\ \forall\ t\geq0$. Here $\openone$ denotes the identity superoperator, corresponding to the trivial situation where the system does not evolve. Whereas if the latter is true, one has that 
\beq
\label{non_neg}
\dot{\mathcal{E}}_{(t)}\neq0,\ \forall\ t\geq0,
\eeq
unless $\mathcal{E}_{(t)}= 0$ for some $t\geq0$. Nevertheless, we know that $\mathcal{E}_{(t)}\neq 0,\ \forall\ t\geq0$, because $\mathcal{E}_{(t)}$ preserves the trace. Thus, the time derivative of $\mathcal{E}_{(t)}$ is always different from zero and (constantly) proportional to $\mathcal{E}_{(t)}$. This allows us to explicitly parametrize the entire semigroup in the exponential form 
\beq
\label{exp}
\mathcal{E}_{(t)}=e^{\mathcal{L}t},
\eeq
$\ \forall\ t\geq0$, where we have introduced the constant linear  map $\mathcal{L}\doteq\dot{\mathcal{E}}_{(0)}\neq0$, the generator of the semigroup. In turn, this parameterization immediately renders the desired  first-order linear differential equation:
\beq
\label{desiredeq}
\dot{\rho}_{(t)}=\mathcal{L}\rho_{(t)},
\eeq
$\ \forall\ t\geq0$. This is the Markovian master equation.

\par Finally, a comment on the assumptions used in the derivation of \eqref{desiredeq} is in place.  For \eqref{first_order}, we implicitly assumed that semigroup $\mathcal{E}$ is continuous and differentiable at all $t\geq0$. Continuity and differentiability are always granted when $\sys$ and $\res$ form an isolated composite system, as the composite dynamics is then governed by a time-independent Hamiltonian. However, the assumptions must be explicitly made in the fully general case. In addition,  in Eq. \eqref{non_neg}, we explicitly used the trace-preservation property of dynamical maps discussed in Sec. \ref{Krausop}. We show now how the other essential property of dynamical maps, complete-positivity, allows one to obtain an explicit form for the generator $\mathcal{L}$.

\subsubsection{Markovian master equation: the Limdbladian}
\label{lindblad}
The superoperator $\mathcal{L}$ generates the quantum dynamical semigroup through the Markovian master equation \eqref{desiredeq}. It can be thought of as the open-system generalization of the Liouvillian generator of the unitary-evolution, given essentially by the commutator between $\rho_{(t)}$ and the Hamiltonian in the Schr\"odinger equation. Here, we derive the most general form of $\mathcal{L}$. We follow a particularly simple approach  similar to that of \cite{preskill98}.

From \eqref{exp}, and for a sufficiently small $dt>0$, we must have ${\cal E}_{(dt)}=1+{\cal L}dt$, so that
\beq
\label{difderiv}
\rho_{(t+dt)}\equiv\mathcal{E}_{(dt)}\rho_{(t)}=\rho_{(t)}+O(dt).
\eeq 
From this, it follows that the most general Kraus decomposition of $\mathcal{E}_{(dt)}$ can be given by a Kraus operator $K_{0}=\openone+O(dt)$ and all other Kraus operators of order $\sqrt{dt}$. Therefore, without loss of generality, we can write the Kraus operators of $\mathcal{E}_{(dt)}$ as
\begin{equation} 
\label{Limbladops}
K_{\mu}\doteq\left\{ 
\begin{array}{ll} 
\openone+(-iH+C)dt & \textrm{for $\mu=0$, }\\ 
\sqrt{\gamma_{\mu}} L_{\mu}\sqrt{dt}  & \textrm{for $0<\mu<d^2_{\sys}$,}
\end{array} \right.
\end{equation} 
where $H$ and $C$ are time-independent Hermitian operators, time-independent operators $L_{\mu}$ have been taken of unit trace-norm,  $\gamma_{\mu}>0$, for all $0<\mu<d^2_{\sys}$, and $d_{\sys}$ is the dimension of the system, assumed to be finite. Kraus operators $K_{\mu}$, for $\mu>0$, describe the possible incoherent transitions that $\sys$ might undergo with probability of order $dt$; and at a rate  $\gamma_{\mu}$. These transitions are also called {\it quantum jumps}, and the operators $L_{\mu}$ quantum jump operators~\cite{carmichael}. In turn, the operator $C$ can be univocally determined using  the normalization condition \eqref{completenessKraus}:
\begin{equation} 
\openone\equiv\sum_{\mu\geq0} K^\dagger_\mu K_\mu= \openone+dt\big(2C+\sum_{\mu>0}\gamma_{\mu}L^{\dagger}_{\mu}L_{\mu}\big),
\end{equation}
which  renders
\begin{equation} 
\label{satisfy}
C=-\frac{1}{2}\sum_{\mu>0}\gamma_{\mu}L^{\dagger}_{\mu}L_{\mu}.
\end{equation}

Next, using \eqref{desiredeq}, \eqref{difderiv}, and the definition of the time derivative, we write
\beq
\label{derivmaster}
\mathcal{L}\rho_{(t)}=
\lim_{dt\to0^+}\frac{\mathcal{E}_{(dt)}\rho_{(t)}-\rho_{(t)}}{dt},
\eeq
and substitute  $\mathcal{E}_{(dt)}\rho_{(t)}$ by its explicit decomposition in terms of Kraus operators \eqref{Limbladops} normalized by \eqref{satisfy}. This, rearranging terms, immediately yields 
\beq
\label{desiredeqLimblad}
\mathcal{L}\rho_{(t)}=-i[H, \rho_{(t)}]+\sum_{\mu>0}\gamma_{\mu}\Big(L_{\mu}\rho_{(t)}L^{\dagger}_{\mu}-\frac{1}{2}\big\{L^{\dagger}_{\mu}L_{\mu},\rho_{(t)}\big\}\Big).
\eeq
The generator of any quantum dynamical semigroup can always be written in this form, referred to as the {\it Lindblad form}. Accordingly, $\mathcal{L}$ is often called the Lindbladian, master equation \eqref{desiredeq} the Lindblad equation, and jump operators $L_{\mu}$ also the Lindblad operators. The first term of \eqref{desiredeqLimblad} is responsible for the coherent part of the dynamics, generated by operator $H$ as a Hamiltonian. The second term accounts for the dissipative part, and it contains two types of contributions. The first one, involving terms of the form $L_\mu\rho_{(t)}L_\mu^\dagger$, is associated to quantum jumps, while the anticommutators $\{L_\mu^\dagger L_\mu,\rho_{(t)}\}$, which stem from the non-Hamiltonian part $C$  of $K_0$ in \eqref{Limbladops}, contribute to a non-unitary evolution of the density operator between quantum jumps \cite{carmichael}.  The coefficients $\gamma_\mu$ play the role of relaxation rates of the open system for its different decay modes.

The assertion that expression \eqref{desiredeqLimblad} is the most general form of the generator of a quantum dynamical semigroup is known as {\it Lindblad theorem}. This was proven in Ref. \cite{Gorini76} for finite-dimensional systems (as in our simple derivation) and in Ref. \cite{Lindblad76} for  general bounded generators. This is not often the case in physical situations: both $H$ and the Lindblad operators can in general be unbounded, as discussed in the examples of Sec. \ref{Examples}. Nevertheless, all known generators of quantum dynamical semigroups can be cast into the Lindblad form \eqref{desiredeqLimblad} \cite{breuer}.  
\subsubsection{The physics behind Markovian versus non-Markovian dynamics: two simple examples}
\label{Examples}

\par As discussed in Sec. \ref{one_parameter}, the Markovian master equation can alternatively be obtained through microscopic derivations. There, instead of taking the the semigroup property \eqref{dynamicalsemigroup} as the basic hypothesis, one starts from a concrete system environment Hamiltonian and works out the reduced system dynamics with the help of some approximations and assumptions. All of these, which are in practice very well satisfied by many quantum-optical systems, have been previously discussed in depth in the literature (see for instance Refs. \cite{Davies76,cohenAP, carmichael,breuer}). Here, we just briefly mention the two most important ones: the {\it Markovian and the Born approximations}. The first one was already mentioned in Sec. \ref{one_parameter}, and consists of neglecting all reservoir-memory effects so as to make the dynamics of $\sys$ local in time. This approximation is to hold in a coarse-grained time regime with temporal resolution $\tau_{\text{cg}}$, and is based on the assumption that the reservoir-correlations decay time $\tau_{\res}$ is much shorter than this resolution. In addition, in order to resolve the system's evolution, its characteristic time scale $\tau_{\sys}$ must be much longer than $\tau_{\text{cg}}$. Altogether, the  approximation is sustained by the Markovian assumption that 
\beq
\label{marvovassump}
\tau_{\res}<<\tau_{\text{cg}}<<\tau_{\sys}.
\eeq
The second one consists of approximating,  at each time $t$, the composite system-reservoir state  as 
\beq
\label{Bornassump}
\rho_{\sys\res (t)}\approx\rho_{(t)}\otimes\rho_{\res},
\eeq
 with $\rho_{\res}$ the state of the reservoir, for the purpose of calculating the density matrix  $\rho_{\sys\res (t+\tau_{cg})}$ in the successive time $t+\tau_{cg}$. The initial factorized-state condition  $\rho_{\sys\res (0)}=\rho_{(0)}\otimes\rho_{\res}$ is implicit. This approximation relies not only on assumption \eqref{marvovassump}, but is also to hold in a weak coupling regime where $\res$, which is infinitely larger than $\sys$, is  affected by $\sys$ negligibly. The idea is that, for weak system-reservoir couplings and short reservoir-correlations decay times, the decay time of any correlation established between system and environment at time $t$ is much smaller than $\tau_{cg}$, so that the initial state for the evolution between $t$ and $t+\tau_{cg}$ can be considered as uncorrelated.  So, as far as its influence on the reduced dynamics of $\sys$ is concerned, the environmental state appears effectively constant.
 
The fact that  the correction to the density matrix in  \eqref{difderiv} is assumed to be proportional to $dt$, implying that the Kraus operators $K_\mu$, with $\mu\not=0$, are proportional to $dt$, as displayed in \eqref{Limbladops},  is also a consequence of the coarse-grained description.  Indeed, this implies, according to  \eqref{unitaryKraus}, that if one monitored continuously the environment, the probability of a quantum jump from the initial state $\ket{0}_R$ to the state $\ket{\mu}_R$ in a time $dt$ should depend linearly on $dt$. But this seems to contradict first-order time-dependent perturbation theory, according to which this probability should be proportional to $(dt)^2$. What happens in fact is that the condition $dt\ge\tau_{cg}\gg\tau_\res$ implies that the transition time $dt$ is sufficiently large so that the Fermi Golden applies, that is, one is in the regime where time-independent rates can be associated to the transitions from $\ket{0}_\res$ to $\ket{\mu}_\res$.  

The two approximations mentioned above are usually applied together, receiving the joint name {\it Born-Markov approximation}. In what follows, we illustrate its  validity with two simple exemplary  situations.


\par {\bf Resonant Jaynes-Cummings interaction: a non-Markovian case}. The Jaynes-Cummings (JC) model describes the coherent interaction between a two-level system, for instance an atom, and a single quantized electromagnetic mode~\cite{jcm1,jcm2}. A single electromagnetic mode is certainly not a complex environment with infinitely many degrees of freedom. However, we force here the identification of it with an effective reservoir $\res$ so as to illustrate when the Born-Markov approximation can fail. The atom is  taken as the main system $\sys$. In the resonant case where the atomic transition frequency  $\omega$ coincides with the frequency of the field, and in the rotating wave approximation \cite{jcm1,jcm2}, the JC Hamiltonian reads
\beq
H^{\text{JC}} = \hbar \omega \left(a^\dagger a +\frac{1}{2}\right) - \hbar \frac{\omega}{2} \sigma^3 -i  \hbar \frac{\Omega_0}{2} \left (\sigma^+ a - \sigma^- a^\dagger\right ).
\eeq
Here the first two terms represent the free energies of the electromagnetic field and the atom, respectively, and the third one the interaction. For simplicity, we have chosen a real coupling constant $\Omega_0$. Operators $a$ and $a^\dagger$ are respectively the annihilation and creation operators corresponding to the electromagnetic field, while $\sigma^+ = \ket{1}\bra{0}$ and its adjoint $\sigma^- = \ket{0}\bra{1}$ denote the raising and lowering operators of the atom, where the computational basis states $\ket{0}$  and $\ket{1}$ are taken as the ground and excited states of the atom, respectively, being eigenstates of $\sigma^3$ with eigenvalues $+1$ and $-1$. In the interaction picture with respect to the free energy $\hbar \omega (a^\dagger a +1/2 - \sigma^3/2)$,  the Hamiltonian is given solely by the interaction term
\beq
\label{JCinteraction}
H^{\text{JC}}_{int} = -i\hbar\frac{\Omega_0}{2}\left (\sigma^+ a - \sigma^- a^\dagger\right ),
\eeq
which describes the processes of excitation of the atom in tandem with the annihilation of one photon (absorption) and decay of the atom accompanied by the creation of a photon (emission). The corresponding unitary evolution, in the interaction picture, is $U_{\sys \res (t)}^{\text{JC}}=\exp[-(\Omega_0 t/2)(\sigma^+ a-\sigma^- a^\dagger)]$. Maps that describe the joint evolution of system and environment can be readily written due to the fact that this evoution separates the Hilbert space into non-communicating sectors. Notice first that the ground state $\ket{0}_\sys \ket{0}_\res$ of no atomic or photonic excitations is a dark state of the evolution. That is, it is an eigenstate of Hamiltonian \eqref{JCinteraction} with eigenvalue zero. Therefore, it is invariant under the unitary evolution. Second, notice that Hamiltonian \eqref{JCinteraction}  connects $\ket{0}_\sys \ket{n}_\res$ only with $\ket{1}_\sys\ket{n-1}_\res$, where $\{\ket{n}_\res\}$ refers to the photonic Fock basis, with $n$ the number of photons in the mode. In particular, the subspace with the field in the vacuum state evolves as
\begin{subequations}
\label{GHZLC}
\begin{align}
\label{JCunitary0}
U_{\sys \res (t)}^{\text{JC}}\ket{0}_\sys\ket{0}_\res & =  \ket{0}_\sys\ket{0}_\res ,\\
\label{JCunitary}
\nonumber
U_{\sys \res (t)}^{\text{JC}}\ket{1}_\sys\ket{0}_\res & =  \cos\left(\frac{\Omega_0}{2} t\right) \ket{1}_\sys\ket{0}_\res \\
& + \sin\left(\frac{\Omega_0}{2} t\right) \ket{0}_\sys\ket{1}_\res\;.
\end{align}
\end{subequations}
 One clearly sees from Eq. \eqref{JCunitary} that, for the initial state $\ket{1}_\sys\ket{0}_\res$, the relevant time of the system dynamics is the time of atomic population inversion (half a Rabi oscillation): $\tau_\sys\approx\frac{\pi}{\Omega_0}$. Now, it is also clear that the time $\tau_\res$ over which the environmental correlations [created by  Rabi oscillations \eqref{JCunitary}] vanish is $\tau_\res\approx\frac{\pi}{2\Omega_0}$, which is of the same order of magnitude as $\tau_\sys$. Needless to say, the state of the reservoir is in addition far from constant during the time $\tau_\sys$. Therefore, the Born-Markov approximation is not expected to hold here. 

\par Indeed, comparing  Eqs. \eqref{JCunitary0} and \eqref{JCunitary} with \eqref{unitaryKraus}, one readily obtains the (time-dependent) Kraus operators of the corresponding dynamical map $\mathcal{E}_{(t)}$ for the evolution of $\sys$:
\begin{eqnarray}
\nonumber
K_0&=&\ket{0}\bra{0}+\cos\Big(\frac{\Omega_0}{2} t\Big)\ket{1}\bra{1}\ \text{and}\\
K_1&=&-i\sin\Big(\frac{\Omega_0}{2} t\Big)\sigma^-.
\label{KrausJC}
\end{eqnarray}
These two Kraus operators satisfy the normalization condition (\ref{completenessKraus}) and provide a complete characterization of the reduced evolution, including coherent Rabi cycles. However, as the reader can immediately check, the semigroup of superoperators  defined by them does not satisfy the Markovian composition rule \eqref{dynamicalsemigroup}.


\par{\bf Spontaneous emission: a Markovian case}. A good model for the environment of a two-level atom is given by an infinite collection of independent harmonic oscillators representing a continuum of electromagnetic modes of frequencies $\omega_k$. As in the previous example, we assume the atom is coupled to every mode via the JC interaction. However, each mode $k$ may now have, apart from a different  natural frequency $\omega_k$, a different coupling constant $\Omega_k$. The Hamiltonian of the composite atom photon-bath system in the rotating wave approximation \cite{jcm1,jcm2} reads 
\begin{eqnarray}
\label{photonic_bath}
\nonumber
H^{\text{PB}}&=&\hbar \sum_k \omega_k \left(a_k^\dagger a_k +\frac{1}{2}\right) - \hbar \frac{\omega}{2} \sigma^3\\
&-&i  \hbar \sum_k \frac{\Omega(\omega_k)}{2} \left (\sigma^+ a_k - \sigma^- a_k^\dagger\right ),
\end{eqnarray}
where the index $PB$ stands for photonic bath. As opposed to \eqref{JCinteraction}, the interaction representation of the photonic-bath Hamiltonian \eqref{photonic_bath} with respect to any non-trivial free energy is not time-independent and does not commute with itself at different times. Nevertheless, the two remarks about the coherent JC case still apply. Namely, first, the ground state $\ket{0}_\sys \ket{0}_\res$, with $\ket{0}_\res$ the vacuum state of the photonic bath, is still a dark state of the evolution. Second, $\ket{1}_\sys \ket{0}_\res$ only couples to $\ket{0}_\sys\ket{1}_\res$, where  $\ket{1}_\res$ represents now a single excitation coherently shared among the infinite field modes according to some distribution that depends on $\Omega(\omega_k)$. The dynamical behavior of the system depends crucially on the dependence of $\Omega(\omega_k)$ on $\omega_k$. The inverse of the width of this function corresponds approximately to the correlation time of the environment.  Therefore, if the width of $\Omega(\omega_k)$ is sufficiently large, so that the condition $\tau_{\res}<<\tau_{\text{cg}}<<\tau_{\sys}$ is verified, the Markovian approximation holds, and one does no longer expects Rabi oscillations, as in the resonant single-mode case.  Under these conditions, environmental memory effects can be safely neglected. We assume this to be the case in the following.

\par This simple heuristics suffices to grasp the main features of the composite unitary evolution, from which the desired reduced map can be obtained. For states initially in the electromagnetic vacuum, it must be
\bea{rcl}
U_{\sys \res (t)}^{\text{PB}}\ket{0}_\sys\ket{0}_\res & = & \ket{0}_\sys\ket{0}_\res\;;\\
U_{\sys \res(t)}^{\text{PB}}\ket{1}_\sys\ket{0}_\res & = & \sqrt{1-p_{(t)}} \ket{1}_\sys\ket{0}_\res + \sqrt{p_{(t)}} \ket{0}_\sys\ket{1}_\res,
\label{APmap}
\eea
where $U_{\sys \res (t)}^{\text{PB}}$ is the evolution operator, in the interaction picture, corresponding to \eqref{photonic_bath}, and  the parameter $0\leq p_{(t)}\leq1$ is the probability of emitting a photon up to time $t$. As before, the Kraus operators of the reduced dynamical map $\mathcal{E}_{(t)}^{\text{AD}}$ are readily given by:

\begin{eqnarray}
\nonumber
&&K^{\text{AD}}_0\doteq\ket{0}\bra{0}+\sqrt{1-p_{(t)}}\ket{1}\bra{1}\ \text{and}\\
 &&K^{\text{AD}}_1\doteq\sqrt{1-p_{(t)}}\ket{0}\bra{0}+\ket{1}\bra{1}.
\label{KrausAP}
\end{eqnarray}
Accordingly, an arbitrary pure state $\ket{\psi} = \alpha \ket{0} + \beta \ket{1}$ evolves as
\begin{eqnarray}
\label{ADquasedef}
\mathcal{E}_{(t)}^{\text{AD}}\ket{\psi}\bra{\psi}&=& |\alpha|^2 + p_{(t)} |\beta|^2\ket{0}\bra{0}\nonumber\\
&+& \alpha \beta^*\sqrt{1- p_{(t)}}\ket{0}\bra{1}\nonumber\\
&+& \alpha^* \beta \sqrt{1- p_{(t)}}\ket{1}\bra{0}\nonumber\\
&+&(1- p_{(t)})|\beta|^2\ket{1}\bra{1}, 
\label{stateAD}
\end{eqnarray}
where the matrix representation in the right-hand side is in the computational basis, as usual. From this, it is clear that one must have the boundary conditions $p(0)=0$ and $\lim_{t\to\infty}p_{(t)}\to 1$. That is, the stationary state is always the ground state.  To  derive the exact behavior of $p_{(t)}$ throughout the dynamics, we consider the composition $[{\mathcal{E}_{(dt)}^{\text{AD}}}]^M$ of $M$ times $\mathcal{E}_{(dt)}^{\text{AD}}$,  with $dt$  an infinitesimal time period such that $M\times dt=t$. Next, we consider the evolution \eqref{stateAD}
of $\ket{\psi}$ again and invoke  the semigroup property \eqref{dynamicalsemigroup}, obtaining
\begin{eqnarray}
\nonumber\mathcal{E}_{(Mdt)}^{\text{AD}}\ket{\psi}\bra{\psi}&\equiv&{\mathcal{E}_{(dt)}^{\text{AD}}}^M\doteq\underbrace{\mathcal{E}_{(dt)}^{\text{AD}} \circ \mathcal{E}_{(dt)}^{\text{AD}}\dots\circ \mathcal{E}_{(dt)}^{\text{AD}}}_M \ket{\psi}\bra{\psi}\\
\nonumber
&=& |\alpha|^2 + \big(1- (1-p_{(dt)})^M\big)|\beta|^2\ket{0}\bra{0}\\
\nonumber
&+& \alpha \beta^*\sqrt{1-p_{(dt)}}^M\ket{0}\bra{1}\\
\nonumber
&+&\alpha^* \beta \sqrt{1-p_{(dt)}}^M\ket{1}\bra{0}\\
&+&(1-p_{(dt)})^M|\beta|^2\ket{1}\bra{1}.
\label{Mdt}
\end{eqnarray}
From Eqs. \eqref{stateAD} and \eqref{Mdt}, one gets $1- p_{(t)}=(1-p_{(dt)})^M$. In addition, from \eqref{non_neg} one knows that $\dot{p}_{(t)}\doteq \gamma\neq0$ for all $t$, with $\gamma$ some real constant. This, together with $p_{(0)}=0$, implies that, for sufficiently small $dt$, one has $p_{(dt)}\approx \gamma dt$. Thus, in the limit $M\to\infty$, and for $dt=t/M$, one gets that $1- p_{(t)}=(1-\gamma dt)^M\to e^{-\gamma t}$. This gives the sought time dependence, $p_{(t)}\equiv1-e^{-\gamma t}$, therefore fully characterizing ${\mathcal{E}_{(t)}^{\text{AD}}}$ for all $t$ through Eqs. \eqref{KrausAP} and \eqref{stateAD}. Notice, in particular, from \eqref{stateAD}, that the amplitude of the excitation $\ket{1}$ is exponentially damped from $\beta$ to 0, at the rate $\gamma$. This explains the reason of the superscript ``AD" in channel ${\mathcal{E}^{\text{AD}}}$, which stands for amplitude damping. As a matter of fact, the exponential amplitude decay is in agreement with the original treatment by Weisskopf and Wigner \cite{weisskopf} of the {\it spontaneous emission} of a photon by a two level atom into the electromagnetic vacuum. Since the latter is nothing but a thermal bath at zero temperature, the proportionality constant  $\gamma$ is typically called zero-temperature dissipation rate. Indeed, in Sec. \ref{Noise_models}, we discuss  channel AD as the zero-temperature case of dissipation into a thermal bath at arbitrary temperature.

\par One should note that the condition $\dot{p}_{(t)}\neq0,\ \forall\ t\geq0$, is implied by the linear dependence with $dt$ in \eqref{difderiv}, which as seen before is a consequence of the transition-rate regime associated with the coarse-grained evolution.

\subsection{Noise models}
\label{Noise_models}
\par We are now in a good position to discuss concrete examples of maps that model physically relevant processes, as well as some of their main  classifications:

\subsubsection{Independent versus collective maps}
\label{indepmaps}
\par In the previous subsections, we have studied the reduced dynamics of the main system $\sys$ without regard to any internal substructure. However, since we are ultimately interested in studying entanglement dynamics, it is crucial to explicitly take into account the fact that the superoperators act on composite systems of $N$ subparts. In this respect, the first distinction we make is between {\it independent} and {\it collective} processes. We use the following notation. Unless otherwise explicitly specified, maps acting on the space $\mathcal{D}(\mathcal{H}_\sys)$ of density operators $\varrho$ of $\sys$ are denoted by $\mathcal{E}$, whereas the ones acting on the space $\mathcal{D}(\mathcal{H}_i)$ of those of the $i$-th subsystem carry in addition the subindex of the corresponding subsystem, as $\mathcal{E}_i$. As we have seen, every physical process can be described by a completely-positive trace-preserving map  $\mathcal{E}$, which admits in turn a Kraus representation as \eqref{krausForm}:
\beq
\label{Kraus_rep2}
\mathcal{E}\rho=\sum_\mu  K_\mu \rho K_\mu^\dag,
\eeq
with  Kraus operators $K_\mu$ with support on $\mathcal{H}_\sys$, and $\mu$ running from 0 to $d_{\sys}\equiv\prod_{i=1}^N d_i$, being $d_i$ the dimension of the $i$-th subsystem's Hilbert space $\mathcal{H}_i$ as usual. It is convenient to explicitly decompose $\mu$ as a multipartite multi-index: $\mu\equiv \mu_1\ ...\ \mu_N$, with  $\mu_i$ running from 0 to $d_i$. 

\par When the Kraus operators of \eqref{Kraus_rep2} admit in turn a decomposition $K_\mu\equiv {K_1}_{\mu_1}\otimes\ ...\ \otimes{K_N}_{\mu_N}$,  where every ${K_i}_{\mu_i}$ has support only on $\mathcal{H}_i$, we say that $\mathcal{E}$ is an {\it  independent map}. That is, the superoperator describes an independent process  if it can be factorized as the tensor product of individual single-partite superoperators, each one acting independently on a different subpart:
\begin{eqnarray}
\label{individual}
\nonumber
\mathcal{E}\rho&=&\sum_{\mu_1\ ...\ \mu_N}  {K_1}_{\mu_1}\otimes\ ...\ \otimes{K_N}_{\mu_N} \rho {K_1}^{\dag}_{\mu_1}\otimes\ ...\ \otimes{K_N}^{\dag}_{\mu_N}\\
&\equiv&\mathcal{E}_1\otimes \mathcal{E}_2 \otimes \ldots
\otimes \mathcal{E}_N\rho. 
\end{eqnarray}
Otherwise, we say that $\mathcal{E}$ is a {\it collective map}. In physical terms, independent maps can be thought of as describing the situations where each subpart is coupled to its own independent environment, whereas collective maps as those where a same environment is coupled to more than one subpart, so that correlations between the subparts may be developed due to the action of the bath. 

\par Naturally, analogously to the case of $k$-separable states, multipartite maps also admit sub-classifications in terms of the multipartitions for which they factorize. For example, in Sec. \ref{Jack} we study maps that decompose as tensor products of two channels acting on two subgroups of neighboring particles. However, for simplicity, unless otherwise explicitly specified, we reserve the term ``independent" for the fully-factorizable case described by \eqref{individual}. On the other hand, it is important to mention that all independent maps are necessarily separable, as defined by \eqref{LOCC}, but a generic separable map needs not factorize  in the form \eqref{individual}. That is, independent maps can not introduce any correlations whatsoever between the subparts, whereas separable maps are allowed to generate correlations and may therefore belong to both the independent or collective classes.

\subsubsection{Pauli maps}
\label{Pauliintro}

\begin{table*}[t!]
\begin{center}
\begin{tabular}{|c|c|c|}
\hline\hline
Channel&Kraus Operators & Associated Physical Process\\
\hline\hline
\begin{tabular}{c}
Generalized \\
Amplitude \\
Damping\\
(GAD)
\end{tabular} 
&
\begin{tabular}{l}
${K}_{0}\equiv\sqrt{\frac{\overline{n}+1}{2\overline{n}+1}}(\ket{0}\bra{0}+\sqrt{1-p}\ket{1}\bra{1})$,\\
${K}_{1}\equiv\sqrt{\frac{\overline{n}+1}{2\overline{n}+1}p}\ \sigma^-$,\\
${K}_{2}\equiv\sqrt{\frac{\overline{n}}{2\overline{n}+1}}(\sqrt{1-p}\ket{0}\bra{0}+\ket{1}\bra{1})$,\\
${K}_{3}\equiv\sqrt{\frac{\overline{n}}{2\overline{n}+1}p}\ \sigma^+$.
\end{tabular} 
&
\begin{tabular}{l}
For $p\equiv 1-e^{-\frac{1}{2}\gamma(2\overline{n}+1)t}$, describes diffusion and dissipation,  in the\\  Born-Markov approximation, with a thermal bath of ave-rage excita-\\tion $\overline{n}$. Constant $\gamma$ is the  zero-temperature dissipation rate, associated\\ to spontaneous emission (channel AD, obtained by setting $\overline{n}=0$).
\end{tabular} 
\\
\hline
\begin{tabular}{c}
Phase \\
Damping \\
 (PD)
\end{tabular} 
&
\begin{tabular}{l}
${K}_{0}\equiv\sqrt{1-p/2}\ \openone$,\\
${K}_{1}\equiv\sqrt{p/2}\ \sigma^3$,\\
\end{tabular} 
&
\begin{tabular}{l}
Also called {\it Dephasing}. Describes the elastic scattering (no excitation\\ exchange) with the environment. With probability $p$, the off-diagonal\\coherences are destroyed. With probability $1-p$, nothing happens.
\end{tabular}
\\
\hline
\begin{tabular}{c}
Phase Flip \\
 (PF)
\end{tabular} 
&
\begin{tabular}{l}
${K}_{0}\equiv\sqrt{1-p}\ \openone$,\\
${K}_{1}\equiv\sqrt{p}\ \sigma^3$,\\
\end{tabular} 
&
\begin{tabular}{l}
Close relative of channel PD. Describes relative $\pi$-phase errors in the\\ computational basis. With probability $p$,
the off-diagonal coherences\\ change sing. With probability $1-p$, nothing happens.
\end{tabular}
\\
\hline
\begin{tabular}{c}
Bit Flip \\
 (BF)
\end{tabular} 
&
\begin{tabular}{l}
${K}_{0}\equiv\sqrt{1-p}\ \openone$,\\
${K}_{1}\equiv\sqrt{p}\ \sigma^1$,\\
\end{tabular} 
&
\begin{tabular}{l}
Same as channel PF but in a rotated basis. Describes relative $\pi$-phase\\ errors in the $X$ eigenbasis. 
\end{tabular}
\\
\hline
\begin{tabular}{c}
Bit-Phase\\ Flip \\
 (BPF)
\end{tabular} 
&
\begin{tabular}{l}
${K}_{0}\equiv\sqrt{1-p}\ \openone$,\\
${K}_{1}\equiv\sqrt{p}\ \sigma^2$,\\
\end{tabular} 
&
\begin{tabular}{l}
Same as channels PF and BF but in yet another basis. Describes rela-\\tive $\pi$-phase errors in the $Y$ eigenbasis. 
\end{tabular}
\\
\hline\begin{tabular}{c}
Depolarizing\\ (D) 
\end{tabular} 
&
\begin{tabular}{l}
${K}_{0}\equiv\sqrt{1-p}\openone$,\\
${K}_{1}\equiv\sqrt{p/3}\ \sigma^1$,\\
${K}_{2}\equiv\sqrt{p/3}\ \sigma^2$,\\
${K}_{3}\equiv\sqrt{p/3}\ \sigma^3$.
\end{tabular} 
&
\begin{tabular}{l}
The most detrimental kind of noise, where all three errors, PF, BF, and\\BPF, can happen. With probability $p$,
the system is depolarized to the\\ maximally mixed state. With probability $1-p$, nothing happens.
\end{tabular}\\
\hline\hline
\end{tabular}
\end{center}
\caption{Brief description of the most popular single-qubit channels. All of them are described in detail in the text.}\label{tabKraus} 
\end{table*}

\par An important family of  separable maps for qubits is that of the {\it Pauli maps}, defined as those where every Kraus operator is proportional to a tensor product of $N$ single-qubit Pauli or identity operators: 
\beq 
\label{PauliKraus}
K_\mu\equiv \sqrt{p_{(\mu_1\ ...\ \mu_N)}}\  {\sigma_1}^{\mu_1} \otimes \ldots  \otimes{\sigma_N}^{\mu_N}\doteq\sqrt{p_{\mu}}\ \sigma^{\mu},
\eeq
with ${\sigma_i}^{0}=\openone_i$, the identity operator on qubit $i$, ${\sigma_i}^{1}=X_i$, ${\sigma_i}^{2}=Y_i$, and ${\sigma_i}^{3}=Z_i$, the three Pauli operators of qubit $i$, as usual, and $p_{(\mu_1\ ...\ \mu_N)}\equiv p_\mu$ any normalized probability distribution. When this distribution factorizes as $p_\mu={p_1}_{\mu_1}\times\ ...\times \ {p_N}_{\mu_N}$, with ${p_i}_{\mu_i}$ a single-partite distribution associated to the $i$-th qubit,  the Pauli map is also an independent map. In addition, if ${p_i}_{\mu_i}$is the same for all $i$, then the composite dynamics is fully characterized by the single-qubit channel. 

\par The best-known examples of single-qubit Pauli channels are the depolarizing (D) and phase-damping (PD)  (or dephasing) channels. Channel D describes the situation where the qubit is probabilistically subject to white noise. That is, it remains untouched with a certain probability $1-p$, or is completely depolarized -- i.e., transformed  into the maximally mixed state $\openone_i/2$ -- with probability $p$. Complete depolarization happens when the qubit is subject to bit-flip ($X_i$), phase-flip ($Z_i$), and bit-phase-flip ($Y_i$) errors with the same probability.  Therefore, the channel is characterized by ${p_i}_{0}=1-p$ and ${p_i}_1={p_i}_2={p_i}_3=p/3$. Channel D describes the most detrimental kind of noise, as not only the coherences are destroyed but also the populations are completely mixed. In turn, channel PD describes the processes where the qubit scatters elastically with the reservoir constituents, which may induce the loss of quantum coherence, with probability $p$, but without any exchange of population in the computational basis (implying that the energy of the system is conserved). The process is thus less detrimental than depolarization, as the (classical) information encoded in the population of the computational basis' elements is preserved. It is characterized by ${p_i}_{0}=1-p/2$, ${p_i}_{1}=0={p_i}_{2}$, and  ${p_i}_{3}=p/2$, so that off-diagonal density-matrix coherence elements vanish  at $p=1$. A close relative of channel PD is the phase-flip  (PF) channel, which, as its name suggests, describes the process by which coherent superpositions of $\ket{0_i}$ and $\ket{1_i}$ are probabilistically subject to a $\pi$-phase shift. It is characterized by ${p_i}_{0}=1-p$, ${p_i}_{1}=0={p_i}_{2}$, and  ${p_i}_{3}=p$, so that the off-diagonal elements change sign  at $p=1$.

\par Other popular single-qubit Pauli channels are for instance the bit-flip (BF) and bit-phase-flip (BPF) channels, which are equivalent to the phase-flip channel but in rotated bases. They are defined respectively by ${p_i}_{0}=1-p$, ${p_i}_{2}=0={p_i}_{3}$, and  ${p_i}_{1}=p$, and ${p_i}_{0}=1-p$, ${p_i}_{1}=0={p_i}_{3}$, and  ${p_i}_{2}=p$. The Kraus operators of all these, and other, single-qubit channels are summarized in table \ref{tabKraus}.

\par Finally, the probability $p$ in all the channels above can be interpreted as a parameterization of time, where $p=0$ refers
to the initial time 0 and $p=1$ to the asymptotic limit $t\rightarrow\infty$, where the system reaches a steady state. A common such parameterization is $p\equiv p_{(t)}\doteq1-e^{-\xi t}$, for some constant decay rate $\xi$. This  choice, which corresponds to a Markovian dynamics, implies that the channels satisfy the Markovian additive composition rule \eqref{dynamicalsemigroup}.

\subsubsection{Independent thermal baths}
\label{Thermaldef}

\par An important example of a non-Pauli single-qubit channel is the generalized amplitude-damping (GAD) channel. It describes the processes of energy diffusion and dissipation, in the Born-Markov approximation, with a thermal bath into which the qubit is individually immersed. It is characterized by the four Kraus operators
\begin{subequations}
\label{Kraus_AmpDamp}
\begin{align}
{K_i}_{0}&\equiv\sqrt{\frac{\overline{n}+1}{2\overline{n}+1}}(\ket{0_i}\bra{0_i}+\sqrt{1-p}\ket{1_i}\bra{1_i}),\\
{K_i}_{1}&\equiv\sqrt{\frac{\overline{n}+1}{2\overline{n}+1}p}\ \sigma_i^-,\\
{K_i}_{2}&\equiv\sqrt{\frac{\overline{n}}{2\overline{n}+1}}(\sqrt{1-p}\ket{0_i}\bra{0_i}+\ket{1_i}\bra{1_i}),\\
{K_i}_{3}&\equiv\sqrt{\frac{\overline{n}}{2\overline{n}+1}p}\ \sigma_i^+,
\end{align}
\end{subequations}
with $\sigma_i^+\doteq\ket{1_i}\bra{0_i}$ and $\sigma_i^-\doteq\ket{0_i}\bra{1_i}$, as usual. Here, $\overline{n}$ is the average number
of excitations in the  bath, $p\equiv p_{(t)}\doteq1-e^{-\frac{1}{2}\gamma(2\overline{n}+1)t}$ is the probability of
the qubit exchanging a quantum with the bath from  time 0 until time $t$, and
$\gamma$ is the zero-temperature dissipation rate already mentioned in Sec. \ref{Examples} when spontaneous emission was discussed. As a matter of fact, channel GAD is nothing but the extension to finite temperature of the purely dissipative AD channel introduced there. Indeed, the AD Kraus operators \eqref{KrausAP} are recovered from \eqref{Kraus_AmpDamp} in the zero-temperature limit $\overline{n}=0$. On the other hand, the purely diffusive case is
obtained in the opposite limit: $\overline{n}\rightarrow\infty$, $\gamma\rightarrow0$, and
$\overline{n}\gamma=\Gamma$, where $\Gamma$ is the diffusion constant. In this limit, channel GAD becomes a Pauli channel, characterized bu  single-qubit probabilities ${p_i}_{0}=\frac{1}{2}(1-p_{(t)}/2+\sqrt{1-p_{(t)}})$, ${p_i}_{1}=\frac{p_{(t)}}{4}={p_i}_{2}$, and  ${p_i}_{3}=\frac{1}{2}(1-p_{(t)}/2-\sqrt{1-p_{(t)}})$. As a matter of fact, in this limit, channel GAD becomes similar to channel D, with both channels having the completely mixed (infinite-temperature thermal) state as the only steady state, for instance.

Table~\ref{tabKraus} summarizes the Kraus operators of all the single-qubit channels discussed in this section, together with a brief description of the physical process associated to each of them.

\section{Theory of open-system dynamics of entanglement: bipartite systems}
\label{IV}

We dive now into the main topic of the review: the dynamics of entanglement under the effect of the environment. The first attempt to study the dynamics of entanglement in open systems of which we have record was done by Yi and Sun, in Ref.~\cite{xxy}. In that article, the authors study the Hilbert-Schmidt distance from a bipartite state $\rho_{AB}(t)$ at time $t$ to the tensor product of its reduced states $\rho_A(t)\otimes \rho_B(t)$. Although this distance is not an entanglement measure, with these pioneering study the authors aimed at capturing some features of the open-sytem entanglement dynamics.  The first real accounts of entanglement dynamics in open systems  of which we have record are due to Rajagopal and Rendell, who analyzed,  in Ref.~\cite{rajagopal}, the dynamics of entanglement for two initially entangled harmonic oscillators under the action of local environments; and to $\dot{\text{K}}$yczkowski and the Horodecki family, who analyzed, in Ref.~\cite{karol0}, the evolution of the average entanglement of formation of  random $2\otimes 2$ states undergoing a sequence of global  unitary evolutions periodically interlaced with a noisy channel. 

Both papers showed that entanglement might vanish at finite times, while coherence, for the same classes of noisy processes, would vanish asymptotically in time. Many distinct cases were discussed in \cite{karol0}: when the unitary part was not present, then,  depending on the noisy process, the average entanglement could vanish asymptotically or at finite time;  when the global unitary was switched on, entanglement revivals could be observed. Since then, many theoretical studies on the dynamics of bipartite entanglement appeared, involving the interaction with local environments \cite{simon02,diosi03,jamroz03,dodd04,duer04,carvalho04,yu04,serafini04,lidar04,hein05,fine05,mintert05b,aravind05,yu06,yu062,santos06,liu06,benatti06,yonac061,ficek2,eberly07,yu:459,yonac:s45,zubairy07,liu07,seligman07,seligman2007,sabrina07,terra01,alejo,ficek08,lopez-2008,marek08,guo08,james08,concentration,hu08,lai08,Paz:220401,paz08,yu08b,xu09,Paz:032102,cavalcanti09,terra02,yu09,mazzola09,zell09,sumanta,papp09,barbosa10,Barreiro10,viviescas10,dur11}, or with collective ones \cite{ficek1,ficek2,ficek08,hor09,adriana10,monz2011}, or yet a combination of both \cite{ficek2}, as discussed in the introduction.

In this section, refraining from an exhaustive compilation of results, we describe some of  the principal aspects of the theory of entanglement evolution in bipartite open systems. We treat the multipartite case in the following section.

\subsection{Two qubits under amplitude damping}
\label{EntDynaAD}

The subtle properties of the dynamics of entanglement, and its striking differences to the dynamics of coherences, are
exhibited in the paradigmatic example of spontaneous atomic decay~\cite{yu04, yu06,yu062}. As discussed in Secs.~\ref{Examples} and \ref{Noise_models}, this dynamics is well approximated by the amplitude damping channel $\mathcal{E}^{\text{AD}}$. Consider now a bipartite system $\sys$, composed of two qubits, $A$ and $B$, initially in state $\ket{\Psi}=\alpha\ket{00}
+\beta\ket{11}$. Under independent AD channels of the same damping strength $p$ on each qubit, the latter state evolves, according to \eqref{ADquasedef}, to the mixed state 
\begin{equation}
\left(\begin{array}{cccc}
|\alpha|^2+p^2|\beta|^2&0&0&(1-p)\alpha\beta^*\\
0&(1-p)p|\beta|^2&0&0\\
0&0&(1-p)p|\beta|^2&0\\
(1-p)\alpha^*\beta&0&0&(1-p)^2|\beta|^2
\end{array}\right)\;,
\label{evolS1S2A}
\end{equation}
\normalsize
with the  matrix written in  the computational basis.

Channel AD  does not create new coherences, the only non-null off-diagonal elements in density matrix \eqref{evolS1S2A} are those of the initial coherence between $|00\rangle$ and $|11\rangle$, damped by the factor $1-p$. One can directly quantify the entanglement of the two qubits throughout their evolution by calculating the concurrence, as given by expression \eqref{concu1}, as a function of $p$, which gives: 
\beq
C(p)=\max\{0,2(1-p)|\beta|(|\alpha|-p|\beta|)\}\,.
\label{concESD}
\eeq
Importantly, in this particular example, the negativity and the concurrence coincide for all $p$ \cite{audenaert,Verstraete01}. Concurrence (\ref{concESD}) features two distinct dynamical regimes:  If (i) $|\beta|\le|\alpha|$, then
entanglement vanishes only asymptotically, i.e. at $p=1$. 
Whereas if (ii) $|\beta |>|\alpha|$, then entanglement vanishes at a finite-time, more precisely at $p=|\alpha/\beta|<1$. 
We will encounter these two disentanglement behaviours repeatedly throughout this review. 
 
Interestingly, since at $p=0$ concurrence \eqref{concESD} depends only on the product of $\alpha$ and $\beta$, and not on themselves, the two different types of decay are consistent with a same initial entanglement. Finite-time disentanglement is also called some times  ``entanglement sudden death'' \cite{yonac061,yu062}. Some intuition of why finite-time disentanglement takes place for $|\beta |>|\alpha|$ can be given for channel AD. Inspecting the joint system-reservoir unitary dynamics \eqref{APmap}, we see that only the excited state $\ket{1}$ couples to the environment. So, the   larger the population $|\beta|^2$ of the excited state, the faster the development of system-environment entanglement, which leads in turn to a faster reduction of system entanglement. In Sec. \ref{TopoAp}, we give in addition a simple geometrical explanation of why this distinction of decay types arises.

Since the joint system-environment state is pure, the rise of system-environment entanglement can be easily verified by calculating the concurrence between system $\sys$ and environment $\res$ through formula  \eqref{concupure}. This yields
\bea{rcl}
\label{CSE}
C_{\sys\res}(p)
&=&2\sqrt{2}|\beta|\sqrt{p(1-p)}\sqrt{1-|\beta|^2 p(1-p)}\;,
\eea
which increases with $|\beta|$. 
Furthermore, the entanglement between each qubit and its own environment increases with $|\beta|^2$:
$C_{\sys_A\res_A}(p)=C_{\sys_B\res_B}(p)= 2 |\beta|^2 \sqrt{p(1-p)}$.

It is also useful to inspect the dynamics of the entanglement between the two environments
\cite{yonac:s45,lopez-2008,alejo}.  At $p=1$, the state of the system (and therefore its entanglement) is completely transferred to the environment~\cite{yonac:s45,lopez-2008}:
\begin{equation}
\left(\alpha\ket{00}+\beta\ket{11}\right)_S\otimes\ket{00}_\res \stackrel{p=1}{\longrightarrow}
\ket{00}_S\otimes\left(\alpha\ket{00}+\beta\ket{11}\right)_\res\;.
\end{equation}
For $p<1$, the entanglement between the two environmental qubits is quantified by the concurrence \cite{alejo} 
\beq
C_{\res_A\res_B}(p)=\max\{0,2p|\beta|[|\alpha|-(1-p)|\beta|]\}\,.
\eeq
Remarkably, this concurrence shows that when system disentanglement occurs at $p<1$, it is accompanied by the appearance of environment entanglement  at 
some $0<p<1$. The latter has some times been called ``entanglement sudden birth'' (ESB), in contraposition with entanglement sudden death'' (ESD). The times for which ESD and ESB occur are  related by  $p_{\text{ESB}}=1-p_{\text{ESD}}$. Thus ESB may occur
before, simultaneously, or after ESD, depending on whether $p_{\text{ESD}}>1/2$, $p_{\text{ESD}}=1/2$, or
$p_{\text{ESD}}<1/2$, respectively. 

To end up with, for baths at non-zero temperature, well approximated by the 
generalized amplitude damping channel, $\mathcal{E}^{\text{GAD}}$, as discussed in Sec. \ref{Thermaldef}, the calculations are somewhat more involved but can still be carried out exactly. For instance, in Ref.~\cite{james08}, it was shown that for a two-qubit system initially in an $X$-state, i.e. with density matrix in the computational basis having non-zero elements only along the diagonals, then, for independent thermal baths at any positive temperature, there is finite-time disentanglement.
Finally, in Ref.~\cite{fine05}, it was shown that the steady-state entanglement of any two-qudit system in touch with independent thermal baths at any non-zero finite temperature always vanishes, for arbitrary interactions between the qudits and the environment. 

\subsection{Two qubits under dephasing}
\label{EntDynaPD}

Another paradigmatic noise channel is the dephasing channel $\mathcal{E}^{\text{PD}}$. In this case the dynamics of entanglement between  two qubits $A$ and $B$, again initially in the  state $\ket{\Psi}=\alpha\ket{00}
+\beta\ket{11}$, is characterized by the concurrence
\beq
\label{concudeph}
C(p)=2 (1-p)|\alpha\beta|\,,
\eeq
which vanishes only at $p=1$, that is, asymptotically in time.

In contrast to the AD case discussed in the previous subsection, 
here the decrease of entanglement in the two-qubit system is accompanied by the generation of genuine four-partite entanglement among the two system qubits and the two environmental ones. For instance, we note that while \eqref{concudeph} decreases with $p$, the total concurrence among all four qubits, as given by \eqref{ConcAndre}, is
\beq
C_{\sys_1\sys_2\res_1\res_2}(p)= |\alpha\beta|\sqrt{4+4 p -p^2}\,,
\eeq
which increases monotonously with $p$. Furthermore, at $p=1$, the joint system-reservoir state evolves towards
\beq
(\alpha|00\rangle+\beta|11\rangle)_\sys|00\rangle_{\res}\stackrel{p=1}{\longrightarrow}\alpha|00\rangle_\sys|00\rangle_\res+\beta|11\rangle_\sys|11\rangle_\res\,.
\eeq
This is 
 a state of the GHZ type, as defined by \eqref{GHZdef0}, which is genuinely four-partite entangled and for which all two-qubit entanglements, between  both system qubits, each of them and its own environmental qubit, and  between both environmental qubits, are null. A detailed analysis  for different initial states was made in \cite{alejo}.

\subsection{Two interacting qubits: entanglement decay and revival}
\label{Interactingdecayandrevival}

\begin{figure}[t]
\begin{center}
\includegraphics[width=\linewidth]{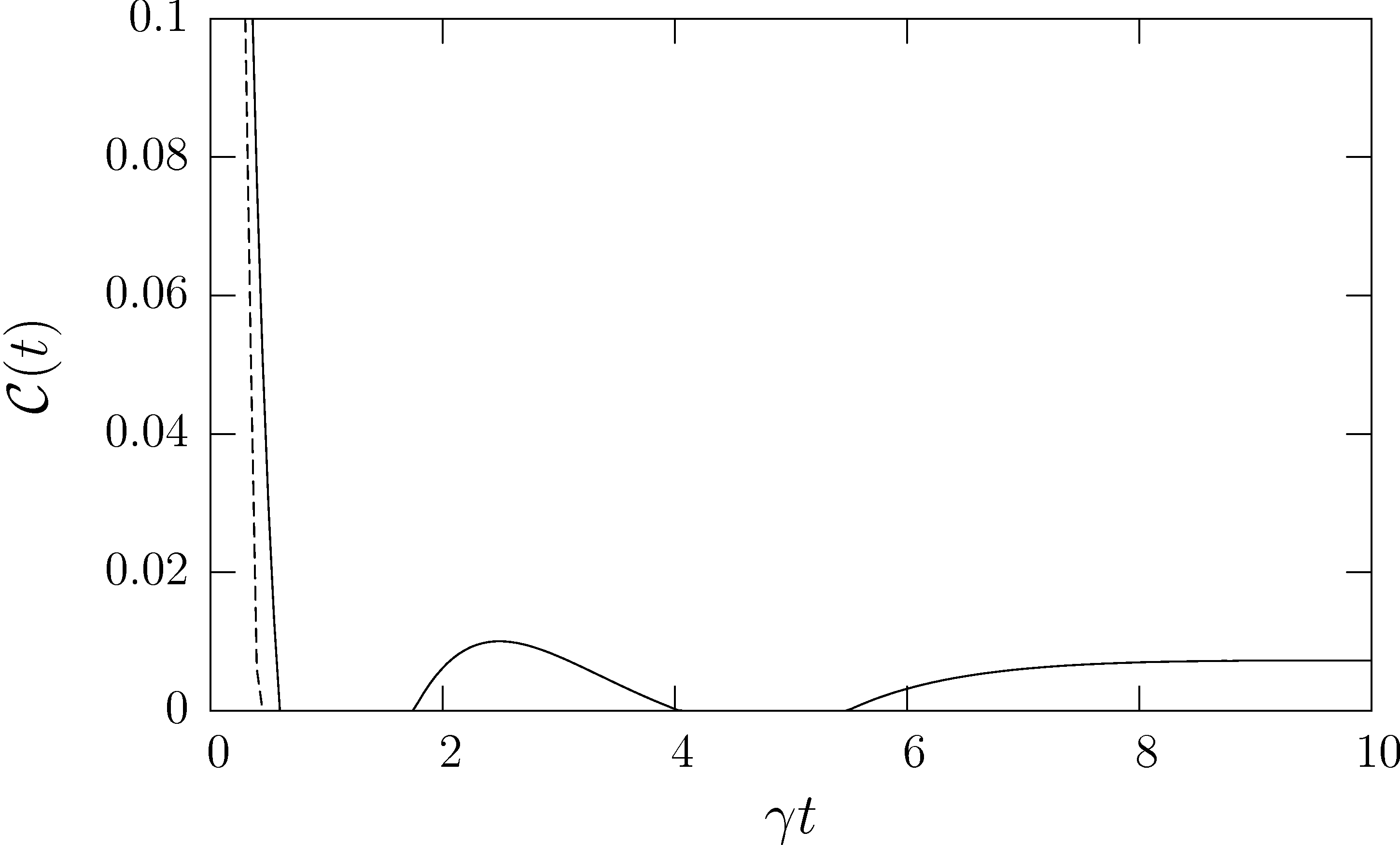}
\caption{Concurrence (solid line) of a pair of two-level atoms interacting via dipole-dipole coupling and under a collective AD environment, as a function of $\gamma t$. The initial state is $\ket{\Psi}=\sqrt{\alpha}|11\rangle+\sqrt{1-\alpha}|00\rangle$, with $\alpha=0.9$. 
The dashed line represents the concurrence of the same initial state but without the dipole-dipole coupling and for local AD environments. For short times both dynamics are similar, with the local coupling prevailing over the global one. For longer times, however, the interaction through the collective environment leads to entanglement revivals. Reprinted figure with permission from Z. Ficek and R. Tana\'s, \href{http://link.aps.org/doi/10.1103/PhysRevA.74.024304}{Phys. Rev. A {\bf 74}, 024304 (2006)}. Copyright (2006) by the American Physical Society.} 
\label{tanasRevival}
\end{center}
\end{figure}

For logical qubits encoded in two-level atoms, their dipole-dipole coupling, as well as the interaction between the atoms' respective environments, can be neglected whenever the distance $r_{12}$ between them is much larger than the resonance wavelength $\lambda$. However,  when $r_{12}\lesssim\lambda$, neither can the atom-atom interaction be neglected nor do the environmental modes resolve the two particles any longer. In this situation, a description based on a common environment becomes necessary. As the two qubits now effectively interact, both directly and via the environment, a competition between entanglement decay and generation takes place, so that entanglement might reappear in the system after vanishing at some finite time.  

Several scenarios where qubit entanglement undergoes revivals have been recently studied in the context of quantum information~\cite{braun02,benatti03,oh06,ficek1,ficek2,das09,sumanta}. However, the effect of dipole-dipole interactions has been previously extensively considered within the framework of superradiance \cite{haroche}, whose first studies trace back to  Dicke's work in the fifties~\cite{dicke}.
Since the baths are collective, decoherence-free subspaces may be found, i.e. some entangled states may be immune against the action of the environment (corresponding to the sub-radiant states of~\cite{dicke}).  

Fig.~\ref{tanasRevival} 
shows results of~\cite{ficek2}, where two two-level atoms are assumed to interact with a collective amplitude damping environment and with each other through dipole-dipole coupling. 
It displays the entanglement dynamics for the initial state $\ket{\Psi}=\sqrt{\alpha}\ket{11}+\sqrt{1 -\alpha}\ket{00}$, with $\alpha=0.9$ and for $r_{12}=\lambda/20$. As usual, $|0\rangle$ and $|1\rangle$ correspond to the atomic ground and excited states, respectively. The entanglement decays and vanishes at finite-time, as for the case of two non-interacting atoms under independent AD discussed in Sec. \ref{EntDynaAD}. However, here, in contrast, entanglement shows revivals. Another exotic example of entanglement creation via the action of both dipole-dipole coupling and collective AD environment takes place when $\alpha=1$. In this case, the initial state is separable, but a sudden birth of system entanglement can take place after some time lag~\cite{ficek2}.

\subsection{Two-qudit isotropic states under local depolarisation or dephasing}
\label{IsoDPD}


Despite the fact that for high-dimensional bipartite systems an exact entanglement evaluation is in general not feasible, there are several interesting families of states  for which relevant results exist. These particular states illustrate  features of entanglement dynamics that can appear also in the general case. 
One such case is that of isotropic states~\cite{isotropic}. These are two-qudit states that fulfil the special symmetry $\varrho_\text{iso}=U\otimes U^* \varrho_\text{iso} U^\dagger\otimes {U^*}^\dagger$, with  $U$ any single-partite unitary operator. The entire family can be parametrized by a single parameter  $0 \le f \le 1$ as
\begin{equation}
\label{isopar}
\varrho_\text{iso}(f) = \frac{1-f}{d^2-1} \big( \openone - \proj{\Phi^+_{d}} \big) + f \proj{\Phi^+_{d}}.
\end{equation}
Parameter $f$ is called the {\it fidelity}, or the {\it singlet fraction}, of $\varrho_\text{iso}(f)$, as it gives the overlap of $\varrho_\text{iso}(f)$ with the maximally entangled state $\ket{\Phi^+_{d}}$. For states \eqref{isopar}, different entanglement measures have been evaluated exactly~\cite{EoFIsoTerhal,chen06c,Rungta03}. For instance, their concurrence reads 
\beq
\label{concuiso}
\mathcal{C}(\varrho_\text{iso}(f)) = \left\{ \begin{array}{cl}
										0&\text{ for } f\le 1/d\\
										(f-1/d)\sqrt{2d/(d-1)}&\text{ for } 1/d\le f\le 1.
										\end{array}\right.
\eeq

Interestingly, as noted by Ann and Jaeger in Ref.~\cite{Ann07-1}, experimentally relevant noise types, as the PD and the D channels map any isotropic state into another isotropic state (of lower singlet fraction). Therefore, the entanglement dynamics of $\varrho_\text{iso}(f)$ can be fully monitored.  
%
A simple but instructive example is the D channel $\mathcal{E}^{D}$ acting on only one of the qudits of $\varrho_\text{iso}(f)$. From \eqref{concuiso}, one straightforwardly calculates the disentanglement time of any isotropic state under $\mathcal{E}^{D}\otimes\openone$. For instance, for the isotropic state of maximal singlet fraction $f=1$, i.e.  the maximally entangled state $\ket{\Phi^+_{d}}$, and for a depolarising strength $p = 1- e^{- d^2 \gamma t}$, the disentanglement time is
%
\beq
t^{\text{D}}_\text{EB} = \ln (d+1)/d^2 \gamma.
\eeq
%
Note that $t^{\text{D}}_\text{EB}$ decreases with the dimension $d$,  which is in accordance with the fact that entanglement tends to become more fragile as the macroscopic limit is approached. The subindex EB in $t^{D}_\text{EB}$ stands for ``entanglement breaking," since the disentanglement time induced by $\mathcal{E}^{\text{D}}\otimes\openone$ on  $\ket{\Phi^+_{d}}$ defines the time at which $\mathcal{E}^{D}$ becomes EB, as discussed in Sec. \ref{EntBreaking}. With the same treatment, one obtains the disentanglement time of isotropic states under $\mathcal{E}^{PD}\otimes\openone$ and therefore also the time at which $\mathcal{E}^{PD}$ becomes EB. 

\subsection{CV systems: two qumodes interacting with local thermal baths}

The dynamics of entanglement for two harmonic oscillators has also been extensively studied in the literature. The dynamics displays some similarities with the case of discrete variables. For example, Refs.~\cite{rajagopal,paris02,serafini04,sabrina07}  show that the entanglement between two qumodes interacting with independent environments may vanish asymptotically or at finite time. Entanglement revivals were also demonstrated for CV systems in the presence of global environments, both in the Markovian~\cite{jakub04,dodd04, benatti06} and non-Markovian~\cite{Ban06,liu07,An007,hammer08} regimes.

\begin{figure}[t]
\begin{center}
\includegraphics[width=\linewidth]{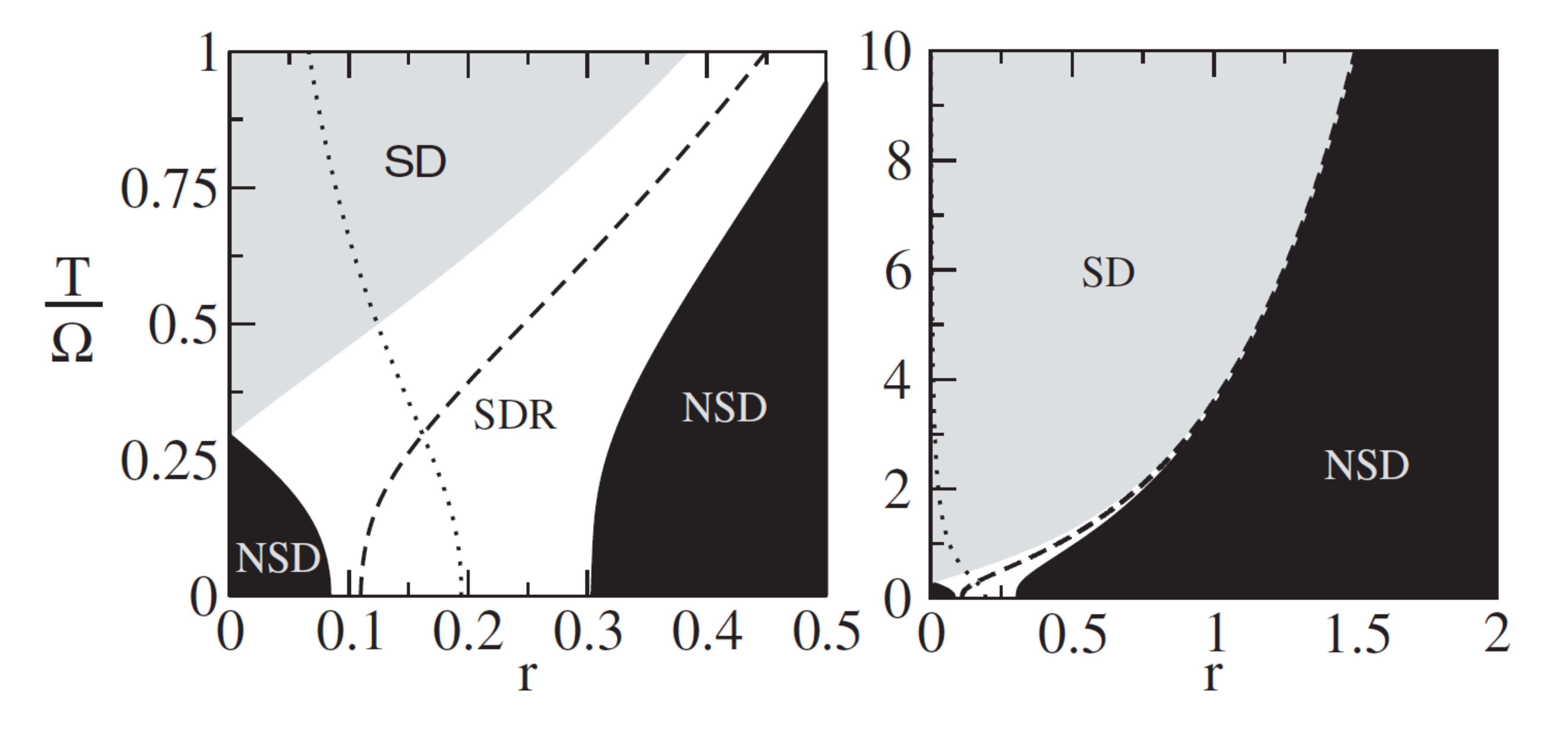}
\caption{Entanglement phase diagram for two oscillators undergoing a quantum Brownian motion under the action of a common environment ($\Omega$ is the renormalized oscillator frequency, here set equal to $1$). Depending on the initial squeezing and temperature, the evolution may induce finite-time disentanglement -- sudden-death (SD) phase --, asymptotic disentanglement or an entangled steady state -- both characterized by the no-sudden-death (NSD) phase --, or a periodic series of entanglement revivals -- sudden-death and revival (SDR) phase. The dotted and dashed lines are related to the  width and position of the different entanglement phases as a function of the temperature (see~\cite{Paz:220401} for details). Reprinted figure with permission from J. P.~Paz  and J. A.~Roncaglia, \href{http://link.aps.org/doi/10.1103/PhysRevLett.100.220401}{Phys. Rev. Lett. {\bf 100}, 220401 (2008)}. Copyright (2008) by the American Physical Society.} 
\label{fig:pazCV}
\end{center}
\end{figure}

In Refs.~\cite{Paz:220401,Paz:032102}, Paz  and Roncaglia drew a general picture of the different ``phases'' of CV entanglement dynamics for two qumodes coupled to the same thermal bath at temperature $T$, and initialized in the two-mode squeezed Gaussian state $e^{-r (a_1^\dagger a_2^\dagger - a_1 a_2)}\ket{0}$, where $r$ is the squeezing parameter, $\ket{0}$ represents the vacuum state, subindices 1 and 2 label the two modes in question, and $a_i$ and $a^{\dagger}_i$, $i=1,2$, are respectively the annihilation and creation operators corresponding to mode $i$. See Fig.~\ref{fig:pazCV}. The coupling to the reservoir is described by the quantum Brownian motion master equation~\cite{Hu92,Martinez13}. Since this dynamics maps Gaussian states onto Gaussian states, the behavior of entanglement for different $r$ and $T$ can be directly assessed from the symplectic eigenvalues, as defined in Sec.~\ref{2.2.1a}, of the covariance matrix of the steady state. Three qualitatively different entanglement phases were identified: (i) Sudden-death (SD) -- the entanglement vanishes at finite-time; (ii) no-sudden-death (NSD) -- the entanglement either decays asymptotically or it remains finite throughout the evolution; and (iii) sudden-death and revival (SDR) --  the entanglement undergoes an infinite series of periodic vanishments and revivals. Also as in the discrete-variable case, the generation of entanglement from two initially separable modes due to the interaction with the common bath is possible.

\subsection{Robust entanglement}

\begin{figure}[t]
\begin{center}
\includegraphics[width=\linewidth]{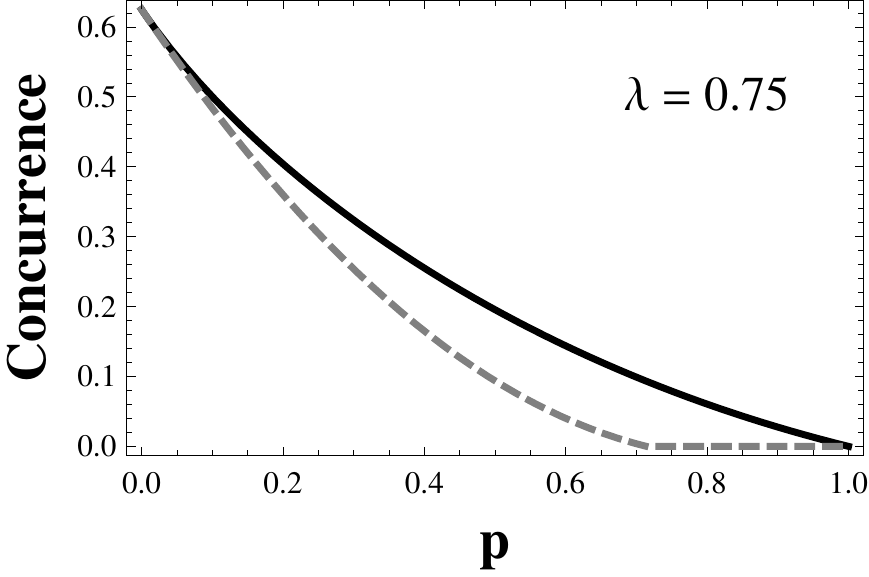}
\caption{Concurrence of two local unitarily equivalent two-qubit states under the same independent AD channels of strength $p$, as a function of $p$. The black  solid line corresponds to the Werner state $\varrho_\rom{Werner}$, whereas the gray dashed line to $\openone\otimes X \varrho_\rom{Werner} \openone\otimes X$. The entanglement of the latter decays faster than that of the Werner state and vanishes at finite time.} 
\label{wernerDiff}
\end{center}
\end{figure}

Distinct initial states with the same amount of entanglement and under the the same type of noise may undergo quite different entanglement dynamics.  This opens the possibility to search for the most robust states, i.e. those whose entanglement decays the slowest. A simple  two-qubit example was studied by Yu and Eberly in Ref.~\cite{yu:459}. As a similar example, we can compare the entanglement dynamics of the Werner state
\bea{l}
\varrho_\rom{Werner} \doteq (1-\lambda)\frac{\openone}{4} + \lambda \ket{\Psi^-}\bra{\Psi^-}\;
\eea
with that of the local-unitary equivalent state
\bea{l}
\openone\otimes X \varrho_\rom{Werner} \openone\otimes X\doteq (1-\lambda)\frac{\openone}{4} + \lambda \ket{\Phi^-}\bra{\Phi^-}\;,
\eea
with $\lambda\in[0,1]$, under independent AD channels. Although both initial states have, by definition, identical entanglements, their entanglement  evolution differs, as shown in Fig.~\ref{wernerDiff} for the exemplary case of $\lambda=0.75$.  Since $\varrho_\rom{Werner}$ has only one excitation shared between both qubits, it gets less affected by the AD noise than $\openone\otimes X \varrho_\rom{Werner} \openone\otimes X$, which possesses two excitations shared between the qubits. Accordingly, the entanglement of the former turns out to be more robust than that of the latter. In conclusion, a local unitary transformation before the noise acts may have a significant positive effect on the entanglement resistance against the noise. 

The two-qudit case was numerically studied by Mintert in Ref. \cite{florian09}. There, he applied lower bounds to the concurrence to search for robust states under independent  PD and AD channels. For  multipartite states, ingenious techniques to find robust states have been obtained~\cite{Froewis11,Froewis12,chaves12b,Ali13}. We discuss these in Secs. \ref{robustGHZdirected}, \ref{robustGHZblockwise}, and \ref{Ali}.

\subsection{Harnessing entanglement dynamics}

Instead of searching for the most robust state, one may try to actively counteract the influence of the environment, so as to slow down the decay of entanglement as much as possible or even to revert it. A variety of  strategies have been proposed~\cite{wiseman:548,doherty:2700,mancini:010304,kwiat01,andre07,andre08,alejo,felix09}. Here we expose two qualitatively different important paradigms for the coherent control  of entanglement. 

\subsubsection{Passive spontaneous-emission based distillation}

Consider a single two-level atom, initially in the state $\alpha \ket{0} +\beta \ket{1}$, interacting with a zero-temperature thermal  bath, where $|0\rangle$ and $|1\rangle$ stand respectively for the ground and excited states, as usual. Furthermore, assume that a perfect detector, able to detect every photon that might eventually be emitted, is placed in the vicinity of the atom. The absence of a click corresponds to a projection onto the vacuum state of the photonic bath. This in turn renders some information about the atom, i.e. it represents a weak measurement of it, and, as such, changes its state. The longer one waits for a click, the higher the probability to project the atom onto $\ket{0}$.

This effect can be employed to distill entanglement  
~\cite{kwiat01,alejo}. One considers two atoms separated enough to approximate their reservoirs as independent. The initial state is $\ket{\Psi_{(0)}} =\alpha\ket{00}+\beta\ket{11}$, with $|\alpha| < |\beta|$
. As before, two detectors continuously monitor each atom's reservoir. Without this monitoring, the dynamics would be ruled by the independent AD channel of strength $p=\exp(-\gamma t)$, but because of the monitoring the system is continuously projected onto a pure state.

Proceeding as in \eqref{APmap} with each atom, we see that, if up to time $t$ no photon is detected by either detector, the  joint system state becomes 
\beq
\ket{\Psi_{(t)} =}\frac{\alpha\ket{00}_\sys+ \beta (1-p)\ket{11}_\sys}{\sqrt{|\alpha|^2 + (1-p)^2|\beta|^2}}.
\eeq
 If at time $t_\text{max} = \gamma^{-1}\ln|\beta/\alpha|$, corresponding to $p_\text{max} = 1-|\alpha/\beta|$, no decay is registered, then the system collapses to the maximally entangled state with $|\alpha|=|\beta|$. The continuous ``no-click" detection reduces the population of $\ket{11}$, balancing it with that of $\ket{00}$
 . In this way, just by locally observing the environment in the vacuum state,  one increases the entanglement of the system. Of course, this scheme succeeds only probabilistically. No click up to time $t_\text{max}$ happens just a fraction $1- [|\alpha|^2 + (1-p)^2|\beta|^2)]$ of the trials. This fraction thus defines the success probability of this particular distillation scheme. An important drawback of the protocol is that maximal entanglement 
 is obtained at a single instant of time. An interesting improvement of this method, also in connection to weak measurements, was proposed in Ref.~\cite{sun10},  where a technique to revert the environment-induced change of entanglement was proposed. The experimental implementation of the latter was reported in \cite{kim11}.

\subsubsection{Active feedback}

\begin{figure}[t]
\begin{center}
\begin{tabular}{cc}
\includegraphics[width=0.35\linewidth]{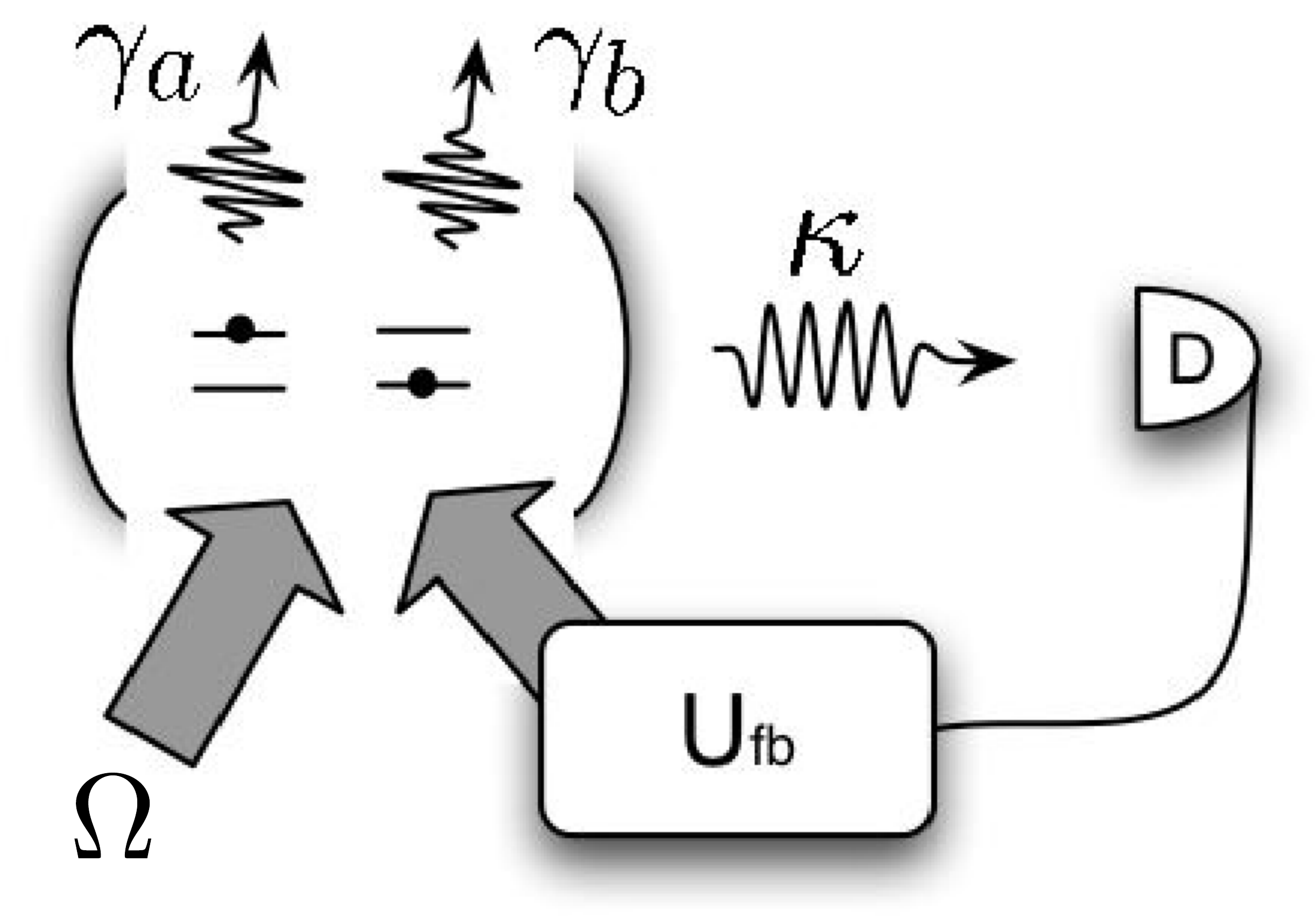}&\includegraphics[width=0.55\linewidth]{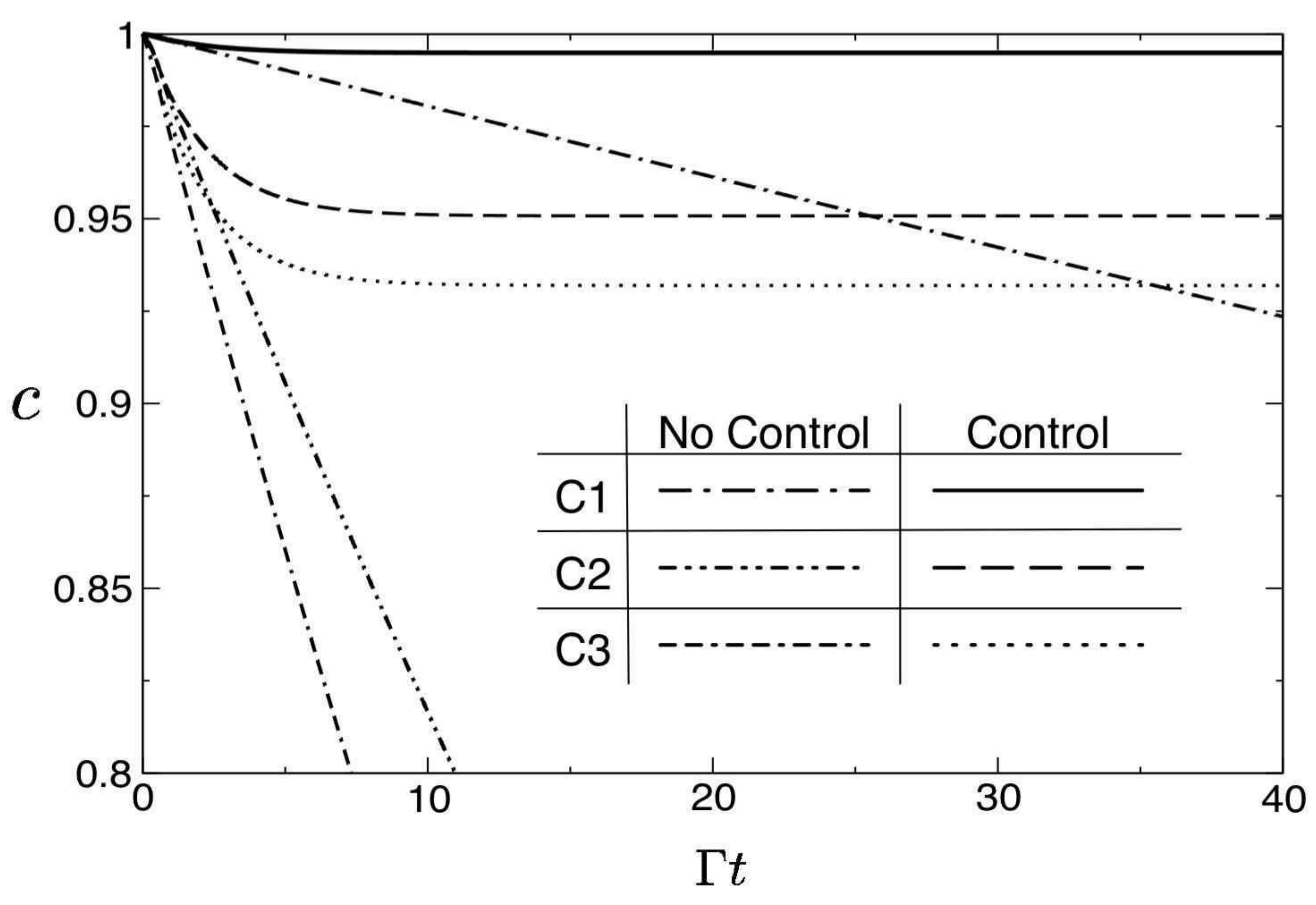}\\
{\bf a)}&{\bf b)}
\end{tabular}
\caption{Slowing down the decay of entanglement of two two-level atoms inside a leaky cavity through active feedback control
. {\bf a)} The atoms' spontaneous emission rates are $\gamma_a$ and $\gamma_b$, while the decay rate of the cavity field is $\kappa$, with $\gamma_a,\gamma_b<<\kappa$. Any photon emitted by the cavity output is  registered by a perfect detector D. This activates the application of the unitary operation $U_\text{fd}$ on one of the atoms. {\bf b)} Entanglement as a function of $\Gamma t$, with and without the feedback control, for the initial state $\ket{\Psi^-}$, $\Omega= 3\Gamma$, and feedback unitary $U_\text{fb}=\openone\otimes \exp(-i \pi X /2)$. The three cases plotted are (C1) $\gamma = 0.001\Gamma$, (C2) $\gamma = 0.01\Gamma$, and  (C3) same as (C2) but with an atomic dephasing rate $\gamma_\text{deph}=\gamma$ additionally taken into account. Reprinted figure with permission from A.~R.~R.~Carvalho and J.~J.~Hope, \href{http://link.aps.org/doi/10.1103/PhysRevA.76.010301}{Phys. Rev. A  {\bf 76}, 010301 (2007)}. Copyright (2007) by the American Physical Society.} 
\label{dede}
\end{center}
\end{figure}

Another possibility is to exploit feedback, i.e. to actively operate on the system  depending on the  outcomes of measurements on its environment. The   idea goes in a similar direction as quantum error-correction. However, here one is mainly interested in the entanglement and not so much in the state itself. An example of such a strategy was proposed  by  Carvalho and Hope in Refs.~\cite{andre07, andre08}. 
They considered the specific setup depicted in  Fig.~\ref{dede} a). There, two non-interacting two-level atoms, $a$ and $b$, are resonantly coupled, with coupling constant $g$, to a cavity, and are simultaneously driven by a classical field of amplitude $\Omega$. The atoms spontaneously decay into the environment with respective decay rates $\gamma_a$ and $\gamma_b$, and into the cavity output mode with decay rate $\kappa$. Both $\gamma_a$ and $\gamma_b$ are assumed much smaller than $\kappa$. So, the dominant decay channel is through the cavity output mode, with an effective decay rate $\Gamma=g^2/\kappa$. This is called the {\it Purcell regime}. The cavity output is in turn continuously monitored by a perfect detector D, which, upon photon click, fires the application of the unitary operation $U_\text{fb}$ on the system. The first difference with the scheme of the previous sub-subsection is that the atoms effectively decay through a collective channel, and the second one is the active application of $U_\text{fb}$.

The best choice of $U_\text{fb}$ requires an optimization, but some simple considerations are sufficient to chose a feedback unitary that preserves high values of entanglement despite of the local and global environments. First note that both the unitary dynamics and the global noise do not mix the symmetric and anti-symmetric subspaces of the two-qubit system. Furthermore, with only the global decay, the maximally entangled state $\ket{\Psi^-}$ is a steady state. The steady-state solution on the symmetric subspace is also entangled, but the overall stationary entanglement turns out to be low ($\sim 0.1$ when quantified by concurrence). When one then considers the local decays, the split between the subspaces is broken. The idea is then to chose a feedback unitary that coherently shifts highly entangled states from the anti-symmetric subspace into the symmetric subspace. This is obtained, for example, by choosing the feedback to act only in a single qubit by taking $U_\text{fb} = \openone\otimes e^{-\frac{i \pi X}{2}}$. As shown in the plot, a significant enhancement of the entanglement robustness is obtained, even when atomic local dephasing  
is additionally considered. 


More recently,  a similar scheme was proposed, in Ref.~\cite{felix09}, but with the application of a time-dependent feedback Hamiltonian instead of a fixed unitary operations. The Hamiltonian is optimized at each instant of time so as to maximize the amount of entanglement. In this way, it is possible to design optimal control strategies for the entanglement dynamics. This is an interesting example within the field of quantum dynamical control, which has recently seen an upsurge of activity~\cite{viola98,viola99,rabitz00,dong10,trail10,lee13}. 

\subsection{An equation of motion for entanglement}

In Refs.~\cite{konrad,tiersch09
} , a dynamical equation for the evolution of the concurrence of two qubits was derived. With this, the entanglement of the system, initially in any pure state, and under the action of any single-sided channel, is totally determined by that of a maximally entangled state undergoing the same dynamics. More precisely, it was shown that
\beq
\mathcal{C}(\openone\otimes\mathcal{E}\ket{\Psi}\bra{\Psi}) = \mathcal{C}(\ket{\Psi})\ \mathcal{C}(\openone\otimes\mathcal{E}\ket{\Phi^+}\bra{\Phi^+})\,,
\label{concMov}
\eeq
for any completely positive single-qubit map $\mathcal{E}$ and pure state $\ket{\Psi}$.

Equation \eqref{concMov} is appealing because of its simplicity. It factorizes into a contribution from the initial state, $\mathcal{C}(\ket{\Psi})$, and another due to the channel, $\mathcal{C}(\openone\otimes\mathcal{E}\ket{\Phi^+}\bra{\Phi^+})$. Indeed, by the Choi-Jamio\l kowsky isomorphism  discussed in Sec. \ref{Krausop}, $\rho_\mathcal{E} \doteq \openone\otimes\mathcal{E}\ket{\Phi^+}\bra{\Phi^+}$ is the dual state of channel $\mathcal{E}$ and therefore contains all the information about it. This is crucial for the efficacy of decomposition \eqref{concMov}. As a corollary of  \eqref{concMov}, one knows that the only single-sided channels $\mathcal{E}$ that map pure entangled states into separable states are the entanglement-breaking channels, that is, those for which the dual state $\rho_\mathcal{E}$ is separable, as discussed in Sec.~\ref{EntBreaking}.

In the general scenario of two-sided channels and initially mixed states, due to the convexity of $C$, one gets instead the upper bound
\beq
\mathcal{C}(\mathcal{E}_1\otimes\mathcal{E}_2 \varrho) \le \mathcal{C}(\varrho)\ \mathcal{C}(\mathcal{E}_1\otimes\openone \ket{\Phi^+}\bra{\Phi^+})\ \mathcal{C}(\openone\otimes\mathcal{E}_2\ket{\Phi^+}\bra{\Phi^+}).
\label{concMovUpper}
\eeq
Remarkably, this bound was shown to be tight in some experimentally relevant cases~\cite{tiersch09}. In turn, this bound has been experimentally verified with  entangled photon pairs~\cite{oswaldo09,xu09}, as discussed in Section~\ref{62}.


Decompositions \eqref{concMov} and \eqref{concMovUpper} can both be generalized to two-qudit systems of arbitrary dimensions $d_1\times d_2$ for the $G$-concurrence~\cite{gour05,gour05b}, and even to multipartite systems for entanglement measures with some particular structures~\cite{gour10, gour12}. The $G$-concurrence is one of the generalizations for $d_1\times d_2$ systems of the two-qubit concurrence \eqref{concu1}, and reduces to the latter when $d_1=d_2=2$. In spite of being an entanglement monotone, it does not detect the entanglement of some entangled states for  $d_1\times d_2>2\times2$. For this reason, it fails to be a well-behaved entanglement measure. However, it has been shown that, for short times, there is a connection between the $G$-concurrence and the usual $C$-concurrence, which may in turn give useful information about the decay of the latter for longer times~\cite{tiersch08}.

\subsection{Geometry of entanglement decay}
An elegant way to extract generic features of entanglement dynamics is to study the geometrical aspects of the set of states $\mathcal{D}(\mathcal{H}_{\sys})$. In the next  sub-subsections we discuss two such aspects.
\subsubsection{Topological view of entanglement trajectories}
\label{TopoAp}


The set of states $\mathcal{D}(\mathcal{H}_{\sys})$ is a closed convex set to which a metric can be assigned~\cite{KarolBook}. Its boundary is defined by all the states for which an arbitrarily small perturbation suffices to take them out of  $\mathcal{D}(\mathcal{H}_{\sys})$. These are all non-full-rank states, defined as those whose rank is smaller than $d^2_\sys$, which is equivalent to having at least one null eigenvalue. An arbitrarily small perturbation maps them into negative operators. In particular, all pure states are rank-1 and therefore non-full-rank. Analogously, the interior of $\mathcal{D}(\mathcal{H}_{\sys})$ is given by the usual definition of the interior of a set: all states around which a ball of some positive radius exists such that every element inside the ball belongs to $\mathcal{D}(\mathcal{H}_{\sys})$. Clearly, these are the full-rank states, as one can always perturb them in any parameter direction by a  quantity  small enough to get another full-rank state. Any state $\varrho\in\mathcal{D}(\mathcal{H}_{\sys})$ can always be expressed as a convex combination of two non-full-rank states or, equivalently, as many rank-1 pure extremal points as the rank of $\varrho$. 

The subset of separable states in  $\mathcal{D}(\mathcal{H}_{\sys})$, in turn, also forms a closed convex set of non-zero volume~\cite{ziczkowski98,VolumeSepState2}. The characterisation of its boundary is of course more intricate as that of $\mathcal{D}(\mathcal{H}_{\sys})$, because apart from containing non-full-rank separable states it also contains full-rank separable states that are arbitrarily close to being entangled. We have already sketched pictorial representations of the geometry of $\mathcal{D}(\mathcal{H}_{\sys})$ and its subset of separable states in Sec. \ref{2.2}, such as for instance the one in Fig. \ref{Geometry}.

Pictorial as they are though, these simple considerations allow one to visualise some  general features of the entanglement dynamics. For instance, one concludes that any dynamic process whose steady state lies within the interior of the set of separable states necessarily induces finite-time disentanglement on all initial states. Because then the intersection between the state trajectory and the border of separability necessarily happens at some finite time $p<1$. This is the case of both the independent channels D and GAD  at infinite temperature, whose unique final state is the maximally mixed state. Furthermore, even for any  non-null finite temperature the steady state of the GAD channel lies in the interior set of separable states, as shown by Yu and Eberly in Ref.~\cite{Yu07}. So finite-time disentanglement is always present there too. 

\par In contrast, when the steady state of the evolution lies at the boundary of the separability set, the dynamic trajectory may never enter the interior of  the separability set. Due to the complexity of $\mathcal{D}(\mathcal{H}_\sys)$, an intuitive picture of when this happens is in general difficult to grasp. For instance, for the independent PD channel acting on any pure entangled state of the form $\ket{\Psi}=\alpha\ket{00}+\beta\ket{11}$, the evolved state is never full rank. Therefore, the entire trajectory from $p=0$ to $p=1$ is tangential to the boundary of $\mathcal{D}(\mathcal{H}_\sys)$. In this case, disentanglement is observed always at $p=1$, as discussed in Sec. \ref{EntDynaPD}. On the contrary, for the independent PD channel composed with independent AD channel, the state is full rank for all $0<p<1$, goes towards the pure state $\ket{00}$, and disentanglement is observed always at $p<1$, as will be discussed in Sec. \ref{Simon} for the multipartite case. However, also for the independent AD alone is the state full rank for all $0<p<1$, as matrix \eqref{evolS1S2A} in Sec. \ref{EntDynaAD} shows, and goes towards $\ket{00}$, but the disentanglement time depends on the initial coefficients $\alpha$ and $\beta$. In Sec. \ref{Simon}, we will see that these generic conclusions remain valid also in the multipartite scenario with initial states of the form $\ket{\Psi}=\alpha\ket{00\hdots0}+\beta\ket{11\hdots1}$.

Of course other types of dynamic trajectories are possible, as for instance those induced by the collective decay of nearby atoms, discussed in Sec. \ref{Interactingdecayandrevival}. There, the dynamic trajectories may repeatedly enter and exit the separability set during the whole evolution.  Another example was given by Fine, Mintert and Buchleitner, in Ref.~\cite{fine05}, who, based solely on geometrical arguments, showed that for any bipartite system interacting with a collective thermal bath, there exists a finite temperature for which the dynamics induces finite-time disentanglement on all initial states. Their argument relies on the fact that, for a sufficiently high temperature, the steady state must be close enough to the maximally mixed state (the steady state for infinite temperature) to also lie within the (finite-volume) set of separable states. A detailed account of the possible trajectories was given by Drummond and Terra Cunha in Refs.~\cite{terra01,terra02}.


\subsubsection{Concentration of entanglement trajectories}
\label{Concentration}
In Ref.~\cite{hayden}, Hayden, Leung, and Winter showed that if one samples pure states at random from the uniform Haar distribution\footnote{The Haar distribution is the uniform probability distribution over the pure state vectors.
Thus, sampling from this distribution renders pure states randomly distributed in a uniform way over all the space of pure states vectors. It defines also a metric, the Haar measure, which is the unique uniform measure on the space of pure state, i.e. the only one invariant under any unitary transformation.}, with very high probability the sampled state is highly entangled in any bipartition. In addition, this probability increases extraordinarily fast with the system size. That is, the distribution of entanglement on the space of pure states concentrates very fast around the maximally entangled states. Following them, in Ref.~\cite{concentration}, Tiersch, de Melo and Buchleitner showed that for entanglement measures that abide by a strong form of continuity known as {\it Lipschitz continuity}, the entanglement trajectories of pure random states under any arbitrary completely positive trace-preserving channel concentrate around the mean entanglement, with the average taken after the evolution under the same map. 

%
\begin{figure}[t]
\label{fig:distributions}
	\centering
	\includegraphics[width=\linewidth]{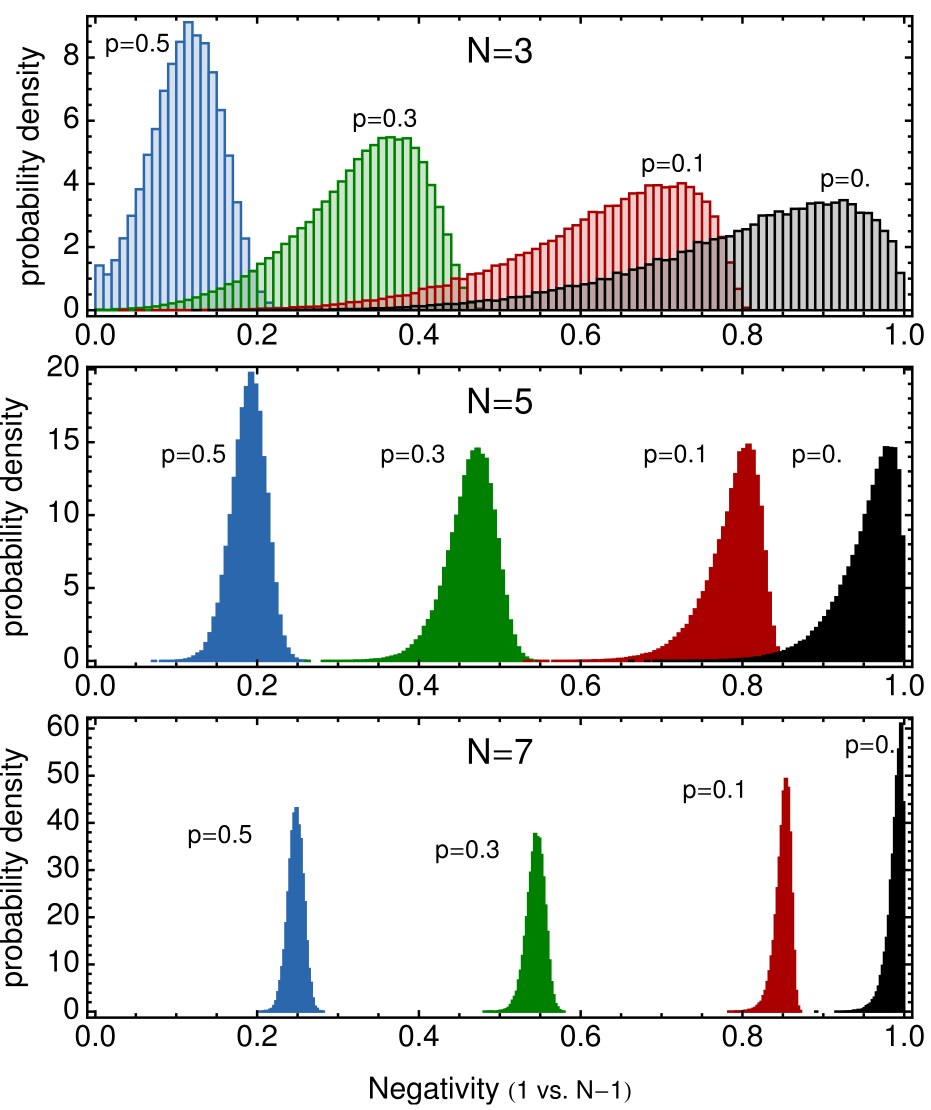}
	\caption{Evolution of the negativity distribution for the bipartition of the first qubit versus the others. For $N=3$ or 5, 100000 states were randomly chosen from the uniform distribution, whereas, for $N=7$, 30000. In all cases all qubits undergo independent PD channels with dephasing strength $p$. For all $p$, the concentration around the mean gets sharper as the number of qubits grows. Adapted from M.~Tiersch, F.~de~Melo and A.~Buchleitner, \href{http://dx.doi.org/10.1088/1751-8113/46/8/085301}{J. Phys. A: Math. Theor.  {\bf 46}, 085301 (2013)}.}	
\end{figure}
%

More precisely, a function $E$ is said to be Lipschitz continuous if there exists a constant $\eta_E>0$, called the {\it Lipschitz constant of $E$}, such that, for all $\varrho,\sigma\in\mathcal{D}(\mathcal{H}_\sys)$,
\beq
\label{LipschitzContDef}
|E(\varrho)-E(\sigma)|\le \eta_E ||\varrho-\sigma||,
\eeq
where ``$||\ ||$" stands for the trace norm, as usual.  The relative entropy of entanglement \eqref{relentro} or the negativity \eqref{Negativity}, studied in Sec. \ref{2.3}, are Lipschitz continuous entanglement measures,  for instance~\cite{MarkusThesis}.

From \eqref{LipschitzContDef}, and by the contraction property of the trace norm, i.e.~the fact that it is non-increasing under any completely-positive trace-preserving channel~\cite{KarolBook}, one immediately gets that
\beq
\label{ineqchannel}
|E(\mathcal{E}(\ket{\Psi}\bra{\Psi})) - E(\mathcal{E}(\ket{\chi}\bra{\chi}))| \le  \eta_E \kappa_\mathcal{E}||\ket{\Psi}\bra{\Psi}-\ket{\chi}\bra{\chi}||;
\eeq
where $\kappa_\mathcal{E}\geq0$ is a  Lipschitz-like constant, but now for the distance between any two states after the action of the channel $\mathcal{E}$
. When the entanglement decay of state $\ket{\chi}$ under $\mathcal{E}$ is known, inequality \eqref{ineqchannel} allows one to estimate the trajectory of the entanglement of $\ket{\Psi}$ under $\mathcal{E}$, and vice versa. Note that the smaller the distance between the initial states, the better  the estimate. In this way, instead of calculating the entanglement of a given state, one can look for a nearby state with some particular symmetry due to which the entanglement evaluation happens to be simpler, and then exploit \eqref{ineqchannel}.  

Furthermore, \eqref{ineqchannel} gives the necessary ingredient to apply Levy's Lemma (see for instance~\cite{ledoux}), which dictates how Lipschitz continuous functions concentrate around their mean when their argument is sampled randomly. When applied to entanglement dynamics, one sees an exponential concentration around the average entanglement at any time and for any trace-preserving CP map.
More precisely, for any $\epsilon>0$ and total system dimension $d$, the authors obtain the bound
%
\begin{equation} 
\label{eq:Concentration}
\Pr \big( \big|E[\mathcal{E}(\ket{\Psi}\bra{\Psi})] - \langle E_\mathcal{E} \rangle\big| > \epsilon \big)
\leq
4e^{ -C \frac{2d-1}{4\eta_E^2 \kappa_{\mathcal{E}}^2} \epsilon^2},
\end{equation}
where $\langle E_\mathcal{E} \rangle := \int d \psi E[\mathcal{E}(\ket{\psi}\bra{\psi})]$ is the entanglement of $\mathcal{E}(\ket{\psi}\bra{\psi})$ averaged over all $\ket{\psi}$ weighted with the uniform Haar measure $d \psi$, and $C$ is a positive constant that can be taken as $(24 \,\pi^2)^{-1}$~\cite{MarkusThesis}. Bound \eqref{eq:Concentration} tells us that the probability that, for a randomly chosen $\ket{\Psi}$, $E[\mathcal{E}(\ket{\Psi}\bra{\Psi})]$ deviates from  $\langle E_\mathcal{E} \rangle$ by more than $\epsilon$, is exponentially small in $d$ and $\epsilon^2$. Thus, the average $\langle E_\mathcal{E} \rangle$ defines the typical entanglement trajectory, i.e. the one that pure random states follow with the highest probability. Moreover, for large enough $d$, 
%
%
all random states follow essentially the same entanglement trajectory, so that knowing the entanglement dynamics of a single typical state is sufficient to infer, with an exponentially small failure probability, that of any other typical state.  Figure \ref{fig:distributions} depicts the evolution of the distribution of negativities for the bipartition of the first qubit versus the rest, for systems of different number of qubits under independent PD channels. A clear tendency to concentrate around the mean as the system dimension grows can be seen there.


\section{Theory of open-system dynamics of entanglement: multipartite systems}
\label{V}
In the multipartite scenario, the system may have arbitrarily many constituents. In such contexts, system isolation is even more strenuous and decoherence effects can by no means  be neglected. On the other hand, in most quantum optical implementations of multipartite quantum systems, the spatial separation among the different particles is such that each subsystem interacts, up to good approximation, locally with its own environment, and that the different environments do not interact with one another. This type of noise is described by independent maps, as defined in Sec.~\ref{indepmaps}, and is certainly the best-understood case. This is the case we mainly cover in this section. In the next section, however, we discuss some examples of experimental studies with collective noise processes. 
 
\par In this section, we have opted for classifying each different main subject into a single subsection. This means that  papers treating different subjects can be repeatedly acknowledged through different subsections, and that chronological order is sometimes violated.  Before diving into the details of multipartite-entanglement decay though, we first briefly introduce some basic definitions of multipartite-entangled states, as well as some basic facts of Pauli maps. The  familiarised reader may skip the following subsection. Finally, throughout the section, unless explicitly specified, we consider exclusively qubit systems.
\subsection{Preliminaries}
\label{prelimi}
In 2000, D\"ur, Vidal, and Cirac realized \cite{duer00b} that the three-qubit W state 
\begin{equation}
\label{Wdef3}
\ket{{\text{W}}_3}\doteq\frac{1}{\sqrt{3}} \big(\ket{001}+\ket{010}+\ket{100}\big)
\end{equation}
 and Greenberger-Horne-Zeilinger (GHZ) state  \cite{ghz89} 
 \begin{equation}
\label{GHZdef3}
\ket{{\text{GHZ}}_3}\doteq\frac{1}{\sqrt{2}} \big(\ket{000}+\ket{111}\big),
\end{equation}
already defined in Sec. \ref{2.1}, cannot be transformed into one another by SLOCC operations, discussed in Sec. \ref{2.2.2}. Furthermore, they realized that any genuinely-tripartite entangled pure state of three qubits can be  transformed via some SLOCC into either one of these two states.   They thus referred to the W and the GHZ types of entanglement as two {\it inequivalent classes} of genuinely-multipartite entanglement. In fact, such fundamental inequivalence makes itself manifest when one considers the reaction of both families to noise. The extremal example is provided when the states are subject to particle-loss noise. Suppose the system, initially in either $\ket{\text{W}}$ or $\ket{\text{GHZ}}$, looses one particle, mathematically described by tracing that particle out. Since both states are symmetric with respect to the exchange of particles, it does not matter which particle is traced out. For the W state, the remaining two qubits end up in $\varrho=\frac{1}{3} \big(\ket{00}\bra{00}+2\ket{\Psi^+}\bra{\Psi^+}\big)$, where $\ket{\Psi^+}=\frac{1}{\sqrt{2}}(\ket{01}+\ket{10})$, as defined in \eqref{Bellstate}. The reduced two-qubit state $\varrho$ is  NPT. Therefore, it is entangled. On the other hand, for the GHZ state, the remaining qubits goes to the separable state $\frac{1}{2} \big(\ket{00}\bra{00}+\ket{11}\bra{11}\big)$.  This simple example illustrates that W-type entanglement bears a sort of built-in robustness as compared to the GHZ type, something that, as we will see  in the forthcoming subsections, is also manifest against other types of noise.  

\par In addition, the number of inequivalent classes of multiparticle entanglement grows very rapidly with the system size (see Ref. \cite{LiLi2012}, where the SLOCC classification for  pure arbitrary $N$-qubit states has been recently worked out, and references therein). This makes the characterization  of multipartite entanglement decay an even harder problem than the bipartite case. In what immediately follows, we introduce only the most popular families of genuinely-multiqubit entangled states.
\subsubsection{GHZ, generalized GHZ, and generalized GHZ-diagonal states}
\label{GHZprelim}
GHZ states  provide simple models of the celebrated Schr\"odinger-cat
state \cite{schrodinger35,leibfried05}, they are also crucial resources for multipartite quantum communication protocols and distributed computing problems \cite{bose98,hillery99,dhondt05}, and  have been experimentally produced with up to 14 trapped ions \cite{monz2011} and up to 8 photons from parametric down-conversion \cite{pan2012,Huang2012}. As mentioned before, GHZ states were originally introduced by Greenberger, Horne, and Zeilinger  in Ref. \cite{ghz89} for three qubits.  As already defined in \eqref{GHZdef0}, they  straightforwardly extend to $N$ qubits as:

\begin{equation}
\label{GHZdef}
\ket{{\text{GHZ}}_N}\doteq\frac{1}{\sqrt{2}} \big( \ket{00\ldots 0} + \ket{11\ldots 1} \big)
\,.
\end{equation}
Any local unitary spin-flip does not alter the class of entanglement. Thus, formally, the superposition $(\ket{k}+\ket{\bar{k}})/\sqrt{2}$,  where $0\leq k\leq 2^{N-1}-1$ is to be understood in binary representation inside the kets, and $\bar{k}\doteq 2^{N}-1-k$ is its bit-wise complement, follows the spirit of the GHZ construction in precisely the same manner. If, in addition, we include possible phase differences of $\pi$, distinguishing states of even and odd parity, we arrive at the GHZ basis, of elements
\begin{equation}
\label{orthoGHZ}
\ket{\psi_k^{\pm}}\doteq(\ket{k}\pm\ket{\bar{k}})/\sqrt{2}.
\end{equation}
Since they form a complete orthonormal basis, any convex combination of them renders a diagonal state in the basis. Nevertheless, the term {\it GHZ-diagonal states} is typically reserved to only a particular form of combination~\cite{duer99, Dur00}:
\begin{eqnarray}
\label{GHZDrho0}
\rho&=&\sum_{k=0}^{2^{N-1}-1} \big( \mathcal{E}_k^+ \ket{\psi_k^{+}}\bra{\psi_k^{+}}+\mathcal{E}_k^-\ket{\psi_k^{-}}\bra{\psi_k^{-}} \big)
\,,
\end{eqnarray}
with $\mathcal{E}_k^+\equiv\mathcal{E}_k^-$ for all $k\neq0$. The coefficients $\mathcal{E}^\pm_k$ represent the probabilities with which each GHZ state \eqref{orthoGHZ} appears in the mixture, and are therefore positive and sum up to one. These states play an important role in the derivation of lower bounds for the entanglement of arbitrary $N$-qubit density matrices, as will be shown in the following. Mixtures \eqref{GHZDrho0} have the peculiarity that their matrix representation in the computational basis has only one non-null off-diagonal coherence element: 
\begin{equation}
\label{DeltaGHZ}
\Delta\doteq\bra{00\ldots 0}\rho\ket{11\ldots 1}=\mathcal{E}_0^+-\mathcal{E}_0^-. 
\end{equation}

\par This inherent symmetry makes these states remarkable for several reasons. First, it allows for a complete characterization of their entanglement and distillability properties.  For all states of the form \eqref{GHZDrho0}, the following hold \cite{duer99, Dur00, Dur01}:
\begin{itemize}
\item ({\it i}) PPT-ness of any bipartition is necessary and sufficient for biseparability in the bipartition;
\item ({\it ii}) biseparability of all bipartitions implies full separability;
\item ({\it iii}) necessary and sufficient for the distillation of a pure maximally entangled state between any two disjoint blocks of qubits is that each and all of the bipartite splits of the $N$ parts for which the two blocks lie on opposite sides of the split are NPT. 
\end{itemize}
An immediate consequence of property ({\it iii}) is that Criterion \ref{criterion:genNdist}, studied in Sec. \ref{multidistil}, which gives a necessary condition for genuine multipartite distillability of arbitrary states, becomes also sufficient for states \eqref{GHZDrho0}. The same happens, of course, for the extension of Criterion \ref{criterion:genNdist} for blockwise $M$-party distillability. For properties ({\it i})-({\it iii}), the evaluation of PPT-ness is crucial. Fortunately, also this evaluation simplifies enormously for these states. Each $1\leq k\leq 2^{N-1}-2$ naturally gives a bipartite split, defined by the set of  zeros versus the that of ones in the binary representation of $k$. With this identification, PPT-ness evaluations for reduce to:  
\begin{equation}
\label{PPTreduce}
\text{bipartition $k$ of state \eqref{GHZDrho0} is PPT }\Leftrightarrow\mathcal{E}_k\geq\Delta/2.
\end{equation}

Second, it holds that \cite{duer99, Dur00,Dur01}
\begin{itemize}
\item ({\it iv}) any arbitrary $N$-qubit density matrix $\varrho$ can be decohered to the form \eqref{GHZDrho0} without changing any of the diagonal elements $\mathcal{E}_k$ or the off-diagonal element $\Delta$ by a sequence of fully local operations. 
\end{itemize}
This property allows one to lower-bound the entanglement and distillability properties of arbitrary states in terms of those of their locally-dephased GHZ-diagonal correspondent. That is, since neither entanglement nor distillability can increase under LOCCs, one arrives at the following criterion \cite{duer99, Dur00,Dur01}.
\begin{criterion}[Multipartite distillability (sufficient)] 
\label{criterion:DistMDist}
If, for an arbitrary $N$-qubit state $\varrho$,  $\mathcal{E}_k>\Delta/2$ for every bipartition $k$ of $M$ given subsets of qubits, with $2\leq M\leq N$, then the state is blockwise $M$-party distillable with respect to the $M$-partition. In addition, if  $\varrho$ is of the form \eqref{GHZDrho0}, then the converse implication is true too.
\end{criterion}

\par A further generalization of \eqref{orthoGHZ} consists of allowing for different amplitudes, which defines the {\em generalized GHZ states} \cite{aolita08,aolitapra09}
\begin{equation}
\label{kalphabeta}
\ket{\psi_k^{\pm}(\alpha,\beta)}\doteq\alpha\ket{k}\pm\beta\ket{\bar{k}},
\end{equation}
with  $\alpha$ and $\beta$  any complex amplitudes such that $|\alpha|^2+|\beta|^2=1$. Because $\alpha$ is  not necessarily equal to $\beta$, one now does not run into redundancies by letting $k$ run from $0$ to $2^{N}-1$, instead of $2^{N-1}-1$ as in \eqref{orthoGHZ} and \eqref{GHZDrho0}. So, in \eqref{kalphabeta}, one has $0\leq k\leq 2^{N}-1$ and $\bar{k}\doteq 2^{N}-1-k$. For each $\alpha$ and $\beta$, the $2^{N+1}$ generalized GHZ states  \eqref{kalphabeta} form an over-complete non-orthogonal basis of ${\mathcal H}_{\sys}$. Arbitrary mixtures of them have in turn  been dubbed {\em generalized GHZ-diagonal states} \cite{aolitapra09}.
\begin{eqnarray}
\label{GGHZDrho0}
\nonumber
\rho&=&\sum_{k=0}^{2^{N}-1} \big( \mathcal{E}_k^+ \ket{\psi_k^{+}(\alpha,\beta)}\bra{\psi_k^{+}(\alpha,\beta)}\\
&+&\mathcal{E}_k^-\ket{\psi_k^{-}(\alpha,\beta)}\bra{\psi_k^{-}(\alpha,\beta)} \big)
\,,
\end{eqnarray}
as they generalize GHZ-diagonal states \eqref{GHZDrho0} ($\mathcal{E}_k^+$ and $\mathcal{E}_k^-$ are no longer necessarily equal for all $k\neq0$) . In contrast to the latter, the computational-basis matrix representation of \eqref{GGHZDrho0} can have all the anti-diagonal elements different from zero. However, as we discuss in Sec. \ref{CompleteGHZ}, analytical scaling laws for the decay under independent noise of the entanglement in their bipartitions can still be derived~\cite{aolitapra09}. Furthermore, their genuine-multipartite entanglement can also be fully characterized, since the necessary Criterion \ref{criterion:BisepGHZ}, discussed in Sec. \ref{BisepCrit}, becomes also sufficient for generalized GHZ-diagonal states \cite{GuehneSeevinck}. Its full $N$-qubit version takes the following form \cite{GuehneSeevinck, Huber10}:

\begin{criterion}[Biseparability of $N$ qubits (GHZ)] 
\label{criterion:BisepGHZN}
If a $N$-qubit state $\varrho$ is biseparable, then 
\begin{equation}
\label{BisepcritN}
\mathfrak{D}^{\ket{\text{GHZ}_N}}(\varrho)\leq\sum_{k=1}^{2^{N-1}-1}\sqrt{\varrho_k\varrho_{\overline{k}}},
\end{equation}
where $\mathfrak{D}^{\ket{\text{GHZ}_N}}(\varrho)\doteq|\bra{00\ldots 0}\varrho\ket{11\ldots 1}|$,  $\varrho_k\doteq\bra{k}\varrho\ket{k}$, and $\varrho_{\overline{k}}\doteq\bra{\overline{k}}\varrho\ket{\overline{k}}$.  In addition, if $\varrho$ is of the form \eqref{GGHZDrho0}, then the converse implication is true too.
\end{criterion}
Notice that all the diagonal matrix elements appear in the sum except for the ones corresponding to 0 ($k=0$)  and $N$ ($k=2^{N}-1$) excitations. As in the three-qubit case of Sec.~\ref{BisepCrit}, the above criterion is valid for all states, and the presence of $\ket{\text{GHZ}_N}$ just makes reference to the fact that inequality \eqref{BisepcritN} is violated specially by states in the vicinity of $\ket{\text{GHZ}_N}$. 
 
Criterion \ref{criterion:BisepGHZN} is more powerful than Criterion \ref{criterion:DistMDist} above in that it addresses multipartite entanglement, instead of distillability addressed by the latter. Both criteria however bear the common advantageous property that only the diagonal elements, plus a single off-diagonal element, of the density matrix are required for their evaluation. This precisely has been the crucial feature that enabled to use these criteria in practice to, for instance, corroborate the presence of both multipartite entanglement and distillability in experimental GHZ states of $N$ as high as 14 \cite{monz2011}, as described in Sec. \ref{67}.

\subsubsection{Basics of independent maps on generalized-GHZ states}
\label{ActionIndependentGHZ}

It is immediate to check that the computational-basis matrix representation of the state $\varrho$ resulting from action of any independent Pauli channel on a  pure state $\ket{\psi_k^{\pm}(\alpha,\beta)}$ satisfies the following: ({\it i}) no off-diagonal element increases its absolute value; and ({\it ii}) every diagonal element $\varrho_k$ is equal to its complementary element $\varrho_{\overline{k}}$, for all $0\leq k\leq 2^{N}-1$. This means that  these single-qubit channels always dephase generalized GHZ states into GHZ-diagonal states\footnote{This is not to be mistaken for property ({\it iv}) of Sec.~\ref{GHZprelim}, for here the values of the diagonal and off-diagonal elements do in general change.}. This simple observation turns out to be of great help in the characterization of GHZ-entanglement dynamics under  physically-relevant types of noise.

\par Any GHZ-diagonal state can be decomposed in the convenient fashion 
\begin{eqnarray}
\label{EntSep}
\rho= \mathcal{E}_{\text{ent}}\varrho_{\text{ent}}+\mathcal{E}_{\text{sep}}\varrho_{\text{sep}}
\,,
\end{eqnarray}
where $\mathcal{E}_{\text{ent}}$ and $\mathcal{E}_{\text{sep}}$ are non-negative probabilities that sum up to 1, $\varrho_{\text{ent}}$ is some entangled state, and $\varrho_{\text{sep}}$ is some separable state diagonal in the computational basis. For the particular cases of depolarizing (D) and phase-damping (PD) channels the calculation is simple~\cite{aolitapra09} and yields in both cases $\mathcal{E}_{\text{ent}}=(1-p)^N$ and $\varrho_{\text{ent}}=\ket{\psi_k^{\pm}(\alpha,\beta)}\bra{\psi_k^{\pm}(\alpha,\beta)}$. 

On the other hand, for the generalized amplitude damping (GAD) channel, whereas  the first statement above is true in any temperature regime, the second is not, as the relation between the populations of $\ket{k}$ and $\ket{\tilde{k}}$ varies with the temperature. Still, it turns out~\cite{aolitapra09} to be possible to decompose any pure state $\ket{\psi_k^{\pm}(\alpha,\beta)}$ under independent GAD noise as in \eqref{EntSep}. The expressions for the corresponding coefficient and state, $\mathcal{E}^{\text{GAD}}_{\text{ent}}$ and $\varrho^{\text{GAD}}_{\text{ent}}$, respectively, are a bit cumbersome but their calculation is detailed in Ref.~\cite{aolitapra09}. 

\subsubsection{Graph and graph-diagonal states}
\label{Graphprelim}
Graph states~\cite{Hein04,Hein_Review} are  quantum states associated to mathematical graphs. They
constitute an important class of genuinely multipartite entangled states with multiple applications in quantum information and communication. The most popular examples are the cluster states, which have been identified as universal resources for measurement-based one-way quantum computation~\cite{Raussendorf01,Briegel09}. However, other members of this family can be used
as codewords for quantum error correction~\cite{Schlingemann01}, to
implement secure quantum communication protocols
\cite{dur05,chen07}, or to simulate the
entanglement distribution of random states~\cite{Dahlsten06}, for instance.
Moreover,  GHZ states \eqref{GHZdef} constitute also a particular example of this general family, as they are local-unitarily equivalent to the graph states associated to both the fully-connected or the star-like graphs~\cite{Hein04,Hein_Review}. For all these reasons, a
great effort has been made both to theoretically understand their
properties~ \cite{Hein04,Hein_Review} and to create
and coherently manipulate them experimentally~\cite{Walther05,Kiesel05,lu07,Chen07b,Vallone08}, including the realization of different graph states of up to 14 ions \cite{monz2011} or 8 photons \cite{Yao2012,pan2012,Huang2012}.

\par A mathematical graph $\mathcal{G}\equiv\mathcal{G}_{(\mathcal{V},\mathcal{E})}\equiv\{\mathcal{V},\mathcal{E}\}$ is defined by a set $\mathcal{V}$ of vertices, or nodes, and a set $\mathcal{E}$ of edges connecting pairs of vertices in $\mathcal{V}$. Examples of such graphs are shown in Figs. \ref{GraphFig} and \ref{Fig2}. To every mathematical graph we associate a physical {\it graph state}, operationally defined for qubits as follows. To each vertex $i\in\mathcal{V}$ we associate a qubit, we initialize all $N$ qubits in
the product state $\ket{{g_{(\mathcal{V})}}_0}\doteq\bigotimes_{i
\in\mathcal{V}} \ket{+_i}$, being
$\ket{+_i}\doteq(\ket{0_i}+\ket{1_i})/\sqrt{2}$,  and, to all pairs
$\{i,j\}$ of qubits  joined by an edge, we apply a
maximally-entangling controlled-$Z$ ($CZ$) gate,
$CZ_{ij}=(\ket{0_i}\bra{0_i}\otimes \openone_j + \ket{1_i}\bra{1_i}\otimes Z_j)\otimes\openone_{\overline{ij}}$, with $\openone_j$ and $\openone_{\overline{ij}}$ the identity operators on qubits $j$ and all qubits but $i$ and $j$, respectively.
The result is the $N$-qubit graph state \cite{Hein04,Hein_Review}
\begin{equation}
\label{graphoperdef}
\ket{\mathcal{G}_{0\ ...\ 0}}\doteq\prod_{\{i,j\} \in \mathcal{E}} CZ_{ij}\ket{{g_{(\mathcal{V})}}_0}.
\end{equation}

\par There exists  an alternative unambiguous definition in terms of their {\it parent Hamiltonians}. Consider the $N$-qubit  Hamiltonian
\begin{equation}
\label{H}
H=-\frac{1}{2}\sum_{i=1}^N \Delta_i S_i,
\end{equation}
where $\Delta_i>0$ are arbitrary coupling strengths (in arbitrary units). Operators 
\begin{equation}
\label{S}
S_i\doteq X_i\otimes\bigotimes_{j\in\mathcal{N}_i} Z_{j},
\end{equation}
with $X_k$ and $Z_k$ the usual Pauli operators acting on qubit $k$, and  $\mathcal{N}_k$ the first neighbours of $k$, i. e.  directly connected to it by some edge,   generate a group with respect to the operator multiplication, called the {\it stabilizer}. We will abuse notation by referring to the generators of the stabilizer group as stabilizer operators. All $N$ stabilizer operators have eigenvalues 1 and $-1$ and commute with each other, so that their $2^N$ common eigenstates form a complete orthonormal basis of the $N$-qubit Hilbert space in question. Thus, since $\Delta_i>0$ for all $i$, Hamiltonian \eqref{H} has a unique ground state. Graph state \eqref{graphoperdef} happens to be the unique common eigenstate of all $N$ stabilizer operators \eqref{S} with eigenvalue 1, and so the unique ground state of Hamiltonian \eqref{H} \cite{Hein04,Hein_Review}. 
The global ground state of the composite Hamiltonian is the ground state of each interaction term of \eqref{H}. Any Hamiltonian with this property is said to be {\it frustration-free}.
The energy difference between the first excited and the ground states of \eqref{H} is in turn given by $\min_{j}{\Delta_j}$, which defines the {\it energy gap} of the Hamiltonian.

Including $\ket{\mathcal{G}_{0\ ...\ 0}}$, we label the $2^N$ orthonormal eigenstates of \eqref{H} by $\ket{\mathcal{G}_{\mu}}$. Subindex $\mu$, without reference to any individual qubit, represents the binary string of individual subindices $\mu\equiv\mu_1\ ...\ \mu_N$, as usual. These states are  related via the local-unitary transformation 
\begin{equation}
\label{graphlocalunit} 
\ket{\mathcal{G}_{\mu}}\doteq\bigotimes_{i=1}^N {Z_{i}}^{\mu_i}\ket{\mathcal{G}_{0\ ...\ 0}},
\end{equation}
so they all possess exactly the same entanglement properties. As mentioned, they define a complete orthonormal basis of $\mathcal{H}_{\sys}$, called the graph-state basis for graph $\mathcal{G}$. Any state diagonal in this basis defines, in turn,  a {\it graph-diagonal state}:
 \begin{eqnarray}
\label{graphdiagonal} 
\rho=\sum_{\mu} p_{\mu}
\ket{\mathcal{G}_{\mu}}\bra{\mathcal{G}_{\mu}},
\end{eqnarray}
where  $p_{\mu}$ is some probability distribution. Any arbitrary $N$-qubit state can always be decohered by some separable map into the form \eqref{graphdiagonal}~\cite{Duer03, Aschauer05}. 

Finally, in Ref. \cite{guehne11b}, G\"uhne {\it et al}. presented a necessary and sufficient criterion for biseparability of four-qubit 1D-graph-diagonal states of a similar form as that of Criteria \ref{criterion:BisepGHZN} and \ref{criterion:BisepWDICKE4}. In addition, the authors showed that, for four-qubit 1D-graph-diagonal, as well as $N$-qubit $Y$-graph-diagonal, states, biseparability and PPT mixture are equivalent. That is, for these family of states, Criterion \ref{criterion:PPTmixtures} of Sec. \ref{2.2.2}, necessary for biseparability, is also sufficient; so that the solid red and dashed blue convex hulls in Fig. \ref{guehne} collapse.

\subsubsection{Basics of Pauli noise on graph-state and generalized-GHZ entanglement}
\label{Sec:ActionPauliGraph}
As mentioned in Sec. \ref{Graphprelim}, every graph state \eqref{graphlocalunit} is, by definition,  eigenstate of generator \eqref{S} with eigenvalues ${s_i}_{\mu}=1$ or $-1$. It follows then that $X_i\ket{\mathcal{G}_{\mu}}\equiv{s_i}_{\mu}X_i.X_i\otimes\bigotimes_{j\in\mathcal{N}_i} Z_{j}\ket{\mathcal{G}_{\mu}}\equiv{s_i}_{\mu}\bigotimes_{j\in\mathcal{N}_i} Z_{j}\ket{\mathcal{G}_{\mu}}$ and $Y_i\ket{\mathcal{G}_{\mu}}\equiv{s_i}_{\mu}Y_i.X_i\otimes\bigotimes_{j\in\mathcal{N}_i} Z_{j}\ket{\mathcal{G}_{\mu}}\equiv{s_i}_{\mu}(-i)Z_{i}\otimes\bigotimes_{j\in\mathcal{N}_i} Z_{j}\ket{\mathcal{G}_{\mu}}$. This implies that any Pauli map $\mathcal{E}$, defined by Kraus operators~\eqref{PauliKraus}, acting on graph states \eqref{graphoperdef}, or graph-diagonal states \eqref{graphdiagonal}, is equivalent to another, modified Pauli map  $\tilde{\mathcal{E}}$. The latter,  in turn, possesses modified Kraus operators $\tilde{K}_\mu$, which are obtained from $K_\mu$ by the substitutions 
\begin{subequations}
\label{subsPauli}
\begin{align}
X_i&\rightarrow\bigotimes_{j\in\mathcal{N}_i} Z_{j},\ \  \text{and}\\
Y_i&\rightarrow Z_i\otimes\bigotimes_{j\in\mathcal{N}_i} Z_{j},
\end{align}
\end{subequations}
 and then regrouping repeated terms. Notice that different  $K_\mu$'s often contribute to a single $\tilde{K}_\mu$. Since  a $Z$  operator acting on any graph (or graph-diagonal) state always yields another graph (or graph-diagonal) state, it is trivial to see that $\tilde{\mathcal{E}}$ takes any initial graph (or graph-diagonal) state to a (another) graph-diagonal state \cite{duer04,hein05}. For example, $\ket{\mathcal{G}_0}$ transforms as
\begin{eqnarray}
\label{ActionPauliGraph} 
\mathcal{E}\proj{\mathcal{G}_0}\equiv\tilde{\mathcal{E}}\proj{\mathcal{G}_0}\doteq\sum_{\mu} \tilde{p}_{\mu}
\ket{\mathcal{G}_{\mu}}\bra{\mathcal{G}_{\tilde{\mu}}},
\end{eqnarray}
where $\tilde{p}_{\mu}$ is the probability corresponding to each $\tilde{K}_\mu$ in the modified map $\tilde{\mathcal{E}}$. The explicit dependence of $\tilde{p}_{\mu}$ with the probabilities of each Kraus operator in  $\mathcal{E}$ can be worked out  and takes a rather simple form (see Sec. II.B.1 of Ref. \cite{chirag10}).

\par In addition, the equivalence also holds  for the generalized-GHZ family: Pauli maps take states \eqref{kalphabeta} and \eqref{GGHZDrho0} into modified generalized GHZ-diagonal states~\cite{aolitapra09}.
\subsubsection{Local dephasing and thermalization as equivalent mechanisms for graph states}
\label{LocalDephisThermalization}
When the Pauli channel corresponds to independent dephasing, a curious effect takes place. Namely, the independent phase-damping channel $\mathcal{E}^{PD}$ always drives graph states towards a thermal state. More precisely,  the thermal graph state $\varrho_T\doteq e^{-H/T}/{\rm Tr}\left(e^{-H/T}\right)$ at temperature $T$ (in units of Boltzmann constant), with $H$ the graph Hamiltonian \eqref{H}, is equivalent to the independently-dephased graph state
\begin{equation}
\label{decoherence1}
\rho_{T}\equiv\mathcal{E}^{PD}\ket{\mathcal{G}_{0\ ...\ 0}}\bra{\mathcal{G}_{0\ ...\ 0}}.
\end{equation}  
Here,  $\ket{\mathcal{G}_{0\ ...\ 0}}$ is the graph state \eqref{graphoperdef} and $\mathcal{E}^{PD}\equiv\mathcal{E}^{PD}_1\otimes\hdots \otimes\mathcal{E}^{PD}_N$, with $\mathcal{E}^{PD}_i$ the single-qubit  PD channel on qubit $i$ with dephasing strength
\begin{equation}
\label{probability}
p_i\equiv\frac{2}{1+e^{\Delta_i/T}}.
\end{equation}
The equivalence was first reported by Raussendorf  {\it et al}., in \cite{Raussendorf05}, and Kay {\it et al}., in \cite{kay06}, for the case of constant couplings $\Delta_i\equiv \Delta$ for all $i$, and later by Cavalcanti {\it et al}., in \cite{cavalcanti10}, for the arbitrary-coupling case. It can be straightforwardly demonstrated by expanding $\rho_{T}$ in the eigenbasis $\{\ket{\mathcal{G}_{\mu}}\}$ of Hamiltonian \eqref{H} and explicitly evaluating $\mathcal{E}^{PD}\ket{\mathcal{G}_{0\ ...\ 0}}\bra{\mathcal{G}_{0\ ...\ 0}}$ in terms of the Kraus decomposition of the PD channel, given in Table~\ref{tabKraus}, and with the help of relationship \eqref{graphlocalunit}.

The equivalence  extends analogously to continuous-variable graph states under independent diffusion along the $q$ quadrature in phase space, which plays the role of the $Z$ quit direction, as reported by Aolita {\it et al}. in \cite{aolita2011}. It constitutes a  connection between the processes of thermalization and independent dephasing.

\subsubsection{W and Dicke states}
\label{W&Dicke}
W-states \eqref{WdefN0} describe $N$ spin-$1/2$ particles with a single excitation coherent and equiprobably shared among all $N$ pairs. They can be used to solve distributed computing problems in quantum networks \cite{dhondt05} and have been experimentally generated with up to $N=8$ trapped ions \cite{haeffner05} and $N=4$ photons \cite{Prevedel09}\footnote{In atomic-ensemble-based quantum memories, single photons are collectively stored among about $10^9$ atoms, in superpositions similar to W states but with non-equal amplitudes. Such quantum memories have been experimentally demonstrated in for instance Refs. \cite{Julsgaard04, choi08,Clausen11}.}. In addition, apart from their robustness against particle loss, they can feature more robust violations (against noise) of Bell inequalities than the GHZ states \cite{SenDe03}. Over the forthcoming subsections, we discuss  how this intrinsic robustness typically also applies for their entanglement.

The three-qubit definition \eqref{Wdef3} extends straightforwardly to the $N$-qubits as  \eqref{WdefN0}:
\begin{equation}
\label{WdefN}
\ket{{\text{W}}_N}\doteq\frac{1}{\sqrt{N}}\sum_{l}\ket{P_l(00\ ...\  01)},
\end{equation}
where $P_l(00\ ...\  01)$ is the $l$-th permutation of the bit string $00\ ...\  01$, and the summation goes over all possible permutations.

\par On the other hand, definition \eqref{Wdef3} can not only be extended to arbitrary $N$, but also to the case of  arbitrary number $k<N$ of excitations. This gives the so-called symmetric Dicke states \cite{Dicke54} of $k$ excitations
\begin{equation}
\label{DickedefN}
\ket{{\text{Dicke}}^k_N}\doteq\frac{1}{{N\choose k}}\sum_{l}\ket{P_l(\underbrace{00\ ... \ 0}_{N-k}\underbrace{11\ ... \ 1}_{k})}.
\end{equation}
Dicke states are simultaneous eigenstates of the squared total angular momentum operator $J^2\doteq(J_1+J_2+J_3)^2$, of maximal eigenvalue $\frac{N}{2}(\frac{N}{2}+1)$, with $J_i\doteq\frac{1}{2}\sum_{j=1}^{N}\sigma^i_j$ (we take $\hbar\equiv1$), and of the total $Z$ angular momentum operator $J_3$, with eigenvalue $ (N-2k)/2$. They have been experimentally produced with up to $N=6$ photons \cite{Prevedel09}. Of course, W states are Dicke states of a single excitation.

Biseparability necessary Criterion \ref{criterion:BisepW}, tailored to test for genuine multipartite entanglement in the vicinity of three-qubit W states, can also be extended to four-qubit W and Dicke states \cite{GuehneSeevinck}. These give the following.
\begin{criterion}[Biseparability of $4$ qubits (W and Dicke)] 
\label{criterion:BisepWDICKE4}
Any $4$-qubit biseparable state $\varrho$ fulfills 
\begin{equation}
\label{BisepcritW4}
\mathfrak{D}^{\ket{{\text{W}}_4}}(\varrho)\leq\sum_{|k|=2}\sqrt{\varrho_{0000}\varrho_{k}}+\sum_{|k|=1}\varrho_{k},
\end{equation}
as well as 
\begin{equation}
\label{BisepcritDicke4}
\mathfrak{D}^{\ket{{\emph{Dicke}}^2_4}}(\varrho)\leq\sqrt{\varrho_{0000}\varrho_{1111}}+\sum_{|k|=1,|l|=3}\sqrt{\varrho_{k}\varrho_{l}}+\frac{3}{2}\sum_{|k|=2}\varrho_{k},
\end{equation}
where $\mathfrak{D}^{\ket{\Psi}}(\varrho)$ is the sum of the absolute values of the off-diagonal elements in the upper triangle of density matrix $\varrho$ for which the corresponding matrix entries in $\ket{\Psi}\bra{\Psi}$ are not null, and where $\varrho_k\doteq\bra{k}\varrho\ket{k}$.
\end{criterion}
Violation of either \eqref{BisepcritW4} or \eqref{BisepcritDicke4} is a sufficient condition for genuine 4-partite entanglement. Condition \eqref{BisepcritW4} is most commonly violated in the vicinity of $\ket{{\text{W}}_4}$, whereas \eqref{BisepcritDicke4} in the vicinity of $\ket{{\text{Dicke}}^2_4}$.

\subsection{Decay of GHZ entanglement}
\label{GHZandmore}

\subsubsection{Entanglement lifetime under local  noise}
\label{Simon}
\par  The first study of the scaling behaviour with $N$ of the entanglement of multipartite systems under independent open-system dynamics was performed in the seminal paper \cite{simon02} by Simon and Kempe. The authors analyzed multiqubit systems under the influence of single-qubit depolarization. They studied the times at which the different bipartitions of W and Dicke states of 3 and 4 qubits,  as well as GHZ states of arbitrary  $N$, become PPT, and the times of survival of entanglement for spin-squeezed states \cite{Kitagawa93,Sorensen01}. 

\par Since channel D drives GHZ states towards GHZ-diagonal states, for which PPT-ness of all bipartitions implies separability, as discussed in Sec. \ref{GHZprelim}, the studies of \cite{simon02} captured the full-separability properties of independently depolarized GHZ states. The authors noticed that the more balanced --i.e. the closer to half-versus-half-- a  bipartite splitting is, the further away from being PPT it is. This can be expressed in terms of the negativity $Neg_{k}(p)$ of a splitting of $k$ versus $N-k$ qubits by the following identity
\begin{eqnarray}
\label{NegPartitions} 
Neg_{1}(p)\leq Neg_{2}(p)\leq \hdots\ \leq Neg_{N/2}(p),
\end{eqnarray}
for all depolarization strengths $p$. Therefore, the critical probability $p_{c}^{\text{D}}$ at which the fully balanced half-versus-half splittings become PPT, gives the disentanglement time, at which the transition from entangled to fully separable takes place. The authors found that such time is always finite ($p_{c}^{\text{D}}<1$) and, curiously, that it grows with $N$.

\par Later on, D\"ur and collaborators performed an exhaustive study of GHZ states under different models of independent decoherence, briefly described in Ref.~\cite{duer04} and elaborated in detail in \cite{hein05}. Those studies focused on genuinely multipartite aspects though, and are discussed in Sec. \ref{Duer}.

\par More recently, Aolita {\it et al}. derived, in Ref. \cite{aolita08}, analytical expressions for the disentanglement times of generalized GHZ states $\ket{\psi_0^{\pm}(\alpha,\beta)}$ under channels D, PD and GAD, for the purely dissipative and fully diffusive regimes, as a function of  $\alpha$, $\beta$ and $N$. Under channel D, these states display finite-time disentanglement for all  $\alpha$ and $\beta$:
\begin{equation}
\label{ESDDC} 
p_{c}^{\text{D}}=1-\big(1+4|\alpha\beta|^{2/N}\big)^{-1/2}.
\end{equation}
For PD, the authors found the disentanglement time
\begin{equation}
\label{ESDPD} 
p_{c}^{\text{PD}}=1,
\end{equation}
for all parameter-regimes. That is,  PD never induces a sudden disappearance of entanglement, it drives the state to separability only in the asymptotic limit $t\to\infty$. On the other hand, channel GAD does not map $\ket{\psi_0^{\pm}(\alpha,\beta)}$ into GHZ-diagonal states, but PPT-ness of any bipartite split can still be proven to imply its biseparability \cite{aolita08}. Furthermore, for the purely dissipative case of AD, PPT-ness of all splits still implies full separability \cite{aolita08}.  Thus, for AD, two regimes for the disentanglement time $p_{c}^{\text{AD}}$ were encountered. For $\alpha\geq\beta$, entanglement vanishes only asymptotically, whereas for $\alpha<\beta$, disentanglement happens always at a finite time:
\begin{equation}
\label{ESD@T=0}
p_{c}^{\text{AD}}=\min\{1,|\alpha/\beta|^{2/N}\}.
\end{equation}
The heuristic explanation for these two different reactions before AD is the same as in the bipartite case discussed in Sec. \ref{EntDynaAD}. Only excited state $\ket{1}$ couples to a zero-temperature reservoir, because all the vacuum can do is absorb excitations from the system. Therefore, the larger the amplitude $\alpha$ of $\ket{00\ldots 0}$, the weaker the effective coupling to the environment and the longer entanglement persists. In contrast, the time $p_{c}^{\text{Diff}}$ at which $\ket{\psi_0^{\pm}(\alpha,\beta)}$ under GAD in the fully diffusive infinite-temperature limit becomes biseparable with respect to all its bipartite splits is finite for all $\alpha$ and $\beta$:
\begin{equation}
\label{ESD@T=infty}
p_{c}^{\text{Diff}}=1+2|\alpha\beta|^{2/N}-\sqrt{1+4|\alpha\beta|^{4/N}}\,.
\end{equation}

\par In addition, An and Kim studied, in Ref. \cite{ba_an09}, generalized GHZ states \eqref{kalphabeta} under the composed action of both PD and AD simultaneously, and for different environment-coupling strengths for each qubit. Also in this case, disentanglement is always found to happen at a finite time.

It is clear from \ref{ESDDC}, \ref{ESD@T=0}, and \ref{ESD@T=infty} that the critical value $p_c$ increases with the number $N$ or qubits. In the first two cases, $p_c\rightarrow1$ when $N\rightarrow\infty$, that is, in this limit the entanglement breaking time becomes infinite. On the other hand, in the diffusive limit, $p_c^{\text{Diff}}\rightarrow 3-\sqrt{5}$ when $N\rightarrow\infty$, so one always have a finite entanglement breaking time. These results imply that, as opposed to coherence, the decay of entanglement does not follow an exponential law.

\par As in the bipartite case, some geometrical intuition of when finite-time disentanglement can arise comes from considering the topological structure of the set of density operators $\mathcal{D}(\mathcal{H}_\sys)$, as discussed in Sec. \ref{TopoAp}. There, we saw that any dynamic process with a  steady state in the interior of the separability set necessarily induces finite-time disentanglement on all initial states, since the intersection between the state trajectory and the separability border must happen at some $p<1$. We saw that this is the case for D and GAD, both at infinite temperature as well as at any finite positive temperature. 

\par We also mentioned that, in contrast, when the steady state lies on the boundary of the separability set, the dynamic trajectory may never enter the interior of  the separability set and an intuitive picture is not so straightforward. The same conclusions drawn in Sec. \ref{TopoAp} for PD and AD acting on the bipartite state $\alpha\ket{00}+\beta\ket{11}$ hold also for the generalized GHZ states $\ket{\psi_0^{+}(\alpha,\beta)}$: On the one hand, for PD the state is never full rank, so that the whole trajectory takes place on the boundary of $\mathcal{D}(\mathcal{H}_\sys)$. So, disentanglement is always asymptotic, at $p=1$. On the other hand, for PD and AD simultaneously, the state is full rank for all $0<p<1$, goes towards pure state $\ket{00\ldots 0}$, and disentanglement happens at $p<1$. However, also for AD alone is the state full rank for all $0<p<1$ and goes towards $\ket{00\ldots 0}$, but the disentanglement time is determined by the initial state.

\par Finally, in the examples of D, AD (for $\beta>\alpha$), and GAD at infinite temperature, where finite-time disentanglement is seen and analytical expressions are available, the disentanglement times \eqref{ESDDC} and \eqref{ESD@T=infty}, as well as the vanishing time of bipartite entanglement \eqref{ESD@T=infty}, all grow with the system's size\footnote{On the other hand, for PD composed with AD, as studied in Ref. \cite{ba_an09}, for the particular case of equal environment-coupling strengths, one can show that the disentanglement time is independent of $N$.}. This could be interpreted as the entanglement robustness growing with $N$. We see, however, in the following subsections, that this is actually far from being true.
\subsubsection{Lifetime of genuine multipartite entanglement under local noise}
\label{Duer} 
D\"ur {\it et al}. provided in  \cite{duer04,hein05} a new insight as to how independently-depolarized GHZ states loose entanglement. The authors found that the least robust bipartitions of one qubit versus the rest become PPT at a critical time that decreases with $N$. This time corresponds to the instant when the state becomes separable with respect to the least stable bipartitions, and therefore sets an upper bound on the lifetime of genuine $N$-partite entanglement. This bound revealed that, in spite of the disentanglement time increasing with $N$, genuine $N$-partite entanglement in GHZ states under local depolarization vanishes at a time that actually decreases with $N$. The exact dependence of this critical $p$ with $N$ is given by the solution of a polynomial equation of order $N$, which can be numerically obtained efficiently.

With Criterion \ref{criterion:BisepGHZN} (Sec. \ref{GHZprelim}), the calculation of the exact times at which GHZ states under independent noise cease to display genuine  $N$-partite entanglement became possible for relevant noise types. For channels D, PD, and AD, for instance, the diagonal elements $\varrho_k$ and the off-diagonal element  $\bra{00\ldots 0}\varrho\ket{11\ldots 1}$ can be calculated explicitly \cite{aolita08} and inequality \eqref{BisepcritN} be checked for. This immediately yields that GHZ states under PD are biseparable only at the asymptotic time \eqref{ESDPD}, in the  steady state, when they become fully separable too. In turn, the time of biseparability of GHZ states under D is again given by the solution of polynomial equation of order $N$, which can be efficiently solved for each $N$. Finally, even though GHZ states under AD are not generalized GHZ-diagonal, and thus the  criterion only gives a necessary condition for biseparability, based on it, G\"{u}hne and Seevinck  derived the exact biseparability time \cite{GuehneSeevinck}: $p_{bs}^{\text{AD}}= (2 ^{N -1}-1)^{-2/N}$.
\subsubsection{Lifetime of genuine multipartite entanglement under global noise}
\label{GlobalDepol}
\begin{figure}
\begin{center}
\includegraphics[width=1\linewidth]{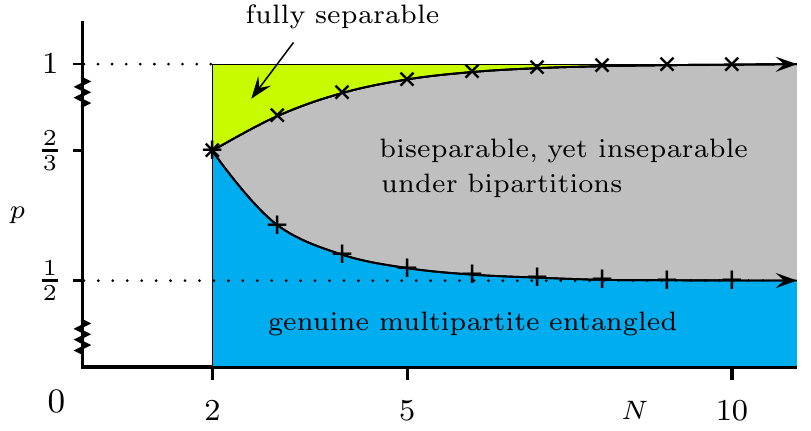}
\caption{
\label{GS}
Complete entanglement classification of $N$-qubit GHZ states under global white noise of strength $p$. The three different regions correspond to markedly different entanglement properties: In the blue region (bottom), for $0\leq p\leq p_{bs}^{\text{GWN}}$, the states are genuine multipartite entangled. In the grey one (middle), for $p_{bs}^{\text{GWN}}\leq p\leq p_{c}^{\text{GWN}}$, the states are no longer genuine multipartite entangled but are still entangled with respect to each and all of their bipartitions. Finally, in the yellow one (top), for $p_{c}^{\text{GWN}}\leq p\leq 1$, the states are fully separable. Even though quantitatively different, the behavior is qualitatively similar to the one observed for independent noise models, in the sense that the full-separability time grows with $N$ whereas the bi-separability one decreases with $N$. From O. G\"uhne and M. Seevinck, \href{http://dx.doi.org/10.1088/1367-2630/12/5/053002}{New J. Phys. {\bf 12}, 053002  (2010)}.}
\end{center}
\end{figure}

Criterion \ref{criterion:BisepGHZN} made it also possible to complete the entanglement classification of GHZ states under the collective depolarizing channel, also called global white noise (GWN). Already in 2000, the critical time for full separability of these states was known \cite{Dur00,Schack00} to be $p_{c}^{\text{GWN}}= (1 +2^{1-N })^{-1}$. With Criterion \ref{criterion:BisepGHZN}, their time of biseparability can be immediately obtained \cite{GuehneSeevinck}: $p_{bs}^{\text{GWN}}=[2(1 -2^{-N })]^{-1}$. The resulting classification is graphically represented in Fig. \ref{GS}, where the three regions of different entanglement properties, whose borders are defined by $p_{c}^{\text{GWN}}$ and $p_{bs}^{\text{GWN}}$, are plotted.

Finally, it is also interesting to note that a method to compute the exact value of the geometric measure of entanglement \eqref{geometricmeasure}
for GHZ and linear-cluster states of arbitrary $N$ under collective dephasing is described in  Ref. \cite{Guehne08}.

\subsubsection{Lifetime of blockwise $M$-party distillability under local Pauli maps}
\label{DuerMpartite}
In Ref.~\cite{hein05}, Hein {\it et al}. also studied the vanishing time of blockwise $M$-party distillable entanglement, with $M<N$, of $N$-qubit GHZ states subject to channel D. For blockwise $M$-party distillability, the system is divided into $M$ blocks of qubits, each of which is treated as a single subpart of larger dimension with respect to which LOCCs are defined, as discussed in Sec. \ref{multidistil}.

Based on Criterion \ref{criterion:DistMDist} (Sec. \ref{GHZprelim}), the authors found that the vanishing time of blockwise $M$-party distillability with respect to any $M$-partitioning coincides with the disentanglement time of the bipartition of the smallest of the $M$ subgroups versus all the rest. For example, for $M$-partitionings where the smallest subgroup is just a single qubit, the vanishing time of blockwise $M$-distillability coincides with the upper bound on the time of biseparability mentioned in the beginning of Sec. \ref{Duer}. This is due to the fact that the bipartitions of one qubit versus the rest are the first ones to disentangle.

\par Apart from channel D, blockwise $M$-distillability of $N$-qubit noisy GHZ states was studied by Bandyopadhyay and Lidar in Ref. \cite{Bandyopadhyay04}, for the case of arbitrary independent Pauli maps. The authors found conditions under which the parity of $N$ can lead to different $M$-distillability properties with respect to those of independent depolarization.

\subsubsection{Full  dynamics of concurrence for few qubits under local noise}
\label{Carvalho}
\par So far we have studied disentanglement times for different types of entanglement, without regard of how entanglement evolves before disentanglement. The first studies that assessed the entire dynamics of entanglement were presented by Carvalho {\it et al}. in Ref. \cite{carvalho04}. There,  GHZ  and W states under the influence of independent thermal baths, at zero and infinite temperature, and purely-dephasing independent reservoirs were considered. An analysis of the entire evolution from $p=0$ to the disentangling time was numerically performed for multiqubit systems of up to  $N=7$, using multipartite concurrence \eqref{ConcAndre} as the entanglement quantifier. The authors calculated decay rates for this concurrence, both in the case of W or GHZ states. For W states, the scaling of the decay rate with $N$ depends on the specific noise type, and its details are discussed in Sec. \ref{Wdecay}. For GHZ states, in contrast, the decay rate grows approximately linearly with $N$, regardless of the specific noise model considered.  This means that, even with disentanglement times that increase with $N$, as discussed in Sec. \ref{Simon}, the concurrence observed in \cite{carvalho04} is exponentially fragile. In Sec. \ref{CompleteGHZ}, we see that, as a matter of fact,  the exponential fragility is actually an intrinsic feature of GHZ entanglement under these local decoherence models, for all entanglement measures. One should note, however, that the decay rate does not faithfully describe the full dynamics of these processes, since the decay is exponential with $N$ but approximately exponential with time only for short times. 
\subsubsection{Full  dynamics of entanglement under local noise}
\label{CompleteGHZ}
\par In Refs.~\cite{aolita08,aolitapra09}, Aolita {\it et al}. studied the entire dynamical evolution of $N$-qubit GHZ entanglement for arbitrary $N$. The authors considered generalized GHZ \cite{aolita08,aolitapra09} and generalized GHZ-diagonal \cite{aolitapra09} states under the three paradigmatic single-qubit channels, D, PD, and GAD at arbitrary temperature. As entanglement measures, they considered the simple  negativity in \cite{aolita08,aolitapra09}, and generic convex entanglement monotones in \cite{aolitapra09}. 

The authors showed  \cite{aolita08}, for all three channels, and for any $\alpha$ and $\beta$ such that $\alpha \beta\neq0$, that the critical probability $p_{\epsilon}$ at which the negativity becomes $\epsilon$ times the initial one, $Neg(p_{\epsilon})=\epsilon Neg(0)$, for any arbitrarily small real $\epsilon>0$, scales as 
\begin{equation}
\label{pcrit}
p_{\epsilon}\approx-\kappa\log(\epsilon)/N,
\end{equation}
in the limit of large $N$, with $\kappa=1$ for channels D and PD, and $\kappa=2$ for GAD. The presence of $\log(\epsilon)$, instead of $\epsilon$, in \eqref{pcrit} shows that the scaling law is sensitive more to the order of magnitude of $\epsilon$ rather than to its exact value.

\par Specifically, the authors showed in \cite{aolita08} that the time at which the entanglement in any bipartition becomes arbitrarily small actually decreases with $N$, independently of finite-time disentanglement happening earlier, later, or not happening at all. For all practical purposes, what matters is not that entanglement does not vanish but that a significant amount of it is left, either to be directly used, or to be distilled without an excessively large overhead in resources. These physical considerations imply that, for multi-particle systems, the amount of entanglement can become too small to be a useful resource long before it vanishes. That is, in general, the disentanglement time is by itself not a valid figure of merit for the robustness of multi-particle entanglement \cite{aolita08}. The entire dynamics of entanglement, and specially the initial-time regime, must be monitored.
  
\begin{figure}
\begin{center}
\includegraphics[width=1\linewidth]{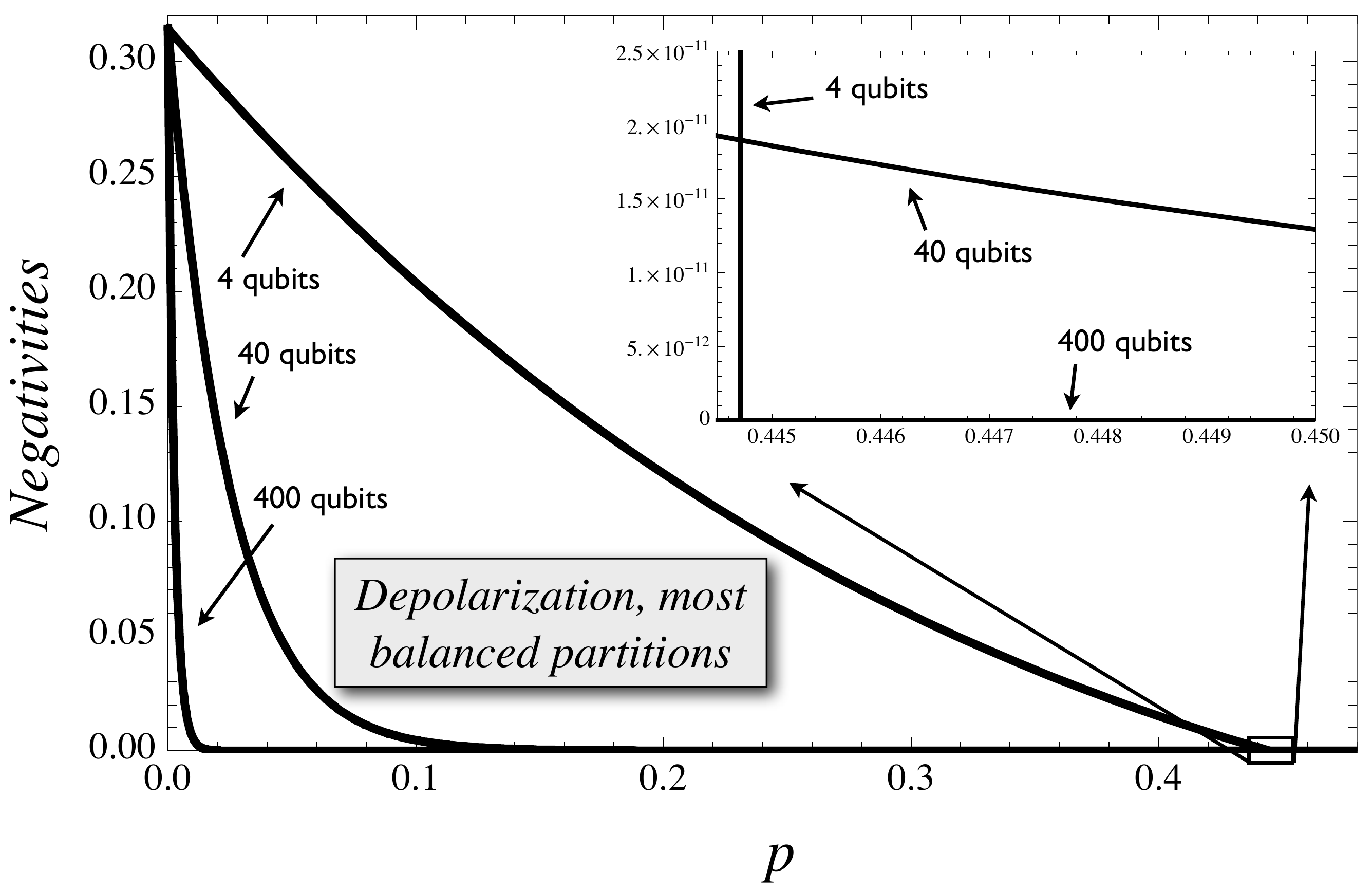}
\caption{
\label{Fig14}
 Negativity of the balanced half-versus-half bipartitions (the most resilient ones) of generalized GHZ states under channel D, as a function of depolarization strength $p$,  for $N=4$, $40$ and $400$. In this graphic $\alpha=1/3$ and $\beta=\sqrt{8}/3$,
but the same behavior is featured for all other parameter choices, and also
for channels PD and GAD. The inset shows a zoom of the region in which
the entanglement of 4 qubits vanishes. Even though the entanglements of both the 40-qubit 
and 400-qubit systems cross the one of 4 qubits and vanish much later, they are many orders of magnitude smaller than their initial value
long before reaching the crossing point. Reprinted figure with permission from L. Aolita \emph{et al.}, \href{http://link.aps.org/doi/10.1103/PhysRevLett.100.080501}{Phys. Rev. Lett. {\bf 100}, 080501 (2008)}. Copyright (2008) by the American Physical Society.}
\end{center}
\end{figure}

\par The situation is illustrated in Fig.~\ref{Fig14}, where the negativity $Neg_{N/2}$ of the most balanced half-versus-half  bipartition is plotted as a function of $p$ and for different $N$. As discussed already in Sec. \ref{GHZprelim}, the negativity captures all the bipartite entanglement content of generalized GHZ states under single-qubit Pauli maps. In addition, even though generalized GHZ states under  GAD are not GHZ-diagonal (so that PPT-ness does not imply biseparability), when all bipartite splittings become PPT, these states can still be proven to be fully separable \cite{aolita08}. So, negativity unambiguously detects the entanglement of the states under scrutiny. The example of Fig.~\ref{Fig14} shows an  initial decay rate that grows with $N$, in agreement with the findings of Carvalho {\it et al}. discussed in Sec. \ref{Carvalho}. This is in remarkable contrast with the behavior of the disentanglement times, which also increase with $N$, as can be seen in the inset of the figure, in agreement with the findings of Simon and Kempe discussed in Sec. \ref{Simon}. Long before disentanglement, the entanglement is the closer to zero the larger  $N$ is. 

\par The observed decay is not a particularity of the negativity. This was shown in Ref. \cite{aolitapra09}, where the authors established analytical upper bounds on entanglement decay, for all convex entanglement measures, and again throughout the noisy evolution. Specifically, for any generalized GHZ-diagonal state $\rho$, they showed that 
\begin{equation}
\label{universalbound}
E(\mathcal{E}\rho) \le (1-p)^N E(\rho),
\end{equation}
for all $0\leq p\leq1$, being $\mathcal{E}$ the D or PD channels acting in each qubit with strength $p$. For pure generalized GHZ states, bound \eqref{universalbound} follows immediately from decomposition \eqref{EntSep}, convexity, and the fact that $E=0$ for all separable states. The extension to generalized GHZ-diagonal states follows in turn from linearity. In addition, bound \eqref{universalbound} is tight, in the sense that there is at least one convex entanglement measure, and a state, that saturates it: For small $p$, or large $N$, negativity $Neg_{N/2}$ for GHZ states is given precisely by $(1-p)^N$  \cite{aolita08,aolitapra09}. This expression clearly shows that, for a fixed value of $p$, the negativity decreases exponentially with $N$.

\par On the other hand, for channel GAD, at  arbitrary temperature (bath population $\bar{n}$), the bounds of \cite{aolitapra09} are restricted to pure generalized GHZ states and are not tight. Again, they follow from decomposition \eqref{EntSep}. Their general form, for $\rho=\ket{\psi_k^{\pm}(\alpha,\beta)}\bra{\psi_k^{\pm}(\alpha,\beta)}$, is given by $E(\mathcal{E}^{GAD}\rho) \leq \mathcal{E}^{GAD}_\text{ent}E(\rho^{GAD}_\text{ent})$; and a simplified version, without the intricate dependence on $\alpha$, $\beta$, and $k$ of $\mathcal{E}^{GAD}_\text{ent}$ and $\rho^{GAD}_\text{ent}$, is
\begin{equation}
\label{universalboundGAD}
E\big(\mathcal{E}^{GAD}\rho\big) \le \Big( 1-\frac{\bar{n}}{2\bar{n}+1}p \Big)^N E_{\text{max}},
\end{equation}
where $E_{\text{max}}$ is the maximal value of the entanglement measure over all states~\cite{aolitapra09}.

\par It is important to emphasize that only the basic requirements of convexity and nullity of $E$ over separable states is necessary for bounds \eqref{universalbound} and \eqref{universalboundGAD} to hold. For this reason, the bounds apply not only to a very general family of entanglement measures, including bipartite and genuinely multipartite, but also to quantifiers of resources for other physical tasks, as for instance non-locality-based \cite{chaves12,chaves12b} or quantum metrology protocols \cite{chaves12b}. Thus, the exponential fragility under local noise appears as an almost universal dynamic property of GHZ-type states. Other families of states do not display such fragility. For example, W states are not exponentially sensitive to local noise, as discussed in Sec. \ref{Wdecay}; and, for random pure states, the negativity of least-balanced splits is numerically observed to violate bound \eqref{universalbound} up to $N=14$, with a violation that increases with $N$ \cite{aolitapra09}. We shall see however, in the next Sec. \ref{robustGHZdirected}, that there are some  particular but very relevant instances where GHZ correlations turn out to be remarkably stable against local decoherence.
\subsubsection{Resistance against local  bit-flip noise}
\label{robustGHZdirected}
In the previous sub-subsection, we have seen that GHZ states are exponentially fragile to  independent channels as D, PD or GAD at any temperature. This turns out not to be the case for independent Pauli noise directed along the transversal $X$ direction, i. e., the single-qubit bit-flip channel $\mathcal{E}^{\text{BF}}$ defined in Sec. \ref{Pauliintro}. In Ref. \cite{borras09}, Borras {\it et al}. numerically observed, up to $N=6$, that GHZ states are considerably more robust against $\mathcal{E}^{\text{BF}}$ than against $\mathcal{E}^{\text{PD}}$. For the particular case of $N=4$, the authors even observed that GHZ states are more robust to $\mathcal{E}^{\text{BF}}$ than to any other channel local-unitary equivalent to it, over a sample of 1000 random local-unitary transformations. This was later formalized by Chaves {\it et al}., in Ref. \cite{chaves12b}, where the following general bound was presented,
\begin{eqnarray}
\label{universalboundBF}
\nonumber 
E\big(\varrho^{\text{BF}}_N\big)&\ge&E\big(\varrho^{\text{BF}}_{N-1}\otimes\ket{0}\bra{0}\big)\\
\nonumber
&\vdots&\\
&\ge& E\big(\varrho^{\text{BF}}_{2}\otimes(\ket{0}\bra{0})^{\otimes N-2}\big).
\end{eqnarray}
Here, $E$ represents an arbitrary entanglement monotone and $\varrho^{\text{BF}}_N\doteq\mathcal{E}^{BF}\ket{\text{GHZ}_N}$. Inequality \eqref{universalboundBF} thus tells that the entanglement of $N$-qubit GHZ states under  independent BF is at least as robust as that of $N-1$-qubit GHZ states under the same noise, for all $N$. It is in a sense a counterpart of inequalities \eqref{universalbound} and \eqref{universalboundGAD}. Consider for instance the $N$-party distilability of $\mathcal{E}^{BF}\ket{\text{GHZ}_N}$. Taking $E$ as any quantifier of two-qubit distillable entanglement, inequality \eqref{universalboundBF} implies that the $N$-party distillability of all GHZ states under independent BF is at least as robust as that of the maximally entangled two-qubit state $\ket{\text{GHZ}_{2}}\equiv\ket{\Phi^+}$ under the same noise, in contrast to \eqref{universalbound} and \eqref{universalboundGAD}. Inequality \eqref{universalboundBF} can also be extended to any graph state \cite{chaves12b}.

In addition, Chaves {\it et al}. observed that the least-balanced-bipartitions negativity of $\mathcal{E}^{BF}\ket{\text{GHZ}_N}$ increases with $N$, approaching the $N$-independent value $1-p$, with $p$  the bit-flip strength, in the limit of large $N$. This behaviour remains even for Pauli channels with small components along directions other than $X$. Apart from entanglement, the authors also analyzed $\mathcal{E}^{BF}\ket{\text{GHZ}_N}$ as resources for high-precision frequency estimation or non-locality based protocols. In both cases they observed an exponential robustness enhancement for bit-flip noise over phase-flip.

Finally, in Ref. \cite{chaves12c}, Chaves {\it et al}. extended these findings to frequency estimation of systems described by a Hamiltonian proportional to $\sigma_z$ and noise proportional to $\sigma_x$.  They showed that, optimizing the pulse duration, a supra-classical precision scaling is possible with GHZ states. The robustness of GHZ states against transversal noise is intuitively connected to the efficacy of the repetition code \cite{nielsen00} to correct bit-flip errors (see also discussion in Sec. \ref{Sec:logicalchannels}). This  culminated in a recent series of papers \cite{Kessler13, Arrad13, Duer13,Ozeri13}, where the same intuition is used to show that, using quantum error-correction, Heisenberg-limited spectroscopy can be attained. That is, for noisy phase evolutions with transversal noise, GHZ states can be error-corrected so as to restore the maximum precision scaling corresponding to a unitary evolution.
\subsubsection{Resistance of blockwise GHZ entanglement for GHZ-encoded blocks against local depolarization}
\label{robustGHZblockwise}
In Refs. \cite{Froewis11,Froewis12}, Fr\"{o}wis and D\"{u}r considered concatenated GHZ states. The name ``concatenated GHZ" is motivated by the notion of making block-wise GHZ states out of blocks that are themselves GHZ states. Specifically, the authors defined concatenated GHZ states as the usual GHZ states \eqref{GHZdef} but where the computational-basis states $\ket{0}$ and $\ket{1}$ are replaced by logical-qubit states $\ket{0}_L$ and $\ket{1}_L$, given by $m$-qubit GHZ states of different parity: $\ket{0}_L\doteq\ket{\psi_0^{+}}$ and $\ket{1}_L\doteq\ket{\psi_0^{-}}$ (using the notation of \eqref{orthoGHZ}). Thus, each $m$-qubit block encodes a logical qubit, and the $N$-logical-qubit GHZ state is composed of a total of $mN$ physical qubits.  

The authors considered the situation where all $mN$ physical qubits undergo the independent channel D, and showed that the coherence, distillability and entanglement at the logical level can be stabilized at the expenses of just a logarithmic overhead at the physical level. More precisely, they showed that even though  all correlations among the logical blocks decay exponentially $N$, as in the usual GHZ states, the decay rate itself decreases exponentially with the block size $m$. So, taking $m=O\big(\log(N)\big)$ physical qubits per logical qubit suffices to ``freeze" the decay at the logical level, i.e. to make the blockwise coherence, distillability and entanglement, or even the usefulness of the states for supra-classical metrology, independent of $N$ for any fixed depolarization strength.

In Sec.~\ref{Sec:logicalchannels}, we derive equivalent logical-encoded noise channels \cite{Kesting13} with which this curious stabilization can be understood as  logical transversal dephasing on logical GHZ states, reducing the problem to that discussed in Sec. \ref{robustGHZdirected}.

\subsection{Decay of graph-state entanglement}
\label{Graphdecay}

\subsubsection{A direct multipartite distillation protocol under local decoherence}
\label{particudist}
\par The first studies on generic graph states under independent decoherence were reported by D\"{u}r {\it et al}. in Ref.~\cite{Duer03}, and later extended in Ref.~\cite{Aschauer05} (see also Sec.~X of Ref. \cite{Hein_Review}). There, a direct genuine-multipartite entanglement distillation protocol  was introduced and theoretically probed in noisy situations. Direct distillation protocols are to be distinguished from those based on bipartite purification, where one distills pure maximally entangled states between all pairs of particles and then re-combine them with LOCCs into a genuinely multipartite entangled pure state. The protocol of \cite{Duer03,Aschauer05} is more efficient -- in terms of conversion rates -- than bipartite distillation methods and outperforms their maximal target-state fidelity when implemented with non-ideal LOCCs. It is applicable to graph states associated to all two-colourable graphs: all those whose vertices can be grouped together into two subgroups (colours) such that every vertex has a different colour from all its neighbours. The two-colour condition restricts the number of edges in the graph, thus simplifying the problem. These include for instance GHZ states, cluster states in all dimensions, and various error-correction codewords. %

The authors focused on the  distillation thresholds, {\it i.e.} the maximal  noise strengths up to which the protocol still purifies, for different models of noise. 
For channel D, they found that the maximal $p$ that linear-cluster (1D-graph) states can tolerate is essentially independent of $N$. Whereas it was found to decrease exponentially with $N$ for GHZ states (associated to star-like and fully connected graphs, with connectivity  $N-1$). 
As discussed in Secs.~\ref{DuerGraph} and \ref{Jack}, these two different behaviours, with respect to this specific graph-state distillation protocol, turn out to represent an  actually rather general feature of graph-state entanglement for several models of local noise. Typically, graph-state entanglement decay rates do not depend on $N$ but rather on the connectivity, or degree, of the associated graph.

\subsubsection{Genuine $N$-partite distillability under local decoherence}
\label{DuerGraph}

\par In the previous subsubsection we discussed the direct distillability of multiparticle-entanglement under a specific distillation protocol, and for the particular case of two-colourable graphs under depolarization. It was also D\"ur and Briegel, in Ref.~\cite{duer04}, and Hein, D\"ur and Briegel, in Ref. \cite{hein05}, who first reported studies on the disentanglement properties of general graph states, and under more general noisy evolutions (see also Sec.~10 of Ref.~\cite{Hein_Review}). There, the authors studied the lifetimes of multipartite distillability, not just with respect to a specific distillation procedure. 

\par Specifically, the authors derived lower and upper bounds on the times at which any entanglement ceases to be distillable. Three different upper bounds were derived for independent thermal baths at zero and infinite temperature, and for independent D, PD, and bit-flip channels. Two of these three upper bounds constitute actually valid upper bounds also on the disentanglement time, as they bound the times at which the considered channels become  entanglement-breaking. On the other hand, a lower bound valid for independent Pauli maps was derived based on an explicit purification protocol, which, in contrast to the one mentioned in Sec.~\ref{particudist}, distills exclusively bipartite maximally-entangled singlet pairs between any two pairs of qubits. We briefly describe this protocol in the end of the sub-subsection. As discussed in Sec.~\ref{multidistil}, the distillation of a maximally-entangled singlet pair between all pairs of qubits
is a necessary and sufficient condition for the distillation of any genuinely multipartite type of pure-state entanglement, including that of the initial graph state. In the context of stability of entanglement as a resource, lower bounds are of special importance, as they set lower limits to the maximal noise strengths that protocols can tolerate. 
\begin{figure}
\begin{center}
\includegraphics[width=1\linewidth]{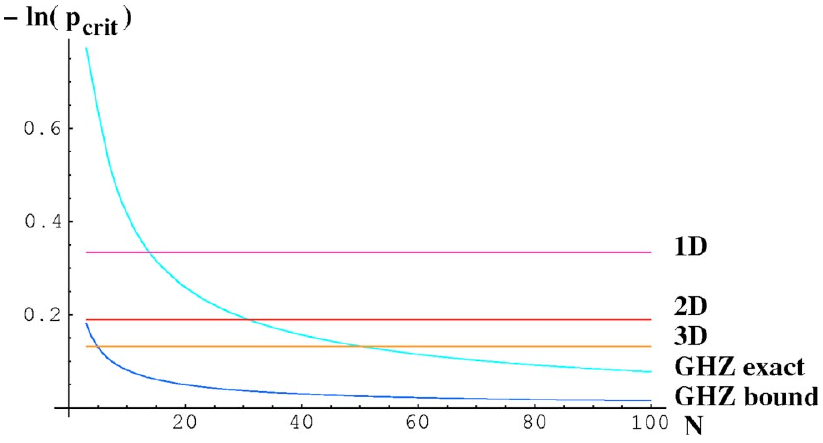}
\caption{
\label{Hein}
Lower bound on the depolarisation strength $p_{crit}$, in log scale, at which locally depolarized GHZ and 1D-, 2D-, and 3D-cluster states cease to be distillable. 
For GHZ states, the exact vanishing time of $N$-party distillability discussed in Sec. \ref{DuerMpartite} is also displayed. For cluster states, the bound decreases with the dimension of the lattice, but are completely independent of the system size $N$. For the GHZ case on the other hand, both curves decrease exponentially with $N$. This  suggests that the robustness of graph-state entanglement does not depend on the system size but only on the connectivity of the associated graph. Reprinted figure with permission from  M. Hein, \emph{et al.}, \href{http://link.aps.org/doi/10.1103/PhysRevA.71.032350}{Phys. Rev. A  {\bf 71} 032350 (2005)}. Copyright (2005) by the American Physical Society.
}
\end{center}
\end{figure}

\par In Fig. \ref{Hein}, the above-mentioned lower bound on the lifetime $p_{crit}$ is plotted as a function of $N$ in logarithmic scale, for initially-pure 1D-, 2D-,  3D-cluster and GHZ states under the influence of independent  D channel. For the GHZ states, in addition, the exact vanishing time of $N$-party distillability, given by the time at which the least robust bipartitions become PPT, as discussed in Sec. \ref{DuerMpartite}, is also shown. For the cluster states, the bound decreases with the dimension of the lattice, and therefore the connectivity of the graph, but are completely independent on $N$. For GHZ states on the other hand, both the bound and the exact disentanglement time strictly decrease with $N$ and tend to zero as $N\longrightarrow\infty$. This suggests a macroscopic robustness of multiparty distillability for graph states with constant degree.  In Sec. \ref{Jack}, we see indeed  that this is a universal reaction of generic graph-state entanglement against the most common types of noise. Graph-state entanglement stability does not depend on the actual size of the system but only on the connectivity of the associated graph.

\par Before switching to the next subject, we stop for a moment to briefly explain the purification protocol considered here, as it is also related to other bounds that we discuss in Sec. \ref{Jack}. As seen in Sec. \ref{Sec:ActionPauliGraph}, any independent Pauli channel $\mathcal{E}$ acting on a graph state is equivalent to that of another separable map $\tilde{\mathcal{E}}$ with Kraus operators composed only of tensor products of $Z$ and identity operators. The application of $\mathcal{E}$ thus clearly commutes with that of local measurements along the $Z$ basis. Now, when all but two qubits in the graph are individually measured in $Z$ basis, the remaining state of the two unmeasured qubits is a pure maximally entangled state \cite{duer04,hein05,Hein_Review}. The protocol is a sequential bipartite procedure, whose essence goes as follows. {\it (i)} Measure all but qubits $\{k,j\}$ of the noisy graph state along the $Z$ basis. This is equivalent to applying the measurements on the pure graph state and then applying $\tilde{\mathcal{E}}$.
{\it (ii)} Trace out all but qubits $\{k,j\}$. The resulting two-qubit state is a noisy entangled state between $k$ and $j$. {\it (iii)} Distill a pure maximally entangled state between qubits $\{k,j\}$ by means of any standard two-qubit purification procedure. {\it (iv)} repeat steps {\it (i)} through {\it (iii)} for all pairs $\{k,j\}$. {\it (v)} Obtain an $N$-qubit pure graph state by locally fusing the distilled maximally entangled pairs. The lifetime of distillability with respect to this particular protocol is given simply by the lifetime of two-qubit NPT-ness after step {\it (ii)} minimized over all pairs $\{k,j\}$. In the case of permutationally invariant states, however, as those used for Fig. \ref{Hein}, the minimization is not required. This lifetime defines the (non-tight) lower bound plotted in the figure.

\subsubsection{Full  dynamics of entanglement under local noise}
\label{Jack}

In the previous sub-subsections, the survival times of entanglement or distillability of decohered graph states were discussed. As discussed in Sec. \ref{CompleteGHZ} for the particular case of GHZ states, survival times on their own are  in general unable to provide a faithful assessment about the state's robustness. The complete evolution must be studied to draw faithful conclusions on entanglement stability. A general framework for the study of the complete dynamics of graph-state entanglement under decoherence was introduced by Cavalcanti {\it et al}., in Ref. \cite{cavalcanti09}, and further developed by Aolita {\it et al}., in Ref. \cite{chirag10}. There, the authors established a systematic method to obtain lower and upper bounds for the entanglement of graph, or (mixed) graph-diagonal, states affected by several models of noise, including both independent and collective arbitrary Pauli maps \cite{cavalcanti09,chirag10}, or independent thermal baths at arbitrary temperature \cite{chirag10}, throughout the evolution. These bounds display the following  properties: 
\begin{itemize}
\item ({\it i}) They apply to all convex LOCC monotones, including of course all convex bipartite or multipartite entanglement quantifiers. 
\item ({\it ii}) They bound the entanglement content of the whole system in terms of that of a significantly smaller subsystem consisting only of those qubits on the boundary of the multipartition considered, requiring no optimization on the full system's parameter space.
\item ({\it iii}) For all Pauli maps,  lower and upper bounds coincide, providing thus the exact entanglement evolution. 
\end{itemize}

\begin{figure}[t]
\begin{center}
\includegraphics[width=0.7\linewidth]{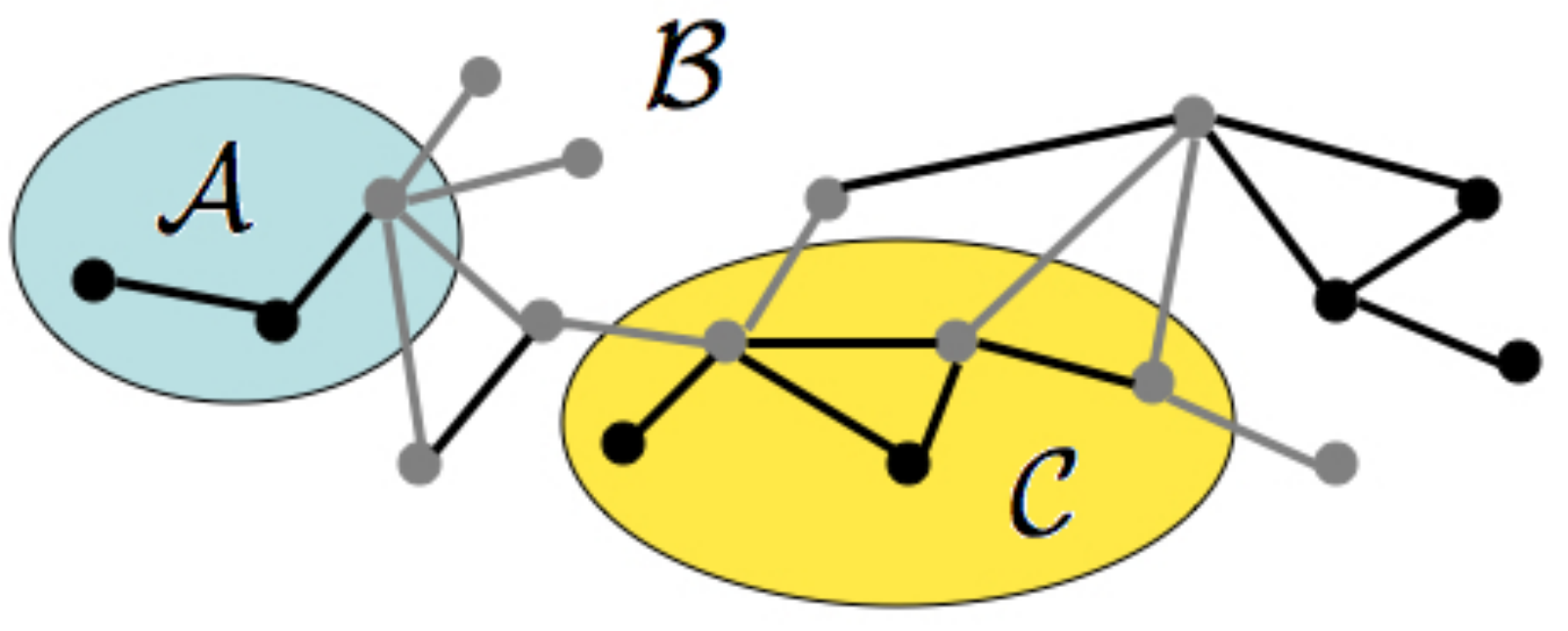}
\caption{\label{GraphFig} Example of a mathematical
graph associated to a physical graph state. An exemplary partition is displayed,  by which the system is split into three
subparts: $\mathcal{A}$ (light blue), $\mathcal{B}$ (white), and $\mathcal{C}$ (yellow). The
vertices and edges in grey correspond to the \emph{boundary
qubits} and the  \emph{boundary-crossing edges}, respectively. Reprinted figure with permission from D. Cavalcanti \emph{et al.}, \href{http://link.aps.org/doi/10.1103/PhysRevLett.103.030502}{Phys. Rev. Let.  {\bf 103}, 030502 (2009)}. Copyright (2009) by the American Physical Society.}
\end{center}
\end{figure}

We  briefly describe the bounds. Consider an arbitrary multipartition of the vertices in a generic $N$-vertex graph $\mathcal{G}_{(\mathcal{V},\mathcal{E})}$. For example, the graph showed in Fig. \ref {GraphFig} has been split into  three parts $\mathcal{A}$, $\mathcal{B}$, and $\mathcal{C}$. Every edge that crosses a subpartition is a \emph{boundary-crossing edge}, and any two vertices connected by such edge are \emph{boundary vertices}. The union of the set $\mathcal{X}$ of all boundary-crossing edges and the set $\mathcal{Y}$ of all boundary
vertices constitutes in turn the {\it boundary graph} $\mathcal{G}_{(\mathcal{Y},\mathcal{X})}\doteq\{\mathcal{Y},\mathcal{X}\}$. An $N$-qubit graph sate $\rho=\ket{{\mathcal{G}_{(\mathcal{V},\mathcal{E})}}_{0}}\bra{{\mathcal{G}_{(\mathcal{V},\mathcal{E})}}_{0}}$ undergoes a noisy physical process, described by map $\mathcal{E}$, during a time $t$. We restrict first to Pauli maps. In this case, as discussed in  Sec. \ref{Sec:ActionPauliGraph}, the evolved state $\rho_{(t)}\equiv\mathcal{E}\rho$ is a graph-diagonal state. We wish to calculate the correlations $E(\rho_{(t)})$  among parts $\mathcal{A}$, $\mathcal{B}$, and $\mathcal{C}$. We refer to $E$ as entanglement, but the only requirements that $E$ is asked to satisfy are convexity and LOCC-monotonicity, discussed in Sec. \ref{2.3.2}. So 
 {\it any convex LOOC-monotone} correlations are also covered by the present formalism. For such calculation, it is  convenient to explicitly factor out all the $CZ$ gates in $\rho_{(t)}$ that do not correspond to a boundary-crossing edge, so that $\rho_{(t)}$, given by \eqref{ActionPauliGraph}, writes as
\begin{eqnarray}
\label{rho_t} 
\nonumber
\rho_{(t)}&=&\bigotimes_{\{i,j\} \in \mathcal{E}/\mathcal{X}}
CZ_{ij}\sum_{\gamma,\delta} \tilde{p}_{\gamma,\delta}
\ket{{G_{(\mathcal{Y},\mathcal{X})}}_\gamma}\bra{{G_{(\mathcal{Y},\mathcal{X})}}_\gamma}\\
&\otimes&\ket{{g_{(\mathcal{V}/\mathcal{Y})}}_\delta}\bra{{g_{(\mathcal{V}/\mathcal{Y})}}_\delta}\bigotimes_{\{k,l\} \in \mathcal{E}/\mathcal{X}}CZ_{kl}.
\end{eqnarray}
Here  we have grouped together all individual indices  $\mu_i$ inside index-string $\mu$ of  \eqref{ActionPauliGraph} corresponding to boundary and non-boundary qubits into two new index-strings: $\gamma$ and $\delta$. Multi-index $\gamma$ accounts for all possible graph states $\ket{{\mathcal{G}_{(\mathcal{Y},\mathcal{X})}}_\gamma}$ associated to the boundary graph, whereas $\delta$ corresponds to all product states $\bigotimes_{j \in\mathcal{Y}}Z^{\delta_j}\ket{{g_{(\mathcal{Y})}}_0}$, with $\ket{{g_{(\mathcal{Y})}}_0}\doteq\bigotimes_{i \in\mathcal{Y}}
\ket{+_i}$. Probability $\tilde{p}_{\gamma,\delta}$  is  the same as $\tilde{p}_{\mu}$ in Eq. \eqref{ActionPauliGraph} with $\mu\doteq(\gamma,\delta)$. For all Pauli maps, its explicit form in terms of the original probabilities in $\mathcal{E}$ is rather simple, and can be found in Sec. II.B.1 of \cite{chirag10}. The key point here is to notice that the  $CZ$ gates explicitly factored out in  \eqref{rho_t} are local unitary operations with respect to the multi-partition of interest, and can therefore be disregarded for the calculation of $E(\rho_{(t)})$:
\begin{eqnarray}
\label{ent}
\nonumber
E(\rho_{(t)})&\equiv&
E\big(\sum_{\gamma,\delta}
\tilde{p}_{\gamma,\delta}\ket{{G_{(\mathcal{Y},\mathcal{X})}}_\gamma}
\bra{{G_{(\mathcal{Y},\mathcal{X})}}_\gamma}\\
&\otimes&\ket{{g_{(\mathcal{V}/\mathcal{Y})}}_\delta}\bra{{g_{(\mathcal{V}/\mathcal{Y})}}_\delta}\big).
\end{eqnarray}

\par Now, since states $\{\ket{{g_{(\mathcal{V}/\mathcal{Y})}}_\delta}\}$ form an orthonormal basis of the Hilbert space of the non-boundary subsystem $\mathcal{V}/\mathcal{Y}$,  a measurement in this basis provides full information about the state of the non-measured boundary subsystem $\mathcal{Y}$. Indeed, for a measurement outcome $\delta$, $\mathcal{Y}$ is projected into the state $\sum_{\gamma}\tilde{p}_{\gamma,\delta}\ket{{G_{(\mathcal{Y},\mathcal{X})}}_\gamma}\bra{{G_{(\mathcal{Y},\mathcal{X})}}_\gamma}$. The mixedness in $\mathcal{Y}$ with respect to $\delta$ is removed. One then typically says that subsystem $\mathcal{Y}$ is {\it flagged} by the measurement outcome of $\mathcal{V}/\mathcal{Y}$. The resulting entanglement is then given by the average entanglement over $\delta$. In turn, since this measurement is an LOCC, the entanglement can only decrease:
\begin{equation}
\label{perfectlower} E(\rho_{(t)})\geq\sum_\delta\tilde{p}_\delta E\Big(\sum_{\gamma}
\tilde{p}_{(\gamma|\delta)}\ket{{G_{(\mathcal{Y},\mathcal{X})}}_\gamma}
\bra{{G_{(\mathcal{Y},\mathcal{X})}}_\gamma}\Big).
\end{equation}
The right-hand side gives the average entanglement over $\delta$, where $\tilde{p}_{\delta}\equiv\sum_{\gamma}\tilde{p}_{\gamma,\delta}$ is the probability of measuring $\delta$ and $\tilde{p}_{(\gamma|\delta)}$ is the conditional probability of finding $\gamma$ given that  $\delta$ takes place. 

\par On the other hand, the convexity of $E$ implies  that \eqref{ent} cannot in turn be greater than the right-hand side of \eqref{perfectlower}. That is, the latter gives at the same time both an upper and a
lower bounds on $E(\rho_{(t)})$, and therefore
\begin{equation}
\label{exact}
E(\rho_{(t)})\equiv\sum_\delta\tilde{p}_\delta E\Big(\sum_{\gamma}
\tilde{p}_{(\gamma|\delta)}\ket{{G_{(\mathcal{Y},\mathcal{X})}}_\gamma}
\bra{{G_{(\mathcal{Y},\mathcal{X})}}_\gamma}\Big).
\end{equation}
This expression reduces the calculation to the average entanglement over a sample of $2^{N_{\mathcal{V}/\mathcal{Y}}}$ states  -- one for each measurement outcome $\delta$ of the non-boundary subsystem -- of  $N_\mathcal{Y}$ (boundary) qubits, with $N=N_\mathcal{Y}+N_{\mathcal{V}/\mathcal{Y}}$. One trades thus an optimization over the entire system space, which contains $O\big(2^{N}\big)$ complex parameters, for at most $2^{N_{\mathcal{V}/\mathcal{Y}}}$ optimisations over $O\big(2^{N_\mathcal{Y}}\big)$ complex parameters each. This directly implies an exponential reduction in the required computer-memory, which, in turn, since each optimisation can be naturally  parallelized, can lead to significant reductions in computing time.

\begin{figure}[t]
\begin{center}
\includegraphics[width=.9\linewidth]{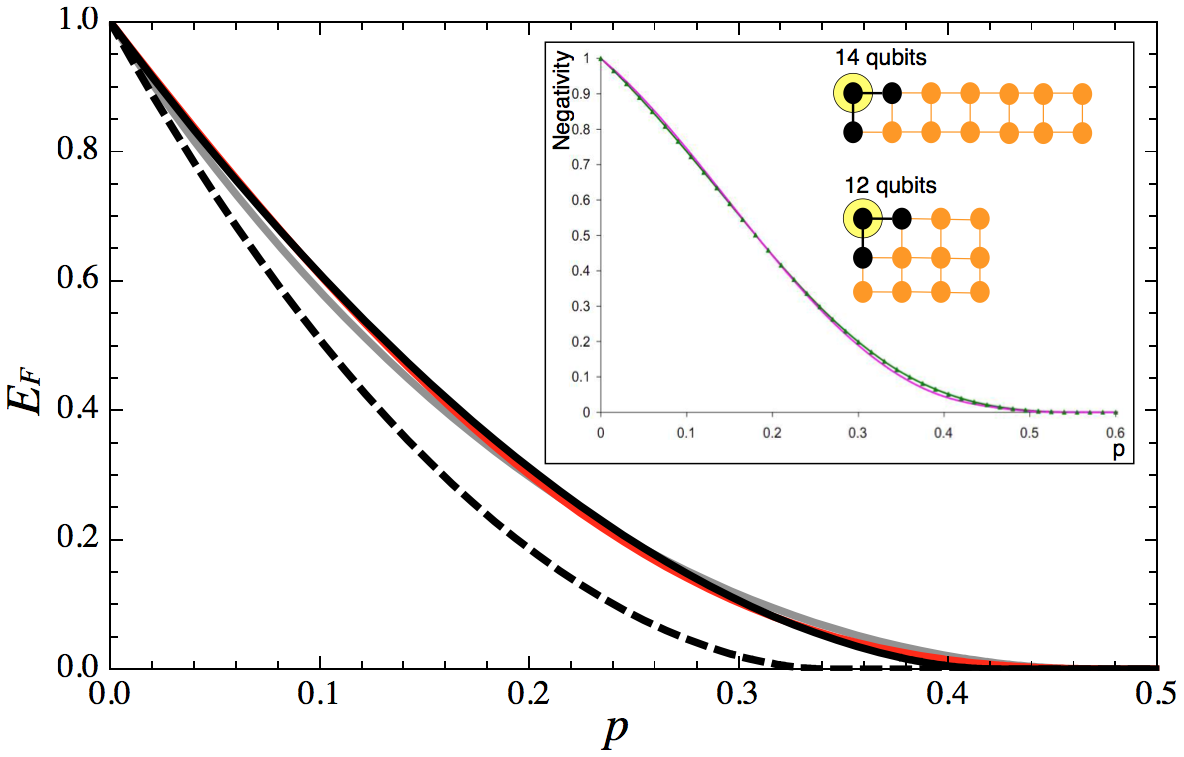}
\caption{ \label{Fig2} Exact entanglement of formation
($E_F$) in the partition of the first particle versus the rest for
linear clusters of 2 (black), 4 (grey) and 7 (red) particles
undergoing independent depolarization as a function of the
noise strength $p$. The lower bound in dashed is obtained by tracing the non-boundary system out and depends therefore only on the two boundary qubits. This shows that, for small $p$, the entanglement in the considered partition is very robust against cluster-length growth. No optimization was required for this plot. From D. Cavalcanti \emph{et al.}, Phys. Rev. Let. {\bf 103}, 030502 (2009). Inset: Exact negativity of the bipartition shown versus $p$, for 14- (green triangles) and 12-qubit (pink solid) rectangular cluster states under independent depolarization. Reprinted figure with permission from L. Aolita \emph{et al.}, \href{http://link.aps.org/doi/10.1103/PhysRevA.82.032317}{Phys. Rev. A, {\bf 82}, 032317 (2010)}. Copyright (2010) by the American Physical Society.}
\end{center}
\end{figure}

\par As an example, in Fig. \ref{Fig2}, the exact value of the entanglement of formation $E_F(\rho_{(t)})$ throughout the evolution is displayed for the bipartition of the first qubit versus the rest for linear cluster states 
under independent D channel as a function of $p$. In this case, since the boundary subsystem is composed only of two qubits, for which the closed formula \eqref{EntofformConcu} for $E_F$ can be applied, \eqref{exact} leads on to $E_F(\rho_{(t)})$ without a single optimisation. In addition, in the inset, the negativity of the bipartition of one qubit versus the rest shown, for twelve- and fourteen-qubit rectangular graph states under independent D, is plotted as a function of $p$. A brute-force negativity calculation would involve diagonalizing a $2^{14} \times 2^{14}= 16384 \times 16384$ matrix  for each value of $p$. Whereas, with the help of \eqref{exact}, each evaluation requires diagonalization of many ($2^{11}$) but very small (dimension $2^{3} \times 2^{3}$) matrices. The realization of the entire negativity calculation, for all $p$ shown, would have been infeasible with the brute-force approach.

Also in Fig. \ref{Fig2}, to assess the robustness of $\rho_{(t)}$ with the system size, a lower bound to $E_F(\rho_{(t)})$ satisfied for all $N$  is displayed in dashed. This is obtained as in \eqref{perfectlower} but directly tracing  the non-boundary qubits out of \eqref{rho_t}, i.e. without accessing any flag information. Then, the mixedness with respect to index $\delta$ remains. The effective noise on the resulting boundary-system state is then described by $\tilde{p}_{\gamma}\equiv\sum_{\delta}\tilde{p}_{\gamma,\delta}$. For  independent or collective D, or independent PD, channels, among other Pauli maps \cite{cavalcanti09}, $\tilde{p}_{\gamma}$ depends only on the boundary subgraph and, at most, its first neighbours, not on the total system. This bound is thus independent on $N$, but introduces in return a decrease in tightness. Still, it allows for a direct assessment of the entanglement robustness merely by  inspection of the graph's connectivity, analogous to the bounds of Fig. \ref{Hein}.  

We notice that, since Pauli maps on graph-diagonal states also render graph-diagonal states, all the arguments exposed so far are also valid for this class of initial mixed states. Furthermore, since any arbitrary state can be dephased towards a graph-diagonal state by means of an LOCC \cite{Aschauer05}, \eqref{exact} can help lower-bound the entanglement content of arbitrary mixed states. 

\par Finally, it is important to mention that, apart from Pauli maps, the whole formalism  extends to other maps $\mathcal{E}$ that do not admit the replacement $\mathcal{E}\leftrightarrow\tilde{\mathcal{E}}$, as described in Sec. \ref{Sec:ActionPauliGraph}, but for which the commutation of $\mathcal{E}$ with the non-boundary $CZ$ gates still yields a separable map. In those cases, despite of in general no longer coinciding, the lower and upper bounds obtained are still highly non-trivial. Remarkably, a crucial family of channels for which this is the case is independent GAD at all temperatures. The bounds for this family are studied in detail in Ref. \cite{chirag10}.

\subsubsection{Blockwise entanglement and effective logical channels for stabilizer-state encoded logical states under local depolarization}
\label{Sec:logicalchannels}

\par In Ref. \cite{hein05}, Hein {\it et al}. studied the decay of logical entanglement in blockwise multipartite entangled states of logical 
 qubits made out of particular stabilizer states, which are local unitarily equivalent to graph states. They considered the situation where all physical qubits in each block undergo independent depolarization of strength $p$. For logical qubits encoded into 5-qubit optimal error-correction codewords \cite{bennett96b,laflamme96}, they showed that, after error-correcting, the physical noise is also described by a D channel but at the logical level, i. e., by a logical channel D acting independently on each block with a logical depolarization strength $p_L$. They showed that, for small $p$, the effective strength $p_L$ is smaller than the physical one $p$. In addition, the authors also considered concatenated encodings with $k$ levels  of concatenation. These consist of logical qubits encoded into codewords made out of also of logical qubits, at an inferior concatenation level, themselves also encoded into codewords, and so an and so forth of to the lowest concatenation level $k$. In this case, always for the 5-qubit encoding,  they obtained that the logical depolarization strength $p_L(k)$ decreases, for small physical strength $p$, doubly-exponentially with $k$.
 
These studies have recently been extended, and further formalized, by Kesting, Fr\"{o}wis, and D\"{u}r, in Ref. \cite{Kesting13}. There, for logical qubits encoded into arbitrary stabilizer-state blocks, and for physical qubits experiencing independent Pauli-channel noise, the authors provide a procedure to derive explicit expressions for the logical effective channel, after error-correcting, acting on the block. They particularize this procedure to two cases. In the first one, they consider arbitrary Pauli noise at the physical level  and logical qubits encoded into the $m$-qubit repetition codewords capable of correcting bit-flip errors. There, the equivalent channel is also a Pauli map at the logical  level, but, interestingly, with the $X$ (bit-flip) and $Y$  (bit-phase-flip) components exponentially damped with $m$, whereas with the $Z$ (phase-flip) component linearly increased with $m$. In the second one, they consider again independent depolarization and a 5-qubit optimal error-correction encoding, and reproduce similar results as in \cite{hein05}.

Since the equivalent channels act on the logical qubits exactly as the corresponding physical channel on physical qubits, one can use them  
to obtain separability or $M$-distillability bounds among different blocks of multipartite blockwise entangled states with all the the known techniques for the case without encodings. When the logical noise strength  is lower than the physical one, one obtains of course higher blockwise entanglement and $M$-distillability lifetimes. 

Importantly, it is interesting to note that, even though the equivalent channels are obtained after error correcting, it is the passive logical encoding, rather than the active error correction, what actually slows down decoherence at the logical level. This is due to the fact that the error-correction step (syndrome measurement and correction) is done locally within each logical block. It is an LOCC with respect to the given $M$-partition and therefore cannot be responsible for the blockwise entanglement or $M$-distillability increases. With this in mind, we see that the effective exponential damping of logical bit-flip errors obtained in \cite{Kesting13} for the repetition code is the origin of the robustness of GHZ states against transversal dephasing  \cite{chaves12b} discussed in Sec. \ref{robustGHZdirected}. In turn,  the robustness of blockwise GHZ entanglement for GHZ encoded logical qubits  \cite{Froewis11,Froewis12}, discussed in Sec. \ref{robustGHZblockwise}, can also be understood in a similar way. The GHZ encoding can be thought of as the $m$-qubit repetition encoding against bit flip, $\ket{0}_{L}\doteq\ket{0}^{\otimes m}$ and $\ket{1}_{L}\doteq\ket{1}^{\otimes m}$, followed by a logical Hadamard rotation $\ket{0}_{L}\to\ket{+}_{L}\doteq\frac{1}{\sqrt{2}}(\ket{0}_{L}+\ket{1}_{L})$ and $\ket{1}_{L}\to\ket{-}_{L}\doteq\frac{1}{\sqrt{2}}(\ket{0}_{L}-\ket{1}_{L})$. The repetition-code encoding implies that physical depolarization acts effectively at the logical level as Pauli channel exponentially dominated by logical $Z$ noise. The logical Hadamard rotation maps the latter into the logical $X$. That is, the total effect of the GHZ encoding is to map the physical D channel to a logical bit-flip channel, against which, as we know from \cite{chaves12b}, logical GHZ states are robust.
 
\subsection{Decay of W- and Dicke-type entanglement}
\subsubsection{Negativity, concurrence, and global entanglement of locally decohered W-states}
\label{Wdecay}

As mentioned in Sec. \ref{Simon}, in Ref. \cite{simon02} Simon and Kempe reported the first preliminary studies on the vanishing times of NPT-ness of W and Dicke states under the independent D channel  for $N= 3$  and 4. 
Later on,  in Ref. \cite{carvalho04}, Carvalho {\it et al}. compared numerically the dynamical evolution of W and GHZ states for $2\leq N\leq 7$, interacting with independent dephasing and thermal (at zero and infinite temperature) baths. The authors considered the multipartite concurrence as the quantifier of entanglement. For both W and GHZ states, they  observed finite-time disentanglement exclusively in the case of independent baths at infinite temperature. For the $W$ states, they observed an approximately linear increase of the damping rate with $N$ for the  infinite-temperature reservoir, and a constant decay rate for the dephasing and zero-temperature thermal reservoirs. This is in contrast to GHZ states, whose damping rates increase with $N$ for all three decoherence models, as we know already. This has been confirmed by  Montakhab and Asadian in Ref. \cite{Montakhab08}, who essentially reproduced the findings of Carvalho {\it et al}. for the global entanglement measure of Meyer and Wallach \cite{meyer02} and up to $N=8$.

\par There is certain intuition behind the behaviour described above. For both pure dissipation (zero-temperature bath) and pure dephasing, the decay rate must be dictated by the total number of excitations in the state. This is essentially due to the fact that the rates of energy dissipation and coherence loss cannot grow faster than proportionally to the number of excitations. This, for W states \eqref{WdefN}, is by definition always equal to 1, regardless of the total number of particles in the system. Therefore, the effect of these two types of noise  on W states is expected to be constant in the system size. This in turn readily implies that W-type entanglement possesses a sort of built-in size-robustness against  several types of decoherence. Pure diffusion (infinite-temperature baths) instead -- as well as its related depolarization -- does not make any distinction between excited or ground states, and therefore its cumulative effect always makes the total decay rate increase without too much regard of the state. In fact, W states under pure diffusion seem to display a dependence of the damping rate with $N$ similar to that of GHZ states  \cite{carvalho04}.

\par Finally, in Ref. \cite{Chaves10}, Chaves and Davidovich studied the robustness  against independent channels PD and AD of W-state-like superpositions as  resources for teleportation of states and the splitting of quantum information between many parties. The authors considered W-states, generalizations of \eqref{WdefN} where each term in the superposition has a different amplitude, and superpositions of the latter with the vacuum. For several cases, the authors obtained analytic expressions, for all $N$, for the negativities, multipartite concurrence and the Meyer-Wallach global entanglement, and compared them with the fidelities for the  protocols corresponding to the different communication tasks. The effect of decoherence on the fidelity corresponding to the splitting of information between many parties was shown to be better described by the Meyer-Wallach global entanglement measure, rather than the negativity associated with the bipartite entanglement.

\subsubsection{Entanglement of Dicke states under global and local decoherence}
\label{Dickedecay}

In Ref. \cite{Guehne08}, G\"{u}hne, Bodoki, and Blaauboer studied the robustness of entanglement of different multipartite-entangled states under the influence of global dephasing. They use the geometric measure of entanglement, $E^G$, defined by Eq. \eqref{geometricmeasure}, to quantify entanglement, and its logarithmic derivative, $\frac{d}{dt}(\ln(E_G(t))\equiv\frac{\frac{d}{dt}E_G(t)}{E_G(t)}$, to quantify the entanglement robustness. The logarithmic derivative captures how fast entanglement decays as compared with how much of it there is left. For four qubits, the authors calculated the exact value of $\partial_t(\ln(E_G(t))$ for globally-dephased GHZ, W, linear-cluster, and symmetric Dicke states. They observed the entanglement in  Dicke states $\ket{{\text{Dicke}}^{2}_4}$, as defined in Eq. \eqref{DickedefN}, to be more robust. 

A full characterization of the dynamics of symmetric Dicke states in the presence of local noise was in turn performed in Ref. \cite{Campbell09} by Campbell, Tame and Paternostro.  More precisely, they  probed the states under independent AD, PD, and D channels. They performed an exhaustive comparison among different tools for the detection of genuine multipartite entanglement after decoherence acts on symmetric Dicke states $\ket{{\text{Dicke}}^{N/2}_N}$. Specifically, they studied the behaviour of $N$-point correlations, of  some state-discrimination techniques \cite{Schmid08},  of collective-spin entanglement witness that require just two local-measurement detection bases \cite{Toth07b}, of a generalized version of the latter \cite{Prevedel09},  of fidelity-based entanglement witnesses,  and of fidelity-based entanglement witnesses plus filtering operations \cite{guehne09}. 
\subsection{Decay of other classes of entanglement}
\label{Moredecay}
\subsubsection{Spin-squeezing under local depolarization}
\label{SpinSqueezing}
\par In Ref.~\cite{simon02}, Simon and Kempe also studied how spin-squeezing \cite{Kitagawa93}, which implies entanglement \cite{Sorensen01}, decays under independent channel D. For some families of initially pure states of arbitrary number of qubits, they calculated the evolution of the spin-squeezing parameters. They found some limiting cases (with asymptotically small initial squeezing) for which spin-squeezing survives up to critical local depolarization strengths as high as $p_{c}=0.29$. 

\subsubsection{Entanglement and distillability of weighted-graph states under local Pauli noise}
\label{Weightedgraph}
\par In Ref.~\cite{hein05}, Hein, D\"{u}r, and Briegel generalized the lower and upper bounds derived in Sec. \ref{DuerGraph} for graph states to the case of weighted graph-states subject to independent Pauli noise. Weighted graph-states are defined analogously to graph states \eqref{graphoperdef} but with non maximally-entangling controlled-phase gates instead of maximally-entangling controlled-$Z$ gates \cite{Hein_Review}. The authors derived for noisy weighted graph-states a lower bound on the distillability lifetime and upper bounds on the entanglement lifetime of the same type as those obtained for graph-states, described in Sec. \ref{DuerGraph}
.
\subsubsection{Few-qubit highly entangled states robust against local bit- or phase-flip noise}
\label{Borras}
\par In Ref. \cite{borras09}, Borras {\it et al}. reported particular forms of 4-, 5- and 6-qubit pure states that are both highly entangled in all the bipartitions and robust against the independent PD or BF channels. The authors numerically compared these states with random pure states, and with W and GHZ states, and found their entanglement decay to be the slowest of all. In particular, for independent dephasing noise, the decay of entanglement in these states is just linear in the dephasing strength. 
\subsubsection{Robustness of multipartite negativity of few-qubit states under local noise}
\label{Ali}
\par In Ref.~\cite{Ali13}, Ali and G\"{u}hne studied the short-to-intermediate time dynamics of the robustness of genuinely multipartite entanglement of several 3- and 4-qubit states under different models of local noise. As a measure of genuinely multipartite entanglement, the authors considered the multipartite negativity $Neg_{\text{Multi}}$  \cite{guehne2011}, defined by Eq. \eqref{MultiNegativity} in Sec. \ref{MultNeg}. They quantified its robustness with its logarithmic derivative $\frac{d}{dt}(\ln(Neg_{\text{Multi}}(t))\equiv\frac{\frac{d}{dt}Neg_{\text{Multi}}(t)}{Neg_{\text{Multi}}(t)}$. As mentioned before, the logarithmic derivative gives the variation of entanglement with time $t$ relative to the amount of  entanglement at time $t$, and therefore is a good figure of merit of the robustness of entanglement independently of how much of it is left.

As the states under scrutiny, they considered 3- and 4-qubit W and GHZ states, the 4-qubit symmetric Dicke state, the 4-qubit singlet, 4-qubit linear-cluster states, 4-qubit weighted graph states and 4-qubit Haar random states, among others. As the GHZ states for all $N$, the 4-qubit symmetric Dicke, singlet and linear-cluster states have all maximal multipartite negativity $Neg_{\text{Multi}}=1/2$. In turn, as models of noise, they considered the independent amplitude damping, phase damping, and depolarizing channels, for noise strengths $p=1-e^{-\gamma t}$, with $0\leq \gamma t\lesssim 0.7$.

For amplitude damping, the authors observed that nearly all states studied display a roughly exponential decay of $Neg_{\text{Multi}}(t)$ with $\gamma t$ (constant logarithmic derivative), with W states possessing the slowest decay rate. The exception to exponential decay was given by GHZ states, whose multipartite negativity features initially an exponential decay but starts decreasing super-exponentially (with the logarithmic derivative itself decreasing) at intermediate times. In contrast, for dephasing, the GHZ states turned out to be the most robust ones, whereas the W states become the least robust ones; with GHZ states displaying an exponential decay and all other states a super-exponential one. Also for depolarization did the GHZ states appear as the most robust of all; and in this case all states showed a super-exponential decay at short times. 

The authors noticed that, except for the case of dephasing, their findings for the W and GHZ states are consistent with the observations of Ref. \cite{carvalho04} for the multipartite concurrence, discussed in Secs. \ref{Carvalho} and \ref{Wdecay}. For the dephasing channel, the results of \cite{carvalho04} and  \cite{Ali13} are opposite. The authors attribute this discrepancy to the possibility that the lower bound used in Ref. \cite{carvalho04} is tighter for W than for  GHZ states. Thus, this bound could be mistaking the former for the latter as the most robust states.

Finally, a remark about the connection between the behaviours of GHZ states reported here and in Ref. \cite{aolita08}, discussed in Sec. \ref{CompleteGHZ}, is in place. From Ref. \cite{aolita08}, it follows that, for sufficiently small $\gamma t$, all bipartite negativities $Neg$ decay approximately exponentially in $\gamma t$. We notice that this is not in contradiction with the super-exponential decay of $Neg_{\text{Multi}}$ at short times observed by Ali and G\"{u}hne for the case of depolarisation. This is simply due to the fact that whereas $Neg$ quantifies bipartite correlations, $Neg_{\text{Multi}}$ exclusively assesses genuinely multipartite ones. As a matter of fact, the results of Ali and G\"{u}hne constitute the first studies, and to our knowledge the only ones up to date, of genuinely multipartite entanglement decay under local noise.

\subsection{Bound entanglement via local noise on multiparticle pure states}
\label{NatMultiBound}
Originally, bound entangled states were meticulously constructed as to provide examples of entangled states from which no entanglement could be distilled -- as for instance the first bipartite entangled PPT states \cite{horodecki98}, or the multipartite examples \cite{bennett99,smolin01} discussed in Sec. \ref{boundmulti}. Later on, bound entanglement was found in a variety of multipartite mixed states, including thermal fermionic systems of up to 12 spins \cite{toth07,patane07} and thermal chains of coupled harmonic oscillators in the thermodynamic limit \cite{Ferraro08,Jack07}. This posteriorly led to the understanding that bound entanglement, rather than a curiosity, is a common feature of multipartite systems that can arise in
natural dynamical processes, more precisely due to the action of local noisy or thermal environments. This is the topic of this section, which has been extensively studied in Refs. \cite{kay06, kay07,Kaszlikowski08, aolita08, aolitapra09, borras09, cavalcanti10, kay10}. Here, we focus mainly on GHZ  and rectangular-cluster states, as described by the proof-of-principle approaches of  \cite{aolita08} and \cite{cavalcanti10}, respectively. Also, we briefly mention the case of random multipartite pure states \cite{aolitapra09, borras09}.
\subsubsection{Bound entanglement in GHZ states under local decoherence}
\label{BoundGHZ}

\par  Aolita {\it et al}. reported in \cite{aolita08} that independent D or GAD channels drive generalized GHZ states \eqref{kalphabeta} towards multipartite bound-entangled states. This is essentially due to the fact that different bipartitions of a GHZ states under local noise  often disentangle at different instants. When the evolution leads each and all of the  bipartitions of one qubit versus the rest to biseparability, the state is necessarily  non-distillable, for no entanglement can be extracted via LOCCs. If in addition the state is still entangled in  other bipartitions, then the state is necessarily  bound entangled. 

\begin{figure}
\begin{center}\includegraphics[width=1\linewidth]{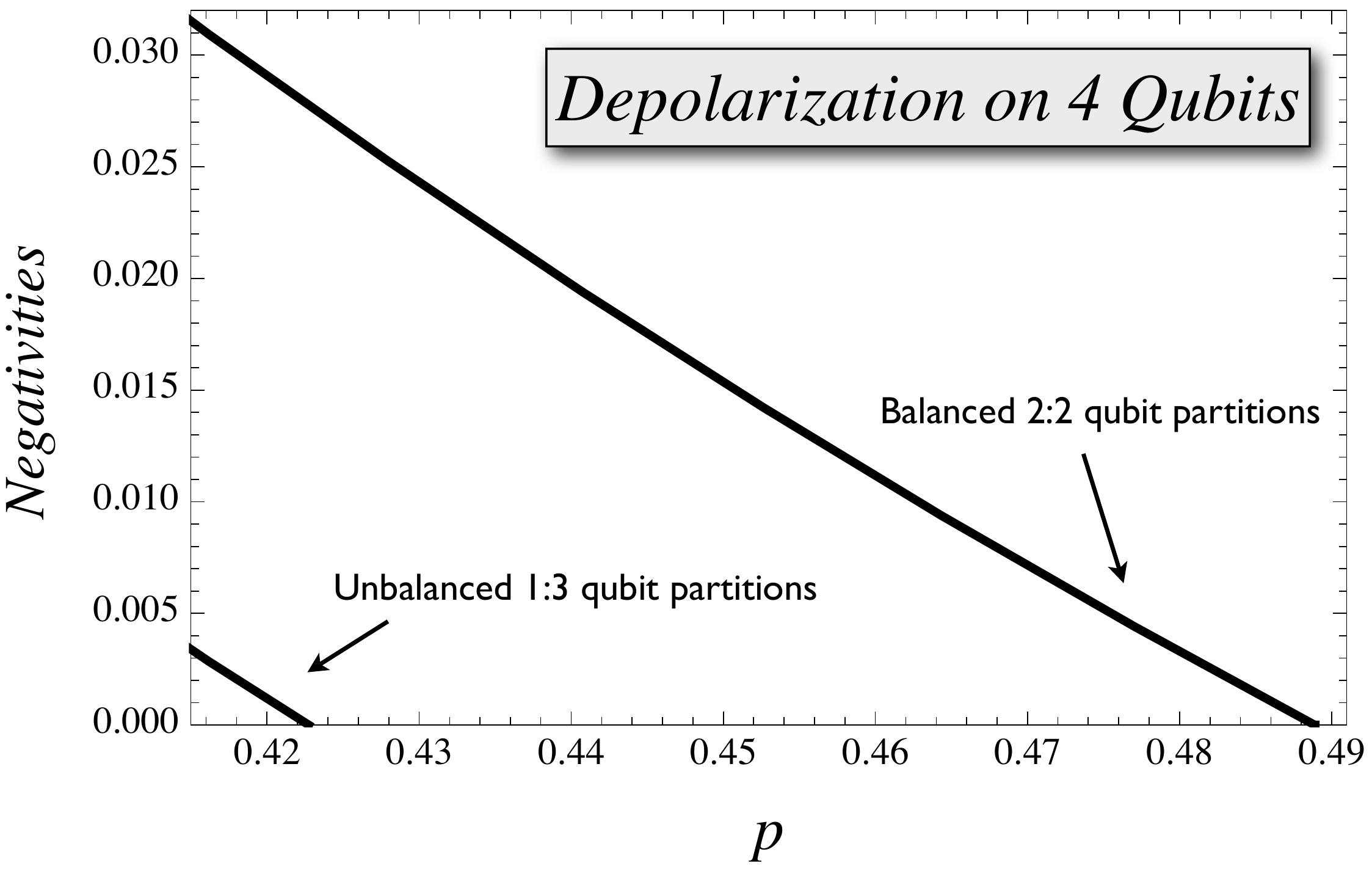}
\caption{Negativity of four-qubit GHZ states under independent depolarization with probability $p$, for the least-balanced $1:3$-qubit bipartitions and the most-balanced $2:2$-qubit ones. Between the instants where both curves vanish the state is bound entangled. A similar behavior is observed with independent thermal baths at any non-null temperature, but the 
effect is not so marked (the smaller the temperature, the weaker the effect). Reprinted figure with permission from L. Aolita {\it et al}., \href{http://link.aps.org/doi/10.1103/PhysRevLett.100.080501}{Phys. Rev. Lett. {\bf 100}, 080501 (2008)}. Copyright (2008) by the American Physical Society.}
\label{Bound}
\end{center}
\end{figure}

\par A four-qubit example is shown in Fig.~\ref{Bound}. There, the $1:3$ negativities vanish before those corresponding to the $2:2$ bipartitions. Between these two events the state is bound entangled. Furthermore, the authors observed a similar behaviour for other values of $\alpha$ and $\beta$ in Eq. \eqref{kalphabeta}, and independent GAD channel at any temperature $T>0$. This is not the case though for channels PD or AD ($T=0$), for which all bipartitions always disentangle simultaneously, as studied in Sec. \ref{CompleteGHZ}. 

\par The noise-strength range of bound entanglement in turn increases with $N$, because, the time of separability of the most-balanced bipartitions grows with $N$, whereas that of the least-balanced ones decreases with $N$  \cite{aolita08}. However, since the entanglement in all bipartitions approaches zero exponentially faster with increasing $N$, the total amount of bound entanglement decreases accordingly with growing system-size. Still, the effect is significant enough to be observed in practice, as we see in Sec. \ref{66}, where a similar approach to the one described here was used for the experimental generation of  bound entangled states. 

\subsubsection{Bound entanglement in random states under local depolarization}
\label{Randomboundent}
The effect described in the previous sub-section for GHZ states appears to be quite general. In Ref.~\cite{borras09}, Borras {\it et al}. numerically observed that, over a sample of 1000 Haar (that is, uniformly distributed) random pure states of $N=6$ qubits, in all cases local depolarization drives the state towards a multipartite bound entangled state at some point in the evolution. This also indirectly follows from Ref. \cite{aolitapra09}, where Aolita {\it et al}. numerically observed that, over a sample of 10000 Haar random pure states of $N=6$ qubits under local depolarization, the time at which the average entanglement in the most-balanced bipartitions vanishes, is smaller than the one at which that in the least-balanced bipartitions does.
\subsubsection{Thermal bound entanglement in graph states under local dephasing.}
\label{Thermalgraph}
In Refs. \cite{kay06, kay07} Kay {\it et al}. calculated the critical temperature of distillability for graph states, which,  due to equivalence \eqref{decoherence1}, yields the critical dephasing strength for distillability of independently dephased graph states. Later, in Ref. \cite{cavalcanti10}, Cavalcanti {\it et al}. showed not only that bound entanglement is always present in independently-dephased graph states of arbitrary size, but also that it is robust against small perturbations in the Hamiltonian couplings. Finally, in Ref. \cite{kay10}, Kay showed that for two-colourable graph states (as for instance all rectangular cluster states) the temperature (local dephasing strength) range of bound entanglement grows with $N$. Here we briefly sketch the argument as presented in \cite{cavalcanti10}, deeply based on  the formalism of Sec. \ref{Jack}.

\begin{figure}
\begin{center}\includegraphics[width=1\linewidth]{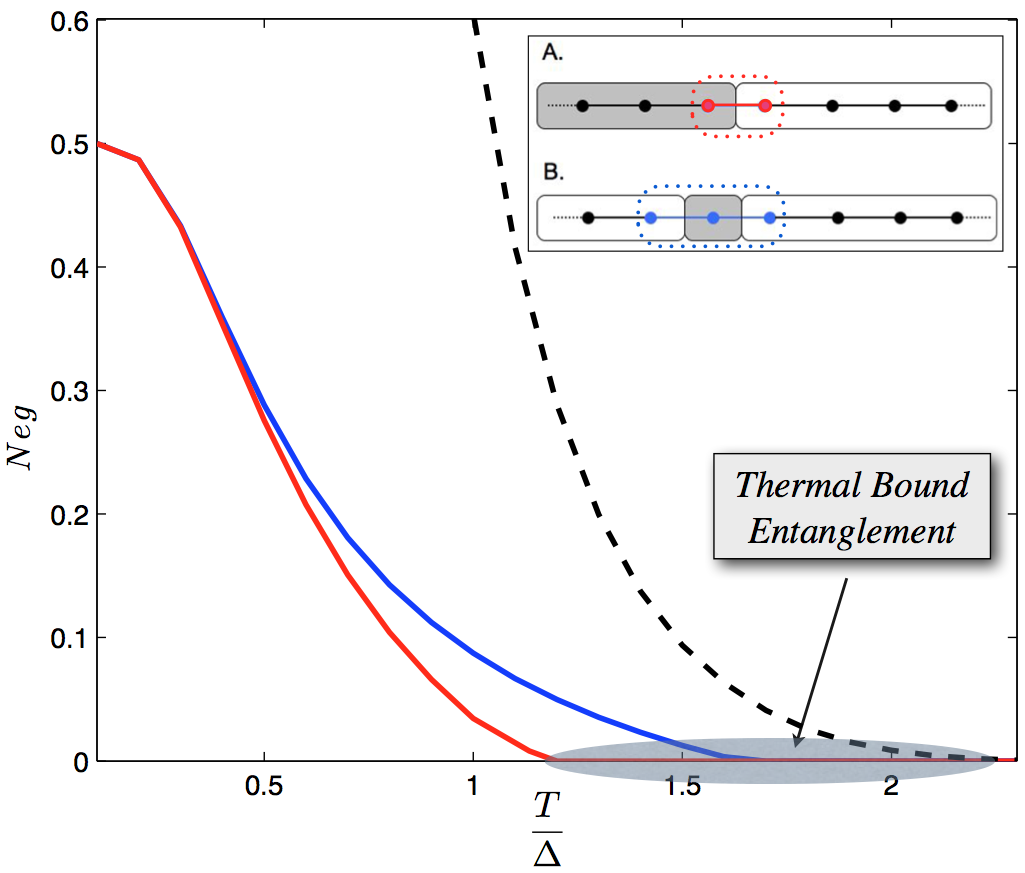}
\caption{Inset: Two possible splittings of a linear cluster (1D graph) into a grey and a white regions. {\bf A.} Two contiguous blocks of nodes are connected by the boundary subsystem in red, consisting of two nodes. {\bf B}. Two non-contiguous blocks of nodes (a single node versus all the others) are connected by the boundary subsystem in blue, consisting of three nodes. For thermal graph states, all the entanglement with respect to any multi-partition is always localized in the corresponding boundary subsystem. Main plot: Negativities of a thermal linear-cluster state as a function of the temperature $T$ (in units of the Hamiltonian gap $\Delta$), for the bipartitions shown in the inset: any two contiguous blocks (red solid) and any qubit versus the rest (blue solid). In addition, in black dashed, the negativity for the even-odd (or ``zig-zag") bipartition of a specimen of $N=12$ qubits is also displayed. The grey-shaded region indicates the temperature range of bound entanglement. The red and blue curves do not depend on $N$. The black dashed one does: its vanishing temperature grows with $N$ \cite{kay10}. Therefore bound entanglement is also present in macroscopic thermal clusters. Adapted from D. Cavalcanti {\it et al}., \href{http://dx.doi.org/10.1088/1367-2630/12/2/025011}{New J. Phys. {\bf 12}, 025011 (2010)}.}
\label{negativity}
\end{center}
\end{figure}

We begin by the simplest case of 1D graphs, represented in the inset Fig. \ref{negativity}, and constant couplings $\Delta_i\equiv \Delta$ in  \eqref{H}, with $\Delta$ the Hamiltonian energy gap.  Consider first a bipartition of the chain into two contiguous blocks, say from qubit $i$ to the left (grey block in inset A) and from qubit $i+1$ to the right (white block in inset A). The system is in thermal equilibrium with a bath at temperature $T$, in the thermal state $\varrho_T\equiv\mathcal{E}^{PD}\big(\ket{\mathcal{G}_{0\ ...\ 0}}\bra{\mathcal{G}_{0\ ...\ 0}}\big)$, as explained in Sec. \ref{LocalDephisThermalization}. Since $\mathcal{E}^{PD}$ is a Pauli map, formula \eqref{exact} for the entanglement in the bipartition applies. More over, for channel $\mathcal{E}^{PD}$ the evaluation is particularly simple, as the channel commutes with all controlled-$Z$ gates. One has then that  $\mathcal{E}^{PD}\equiv\tilde{\mathcal{E}}^{PD}$ and immediately obtains that $\tilde{p}_{\gamma,\delta}\equiv
\tilde{p}_{\gamma}\tilde{p}_{\delta}$. That is, the summations over $\gamma$ and $\delta$ in \eqref{exact} become independent, and only the mixing due to the independent dephasing on the boundary subgraph, depicted in blue in inset A of the figure, survives.

As a result, the total entanglement between any two contiguous blocks is equal to that of a two-qubit thermal graph state at the same temperature. Thus, the critical temperature ${T^c}_{2}$ for which the entire thermal cluster becomes separable with respect to all contiguous-block bipartitions is immediately obtained \cite{kay06,cavalcanti10},
\begin{equation}\label{Tc1}
{T^c}_{2}(\Delta)=\frac{-\Delta}{\ln(\sqrt{2}-1)}\approx 1.1 \Delta.
\end{equation}
Above this temperature, the $N$-qubit thermal state $\varrho_T$ is non-distillable, as for any two qubits a contiguous-block bipartition can be found in which each particle lies on a different side of the partition and is therefore separable from the other, so that  not even entanglement between any two qubits can be extracted by LOCCs.

On the other hand, with a similar reasoning as in the previous subsection, one finds a range of  temperatures $T\geq {T^c}_{2}(\Delta)$  such that $\varrho_T$ is non-separable. Then $\varrho_T$ is necessarily bound entangled. This range of temperatures comes from considering non-contiguous bipartitions of the chain as that of an $i$-th qubit versus all the rest (grey versus white regions in inset B of Fig. \ref{negativity}). In this case, the boundary system corresponds to a three-qubit thermal system, represented in blue in inset B, whose negativity vanishes at a higher temperature. Furthermore, the two-colourable ``zig-zag" bipartition is always the last one to become PPT, and its corresponding critical temperature grows with $N$ \cite{kay10}. Similar results are obtained for higher dimensional clusters and  for unequal Hamiltonian couplings \cite{cavalcanti10,kay10}.

\par To end up with, two very  important comments are in place. First, since  $\tilde{p}_{\gamma,\delta}=\tilde{p}_{\gamma}\tilde{p}_{\delta}$, and since  the boundary subgraph does not change when the cluster's size  varies, none of the latter results depends at all on $N$. This implies that bound entanglement  is present in thermal graphs of arbitrary size, in particular also  in macroscopic specimens. This complements the discoveries for bosonic systems by Ferraro {\it et al}. \cite{Ferraro08,Jack07}. Second, the latter, together with the robustness of bound entanglement against changes in the Hamiltonian couplings, suggests that thermal bound entanglement, more than a singularity of a peculiar interaction pattern, might be a rather typical feature,  very likely to be present in many other systems of strongly-interacting particles.

\section{Experiments on open-system dynamics of entanglement}
\label{6}

\begin{figure}[t]
\begin{center}
\includegraphics[width=\linewidth]{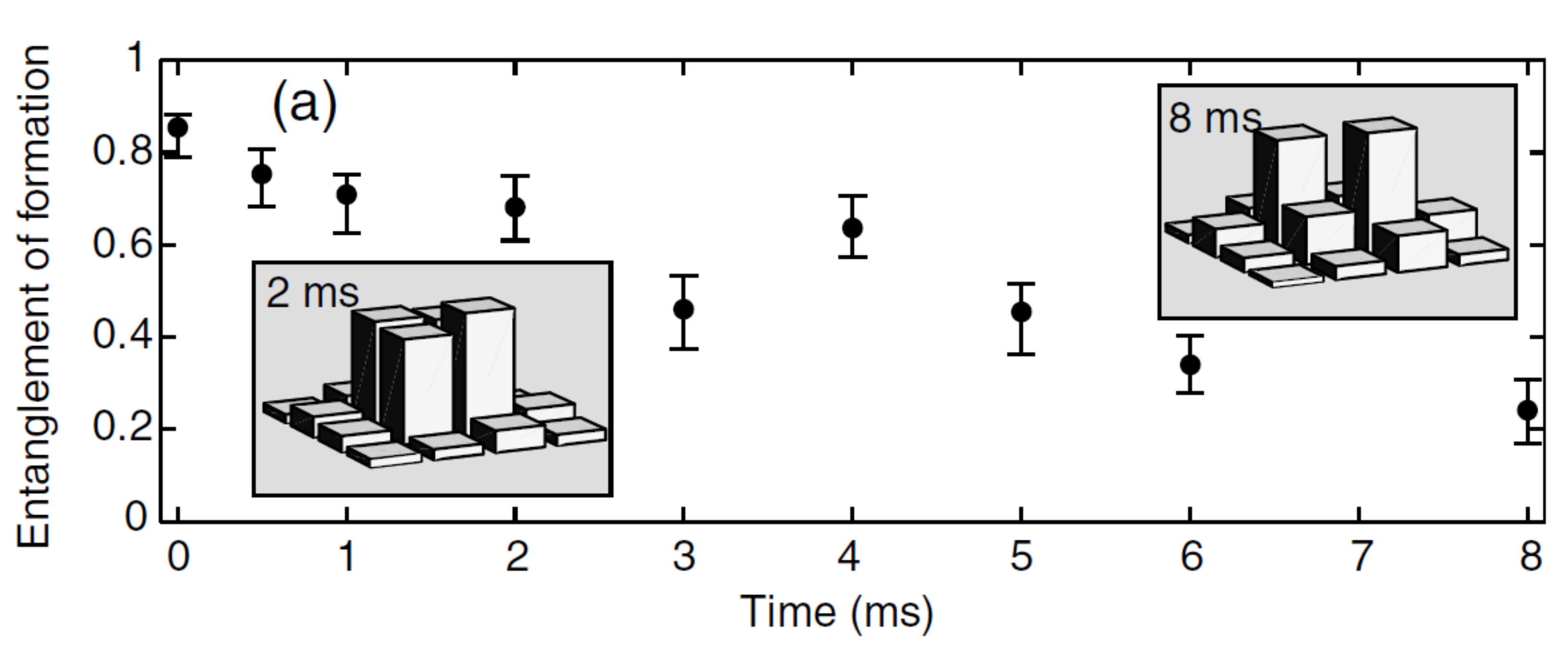}
\caption{Entanglement evolution of $|\psi^+\rangle$ encoded into two trapped 
ions~\cite{roos04b}. Each black dot amounts to a full tomography of the two ions' state, implying several realizations for each experimental point. After the density matrix is determined, the entanglement of formation is evaluated. The density matrix displayed in the left inset ($t=2ms$), in the computational basis $\{\ket{00}_\sys,\ket{01}_\sys,\ket{10}_\sys,\ket{11}_\sys\}$, is to be compared to the one shown in the right inset, ($t=8ms$). Clearly the off-diagonal elements, the coherences, decay due to the interaction with the environment, implying the decay of entanglement -- not necessarily both features decay at the same speed. Reprinted figure with permission from C. F. Ross {\it et al}., \href{http://link.aps.org/doi/10.1103/PhysRevLett.92.220402}{Phys. Rev. Lett. {\bf 92}, 220402 (2004)}. Copyright (2004) by the American Physical Society. } 
\label{roos}
\end{center}
\end{figure}

 Practically all experiments that aim to create entangled states for any given purpose indirectly assess the entanglement dynamics. The simple preparation of an entangled state and its later measurement is already probing the dynamics of entanglement due to the unavoidable influence of the natural environment. For instance, an early experiment 
where this procedure was implemented was reported by C.~F.~Roos \emph{et~al.}, in Ref.~\cite{roos04b}. There, the four Bell-states were deterministically prepared in the energy levels of two trapped ions, and were left to evolve exclusively under the detrimental action of their natural decoherence processes, dominated by spontaneous emission. At different points in time, a full tomography was performed and an entanglement measure evaluated, in this case the entanglement of formation (see Fig.~\ref{roos}). An overall decay of entanglement was clearly observed.

Nevertheless, to study in detail the many aspects of entanglement dynamics, as introduced in the earlier sections, a remarkable control of the interaction between system and environment is necessary. Experiments with such level of control are not so frequent. Bellow we analyze some aspects of the few experiments explicitly designed to probe  open-system entanglement dynamics. In most cases, the entanglement dynamics is actually not studied as a function of time, but rather of the relevant noise strength of the experiment, which can be varied in a controlled way. In previous sections we identified this parameter with a noise probability $p$. Here,  this may represent the thickness of a quartz slab that attenuates an optical beam, or the amount of dephasing introduced between two atomic states, or yet the coupling between polarization and spatial degrees of freedom of a photon. Changing the value of this parameter is equivalent to changing the time duration of the noisy process, and therefore the corresponding investigation translates directly into a time-dependent analysis.

\subsection{Observing finite-time disentanglement}
\label{61}

\begin{figure}[t!hf]
\begin{center}
\includegraphics[width=.9\linewidth]{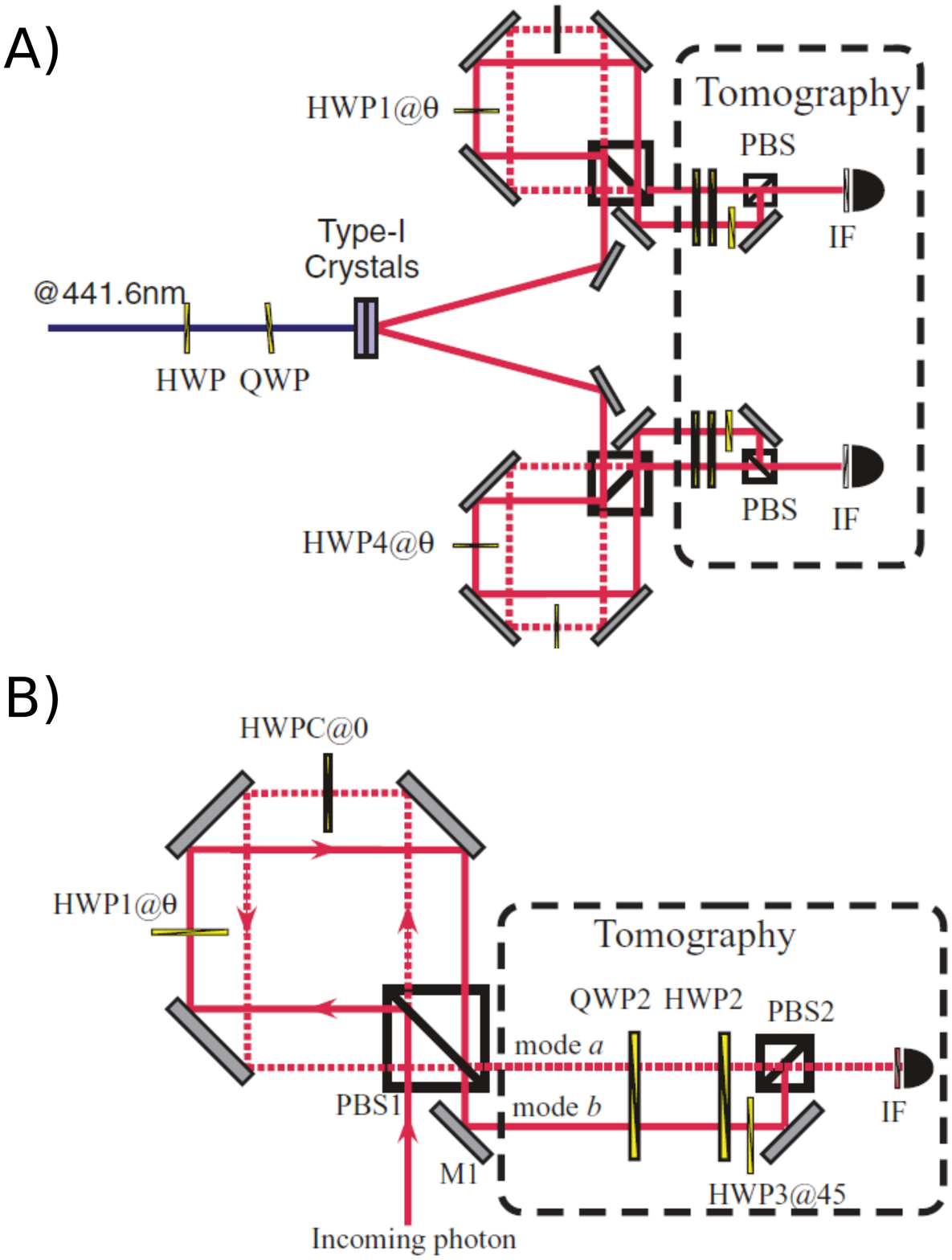}
\caption{Experimental setup for probing the open-system dynamics of entanglement. Two adjacent nonlinear crystals pumped by a continuous laser source generate pairs of polarization-entangled photons. Each photon is sent through an interferometer, where a polarized beam splitter sends orthogonal polarizations into different propagation modes, which act as environment for the photon polarizations. Quantum tomography on the outgoing field allows the reconstruction of the polarization state and the calculation of its concurrence. From M. F. Almeida {\it et al}., \href{http://dx.doi.org/10.1126/science.1139892}{Science {\bf 316}, 579 (2007)}. Reprinted with permission from AAAS. } 
\label{setupexp}
\end{center}
\end{figure}

The first report of an experimental study of the full entanglement dynamics under a controlled environment is due to Almeida {\it et al}.~\cite{almeida07}. There, the finite-time disentanglement of two qubits, as described in Sec.~\ref{EntDynaAD}, was verified.  The corresponding setup is shown in Fig.~\ref{setupexp}.

A twin-photon pair entangled in the polarization degrees of freedom is produced by a down-conversion setup. The produced two-photon state is close to $|\psi\rangle=\alpha|HH\rangle+\beta|VV\rangle$, where $H$ and $V$ stand respectively for the horizontal and vertical linear polarization states, and $\alpha$ and $\beta$ are complex coefficients (with $|\alpha|^2+|\beta|^2=1$) that depend on the orientation of the polarization of the pumping laser. 
Each photon is sent through separate interferometers, which play the role of independent environments. The interferometers couple the $H$ and $V$ polarizations to  different propagation modes. This is done with a polarized beam splitter (PBS) and an arrangement of mirrors, as shown in Fig.~\ref{setupexp} B). More precisely, a horizontally polarized photon is transmitted through the PBS, and propagates in the interferometer mode ``0,''  represented by a red dashed line. A vertically polarized photon is reflected by the PBS into the interferometer mode ``1,''  represented by a red dashed line. If its polarization remains the same, then it is reflected again by the PBS, emerging from the interferometer in mode ``0,'' thus following the same output path as the incoming horizontally-polarized photon. On the other hand, if its polarization is rotated by the half-wave plate (HWP) inserted inside the interferometer into propagation mode ``1,''  so that $|V\rangle\rightarrow \cos(2\theta)|V\rangle + \sin(2\theta) |H\rangle$, where $\theta$ is the angle of the plate, the horizontal component will emerge from the interferometer in mode ``1,'' while the vertical component will emerge along mode ``0,'' as before. Modes ``0'' and ``1'' represent two different states of the environment.

Through this procedure, the following transformation is implemented:
\begin{eqnarray}
\label{transfoequiv}
&&|H\rangle|0\rangle\rightarrow|H\rangle|0\rangle\nonumber\\
&&|V\rangle|0\rangle\rightarrow \sqrt{1-p}|V\rangle|0\rangle+\sqrt{p}|H\rangle|1\rangle\,,
\end{eqnarray}
where $p=\cos2\theta$. Identifying $|H\rangle$ and $|V\rangle$ with the ground and excited states of a two-level atom, respectively, we see that this map can be used to simulate different types of dynamical behaviours. For instance, for $p=1-e^{-\gamma t}$ it is equivalent to transformation \eqref{APmap} of Sec.~\ref{Examples}, which models the spontaneous decay of an atom, inducing the independent amplitude-damping channel \eqref{KrausAP}. On the other hand, for $p=\sin^2(\Omega t/2)$ it is equivalent to \eqref{GHZLC}, which models the resonant Jaynes-Cummings dynamics. In the first case, vanishing of entanglement for a value of $p<1$ corresponds to finite-time disentanglement. Whereas, for the resonant Jaynes-Cummings model, the oscillation of $p$ leads to the disappearance and subsequent revival of entanglement at times corresponding to half Rabi cycles, as expected. Equation \eqref{transfoequiv} makes clear that, for the processes considered above, the environment can be described as a single qubit.

The two emerging paths are reunited incoherently with the help of a beam splitter, as shown in Fig.~\ref{setupexp}B) before the photon hits a detector. This amounts to tracing out the environment. Half-wave and quarter-wave plates placed on the exiting paths allow a quantum tomographic reconstruction of the emerging polarization state.

\begin{figure}[t]
\begin{center}
\includegraphics[width=\linewidth]{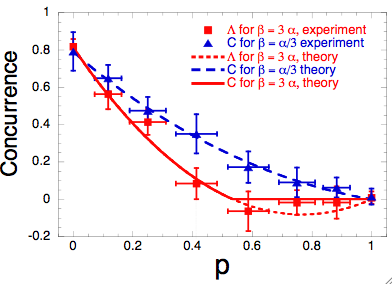}
\caption{Concurrence as a function of the transition probability $p$. Two different initial states are considered, close to $\alpha|00\rangle+\beta|11\rangle$, with the same initial concurrence, but differing on the relation between $|\alpha|$ and $|\beta|$: either $|\alpha|=\sqrt{3}|\beta|$ (triangles) or $|\beta|=\sqrt{3}|\alpha|$ (squares). The points correspond to experimental data for the quantity $\Lambda$ defined by (\ref{concu1})(b), obtained by tomographic reconstruction of the twin-photon state for each value of $p$. According to \eqref{concu1}(a), concurrence coincides with $\Lambda$ when $\Lambda\ge 0$ and  is equal to zero when $\Lambda<0$. The dashed lines correspond to the value of $\Lambda$ obtained by applying the amplitude map to the actual initial state produced in the experiment. Adapted from M. F. Almeida {\it et al}., \href{http://dx.doi.org/10.1126/science.1139892}{Science {\bf 316}, 579 (2007)}. Reprinted with permission from AAAS. } 
\label{concuexp}
\end{center}
\end{figure}

Applying this procedure to a photon pair detected in coincidence, as sketched in Fig.~\ref{setupexp}A), one is able to reconstruct the two-qubit state and calculate for instance the corresponding concurrence. The experimental results for the concurrence are shown in Fig.~\ref{concuexp}, and clearly demonstrate the (in general) non-exponential decay of this quantity. In particular, the disappearance of entanglement for $p<1$  (corresponding, for a decaying atom, to a finite time) is verified when the probability of the excited-state component is larger than that of the ground state ($|\beta|>|\alpha|$).

\begin{figure}[t]
\begin{center}
\includegraphics[width=\linewidth]{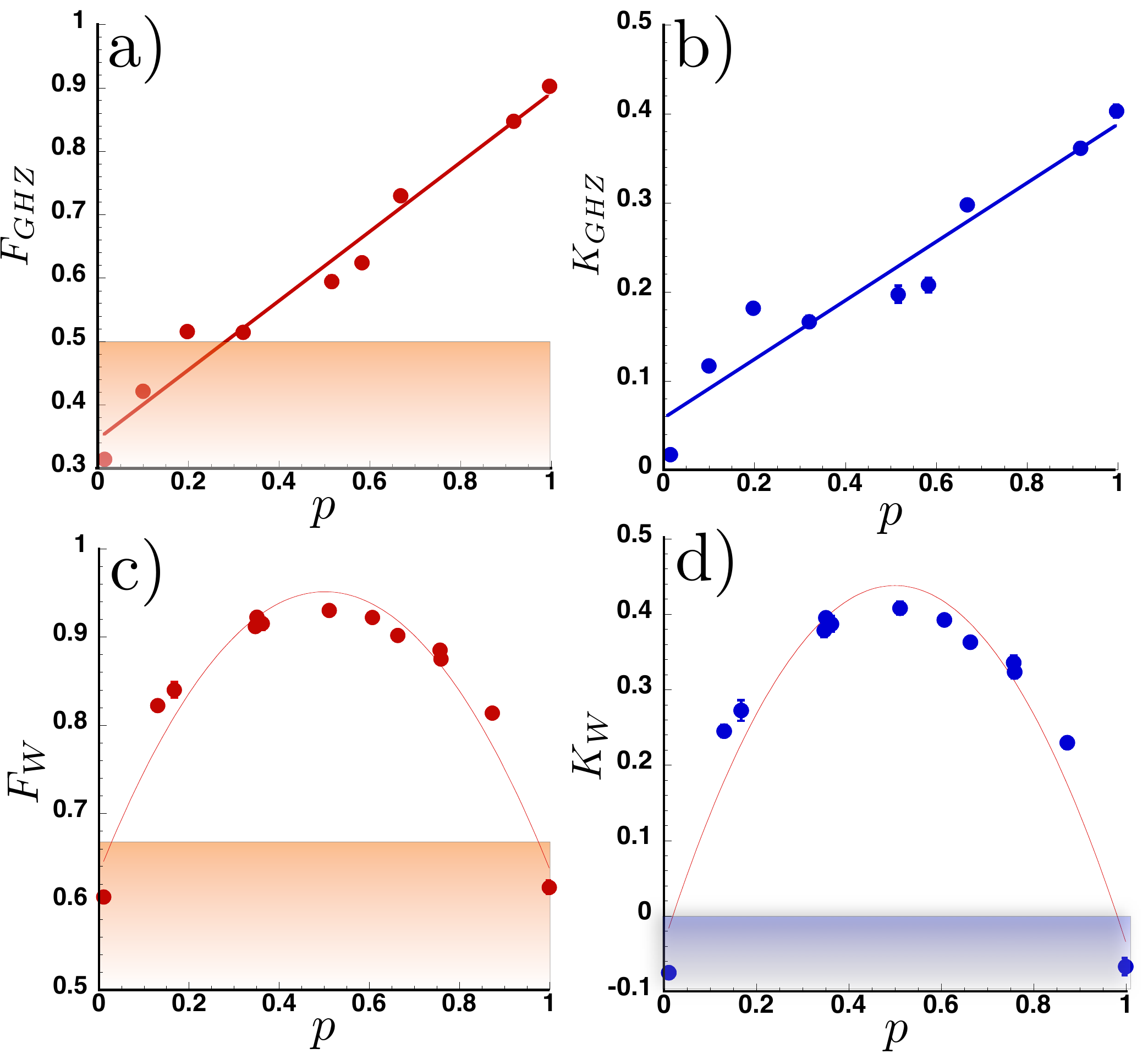}
\caption{Indicators of genuine multipartite entanglement between a two-qubit system and a one-qubit environment as a function of the noise strength $p$, referring to Table \ref{tabKraus}: (a) Fidelity $F_{GHZ}$ with respect to the GHZ state and PD channel; $F_{GHZ}\geq1/2$ indicates genuine tripartite entanglement. (b) Violation $K_{GHZ}$ of Criterion \ref{criterion:BisepGHZ} and PD channel; $K_{GHZ}>0$ implies genuine tripartite entanglement. (c) Fidelity $F_{W}$ with respect to the W state and AD channel; $F_{W}\geq2/3$ indicates genuine tripartite entanglement. (d) Violation $K_{W}$ of Criterion \ref{criterion:BisepGHZ} and AD channel; $K_{W}>0$ guarantees genuine tripartite entanglement. The lines are fittings to the experimental data. Reprinted figure with permission from O. J. Far\'\i as {\it et al}., \href{http://link.aps.org/doi/10.1103/PhysRevLett.109.150403}{Phys. Rev. Lett. {\bf 109}, 150403 (2012)}. Copyright (2012) by the American Physical Society.} 
\label{emergence}
\end{center}
\end{figure}

If, instead of reuniting the two emerging paths, as shown in Fig.~\ref{setupexp} B), one measures the population of each path and also their coherence by letting them go through another interferometer, one is able to reconstruct the full state of system plus environment. This was done by Jim\'enez-Far\'ias {\it et al}. in \cite{osvaldo2012}, by using an interferometric setup that allowed the implementation of several decoherence channels and full access to all environmental degrees of freedom. In particular, this setup allows the quantum tomography of the joint polarization-path degrees of freedom, thus exhibiting the flow of entanglement from the initial two-partite system towards the environment, as discussed in Section \ref{EntDynaAD}. It was shown in \cite{osvaldo2012} that, when a qubit from an entangled pair interacts with the environment, which as in Eq.~(\ref{transfoequiv}) may be described by a qubit, the initial bipartite entanglement gets redistributed into bipartite and genuine tripartite entanglements. This complements the theoretical results of \cite{lopez-2008} concerning the entanglement of two independent environments coupled to two qubits of an initially entangled pair. Although ideally the state corresponding to system plus environment should be pure, this is of course not true in the experiment, due to uncontrollable noise and uncertainties in the preparation of the initial state. In \cite{osvaldo2012}, emergence of genuine multipartite entanglement of the W and GHZ types, between system (polarization degrees of freedom) and environment (spatial modes), was identified through the use of both fidelity-based witnesses, corresponding to \eqref{recipeopt}, and Criteria \ref{criterion:BisepGHZ} and \ref{criterion:BisepW}, discussed in Sec. \ref{BisepCrit}, and proposed in Ref.~\cite{GuehneSeevinck}.  Figure \ref{emergence} displays the experimental results obtained for the amplitude damping (AD) and phase-damping (PD) channels. In this picture, $K_{GHZ}$ and $K_{W}$ correspond respectively to the differences between the left-hand side and the right-hand side of Eqs.~\eqref{BisepcritGHZ} and \eqref{BisepcritW}, so that positive values of $K_{GHZ}$ and $K_{W}$ signal the existence of genuine tripartite entanglement of the GHZ or W type, respectively. In the figure, we can see  that Criteria \ref{criterion:BisepGHZ} and \ref{criterion:BisepW} identify genuine tripartite entanglement over a wider range of values of $p$ than the fidelity-based witnesses.

Finite-time disappearance of entanglement was also observed by Laurat {\it et al}. \cite{Laurat}. In this realization, an effective two-qubit entangled state, where a single collective atomic excitation is coherently shared between two separate atomic clouds, was  prepared probabilistically in a heralded fashion, with a setup based on the Duan-Lukin-Cirac-Zoller (DLCZ) scheme \cite{DLCZ}. 
While stored in the atomic clouds, entanglement is exposed to the action of the environment, and as such, it decays. After an interval $\tau$, the remaining entanglement is measured by mapping the respective atomic states back  into a photonic state, with the help of read pulses. The tomography of the photonic state, and further entanglement evaluation displayed the vanishing of entanglement for a finite value of $\tau$~\cite{Laurat}. In spite of the usually low values of entanglement obtained in this type of setup, it represents an important step towards quantum networks~\cite{kimbleQN}, since it allows for storage and retrieval of quantum information in a heralded fashion.

\subsection{Verifying equations of motion of entanglement}
\label{62}

\begin{figure}[t]
\begin{center}
\includegraphics[width=\linewidth]{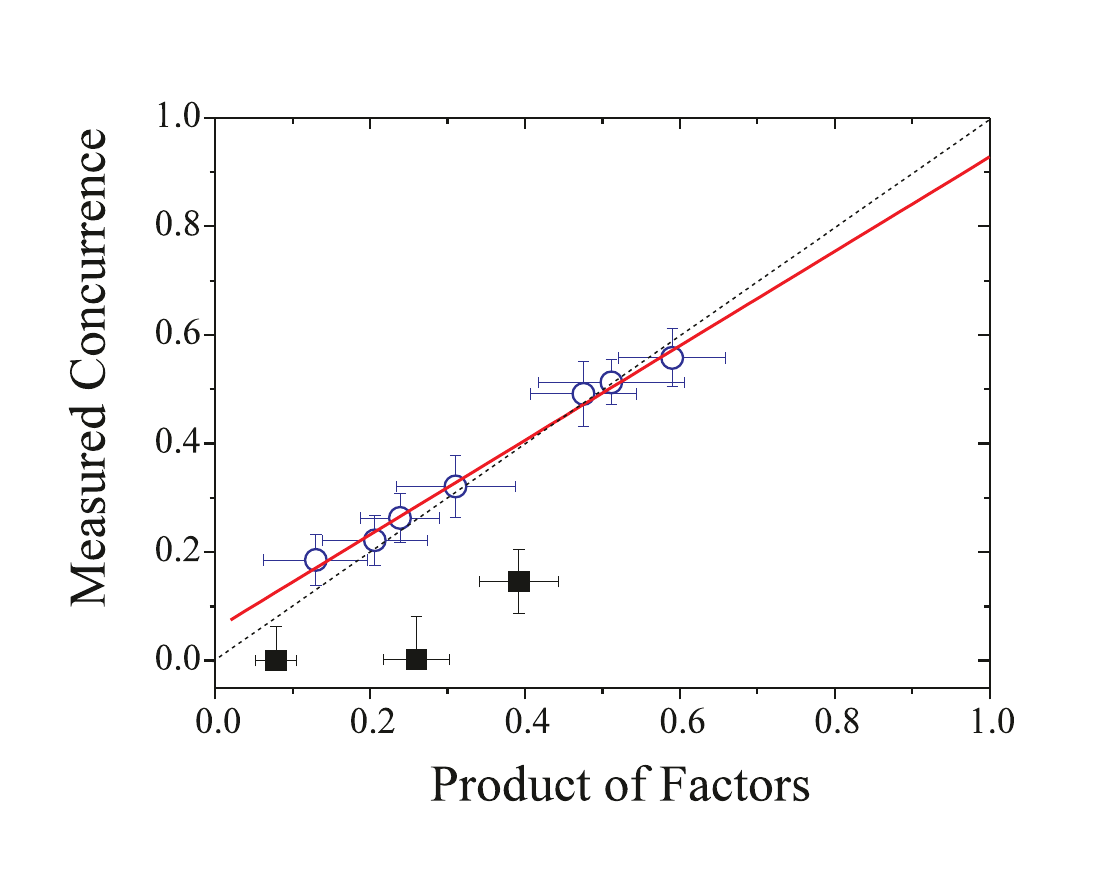}
\caption{Experimental test of the dynamical law for the evolution of entanglement, for one-sided channels. Comparison between measured concurrence and the product of factors on the right side of Eq.(\ref{concMovUpper}). The circles and squares correspond respectively to quasi-pure and mixed initial states. The dashed line has unit slope and the solid line is a linear fit to the circles (slope $\sim 0.87$). From O. Jim\'enez-Far\'\i as {\it et al}., \href{http://dx.doi.org/10.1126/science.1171544}{Science {\bf 324}, 1414 (2009)}. Reprinted with permission from AAAS.} 
\label{dynamicsexp}
\end{center}
\end{figure}

The strategy displayed in Fig.~\ref{setupexp} is also useful to test the dynamical law of entanglement given by Eq.~(\ref{concMov}). 
The first experimental test was implemented by applying the channel shown in Fig.~\ref{setupexp} B) to one of the photons of the entangled pair \cite{farias}. The first factor on the right-hand side of (\ref{concMov}) is determined by tomography of the input state, and subsequent evaluation of its concurrence; whereas the second one by process tomography of the single-qubit channel ${\cal E}$, and subsequent evaluation of the concurrence of the state $\openone\otimes{\cal E}|\Phi^+\rangle\langle\Phi^+|$. Process tomography of ${\cal E}$ can be done with a laser field sent though the channel with four different incoming polarizations, and measuring the corresponding polarizations of the outgoing field. This suffices to determine all the Kraus operators of the channel \cite{nielsen00}. 

An alternative to direct process tomography of ${\cal E}$ would be to perform state tomography of $\openone\otimes{\cal E}|\Phi^+\rangle\langle\Phi^+|$. This state, by the Choi-Jamio\l kowsky isomorphism~\eqref{CJDualism}, discussed in Sec.~\ref{Krausop}, unambiguously determines channel ${\cal E}$. However, a maximally entangled state cannot be precisely produced in the lab. Furthermore, the direct process tomography, since it uses intense laser beams, yields a signal-to-noise ratio much larger than the one corresponding to measuring the effect of the channel on a maximally-entangled state.

Figure~\ref{dynamicsexp} displays the experimental results for states that are initially close to pure states, and also for initially mixed states, undergoing the amplitude-damping channel. A very good agreement between the experimental data and the entanglement equation of motion~(\ref{concMov}) is obtained for pure states -- the independently measured $l.h.s.$ and $r.h.s.$ of Eq.~(\ref{concMov}) agree, within experimental errors, for all the points representing initially pure states (circles), as depicted by the dashed line with unit slope in Fig.~\ref{dynamicsexp}. For mixed states (squares), the inequality (\ref{concMovUpper}) applies, and the experimental points are found below the unit slope curve as expected. A fit to the experimental data gives a slope of $\sim 0.87$, taken into account only the initially quasi-pure states (circles).

\begin{figure}[t]
\begin{center}
\includegraphics[width=\linewidth]{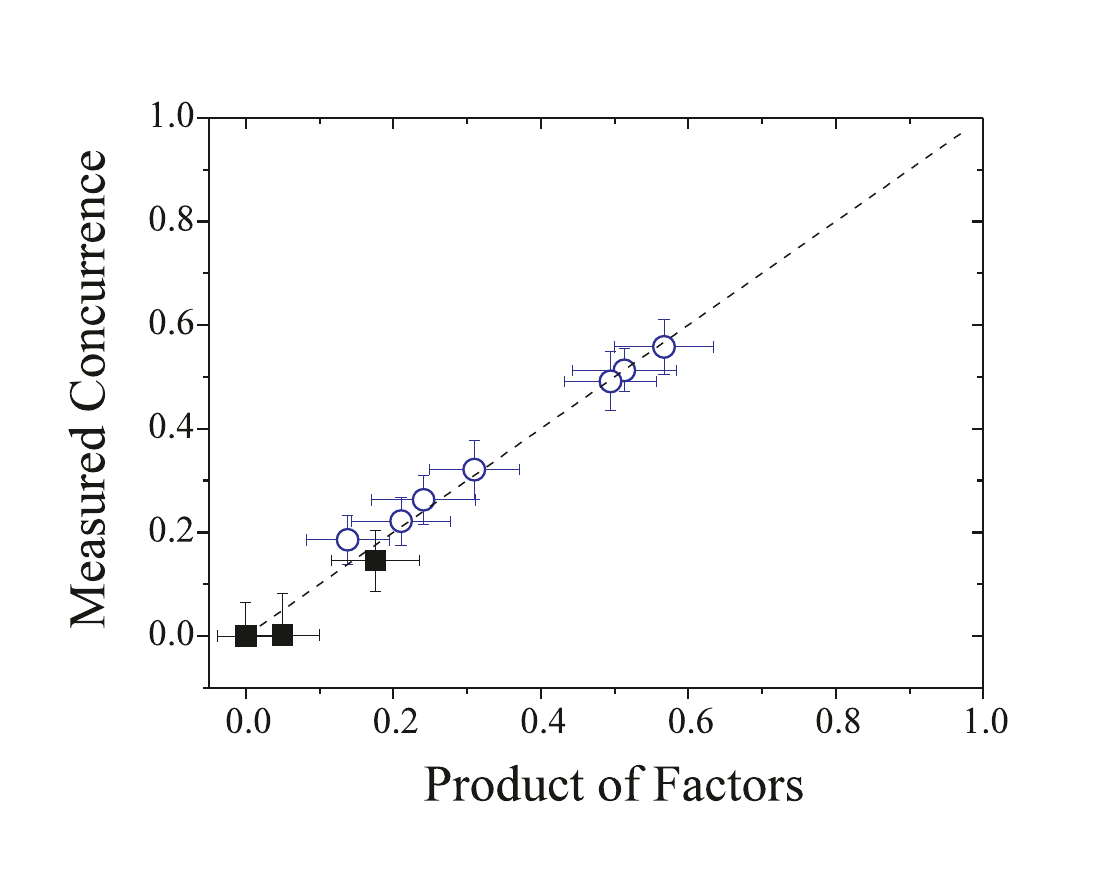}
\caption{Experimental test of the generalized dynamical law for the evolution of entanglement, for one-sided channels. Comparison between measured concurrence and the product of factors on the right side of Eq.(\ref{general}). The circles and squares correspond respectively to quasi-pure and mixed initial states. The slope of the line fitting all the data set is now $\sim 0.97$. From O. Jim\'enez-Far\'\i as {\it et al}., \href{http://dx.doi.org/10.1126/science.1171544}{Science {\bf 324}, 1414 (2009)}. Reprinted with permission from AAAS.} 
\label{dynamicsexp2}
\end{center}
\end{figure}

An extension of  equation of motion~(\ref{concMov}) to the case of initially mixed states was proposed in Ref.~\cite{farias}. It is based on a relation between generic states and pure states undergoing trace-preserving channels, in a similar fashion as the Choi-Jamio\l kowsky state-channel duality. This is expressed by the fact that for every normalised state $\varrho\in\mathcal{D}(\mathcal{H}_{\sys}\otimes\mathcal{H}_{\sys})$ there is always a completely positive trace-preserving channel $\mathcal{E}_{\varrho}$, acting on $\mathcal{D}(\mathcal{H}_{\sys})$, and a pure state $\ket{\Psi_{\varrho,\mathcal{E}_{\varrho}}}\in\mathcal{D}(\mathcal{H}_{\sys}\otimes\mathcal{H}_{\sys})$, such that $\varrho=\openone\otimes\mathcal{E}_{\varrho}\proj{\Psi_{\varrho,\mathcal{E}_{\varrho}}}$. Notice that $\ket{\Psi_{\varrho,\mathcal{E}_{\varrho}}}$ is in general not maximally entangled, otherwise the statement would hold only for matrices $\varrho$ with maximally mixed reduced states, as discussed in Sec. \ref{Krausop} after \eqref{CJDualism}. In addition, $\mathcal{E}_{\varrho}$ and $\ket{\Psi_{\varrho,\mathcal{E}_{\varrho}}}$ are not unique. There is an entire family of channels $\mathcal{E}'_{\varrho}$ and $\ket{\Psi_{\varrho,\mathcal{E}'_{\varrho}}}$, all connected by local unitary transformations, that yields the same $\varrho$. Nevertheless, since entanglement is invariant under local unitaries, this is not an issue for the following purposes.

Since $\mathcal{C}\left[(\openone\otimes\mathcal{E})\varrho\right]=\mathcal{C}\left[(\openone\otimes\mathcal{E})(\openone\otimes\mathcal{E}_{\varrho})\proj{\Psi_{\varrho,\mathcal{E}_{\varrho}}}\right]$, and $\proj{\Psi_{\varrho,\mathcal{E}_{\varrho}}}$ is a pure state, it follows from Eq. (\ref{concMov}) that
\begin{equation}
\label{general}
\mathcal{C}\left[(\openone\otimes\mathcal{E})\varrho\right]=\mathcal{C}(\proj{\Psi_{\varrho,\mathcal{E}_{\varrho}}}) \; \mathcal{C}\left[(\openone\otimes\mathcal{E}\circ\mathcal{E}_{\varrho})|\Phi^+\rangle\langle\Phi^+|\right]\,.
\end{equation}
This is the generalization of equation of motion~(\ref{concMov}) to initially mixed states. One should note however that, in order to use this expression to get the final concurrence, it is not sufficient to know the channel (through process tomography) and the initial concurrence, as before: one needs to tomographically reconstruct the initial state, in order to get $\proj{\Psi_{\varrho,\mathcal{E}_{\varrho}}}$ and $\mathcal{E}_{\varrho}$. This is still advantageous however from the experimental point-of-view, since one still avoids the tomographic reconstruction of the final state, when the signal-to-noise ratio is decreased by the action of the channel. 

Figure \ref{dynamicsexp2} displays the comparison between the measured final concurrence and the product of factors on the right side of (\ref{general}), for initial pure and mixed states. The experimental points now fall nicely on a straight line with slope almost one ($\sim 0.97$), within a very good approximation, even when the initially mixed states (squares) are taken into account.

\subsection{Non-Markovian entanglement dynamics: decay and revival}
\begin{figure}[t!]
\begin{center}
\includegraphics[width=0.7\linewidth]{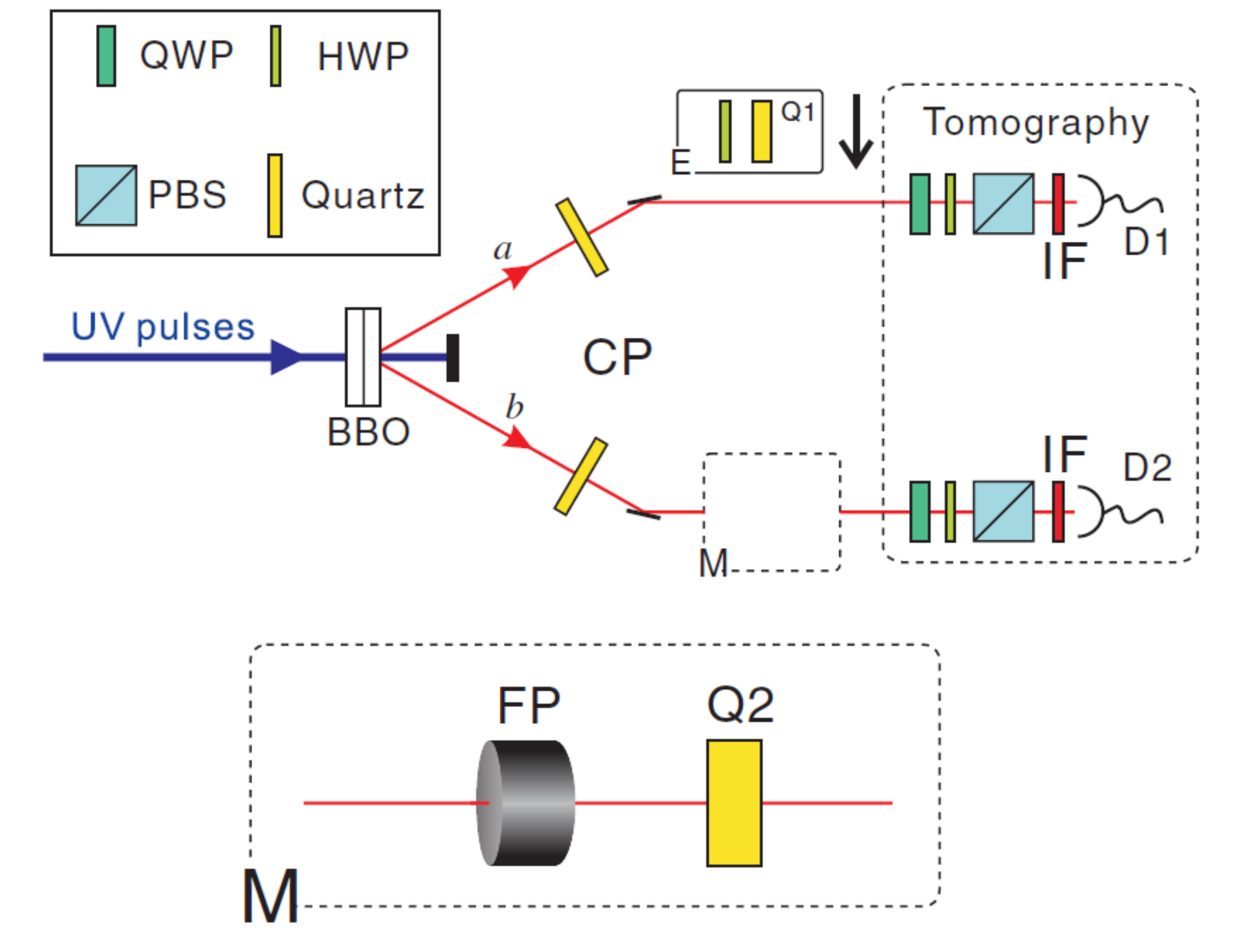}\\ \includegraphics[width=0.7\linewidth]{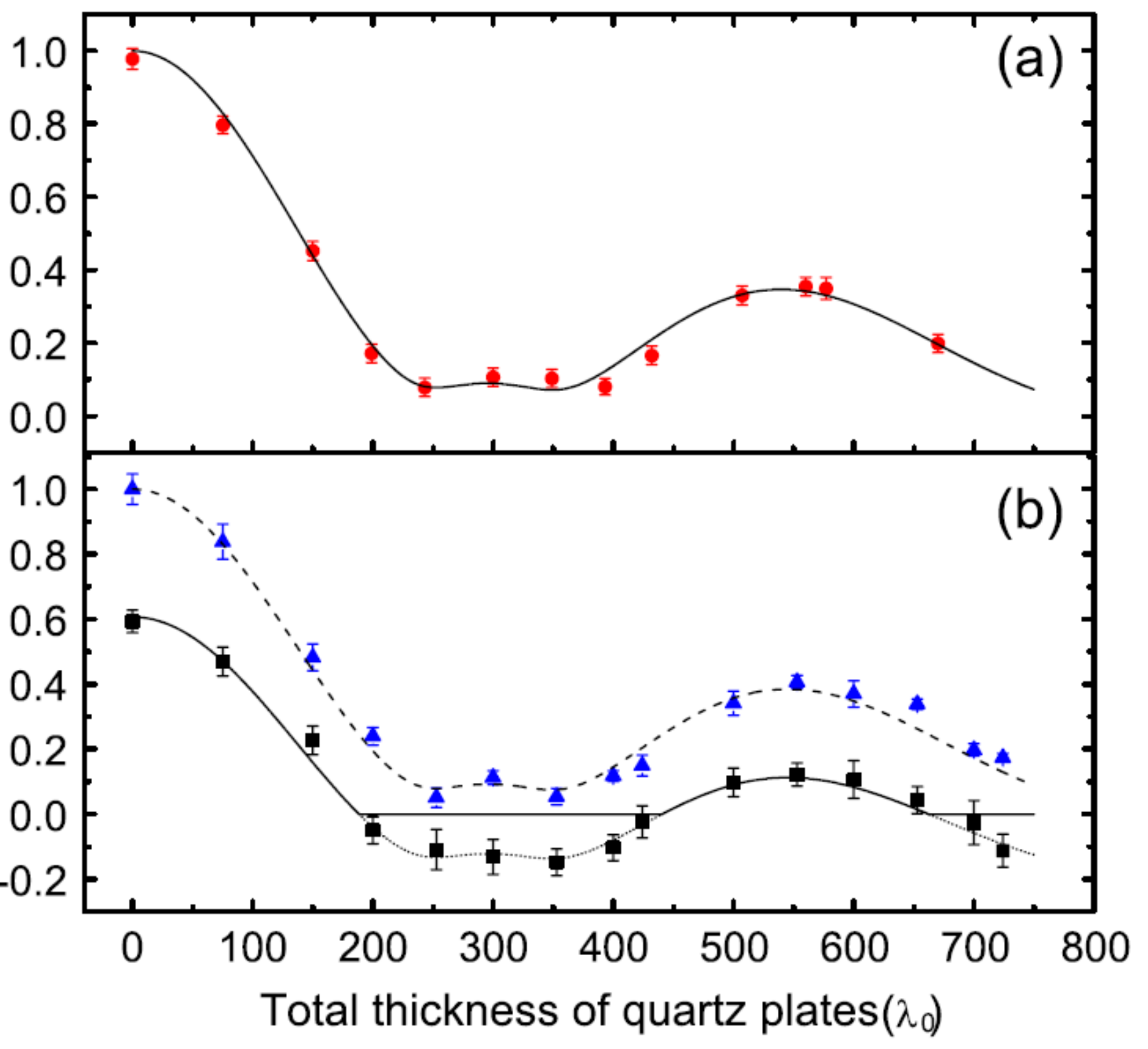}
\caption{Entanglement dynamics: decay  and revival. Top panel (experimental setup): Polarization-entangled states are created by type I parametric down conversion in two BBO crystals, and path information is erased by thin quartz plates (CP). The photon taking the upper mode may undergo decoherence in different basis by the introduction of a QWP and a quartz crystal. On the lower mode, the photon experiences a non-markovian environment induced by a Fabry-Perot cavity followed by a quartz crystal. At the end, a full tomography is carried out. Bottom panel (results): The dots on the top show the experimentally observed evolution of the concurrence for an initially maximally entangled state for different sizes of the quartz crystal -- a non-monotonic behavior of entanglement is observed. The solid line is the theoretical prediction for the concurrence. On the bottom, the experimental lower square dots, corresponding to the quantity $\Lambda$ defined in \eqref{concu1}, show the evolution of entanglement for an initially non-maximally entangled state. The solid line represents the theoretical fit of concurrence, which is set to 0 whenever $\Lambda$ becomes negative. The upper curve in the bottom panel displays the degree of polarization of the photonic state, which is the magnitude of the corresponding Bloch vector. The entanglement in this state vanishes  at a finite time first and revives then at a latter time (time is parametrized by the length $L$ of the quartz crystal). Reprinted figure (adapted) with permission from Jin-Shi Xu {\it et al}. \href{http://link.aps.org/doi/10.1103/PhysRevLett.104.100502}{Phys. Rev. Lett. {\bf 104}, 100502 (2010)}. Copyright (2010) by the American Physical Society. \label{jinshi}} 
\end{center}
\end{figure}

With a different experimental setup, shown in Fig.~\ref{jinshi}, Jin-Shi Xu \emph{et al.} \cite{jinshi02} also verified the dynamical law for entanglement evolution~\eqref{concMov}. The nice thing about this setup is that the introduction of a Fabry-Perot cavity into one of the modes allows for the realization of non-Markovian dynamics. 

The principle works as follows: suppose that a single polarization-qubit, initially in the state $(\ket{H}+\ket{V})/\sqrt{2}$,  passes through a crystal of quartz. Due to the birefringence of the quartz, different frequencies of the laser field (which is not exactly monochromatic) acquire different phases and the state evolves to $(\proj{H}+\proj{V} +\kappa \ket{H}\bra{V} + \kappa^* \ket{V}\bra{V})/2$. Usually, the distribution of frequencies of the pump laser field is well approximated by a Gaussian distribution, and thus  $\kappa = \int f(\omega)\exp(i \alpha \omega) d\omega$ with $f(\omega) = (2/\sqrt{\pi} \proj{\Psi_{\varrho,\mathcal{E}_{\varrho}}})\exp[-4(\omega-\overline{\omega})]$ ($\proj{\Psi_{\varrho,\mathcal{E}_{\varrho}}}$ is the distribution variance, and $\overline{\omega}$ is the laser central line), and $\alpha \propto L$ the thickness of the quartz slab. In this case, $\kappa$ presents an exponential decay as function of $L$. Now, if before the quartz plate the photon passes through a Fabry-Perot cavity, only a finite number of the frequencies are transmitted -- a non-Markovian dynamics is established. As expected from a finite dimensional system, the coherences may refocus and $\kappa$ may increase for some values of $L$.

With this construction, Jin-Shi Xu \emph{et al.}~were not only able to verify the entanglement equation of motion~\cite{xu09}, but also were able to observe the revival of entanglement from a separable state~\cite{jinshi02}. Some of these results are illustrated in Fig.~\ref{jinshi}.

\subsection{Probing entanglement robustness with decoherence-free subspaces}
\label{63}


The quest for robust entanglement has stimulated experiments in different frameworks, in particular for instance in trapped ions~\cite{kielpinski01,haeffner05b,roos06} and linear optics~\cite{kwiat00,yamamoto}. In~\cite{haeffner05b}, H\"affner {\it et al}.  probed the robustness of decoherence-free subspaces when entanglement is created between two nearby trapped Ca$^+$ ions. In order to eliminate the effects of local spontaneous emission, the qubits are encoded into Zeeman sub-levels of the S$_{1/2}$ ground state. The main source of decoherence that remains are the fluctuations of the trapping fields, which are common to both ions. In this regime, entanglement formed by superpositions of states with the same energy, such as $(\ket{01}\pm\ket{10})/\sqrt{2}$, is in principle decoupled from the environment and therefore robust. These superpositions define a decoherence-free subspace with respect to this type collective noise.

\begin{figure}[t]
\begin{center}
\includegraphics[width=\linewidth]{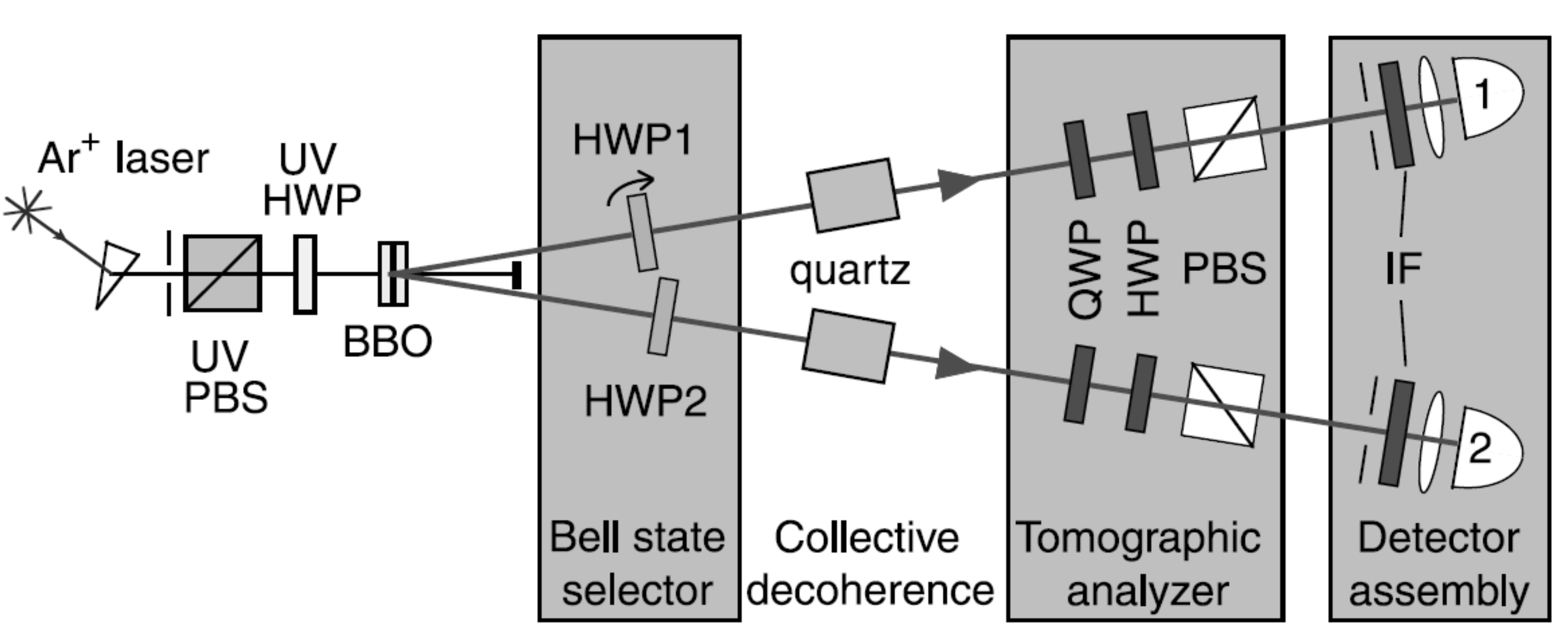}\\
\includegraphics[width=\linewidth]{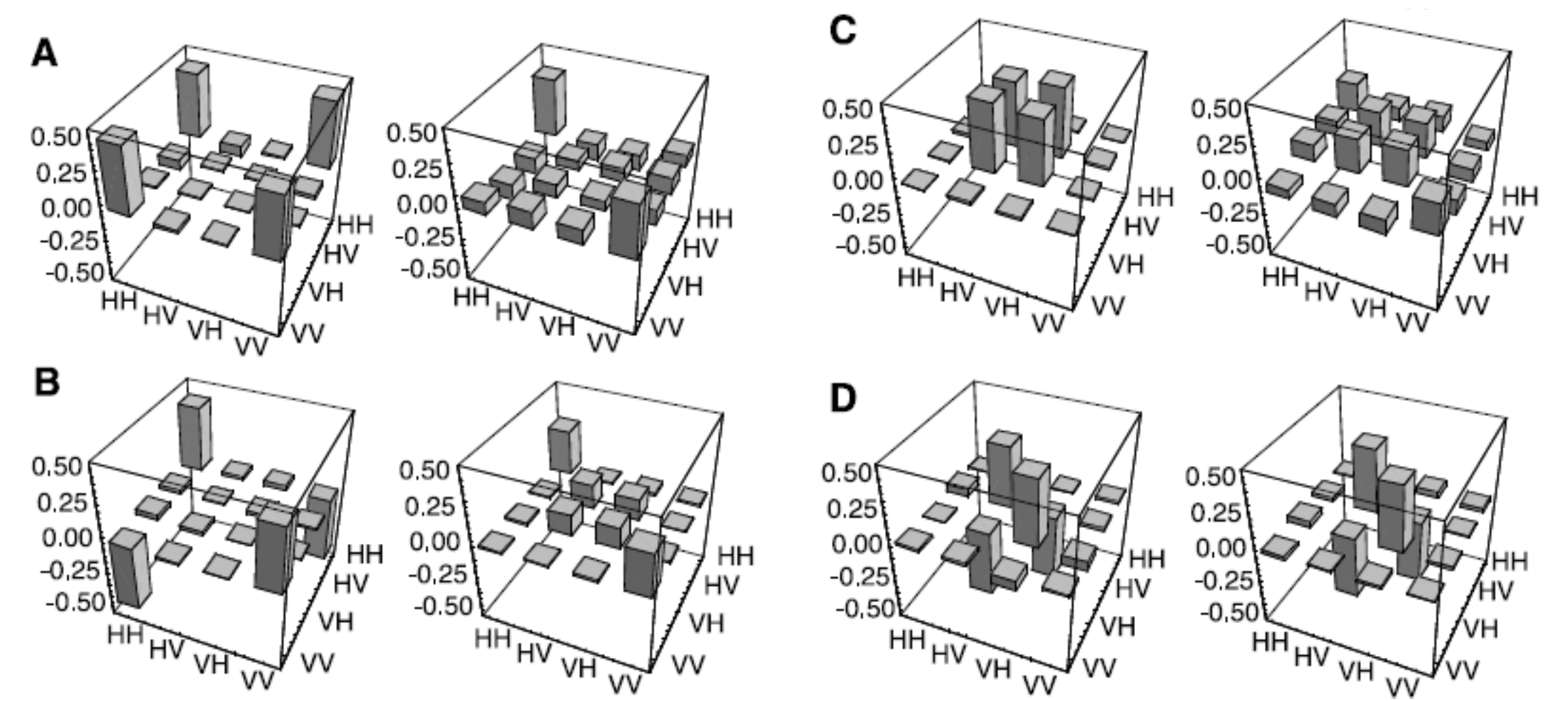}\\
\caption{Experimental arrangement  to investigate polarization decoherence-free subspaces.  Top panel: experimental setup. Polarization-entangled photons are produced at the nonlinear crystals (BBO). Half-waveplates (HWP1 and HWP2) are used to prepare the four Bell states. The decohering elements are separate slabs of quartz whose thicknesses are correlated to mimic collective decoherence. The final quarter-waveplate (QWP) and half-waveplate (HWP) in each arm, along with polarizing beam splitters (PBS), enable tomographic reconstruction of the polarization density matrix.  Bottom panel: results. From A to D, tomographic reconstructions are shown before (left) and  after (right) the action of decoherence, for the cases of $\ket{\Phi^+},\,\ket{\Phi^-},\,\ket{\Psi^+}, \text{and}\, \ket{\Psi^-}$ respectively. State $\ket{\Psi^-}$ is the most robust  among the four. From P. G. Kwiat {\it et al}. \href{http://dx.doi.org/10.1126/science.290.5491.498}{Science {\bf 290}, 498 (2000)}. Reprinted with permission from AAAS.} 
\label{kwiat}
\end{center}
\end{figure}

To measure the robustness  within the decoherence-free subspace, the authors generated the state $\ket{\Psi^+(\phi)} = (\ket{01}+e^{i\phi}\ket{10})/\sqrt{2}$, and let it interact with the environment for some time. In principle, concurrence could be calculated by tomographical reconstruction of the two-qubit density matrix. However, this process requires many experimental cycles (of the order of 1000). A fidelity-based entanglement witness was then employed,  corresponding to \eqref{recipeopt} for the two-qubit case. The ions are still entangled if the fidelity $F=\bra{\Psi^+(\phi)}\varrho\ket{\Psi^+(\phi)}
$ between the initial state $\ket{\Psi^+(\phi)}$ and the evolved state $\varrho$ is higher than $1/2$.   Due to a magnetic field gradient, which lifts the degeneracy between the energy levels of states $\ket{01}$ and $\ket{10}$, the phase $\phi$ undergoes a deterministic evolution, which can be determined and corrected, in such a way that the fidelity becomes $F=(\varrho_{01,10}+\varrho_{10,01})/2+|\varrho_{01,10}|$, which satisfies the bound 
$F\ge F_{\rm min}\doteq2|\varrho_{01,10}|$. Through the measurement of the single off-diagonal element $\varrho_{01,10}$
, entanglement was witnessed for up to 20s, which is to be compared with the 1s coherence decay time when the qubits are  encoded outside the S$_{1/2}$ ground states and local spontaneous emission is present. The deviation from a perfect decoherence-free subspace was mainly attributed to fluctuations of the magnetic-field gradient across the trap.

In Ref.~\cite{kwiat00}, Kwiat {\it et al}. investigated the robustness of photonic polarization entanglement with a setup based also on decoherence-free subspaces. Entangled photon pairs were produced by spontaneous parametric down-conversion, and submitted to controllable decoherence as the photons pass through thick, adjustable birefringent elements.  These elements add  phases to the photon states that can be correlated, mimicking collective random phases. Through tomographic analysis of the final two-photon state, it is observed that the singlet state $(|HV\rangle - |VH\rangle)/\sqrt{2}$ is considerably more stable than the other Bell states.  The experimental setup and results are shown in Fig.~\ref{kwiat}.

\subsection{Continuous variables: tripartite entanglement dynamics and robustness}
\label{64}

\begin{figure}{t}
\begin{center}
\includegraphics[width=\linewidth]{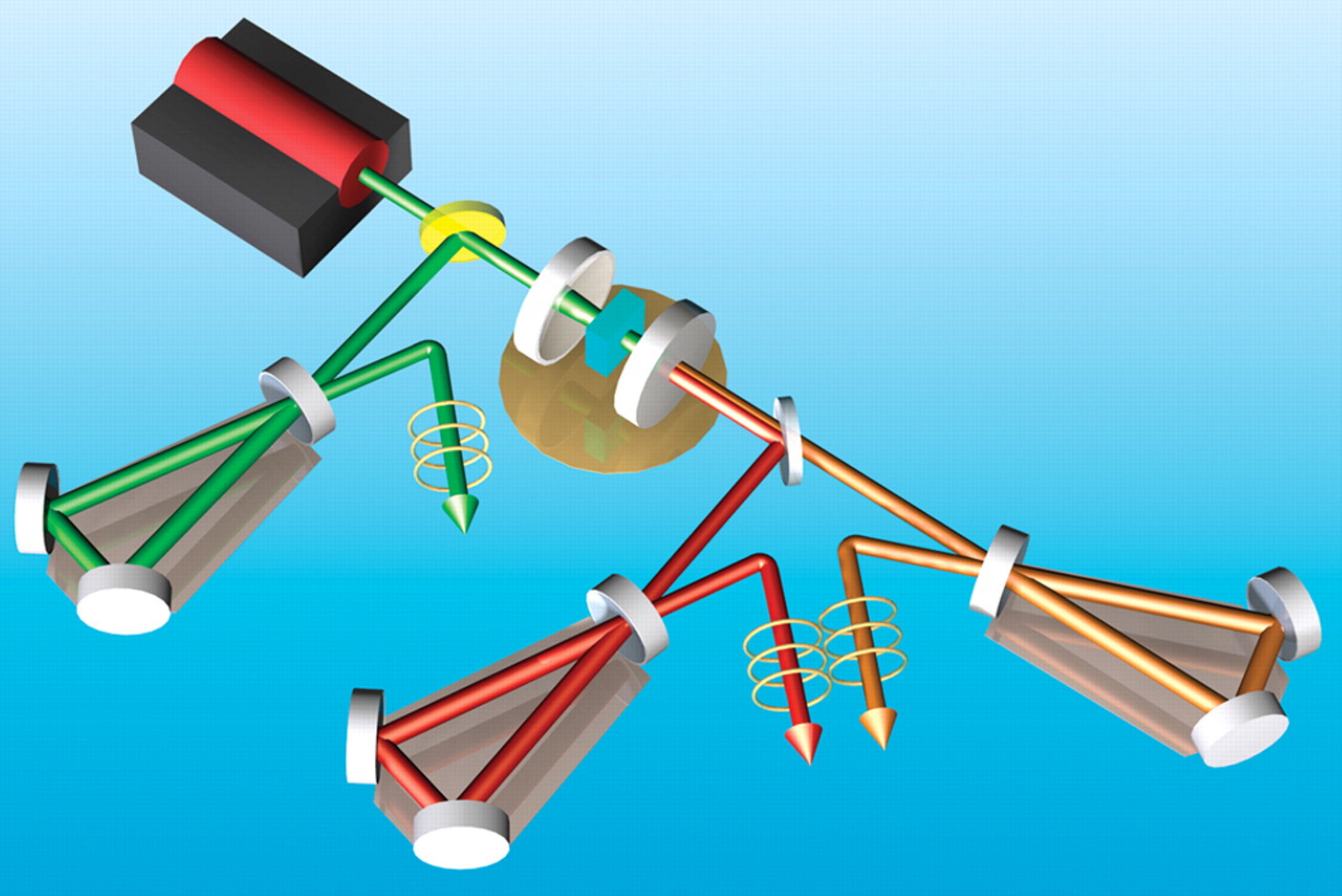}
\caption{Setup for investigation of continuous-variables tripartite entanglement. The optical parametric oscillator (blue, between two mirrors at the center of the picture) is pumped by a coherent light source, producing three light fields of different wavelengths (the reflected one, coloured green, plus two transmitted beams, coloured red and orange). The three beams are sent into near-resonant empty cavities, and then detected on high-quantum efficiency photodetectors. This allows, upon scanning the frequency of the cavity, for the measurement of quadrature fluctuations. From A. S. Coelho {\it et al}., \href{http://dx.doi.org/10.1126/science.1178683}{Science {\bf 326}, 823 (2009)}. Reprinted with permission from AAAS.}
\label{nussenzveig1}
\end{center}
\end{figure}

In Ref.~\cite{nussenzveig09}, Coelho {\it et al}. reported the disentanglement of a tripartite continuous-variable entangled state. 
The authors used an optical parametric oscillator (OPO) to generate entanglement between three bright beams of light with different wavelengths (see Fig.~\ref{nussenzveig1}). The beams are entangled in the amplitude and phase quadrature components. The experiment analysed the robustness of the entanglement against losses, and demonstrated that disentanglement may occur for finite channel losses,  corresponding to finite-time disentanglement.

\begin{figure}{t}
\begin{center}
\includegraphics[width=\linewidth]{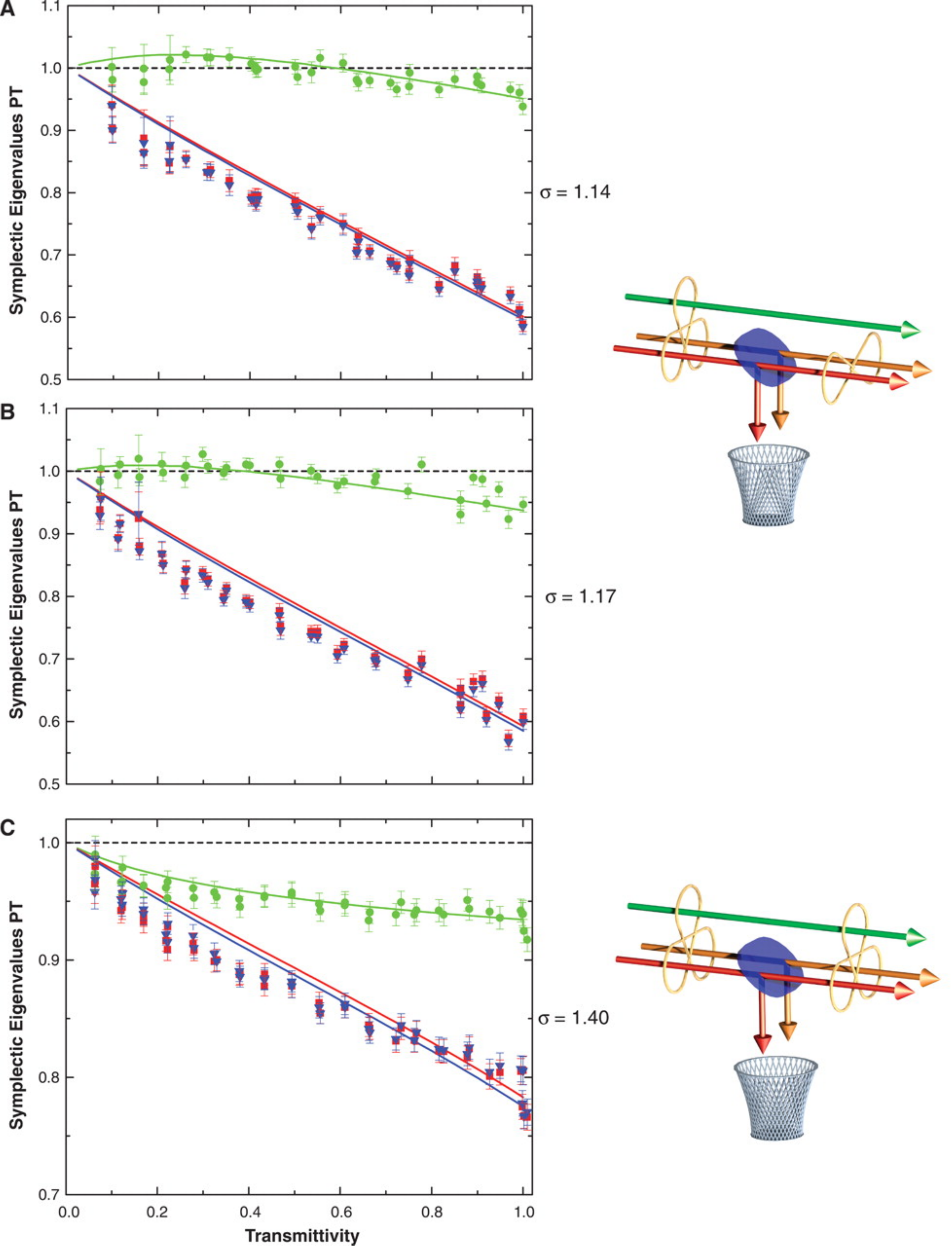}
\caption{Biseparability at finite losses. Symplectic eigenvalues of the measured covariance matrices as functions of the transmissivity of the twin modes, for three different values of the relative pumping power $\sigma$. The green circles represent the eigenvalues corresponding to the pump mode, the red squares those of the signal, and the blue triangles the ones of the idler. When any of these is greater or equal to 1, the associated mode is separable from the other two. In A ($\sigma = 1.14$) and B ($\sigma = 1.17$),  the pump beam becomes separable from the twins for finite losses (transmittance near 0.6 and 0.4, respectively). The situation is pictorially sketched on the right-hand side of the figure, where golden ribbons symbolise the presence of entanglement. 
In C ($\sigma = 1.40$),  all three fields remain inseparable until zero transmissivity, as sketched again on the right-hand side. 
From A. S. Coelho {\it et al}., \href{http://dx.doi.org/10.1126/science.1178683}{Science {\bf 326}, 823 (2009)}. Reprinted with permission from AAAS.}
\label{nussenzveig2}
\end{center}
\end{figure}

Above the oscillation threshold, the OPO, pumped by a coherent light source, generates narrow-band and tunable bright twin beams, with strong intensity correlations among them. Furthermore, in order to produce twin photons, a pump photon must be annihilated, which implies anti-correlations between the reflected pump intensity and the sum of the intensities of the twin beams. Besides, the sum of the frequencies of the twin beams must match the pump frequency. This frequency constraint implies a constraint for the phase variations (or fluctuations) of the three fields. The phase fluctuations of the twin beams should be anti-correlated, and their sum should be correlated to the phase fluctuations of the pump. These correlations translate into entangled amplitude and phase quadrature components. The amplitude difference  and the phase sum play respectively the roles of momentum difference and position sum in EPR's experiment.

After the generation of the twin entangled beams, they are separated by their polarization, and their quadrature fluctuations are measured, together with those of the reflected pump beam, through the use quasi-resonant empty cavities followed by high quantum-efficiency photodetectors. This allows for the complete reconstruction of the covariance matrix for the three beams, as well as the measurement of higher-order quadrature correlations. The authors measured up to  tenth moment correlations, and established, up to excellent approximation, that the three-field state is Gaussian. Therefore, Criteria \ref{criterion:Duan} and \ref{criterion:PPTCV} of Sec. \ref{2.2.1a} provide necessary and sufficient conditions for biseparability in the different bipartitions. For instance, as explained in detail in Sec.~\ref{2.2.1a}, a multi-mode Gaussian state is biseparable with respect to a bipartition of any one mode versus the rest iff the symplectic eigenvalues of its covariance matrix corresponding to that mode are greater than or equal to 1.


In the experiment, the smallest symplectic eigenvalues associated to each mode were obtained for several values of $\sigma=P/P_{th}$, where $P$ is the pump power and  $P_{th}$ the  oscillation-threshold pump power.  With this setup, one can observe the dependence of entanglement  on controlled linear losses imposed on the twin beams, by placing variable attenuators just before the corresponding photodetectors (pump beam losses had little effect on the symplectic eigenvalues).  Figure~\ref{nussenzveig2} displays the experimental results for three different values of the pump power, as a function of the transmissivity  of the attenuators.  For high transmissivities (low losses) all the symplectic eigenvalues are smaller than one, implying full inseparability of the three fields  -- the resulting state is thus an entangled state of three bright beams of light.  For smaller pump powers, the pump beam becomes separable from the twin beams, for finite values of the transmittivity, even though the squeezing of individual beams vanishes only asymptotically as a function of losses. This is reminiscent of finite-time disentanglement, but in the continuous-variable case and with respect to a particular bipartition. As the pump power increases, the entanglement signature is monotonically reduced, but the three fields remain inseparable until the transmissivity is zero (total loss), thus entanglement, even though smaller, is more robust in this case.  

A further development of this study, involving the disentanglement between the twin beams at finite losses, was reported in Ref. \cite{barbosa10}.

\subsection{Multiqubit entanglement dynamics: four qubits}
\label{65}

\begin{figure}[t]
\begin{center}
\includegraphics[width=\linewidth]{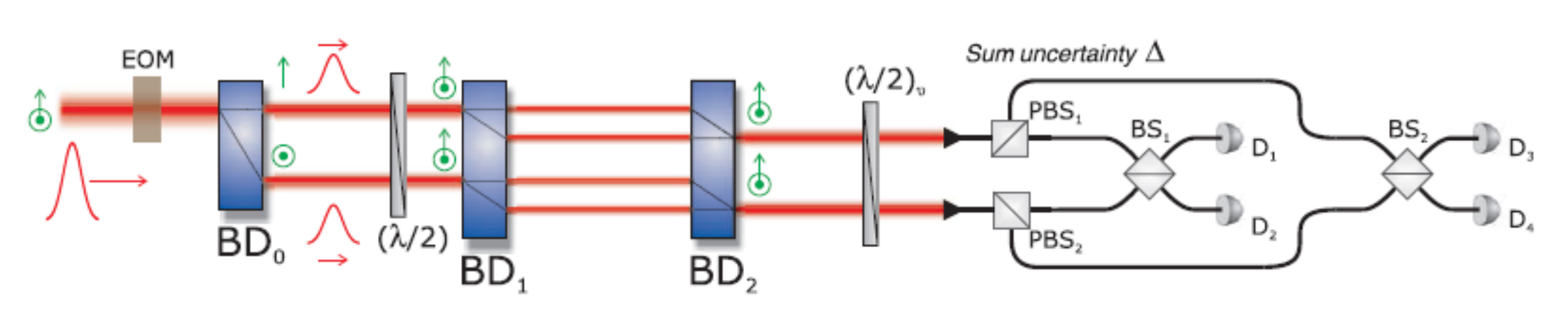}\\
\includegraphics[width=\linewidth]{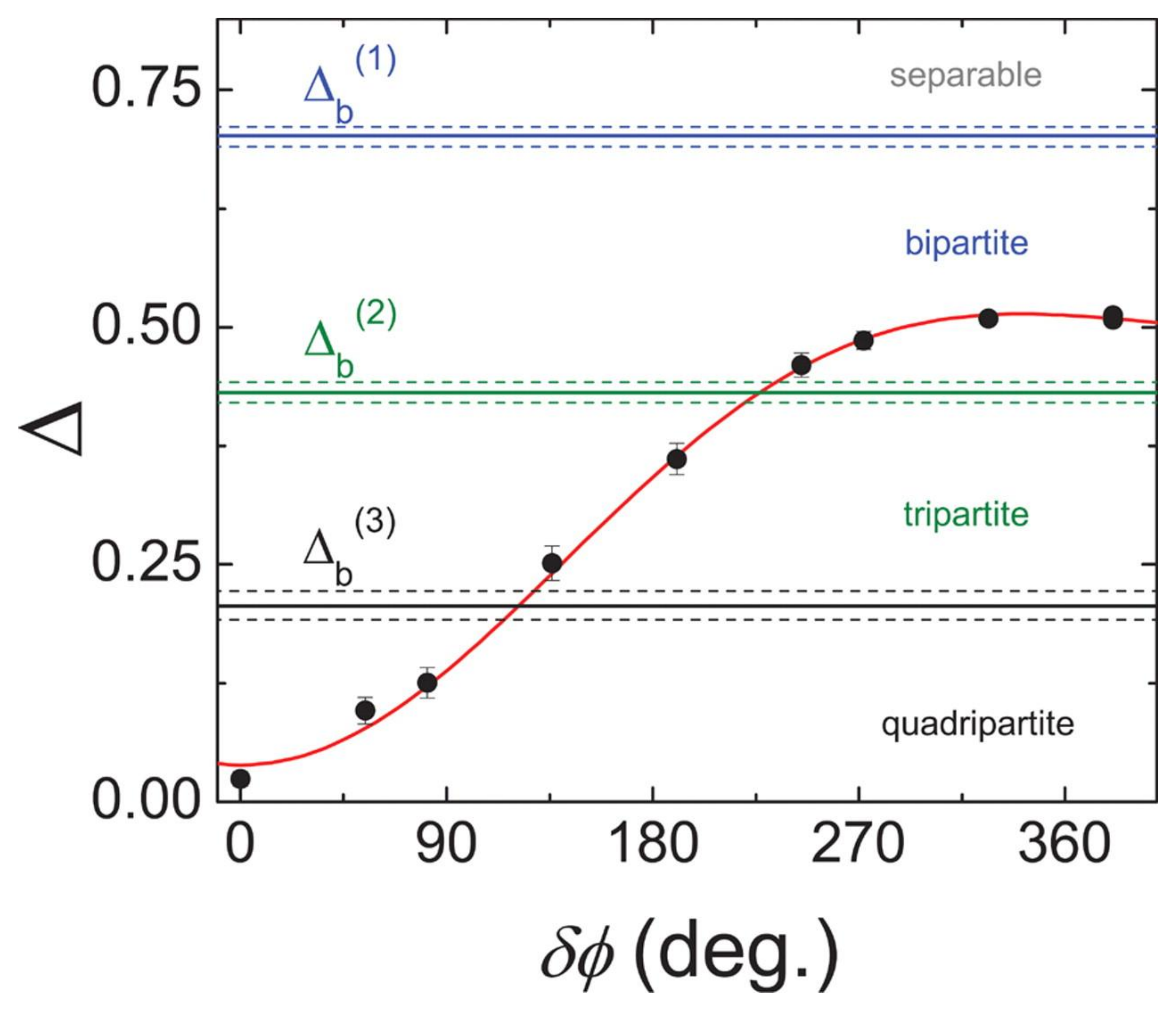}
\caption{Genuine multi-qubit entanglement decay. Top panel (experimental setup): The beam displacers BD$_0$ and BD$_1$ create a quadripartite mode entangled state by splitting  a single photon into four optical modes. This state is then measured in  orthonormal basis \eqref{Worthogonal} and the  uncertainty $\Delta$ is evaluated.  Bottom panel (results):  $\Delta$ as function of the dephasing strength $\delta\phi$  signals the trajectory of correlations across the different shells of genuine $k$-partite entanglement. The horizontal solid lines indicate the boundaries $\Delta_b^{(k)}$ for  genuine $k$-entanglement, and the dashed lines indicate the uncertainty on these boundaries due to the possibility of two or more photons in the system. From S. B. Papp {\it et al}., \href{http://dx.doi.org/10.1126/science.1172260}{Science {\bf 324}, 764 (2009)}. Reprinted with permission from AAAS.}  
\label{papp}
\end{center}
\end{figure}

The first experiment to explicitly address multi-qubit entanglement decay was reported by Papp {\it et al}. in Ref.~\cite{papp09}. There, a single heralded photon from a cloud of Cs atoms is coherently split among the four modes of an optical interferometer (see Fig.~\ref{papp}) to create the W-like state
\bea{rl}
\ket{W} =  \frac{1}{2}\left[ (\ket{1000}+ \right. & e^{i\phi_1}\ket{0100})+ \\
&\left. e^{i\phi}(\ket{0010}+ e^{i\phi_2}\ket{0001})\right].
\eea
The relative phase $\phi$ can be adjusted by an electro-optic modulator (EOM), while $\phi_1$ and $\phi_2$ are kept fixed throughout the experiment. Driving the EOM with a fluctuating voltage induces a controlled dephasing process between the subspaces spanned by $\ket{1000}$ and $\ket{0100}$ and by $\ket{0010}$ and $\ket{0001}$. The amplitude of this phase noise is $\delta\phi$: $\delta\phi=0$ means no dephasing, and a very well defined phase difference, whereas $\delta\phi=360^\circ$ describes the case where the phase is totally random.

Due to the dephasing, the initially genuine quadripartite entanglement decays first into tripartite and then into bipartite entanglement. See the bottom panel of Fig.~\ref{papp}. To resolve among the different types of entanglement, the authors borrow a powerful idea from quantum metrology: there are  genuine $N$-partite  entangled states that allow one to measure observables more precisely than any state without genuine $N$-partite entanglement. With this in mind, a figure of merit based on the variance of four-qubit unidimensional orthogonal projectors $\{M_1, M_2, M_3, M_4\}$, such that they spam the single-excitation subspace, is introduced
\beq
\Delta = \sum_{i=1}^4 \langle M_i^2\rangle - \langle M_i \rangle^2= 1 - \sum_{i=1}^4  \langle M_i \rangle^2.
\eeq
For states without genuine $k$-partite entanglent, with $k\leq 4$,  this quantity is lower bounded by $\Delta_b^{(k)}$. In this way, if a value of $\Delta$ smaller than $\Delta_b^{(k)}$ is obtained, one is sure to have a state that has genuine $(k+1)$-partite entanglement. The specific projectors used in this experiment were $M_i = \proj{W_i}$, with
\bea{c}
\label{Worthogonal}
\ket{W_1}=\frac{1}{2}\left( \ket{1000} + e^{i \beta_1}\ket{0100} +e^{i \beta_2}\ket{0010}+e^{i \beta_3}\ket{0100}\right),\\
\ket{W_2}=\frac{1}{2}\left( \ket{1000} - e^{i \beta_1}\ket{0100} - e^{i \beta_2}\ket{0010}+e^{i \beta_3}\ket{0100}\right),\\
\ket{W_3}=\frac{1}{2}\left( \ket{1000} - e^{i \beta_1}\ket{0100} +e^{i \beta_2}\ket{0010}-e^{i \beta_3}\ket{0100}\right),\\
\ket{W_4}=\frac{1}{2}\left( \ket{1000} + e^{i \beta_1}\ket{0100} - e^{i \beta_2}\ket{0010}-e^{i \beta_3}\ket{0100}\right),
\eea
where the phases $\beta_i$ were adjusted so as to minimize $\Delta$.

This procedure resembles the one based on entanglement witnesses, as it renders a sufficient criteria for  genuine $k$-partite entanglement. However, since $\Delta$ cannot be written as the expectation value of an observable, it rather defines a  {\it non-linear entanglement witness}. In any case, the dynamics through the different layers of genuine $k$-partite entanglement can be inferred as function of the noise strength $\delta\phi$.   
 
 \subsection{Dynamical generation of bound states} 
 \label{66}
 
 \begin{figure}[t]
\begin{center}
\includegraphics[width=\linewidth]{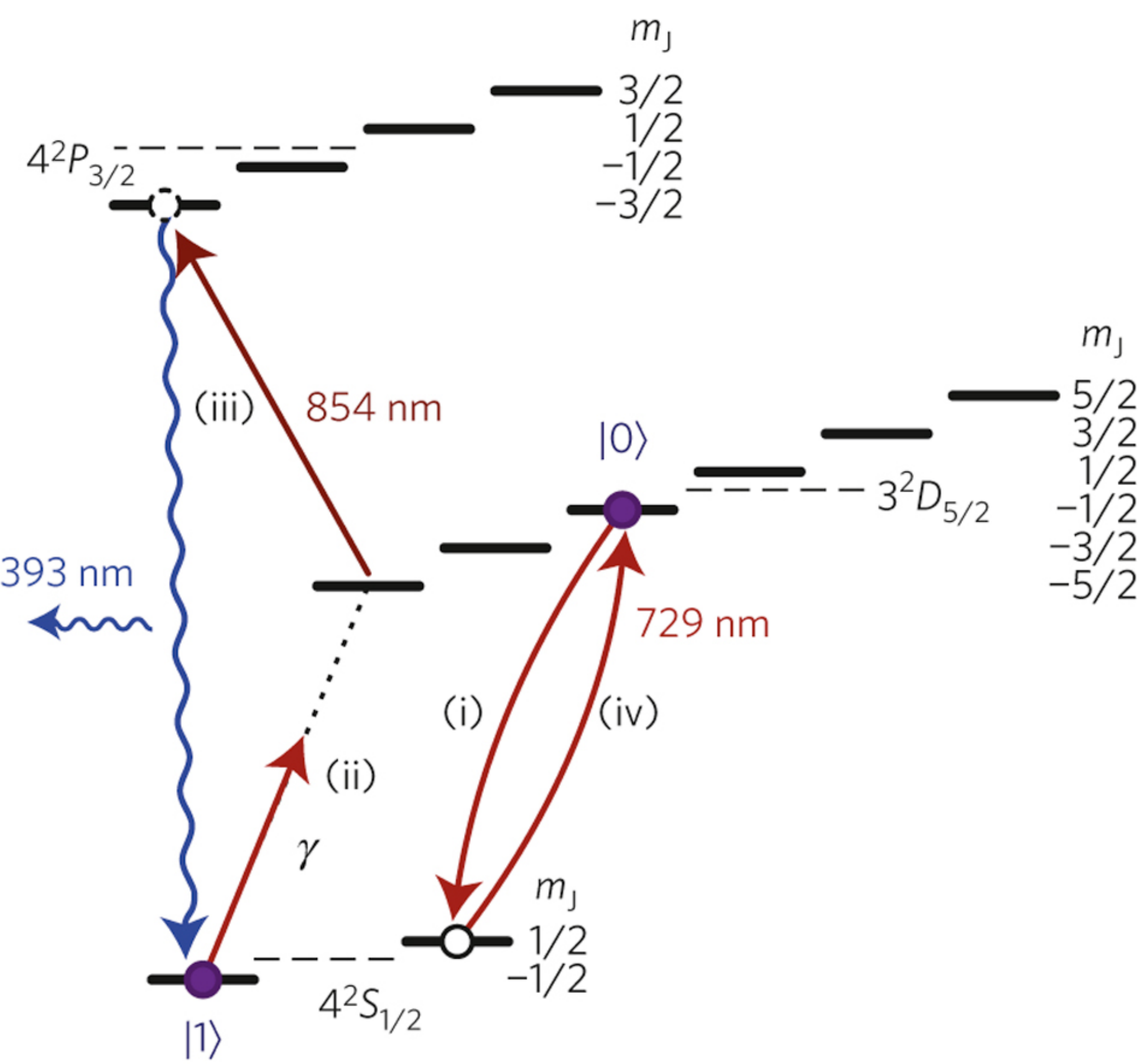}
\end{center}
\caption{Zeeman-level structure of ${}^{40}$Ca$^+$ and pulse-sequence for the implementation of tunable dephasing. Local dephasing is realized by carrying out the following sequence independently on each ion:  ({\it i}) hiding of $|0\rangle$,  ({\it ii}) transfer of the population in $|1\rangle$ into the superposition $\sqrt{1-\gamma}|1\rangle+\sqrt{\gamma}|D_{5/2}(m=-5/2)\rangle$,  ({\it iii}) optical pumping of the population in $D_{5/2}(m=-5/2)$ into $P_{3/2}(m=-3/2)$ and subsequent decay back to $|1\rangle$ by spontaneous emission, and finally ({\it iv}) restoring the initially hidden population back to $|0\rangle$. From J. T. Barreiro {\it et al}., \href{http://dx.doi.org/10.1038/nphys1781}{Nature Physics {\bf 6}, 943 (2010)}. }
\label{barreiro0}
\end{figure}

In  a following experiment reported in Ref.~\cite{Barreiro10}, Barreiro {\it et al}.~go a step further than in the previous subsection and demonstrate a rich dynamical behaviour of four initially entangled qubits embedded in independent decohering environments, including the appearance of bound entanglement due to the action of the local environments. Bound entanglement arises here because of a similar mechanism to the one described by Fig.~\ref{Bound} of Sec.~\ref{BoundGHZ}, for the case of GHZ states. Different bipartitions of an initially entangled $N$-qubit state may become, under the action of individual environments, separable at different times, implying that undistillability and entanglement coexist, before the state  completes its disentanglement. It is however important to distinguish that the initial entangled state prepared in \cite{Barreiro10} is close to the Smolin state \eqref{Smolin}, which is already itself bound entangled, as discussed in Sec.~\ref{boundmulti}. The setup in \cite{Barreiro10} comprises  four $\,^{40}\rm{Ca}^+$ ions confined in a linear Paul trap. In each ion a  qubit is encoded in the Zemman levels as $\ket{0}\doteq \ket{D_{5/2} (m=-1/2)}$ and $\ket{1}\doteq \ket{S_{1/2} (m=-1/2)}$, as shown in Fig.~\ref{barreiro0}. 
 
 A controlled dephasing channel is applied locally to each qubit, through a sequence of operations illustrated in Fig.~\ref{barreiro0}:  Initially, ({\it i}) the population of the population of state $\ket{0}$ is hidden by a full coherent transfer into the $S_{1/2}(m = 1/2)$ level. Then, ({\it ii}) the population in $\ket{1}$  is partially transferred to a superposition between the state $\ket{1}$ and an auxiliary level $D_{5/2}(m=-5/2)$. This is followed by ({\it iii}) optical pumping of $D_{5/2}(m=-5/2)$ to the excited state $P_{3/2}(m=-3/2)$, with the subsequent spontaneous decay back to $|1\rangle$. Finally, ({\it iv}) the population hidden initially is restored back to $\ket{0}$. The photon scattered in step ({\it iii})  carries partial information about the qubit state into the environment. This implements an effective dephasing channel, with dephasing strength given by the probability of photon emission, which is in turn given by probability $0\le\gamma\le 1$ of transferring all the initial population from $\ket{1}$ to $D_{5/2}(m=-5/2)$ in ({\it ii}). 
 
 \begin{figure}[t]
\begin{center}
\includegraphics[width=\linewidth]{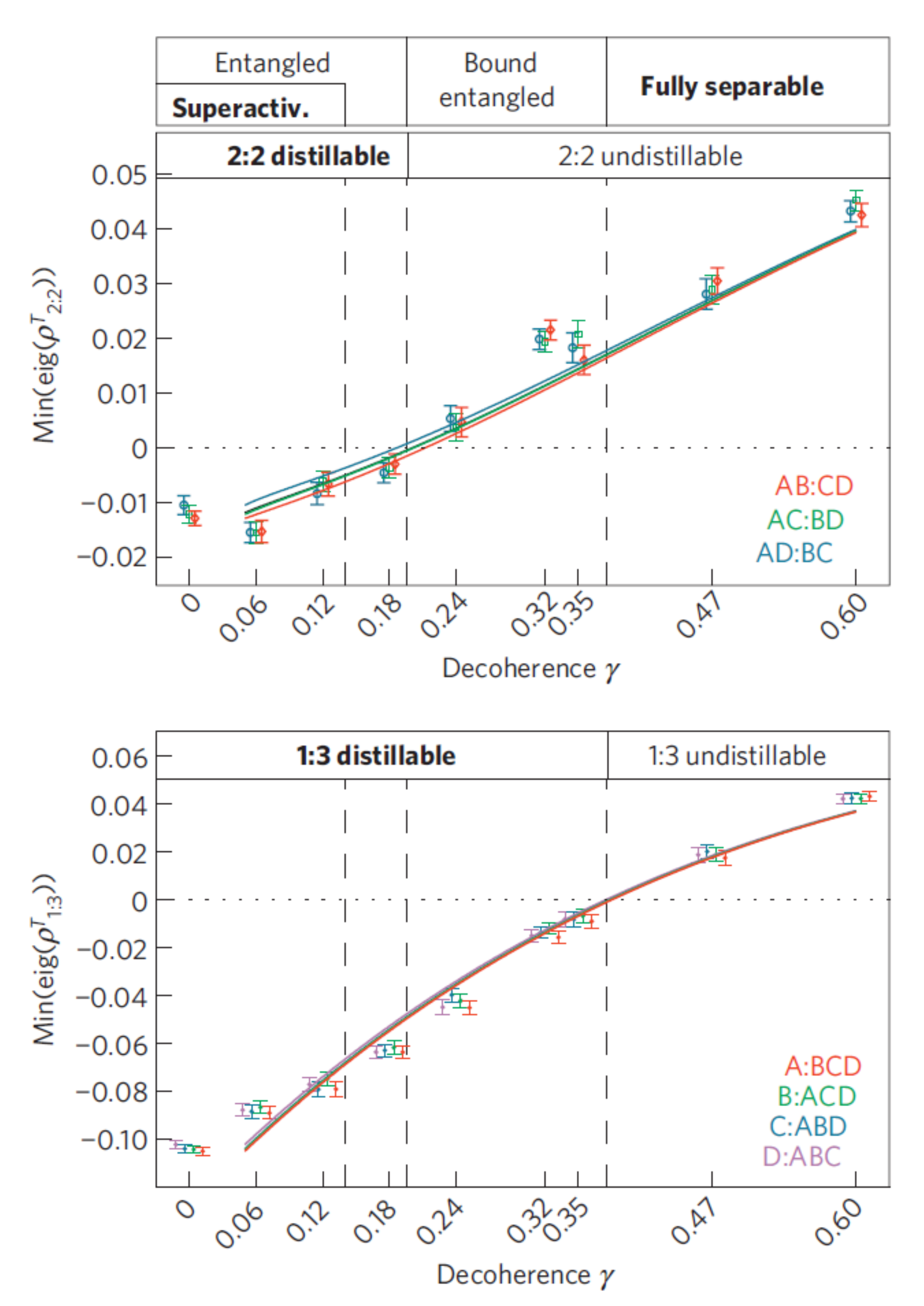}
\end{center}
\caption{Decay of entanglement through its different regimes. The smallest eigenvalues of the partially transposed experimental density matrix with respect to all the bipartitions are plotted as functions of the dephasing strength $\gamma$. The upper panel corresponds to all the bipartitions of 2 versus 2 qubits; whereas the lower one to those of 1 versus 3 bipartition. The dashed vertical lines delimit different regimes of multipartite entanglement. At $\gamma=0$ the state is capable of a CHSH-inequality violation (not plotted). As $\gamma$ increases, the state goes first beyond the threshold for superactivation, and then, when the 2:2 eigenvalues become zero, beyond that of distillability. The latter  demonstrates the dynamical generation of bound entanglement. Finally, when 1:1 eigenvalues become zero, the state reaches full separability. Adapted from J. T. Barreiro {\it et al}., \href{http://dx.doi.org/10.1038/nphys1781}{Nature Physics {\bf 6}, 943 (2010)}. }
\label{barreiro}
\end{figure}

As mentioned, the initial 4-qubit state is close to the Smolin state, and, as the latter, violates a CHSH-type Bell inequality and enables entanglement superactivation \cite{shor03}. Entanglement superactivation is the most extreme example of superadditivity of distillable entanglement, whereby two copies of a multipartite bound entangled state (i.e., with zero distillable entanglement each) tensored together possess positive distillable entanglement. Nevertheless, in contrast to the Smolin state, which is separable in all 2:2 bipartitions and entangled in all 1:3 bipartitions, the initial state prepared is entangled in the 1:3 bipartitions and also slightly entangled in the 2:2 ones.

As shown in Fig.~\ref{barreiro}, the different regimes of entanglement of the initial state under local dephasing are monitored. For $0\le \gamma < 0.06$ the state violates a CHSH-like inequality (not shown) and can be super activated. Between $0.06 \le \gamma < 0.12$, CHSH violation ceases but superactivation is still possible.  In the region $0.21\le\gamma<0.35$ the state is bound entangled. 
Finally, at $\gamma=0.35$ the state reaches full separability.

\subsection{Dynamics of  14-qubit GHZ entanglement}
\label{67}

 \begin{figure}[t]
\begin{center}
\includegraphics[width=\linewidth]{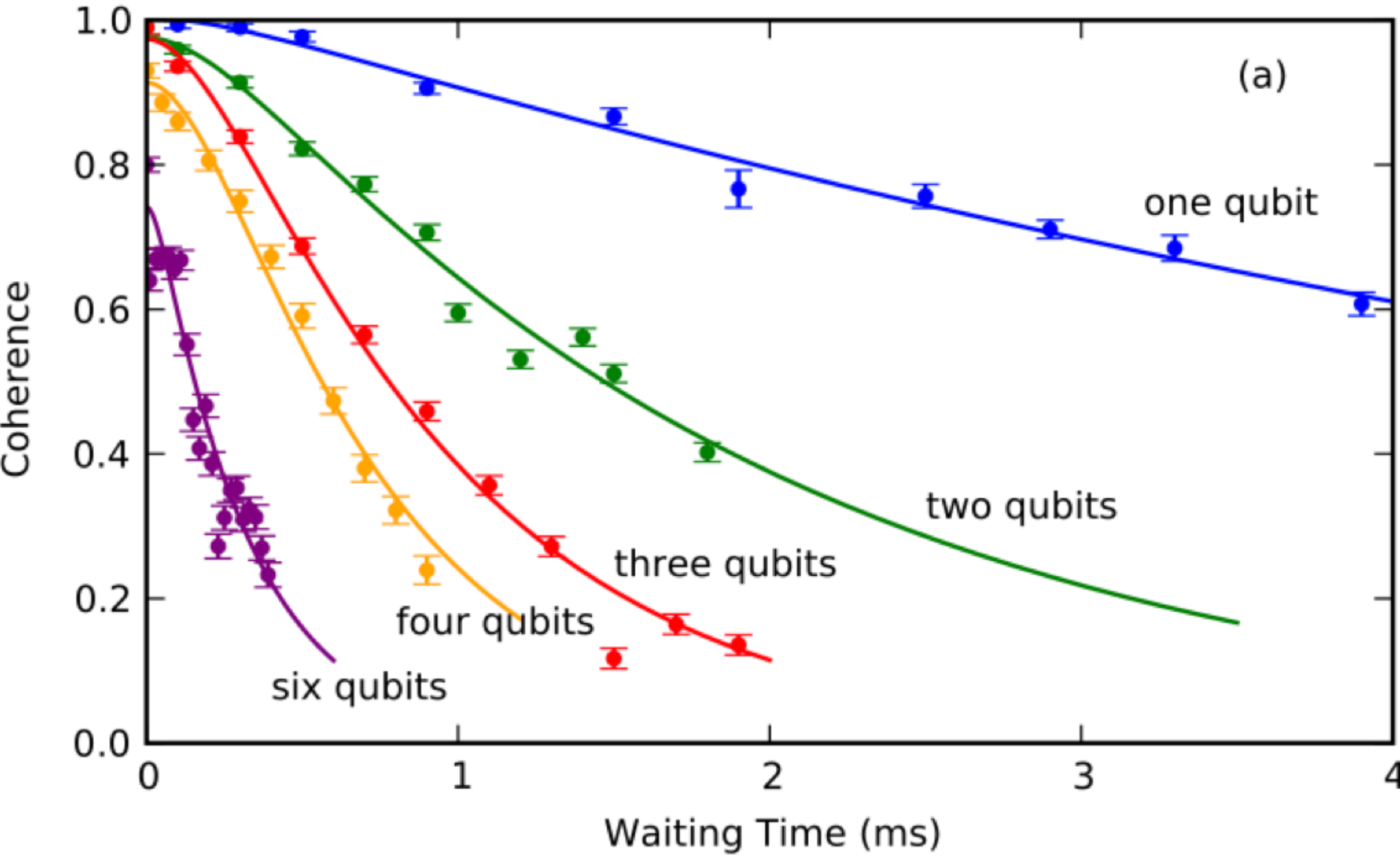}
\end{center}
\caption{Coherence as a function of time of a single-qubit superposition (blue) and GHZ states of 2 (green), 3 (red), 4(orange), and 6 (purple) qubits. 
These results are consistent with an exponential decay rate proportional to the squared number of ions. Reprinted figure with permission from T. Monz {\it et al}., \href{http://link.aps.org/doi/10.1103/PhysRevLett.106.130506}{Phys. Rev. Lett. {\bf 106}, 130506 (2011)}. Copyright (2011) by the American Physical Society.}
\label{nsquare}
\end{figure}

In Ref.~\cite{monz2011}, Monz \emph{et al}. reported the creation of GHZ states of up to 14 ion-qubits. The system is the same as in the previous subsection, a string of ${}^{40}$Ca$^+$ ions confined in a linear Paul trap. The qubit encoding is again given by the ground state  $|1\rangle\equiv S_{1/2}(m=-1/2)$ and the metastable level $|0\rangle\equiv D_{5/2}(m=-1/2)$, shown in Fig.~\ref{barreiro0}. First, the internal degree of freedom is optically pumped to the initial state $|1\dots 1\rangle$, and the motional degree is cooled to the ground state by Dopler and side-band cooling. Then, the ions are subject to  a collective M\o lmer-S\o rensen entangling interaction~\cite{sorensen00,benhelm08}, which takes $|1\dots 1\rangle$ to the GHZ state. 


The authors measure the populations, in the computational basis, of the experimentally created state $\rho$, as well as the coherence $\Delta\doteq\bra{00\ldots 0}\rho\ket{11\ldots 1}$. As discussed in Sec.~\ref{GHZprelim}, these matrix elements suffice to assess the genuine multipartite entanglement and multiparty distillability properties of noisy GHZ states. Indeed, with these, the presence of genuinely multipartite entanglement is corroborated by three different approaches. $N$-qubit distillability is confirmed with Criterion \ref{criterion:DistMDist} of Sec.~\ref{GHZprelim}. Whereas genuine $N$-qubit entanglement is confirmed both through Criterion \ref{criterion:BisepGHZN}, also in Sec.~\ref{GHZprelim}, and through fidelity-based entanglement witness \eqref{recipeopt}, of Sec.~\ref{2.2.2}.
The populations are measured by direct fluorescence detection, and the off-diagonal element $\Delta$ by measurements of parity oscillations induced by collective rotations on the state.

Up to $N=8$ qubits, the authors monitor the decay of $\Delta$ over the time between the state generation and read-out. Fig.~\ref{nsquare} displays the observed decay of $\Delta$, which is consistent with an exponential decay of  with decay rate proportional to $N^2$. This is in contrast to the decay rate proportional to $N$ typical of local noise processes, as described in Sec.~\ref{CompleteGHZ}. This quadratic decay rate was shown to stem from the presence of correlated phase noise, caused by collective fluctuations of the magnetic field. Correlated noise appears, in some form or another, in many experimental platforms. The fast coherence decay it can induce imposes stringent constraints on the scalability of GHZ states to large $N$.

\section{Conclusion}
\label{VII}

From a mesmerizing quantum phenomenon to a physical resource for quantum-information processing, entanglement has followed a peculiar trajectory since it was discussed in the 1935 paper by Einstein, Podolsky, and Rosen, and in the 1935 and 1936 papers by Schr\"odinger. Quantum correlations acquired a new status with the 1964 paper by John S. Bell, and were recognized in the 1990's as  resources for quantum computation and communication. These potential applications motivated a deeper analysis of the mathematics of entanglement, and of the possible deleterious effects of the environment.

The amount of insight gained in the last two decades on this subtle property of the quantum world is enormous. The huge reference section of this paper is an evidence of the collective effort of theoretical and experimental physicists, computer scientists, and mathematicians devoted to tackle different aspects of entanglement. 

In this review, we have focused on the dynamics of entanglement of systems that interact with different kinds of environments. This specific problem is of utmost importance within the vast domain of studies of quantum correlations, since it is directly connected to the analysis of the robustness of quantum computing, quantum simulations, quantum metrology, and quantum communication protocols. In addition, it turns out to be related to several fundamental questions in quantum theory, {as we hope is clear from the previous sections. 

Nevertheless, important questions connected to entanglement dynamics still remain as open challenges or promising research directions. Some of them  are summarized in the list below.

\begin{itemize}

\item{\it Dynamics of entanglement between macroscopic subsystems}. This problem is intrinsically related to  recent experimental efforts that intend to push quantum coherence and entanglement to macroscopic scales \cite{marshall03,lvovsky2013,bruno,groblacher}. These experiments could lead to stringent tests of quantum mechanics and its classical limit \cite{leggett02}, and might be able to probe subtle decoherence effects of the gravitational field \cite{joos,marshall03,kleckner08}. Assessment of the dynamical signature of different sources of decoherence might help to identify the contribution of gravitation to decoherence.

\item{\it Further effective logical channels}. Effective noise channels at the logical level have been studied \cite{Kesting13} for logical qubits encoded into some stabilizer-state error-correction codewords under local Pauli noise. An interesting direction to explore is the investigation of effective channels for logical encodings relevant to current quantum computing or communication experiments. These may be for instance few-qubit decoherence-free subspaces under realistic noise models for ion-trap~\cite{aolita07ions,Monz09} or superconducting-circuit~\cite{Wu08,Xue09} architectures, or photonic encodings involving transverse spatial modes, useful for misalignment-immune quantum communication \cite{aolita07quantcom,souza08,Dambrosio12,Vallone14}, ultra-sensitive rotational sensors~\cite{Fickler12, DAmbrosio13}, or high-dimensional qudit processing~\cite{Marrucci11,Fickler12, DAmbrosio13,Flicker14}, under atmospheric turbulence or other spatial perturbations~\cite{Dambrosio12,Krenn14,Ozzy14}.

\item{\it Equation of motion for average entanglements}. It was shown in Ref.~\cite{concentration} that, as the number of parts grows, the entanglement trajectories of initially Haar-random  pure states under noise concentrate around that of the average entanglement. An interesting open question is to determine an equation of motion for the Haar-average  entanglement, as well as for the width of the distribution or even higher moments thereof, given a particular noise model. The uniform Haar distribution has the advantage of readily allowing for mathematical tools such as Levy's Lemma. However, it would be very interesting to aim at other distributions of  pure states from which one can, contrary from the Haar distribution, sample efficiently. This may potentially render more realistic insights about typical entanglement dynamics in natural systems.

\item{\it Dynamics of entanglement for identical particles.} Entanglement dynamics for systems composed of indistinguishable particles, such as Bosons, Fermions or Anyons, for instance, remains a largely unexplored field. In spite of recent efforts~\cite{Schliemann01, eckert02, Ghirardi04, Benatti20121304, tichy2013}, the lack of accepted entanglement criteria and measures for such systems prevents a better understanding of this problem. A great obstacle is that a resource theory based on LOCC's is no longer suitable: Due to the necessary symmetrization  or anti-symmetrization of identical particles, the notion of local operations does not apply anymore.


\item{\it Dissipative stabilization of entanglement.} Recent developments point to new directions in this fascinating field. The stabilization of entanglement through engineered dissipation has been demonstrated in recent experiments \cite{devoret,lin}. This technique, which stems from \cite{poyatos} -- see also \cite{andre01} --,  may have applications in dissipative state preparations, quantum computations and dissipatively-driven quantum phase transitions  \cite{kraus08,diehl08,verstraete09,Eisert10, barreiro11}.

\item{\it Entanglement dynamics for improved quantum metrology.} The entangled states that have been proposed  \cite{bollinger,dowling02} for attaining the ultimate quantum precision limit in the estimation of transition frequencies and phases are very fragile against decoherence. The problem of increasing the precision in quantum metrology is thus closely related to the search of procedures to increase the robustness of entanglement. Recent progress has led to new estimates of precision limits in the presence of noise \cite{escher}, and new strategies to increase the robustness against particular models of local noise of entangled states used for quantum metrology have recently appeared, both passively \cite{chaves12b,chaves12c} and exploiting error-correction \cite{Kessler13, Arrad13, Duer13,Ozeri13}. This venue of research may lead to interesting results in the years to come.

\item{\it Scaling laws for the decay of genuinely multipartite entanglement}. The recently introduced multipartite negativity  \cite{guehne2011}, sensitive only to genuinely multipartite entanglement, which allows for a quite practical evaluation through semidefinite programming, has provided a new tool to study the robustness of genuine-multipartite entanglement \cite{Ali13}. A remaining challenge in this direction is the derivation of analytical scaling laws, in the sense of \cite{aolita08}, for the decay of genuinely multipartite entanglement, either for the multipartite negativity or for other genuinely multipartite measures yet to be conceived.

\item{\it Scaling laws for entanglement decay in many-body-system quantum simulations}. An ambitious but very interesting question is whether one can come up with (at least approximate) analytical scaling laws for entanglement decay \cite{aolita08}, either in ground and thermal states, or in non-equilibrium (quenched) systems, for Hubbard or spin models, in their different phases, under realistic noise types present in physical implementations, such as for instance atoms in optical lattices \cite{Bloch08, bloch2012,cirac2012}.

\item{\it Entanglement propagation in many-body systems}. This theme is relevant to many fundamental questions in physics, such as quantum transport, localization, and thermalization.  A toy example, discussed in Sec's. \ref{EntDynaAD}, \ref{EntDynaPD} and \ref{61}, is the flow of entanglement from an initial two-qubit entangled system into the corresponding environment. An important problem is the determination of the maximum speed of propagation of information, and how fast different parts of the system become correlated \cite{lieb72,hastings04,sims,braviy06,arealaw,eisert13}. The connection between entanglement propagation and thermalization, and the effect of noisy environments  on the propagation, are still  largely unexplored subjects \cite{eisert13}. For noiseless evolutions and short-range interactions, constant-velocity bounds were derived by Lieb and Robinson \cite{lieb72}. A generalization of this bound to Markovian quantum evolutions was derived in \cite{Poulin}. However, in spite of recent progress \cite{hauke13}, this is still an open problem for long-range interactions, even in the absence of noise. Recent experiments \cite{roos13,jurcevic14} have allowed a detailed study of this question, and constitute implementations of quantum simulations that are helpful to investigate still unknown properties of entanglement propagation.

\end{itemize}

Further progress in the field is however intrinsically constrained by advance in the problem of quantifying entanglement, in its different varieties, for  multipartite systems, which, in spite of the  impressive recent progress here described, remains far from closed. All in all, Schr\"odinger's predicament in his famous Naturwissenschaften paper \cite{schrodinger35} still holds true: entanglement ``keeps coming back to haunt us.''

\begin{acknowledgements}

The views expressed in this review were greatly influenced by many of our colleagues. In particular, we  would specially like to thank Antonio Ac{\'{\i}}n,  Fernando Brand\~{a}o, Michel Brune, Andreas Buchleitner, Daniel Cavalcanti, Rafael Chaves, Joseph H. Eberly, Bruno M. Escher, Marcelo Fran\c ca, Serge Haroche, Wolfgang D\"{u}r, Jens Eisert, Alessandro Ferraro, Otfried G\"{u}hne, Thomas Konrad, Barbara Kraus, Ruynet L. de Matos Filho, P\'erola Milman, Florian Mintert, Jean-Michel Raimond, Augusto Roncaglia, Paulo Henrique Souto Ribeiro, Markus Tiersch,  Stephen P. Walborn, and Nicim Zagury.

We would also like to thank institutions that were crucial for the production of this paper. L.A. would like to thank his former affiliation ICFO - the Institute of Photonic Sciences (Barcelona, Spain), where significant parts of this review were written; and  the ``Centro de Ciencias de Benasque Pedro Pascual" (Benasque, Spain), for repeatedly hosting crucial discussions about the topics here reviewed.  F.d-M. would like to thank the Freiburg University, the Catholic University of Leuven, and the Amsterdam Center for Mathematics and Computer Sciences, where part of this work was written. L.A. and F.d-M. would like to thank the Quantum Optics and Quantum Information group at the Federal University of Rio de Janeiro for hosting them in various occasions, during which the writing of this review was greatly pushed forward.

This work was partially funded by the German Alexander von Humboldt Foundation, the Belgian Interuniversity Attraction Poles Programme P6/02, the Netherlands Organization for Scientific Research, the Spanish MICIIN through a ``Juan de la Cierva" grant and the European Union under Marie Curie IEF No 299141, the Brazilian National Institute for Science and Technology on Quantum Information, and the Brazilian funding agencies CNPq, Faperj and CAPES. 
\end{acknowledgements}

%
\bibliographystyle{unsrt}
\bibliography{Master_Bib6}

\begin{thebibliography}{100}

\bibitem{epr35}
A.~Einstein, B.~Podolsky, and N.~Rosen.
\newblock Can {Q}uantum-{M}echanical {D}escription of {P}hysical {R}eality {B}e
  {C}onsidered {C}omplete?
\newblock {\em Phys. Rev.}, 47:777, 1935.

\bibitem{schrodinger35}
E.~Schr{\"{o}}dinger.
\newblock Die gegenw{\"{a}}rtige situation in der quantenmechanik.
\newblock {\em Die Naturwissenschaften}, 23:807--812, 1935.

\bibitem{schrodinger2}
E.~Schr{\"{o}}dinger.
\newblock Discussion of probability relations between separated systems.
\newblock {\em Proceedings of the Cambridge Philosophical Society},
  31:555--563, 1935.

\bibitem{schrodinger3}
E.~Schr{\"{o}}dinger.
\newblock Discussion of probability relations between separated systems.
\newblock {\em Proceedings of the Cambridge Philosophical Society},
  32:446--451, 1936.

\bibitem{bellepr64}
J.~S. Bell.
\newblock On the {E}instein-{P}odolsky-{R}osen paradox.
\newblock {\em Physics}, 1:195--200, 1964.

\bibitem{bell04}
J.~S. Bell.
\newblock {\em Speakable and Unspeakable in Quantum Mechanics}.
\newblock Cambridge, New York, 2004.

\bibitem{ekert91}
A.~K. Ekert.
\newblock Quantum cryptography based on {B}ell{'}s theorem.
\newblock {\em Phys. Rev. Lett.}, 67:661, 1991.

\bibitem{bennett92a}
C.~H. Bennett and S.~J. Weisner.
\newblock Communication via one- and two-particle operators on
  {E}instein-{P}odolsky-{R}osen states.
\newblock {\em Phys. Rev. Lett.}, 69:2881, 1992.

\bibitem{bennett93}
C.~H. Bennett, G.~Brassard, C.~Crepeau, R.~Jozsa, A.~Peres, and W.~K. Wootters.
\newblock Teleporting an {U}nknown {Q}uantum {S}tate via {D}ual {C}lassical and
  {E}instein-{P}odolsky-{R}osen channels.
\newblock {\em Phys. Rev. Lett.}, 70:1895--1899, 1993.

\bibitem{shor97}
P.~W. Shor.
\newblock Polynomial-{T}ime {A}lgorithms for {P}rime {F}actorization and
  {D}iscrete {L}ogarithms on a {Q}uantum {C}omputer.
\newblock {\em SIAM J. Sci. Statist. Comput.}, 26:1484--1509, 1997.

\bibitem{steane98}
A.~M. Steane.
\newblock Quantum computing.
\newblock {\em Rep. Prog. Phys.}, 61:117--173, 1998.

\bibitem{bennetnature}
C.~H. Bennett and B.~D. DiVicenzo.
\newblock Quantum information and computation.
\newblock {\em Nature}, 404:247--255, 2000.

\bibitem{nielsenchuang}
M.~A. Nielsen and I.~L. Chuang.
\newblock {\em Quantum Computation and Quantum Information}.
\newblock Cambridge University Press, 2000.

\bibitem{bouwmeester00}
D.~Bouwmeester, A.~Ekert, and A.~Zeilinger, editors.
\newblock {\em The Physics of Quantum Information}.
\newblock Springer Verlag, Berlin, 2000.

\bibitem{einsteinborn}
M.~Born.
\newblock {\em The Born-Einstein Letters: Friendship, Politics and Physics in
  Uncertain Times}.
\newblock Macmillan, 2005.

\bibitem{mattle}
K.~Mattle, H.~Weinfurter, P.~G. Kwiat, and A.~Zeilinger.
\newblock Dense coding in experimental quantum communication.
\newblock {\em Phys. Rev. Lett.}, 76(4656-4659), 1996.

\bibitem{davidovich0}
L.~Davidovich, N.~Zagury, M.~Brune, J.~M. Raimond, and S.~Haroche.
\newblock Teleportation of an atomic state between two cavities using nonlocal
  microwave fields.
\newblock {\em Phys. Rev. A}, 50:R895--898, 1994.

\bibitem{zeilinger97}
D.~Bouwmeester, J.~W. Pan, K.~Mattle, M.~Eibl, H.~Weinfurter, and A.~Zeilinger.
\newblock Experimental quantum teleportation.
\newblock {\em Nature}, 390:575--579, 1997.

\bibitem{boschi}
D.~Boschi, S.~Branca, F.~De Martini, L.~Hardy, and S.~Popescu.
\newblock Experimental realization of teleporting an unknown pure quantum state
  via dual classical and {E}instein-{P}odolsky-{R}osen channels.
\newblock {\em Phys. Rev. Lett.}, 80:1121--1125, 1998.

\bibitem{Riebe04}
M.~Riebe, H.~H\"affner, C.~F. Roos, W.~H{\"a}nsel, J.~Benhelm, G.~P.~T.
  Lancaster, T.~W. K{\"o}rber, C.~Becher, F.~Schmidt-Kaler, D.~F.~V. James, and
  R.~Blatt.
\newblock Deterministic quantum teleportation with atoms.
\newblock {\em Nature}, 429:734, 2004.

\bibitem{gisin02}
N.~Gisin, G.~Ribordy, W.~Tittel, and H.~Zbinden.
\newblock Quantum cryptography.
\newblock {\em Rev. Mod. Phys.}, 74:145, 2002.

\bibitem{Barrett05a}
J.~Barrett, L.~Hardy, and A.~Kent.
\newblock No signalling and quantum key distribution.
\newblock {\em Phys. Rev. Lett.}, page 010503, 2005.

\bibitem{Acin2006a}
A.~Ac{\'{i}}n, N.~Gisin, and L.~Masanes.
\newblock From {B}ell's {T}heorem to {S}ecure {Q}uantum {K}ey {D}istribution.
\newblock {\em Phys. Rev. Lett.}, 97:120405, 2006.

\bibitem{Acin07}
A.~Ac{\'{i}}n, N.~Brunner, N.~Gisin, S.~Massar, S.~Pironio, and V.~Scarani.
\newblock Device-independent security of quantum cryptography against
  collective attacks.
\newblock {\em Phys. Rev. Lett.}, 98(230501), 2007.

\bibitem{Colbeck07}
R.~Colbeck.
\newblock {\em Quantum and relativistic protocols for secure multi-party
  computation}.
\newblock PhD thesis, University of {C}ambridge, 2007.

\bibitem{Pironio10}
S.~Pironio, A.~Ac{\'{i}}n, S.~Massar, A.~B. de~la Giroday, D.~Matsukevich,
  P.~Maunz, S.~Olmschenk, D.~Hayes, L.~Luo, T.~Manning, and C.~Monroe.
\newblock Random {N}umbers {C}ertified by {B}ell's {T}heorem.
\newblock {\em Nature}, 464:1021, 2010.

\bibitem{Colbeck12}
R.~Colbeck and R.~Renner.
\newblock Free randomness can be amplified.
\newblock {\em Nat. Phys.}, 8:450, 2012.

\bibitem{Gallego13}
R.~Gallego, L.~Masanes, G.~de~la Torre, C.~Dhara, L.~Aolita, and A.~Ac\'{i}n.
\newblock Full randomness from arbitrarily deterministic events.
\newblock {\em Nat. Commun.}, 4:2654, 2013.

\bibitem{Brandao13}
F.~G. S.~L. Brand{\~{a}}o, R.~Ramanathan, A.~Grudka, K.~Horodecki,
  M.~Horodecki, and P.~Horodecki.
\newblock Robust device-independent randomness amplification with few devices.
\newblock {\em ArXiv:1310.4544}, 2013.

\bibitem{Braunstein1996}
S.~L. Braunstein, C.~M. Caves, and G.~J. Milburn.
\newblock Generalized uncertainty relations: Theory, examples, and {L}orentz
  invariance.
\newblock {\em Ann. Phys.}, 247:135--173, 1996.

\bibitem{leibfried04}
D.~Leibfried, M.~D. Barrett, T.~Schaetz, J.~Britton, J.~Chiaverini, W.~M.
  Itano, J.~D. Jost, C.~Langer, and D.~J. Wineland.
\newblock Toward {H}eisenberg-limited spectroscopy with multiparticle entangled
  states.
\newblock {\em Science}, 304:1476--1478, 2004.

\bibitem{escher}
B.~M. Escher, R.~L. de~Matos~Filho, and L.~Davidovich.
\newblock General framework for estimating the ultimate precision limit in
  noisy quantum-enhanced metrology.
\newblock {\em Nat. Phys.}, 7:406--411, 2011.

\bibitem{GLM2011:222}
V.~Giovannetti, S.~Lloyd, and L.~Maccone.
\newblock Advances in quantum metrology.
\newblock {\em Nat. Photon.}, 5(4):222--229, 2011.

\bibitem{feynman82}
R.~P. Feynman.
\newblock Simulating {P}hysics with {C}omputers.
\newblock {\em International Journal Of Theoretical Physics}, 21:467--488,
  1982.

\bibitem{lloyd96}
S.~Lloyd.
\newblock Universal quantum simulators.
\newblock {\em Science}, 273:1073--1078, 1996.

\bibitem{Jaksch98}
D.~Jaksch, C.~Bruder, J.~I. Cirac, C.~W. Gardiner, and P.~Zoller.
\newblock Cold {B}osonic {A}toms in {O}ptical {L}attices.
\newblock {\em Phys. Rev. Lett.}, 81:3108--3112, 1998.

\bibitem{Greiner02}
M.~Greiner, O.~Mandel, T.~Esslinger, T.~W. H{\"a}nsch, and I.~Bloch.
\newblock Quantum {P}hase {T}ransition from a {S}uperfluid to a {M}ott
  {I}nsulator in a {G}as of {U}ltracold {A}toms.
\newblock {\em Nature}, 415:39--44, 2002.

\bibitem{Bloch08}
I.~Bloch, J.~Dalibard, and W.~Zwerger.
\newblock {M}any-{B}ody {P}hysics with {U}ltracold {G}ases.
\newblock {\em Rev. Mod. Phys.}, 80:885, 2008.

\bibitem{cirac2012}
J.~I. Cirac and P.~Zoller.
\newblock Goals and opportunities in quantum simulation.
\newblock {\em Nature Physics}, 8:264--266, 2012.

\bibitem{bloch2012}
I.~Bloch, J.~Dalibard, and S.~Nascimb{\`e}ne.
\newblock Quantum simulations with ultracold quantum gases.
\newblock {\em Nat. Phys.}, 8:267--276, 2012.

\bibitem{blatt2012}
R.~Blatt and C.~F. Roos.
\newblock Quantum simulations with trapped ions.
\newblock {\em Nat. Phys.}, 8:277--284, 2012.

\bibitem{guzik2012}
A.~Aspuru-Guzik and P.~Walther.
\newblock Photonic quantum simulators.
\newblock {\em Nat. Phys.}, 8:285--291, 2012.

\bibitem{houck2012}
A.~A. Houck, H.~E. T{\"u}reci, and J.~Koch.
\newblock On-chip quantum simulation with superconducting circuits.
\newblock {\em Nat. Phys.}, 8:292--299, 2012.

\bibitem{AA11}
S.~Aaronson and A.~Arkhipov.
\newblock The {C}omputational {C}omplexity of {L}inear {O}ptics.
\newblock In {\em Proceedings of {A}{C}{M} {S}ymposium on the {T}heory of
  {C}omputing, STOC}, page 333. Association for Computing Machinery, New York,
  2011.

\bibitem{arno}
A.~Rauschenbeutel, G.~Nogues, S.~Osnaghi, P.~Bertet, M.~Brune, J.~M. Raimond,
  and S.~Haroche.
\newblock Step-by-step engineered multiparticle entanglement.
\newblock {\em Science}, 288:2024--2028, 2000.

\bibitem{Kuhn02}
A.~Kuhn, M.~Hennrich, and G.~Rempe.
\newblock Deterministic single-photon source for distributed quantum
  networking.
\newblock {\em Phys. Rev. Lett.}, 89:067901, 2002.

\bibitem{McKeever04}
J.~McKeever, A.~Boca, A.~D. Boozer, and J.~R.~Buck nad H.~J.~Kimble.
\newblock Experimental realization of a one-atom laser in the regime of strong
  coupling.
\newblock {\em Nature}, 425:268, 2003.

\bibitem{keller04}
M.~Keller, B.~Lange, K.~Hayasaka, W.~Lange, and H.~Walther.
\newblock Continuous generation of single photons with controlled waveform in
  an ion-trap cavity system.
\newblock {\em Nature}, 431:1075--1078, 2004.

\bibitem{Wilk08}
T.~Wilk, S.~C. Webster, A.~Kuhn, and G.~Rempe.
\newblock Single-{A}tom {S}ingle-{P}hoton {Q}uantum {I}nterface.
\newblock {\em Science}, 317:488, 2008.

\bibitem{blinov}
B.~B. Blinov, D.~L. Moehring, L.~M. Duan, and C.~Monroe.
\newblock Observation of entanglement between a single trapped atom and a
  single photon.
\newblock {\em Nature}, 428:153--157, 2004.

\bibitem{Julsgaard04}
B.~Julsgaard, J.~Sherson, J.~I. Cirac, J.~Fiurasek, and E.~S. Polzik.
\newblock Experimental demonstration of quantum memory for light.
\newblock {\em Nature}, 432:482, 2004.

\bibitem{choi08}
K.~S. Choi, H.~Deng, J.~Laurat, and H.~J. Kimble.
\newblock Mapping photonic entanglement into and out of a quantum memory.
\newblock {\em Nature}, 452:72--75, 2008.

\bibitem{Clausen11}
C.~Clausen, I.~Usmani, F.~Bussi\'eres, N.~Sangouard, M.~Afzelius,
  H.~de~Riedmatten, and N.~Gisin.
\newblock Quantum storage of photonic entanglement in a crystal.
\newblock {\em Nature}, 469:508--511, 2011.

\bibitem{sackett00}
C.~A. Sackett, D.~Kielpinski, B.~E. King, C.~Langer, V.~Meyer, C.~J. Myatt,
  M.~Rowe, Q.~A. Turchette, W.~M. Itano, E.~J. Wineland, and C.~Monroe.
\newblock Experimental entanglement of four particles.
\newblock {\em Nature}, 404:256, 2000.

\bibitem{leibfried05}
D.~Leibfried, E.~Knill, S.~Seidelin, J.~Britton, R.~B. Blakestad,
  J.~Chiaverini, D.~B. Hume, W.~M. Itano, J.~D. Jost, C.~Langer, R.~Ozeri,
  R.~Reichleand, and E.~J. Wineland.
\newblock Creation of a six-atom {`}schr{\"{o}}dinger cat{'} state.
\newblock {\em Nature}, 438:639, 2005.

\bibitem{haeffner05}
H.~Haffner, W.~Hansel, C.~F. Roos, J.~Benhelm, D.~Chek-al kar, M.~Chwalla,
  T.~Korber, U.~D. Rapol, M.~Riebe, P.~O. Schmidt, C.~Becher, O.~Guhne, W.~Dur,
  and R.~Blatt.
\newblock Scalable multiparticle entanglement of trapped ions.
\newblock {\em Nature}, 438:643--646, 2005.

\bibitem{monz2011}
T.~Monz, P.~Schindler, J.~T. Barreiro, M.~Chwalla, D.~Nigg, W.~A. Coish,
  M.~Harlander, W.~H{\"a}nsel, M.~Hennrich, and R.~Blatt.
\newblock 14-qubit entanglement: Creation and coherence.
\newblock {\em Phys. Rev. Lett.}, 106:130506, 2011.

\bibitem{pan00}
J.~W. Pan, D.~Bouwmeester, M.~Daniel, H.~Weinfurter, and A.~Zeilinger.
\newblock Experimental test of quantum nonlocality in three-photon
  {G}reenberger-{H}orne-{Z}eilinger entanglement.
\newblock {\em Nature}, 403:515, 2000.

\bibitem{zhao}
Z.~Zhao, Y-A. Chen, A-N. Zhang, T.~Yang, H.~J. Briegel, and J-W. Pan.
\newblock Experimental demonstration of five-photon entanglement and
  open-destination teleportation.
\newblock {\em Nature}, 430:54--58, 2004.

\bibitem{Walther05}
P.~Walther, K.~J. Resch, T.~Rudolph, E.~Schenck, H.~Weinfurter, V.~Vedral,
  M.~Aspelmeyer, and A.~Zeilinger.
\newblock Experimental one-way quantum computing.
\newblock {\em Nature}, 434:169, 2005.

\bibitem{lu07}
C.~Y. Lu, X.~Q. Zhou, O.~G{\"{u}}hne, W.~B. Gao, J.~Zhang, A.~G{\"{o}}bel
  Z.~S.~Yuan, T.~Yang, and J.~W. Pan.
\newblock Experimental entanglement of six photons in graph states.
\newblock {\em Nat. Phys.}, 3:91, 2007.

\bibitem{Prevedel07}
R.~Prevedel, P.~Walther, F.~Tiefenbacher, P.~B{\"o}hi, R.~Kaltenbaek,
  T.~Jennewein, and A.~Zeilinger.
\newblock High-speed linear optics quantum computing using active feed-forward.
\newblock {\em Nature}, 445:65, 2007.

\bibitem{Pino08}
G.~Vallone, E.~Pomarico, F.~De Martini, and P.~Mataloni.
\newblock Active {O}ne-{W}ay {Q}uantum {C}omputation with {T}wo-{P}hoton
  {F}our-{Q}ubit {C}luster {S}tates.
\newblock {\em Phys. Rev. Lett.}, 100:160502--160506, 2008.

\bibitem{Prevedel09}
R.~Prevedel, G.~Cronenberg, M.~S. Tame, M.~Paternostro, P.~Walther, M.~S. Kim,
  and A.~Zeilinger.
\newblock {E}xperimental {R}ealization of {D}icke {S}tates of up to {S}ix
  {Q}ubits for {M}ultiparty {Q}uantum {N}etworking.
\newblock {\em Phys. Rev. Lett.}, 103:020503 -- 020507, 2009.

\bibitem{Ceccarelli09}
R.~Ceccarelli, G.~Vallone, F.~De Martini, P.~Mataloni, and A.~Cabello.
\newblock Experimental {E}ntanglement and {N}onlocality of a {T}wo-{P}hoton
  {S}ix-{Q}ubit {C}luster {S}tate.
\newblock {\em Phys. Rev. Lett.}, 103:160401, 2009.

\bibitem{tenqubitpan}
W.-B. Gao, C.-Y. Lu, X.-C. Yao, P.~Xu, O.~G{\"u}hne, a.~Goebel, Y.-A. Chen,
  C.-Z. Peng, Z.-B. Chen, and J.-W. Pan.
\newblock Experimental demonstration of a hyper-entangled ten-qubit
  {S}chr{\"o}dinger cat state.
\newblock {\em Nat. Phys.}, 6:331--335, 2010.

\bibitem{pan2012}
X-C. Yao, T-X. Wang, P.~Xu, H.~Lu, G-S. Pan, X-H. Bao, C-P., C-Y. Lu, Y-A.
  Chen, and J-W. Pan.
\newblock Observation of eight-photon entanglement.
\newblock {\em Nature Photon.}, 6:225--228, 2012.

\bibitem{Yao2012}
X.-C. Yao, T.-X. Wang, H.-Z. Chen, W.-B. Gao, A.~G. Fowler, R.~Raussendorf,
  Z.-B. Chen, N.-L. Liu, C.-Y. Lu, Y.-J. Deng, Y.-A. Chen, and J.-W. Pan.
\newblock Experimental demonstration of topological error correction.
\newblock {\em Nature}, 482:489--494, 2012.

\bibitem{Huang2012}
Y.-F. Huang, B.-H. Liu, L.~Peng, Y.-H. Li, L.~Li, C.-F. Li, and G.-C. Guo.
\newblock Experimental generation of an eight-photon
  {G}reenberger--{H}orne--{Z}eilinger state.
\newblock {\em Nat. Commun.}, 2:546, 2012.

\bibitem{julsgaard01}
B.~Julsgaard, A.~Kozhekin, and E.~S. Polzik.
\newblock Experimental long-lived entanglement of two macroscopic objects.
\newblock {\em Nature}, 413:400--403, 2001.

\bibitem{Choi10}
K.~S. Choi, A.~Goban, S.~B. Papp, S.~J. van Enk, and H.~J. Kimble.
\newblock Entanglement of spin waves among four quantum memories.
\newblock {\em Nature}, 468:412--416, 2010.

\bibitem{Majer07}
J.~Majer, J.~M. Chow, J.~M. Gambetta, J.~Koch, B.~R. Johnson, J.~A. Schreier,
  L.~Frunzio, D.~I. Schuster, A.~A. Houck, A.~Wallraff, A.~Blais, M.~H.
  Devoret, S.~M. Girvin, and R.~J. Schoelkopf.
\newblock Coupling {S}uperconducting {Q}ubits via a {C}avity {B}us.
\newblock {\em Nature}, 449:443, 2007.

\bibitem{Matthews09}
J.~C.~F. Matthews, A.~Politi, A.~Stefanov, and J.~L. O'Brien.
\newblock Manipulation of multiphoton entanglement in waveguide quantum
  circuits.
\newblock {\em Nat. Photon.}, 3:346, 2009.

\bibitem{susan}
S.~Ghosh, T.~F. Rosenbaum, G.~Aeppli, and S.~N. Coppersmith.
\newblock Entangled quantum state of magnetic dipoles.
\newblock {\em Nature}, 48:48--51, 2003.

\bibitem{schrodinger26a}
E.~Schr{\"{o}}dinger.
\newblock The continuous transition from micro to macro-mechanics.
\newblock {\em Die Naturwissenschaften}, 28:664--666, 1926.

\bibitem{glauber1}
R.~J. Glauber.
\newblock The quantum theory of optical coherence.
\newblock {\em Phys. Rev.}, 130:2529--2539, 1963.

\bibitem{glauber2}
R.~J. Glauber.
\newblock Coherent and incoherent states of the radiation field.
\newblock {\em Physical Review}, 131:2766--2788, 1963.

\bibitem{vonneumann32}
J.~von Neumann.
\newblock {\em Matematische Grundlagen der Quantenmechanic}.
\newblock Springer, Berlin, 1932.

\bibitem{wheeler83}
J.~A. Wheeler and W.~Hubert Zurek, editors.
\newblock {\em Quantum Theory and Measurement}.
\newblock Princeton University Press, Princeton, NJ, 1983.

\bibitem{caldeira}
A.~O. Caldeira and A.~J. Leggett.
\newblock Influence of damping on quantum interference: An exactly soluble
  model.
\newblock {\em Phys. Rev. A}, 31:1059--1066, 1985.

\bibitem{joos}
E.~Joos, H.~D. Zeh, C.~Kiefer, D.~Giulini, J.~Kupsch, and I.~O. Stamatescu.
\newblock {\em Decoherence and the Appearance of a Classical World in Quantum
  Theory}.
\newblock Springer, 2003.

\bibitem{paz01}
J.~P. Paz and W.~H. Zurek.
\newblock Environment-induced decoherence and the transition from quantum to
  classical.
\newblock In R.~Kaiser, C.~Westbrook, and F.~David, editors, {\em Course 8 of
  Les Houches Lectures Session LXXII: Coherent Atomic Matter Waves}, page 533,
  Berlin, 2001. Springer.

\bibitem{zurek:715}
W.~H. Zurek.
\newblock Decoherence, einselection, and the quantum origins of the classical.
\newblock {\em Rev. Mod. Phys.}, 75(3):715, 2003.

\bibitem{schlosshauer07}
M.~A. Schlosshauer.
\newblock {\em {Decoherence and the quantum-to-classical transition}}.
\newblock The frontiers collection. Springer, 2007.

\bibitem{davidovich1}
L.~Davidovich, M.~Brune, J.~M. Raimond, and S.~Haroche.
\newblock Mesoscopic quantum coherences in cavity qed: Preparation and
  decoherence monitoring schemes.
\newblock {\em Phys. Rev. A}, 53:1295--1309, 1996.

\bibitem{davidovich2}
L.~Davidovich, A.~Maali, M.~Brune, J.~M. Raimond, and S.~Haroche.
\newblock Quantum switches and nonlocal microwave fields.
\newblock {\em Phys. Rev. Lett.}, 71:2360--2363, 1993.

\bibitem{enscat}
M.~Brune, E.~Hagley, J.~Dreyer, X.~Ma{\^{i}} tre, A.~Maali, C.~Wunderlich,
  J.~M. Raimond, and S.~Haroche.
\newblock Observing the progressive decoherence of the {``}meter{''} in a
  quantum measurement.
\newblock {\em Phys. Rev. Lett.}, 77:4887--4890, 1996.

\bibitem{myatt2}
C.~J. Myatt, B.~E. King, Q.~A. Turchette, C.~A. Sackett, D.~Kielpinski, W.~M.
  Itano, C.~Monroe, and D.~J. Wineland.
\newblock Decoherence of quantum superpositions through coupling to engineered
  reservoirs.
\newblock {\em Nature}, 403:269--273, 2000.

\bibitem{Deleglise08}
S.~Del\'eglise, I.~Dotsenko, C.~Sayrin, J.~Bernu, M.~Brune, J.-M. Raimond, and
  S.~Haroche.
\newblock Reconstruction of non-classical cavity field states with snapshots of
  their decoherence.
\newblock {\em Nature}, 455:510, 2008.

\bibitem{leggett02}
A.~J. Leggett.
\newblock Testing the limits of quantum mechanics: motivation, state of play,
  prospects.
\newblock {\em Journal of Physics: Condensed Matter}, 14(15):R415, 2002.

\bibitem{lee11}
K.~C. Lee, M.~R. Sprague, B.~J. Sussman, J.~Nunn, N.~K. Langford, X.-M. Jin,
  T.~Champion, P.~Michelberger, K.~F. Reim, D.~England, D.~Jaksch, and I.~A.
  Walmsley.
\newblock Entangling macroscopic diamonds at room temperature.
\newblock {\em Science}, 334(6060):1253--1256, 2011.

\bibitem{demartini}
F.~De~Martini, F.~Sciarrino, and C.~Vitelli.
\newblock Entanglement test on a microscopic-macroscopic system.
\newblock {\em Phys. Rev. Lett.}, 100:253601, Jun 2008.

\bibitem{lvovsky2013}
A.~I. Lvovsky, R.~Ghobadi, A.~Chandra, A.~S. Prasad, and C.~Simon.
\newblock Observation of micro-macro entanglement of light.
\newblock {\em Nat Phys}, 9(9):541--544, 09 2013.

\bibitem{bruno}
N.~Bruno, A.~Martin, P.~Sekatski, N.~Sangouard, R.~T. Thew, and N.~Gisin.
\newblock Displacement of entanglement back and forth between the micro and
  macro domains.
\newblock {\em Nat Phys}, 9(9):545--548, 09 2013.

\bibitem{gigan06}
S.~Gigan, H.~R. Bohm, M.~Paternostro, F.~Blaser, G.~Langer, J.~B. Hertzberg,
  K.~C. Schwab, D.~Bauerle, M.~Aspelmeyer, and A.~Zeilinger.
\newblock Self-cooling of a micromirror by radiation pressure.
\newblock {\em Nature}, 444(7115):67--70, 11 2006.

\bibitem{arcizet}
O.~Arcizet, P.~F. Cohadon, T.~Briant, M.~Pinard, and A.~Heidmann.
\newblock Radiation-pressure cooling and optomechanical instability of a
  micromirror.
\newblock {\em Nature}, 444(7115):71--74, 11 2006.

\bibitem{kleckner}
D.~Kleckner and D.~Bouwmeester.
\newblock Sub-kelvin optical cooling of a micromechanical resonator.
\newblock {\em Nature}, 444(7115):75--78, 11 2006.

\bibitem{rajagopal}
A.~K. Rajagopal and R.~W. Rendell.
\newblock Decoherence, correlation, and entanglement in a pair of coupled
  dissipative oscillators.
\newblock {\em Phys. Rev. A}, 63:022116, 2001.

\bibitem{karol0}
K.~{\.{Z}}yczkowski, P.~Horodecki, M.~Horodecki, and R.~Horodecki.
\newblock Dynamics of quantum entanglement.
\newblock {\em Phys. Rev. A}, 65:012101, 2001.

\bibitem{duan00}
L.~M. Duan, G.~Giedke, J.~I. Cirac, and P.~Zoller.
\newblock Inseparability criterion for continuous variable systems.
\newblock {\em Phys. Rev. Lett.}, 84:2722--2725, 2000.

\bibitem{simon02}
C.~Simon and J.~Kempe.
\newblock Robustness of multiparty entanglement.
\newblock {\em Phys. Rev. A}, 65:052327, 2002.

\bibitem{diosi03}
L.~Di{\'{o}}si.
\newblock Progressive decoherence and total environmental disentanglement.
\newblock In F.~Benatti and R.~Floreanini, editors, {\em Irreversible Quantum
  Dynamics}, Lecture Notes in Physics. Springer, Berlin, Berlin, 2003.

\bibitem{jamroz03}
L.~Jak{\'{o}}bczyk and A~Jamr{\'{o}z}.
\newblock Entanglement and nonlocality versus spontaneous emission in two-atom
  systems.
\newblock {\em Phys. Lett. A}, 318:318, 2003.

\bibitem{dodd04}
P.~J. Dodd and P.~J. Halliwell.
\newblock Disentanglement and decoherence by open system dynamics.
\newblock {\em Phys. Rev. A}, 69:052105, 2004.

\bibitem{duer04}
W.~D{\"{u}}r and H.~J. Briegel.
\newblock Stability of macroscopic entanglement under decoherence.
\newblock {\em Phys. Rev. Lett.}, 92:180403, 2004.

\bibitem{ficek1}
R.~Tana\'{s} and Z.~Ficek.
\newblock Entangling two atoms via spontaneous emission.
\newblock {\em J. Opt. B: Quantum. Semiclass. Opt.}, 6:S90, 2004.

\bibitem{carvalho04}
A.~R.~R. Carvalho, F.~Mintert, and A.~Buchleitner.
\newblock Decoherence and multipartite entanglement.
\newblock {\em Phys. Rev. Lett.}, 93:230501, 2004.

\bibitem{yu04}
T.~Yu and J.~H. Eberly.
\newblock Finite-time disentanglement via spontaneous emission.
\newblock {\em Phys. Rev. Lett.}, 93:140404, 2004.

\bibitem{serafini04}
A.~Serafini, F.~Illuminati, M.~G.~A. Paris, and S.~De~Siena.
\newblock Entanglement and purity of two-mode gaussian states in noisy
  channels.
\newblock {\em Phys. Rev. A}, 69(2):022318, 2004.

\bibitem{lidar04}
S.~Bandyopadhyay and D.~A. Lidar.
\newblock Entangling capacities of noisy two-qubit hamiltonians.
\newblock {\em Phys. Rev. A}, 70(1):010301, Jul 2004.

\bibitem{hein05}
M.~Hein, W.~D{\"{u}}r, and H.~J. Briegel.
\newblock Entanglement properties of multipartite entangled states under the
  influence of decoherence.
\newblock {\em Phys. Rev. A}, 71:032350, 2005.

\bibitem{fine05}
B.~V. Fine, F.~Mintert, and A.~Buchleitner.
\newblock Equilibrium entanglement vanishes at finite temperature.
\newblock {\em Phys. Rev. B}, 71(15):153105, Apr 2005.

\bibitem{mintert05b}
F.~Mintert, A.~R.~R. Carvalho, M.~Ku{\'{s}}, and A.~Buchleitner.
\newblock Measures and dynamics of entangled states.
\newblock {\em Phys. Rep.}, 415:207, 2005.

\bibitem{aravind05}
D.~Tolkunov, V.~Privman, and P.~K. Aravind.
\newblock Decoherence of a measure of entanglement.
\newblock {\em Phys. Rev. A}, 71(6):060308, Jun 2005.

\bibitem{yu06}
T.~Yu and J.~H. Eberly.
\newblock Quantum open system theory: {B}ipartite aspects.
\newblock {\em Phys. Rev. Lett.}, 97:140403, 2006.

\bibitem{yu062}
T.~Yu and J.~H. Eberly.
\newblock Sudden death of entanglement: {C}lassical noise effects.
\newblock {\em Opt. Commun.}, 264:393--397, 2006.

\bibitem{santos06}
M.~F. Santos, P.~Milman, L.~Davidovich, and N.~Zagury.
\newblock Direct measurement of finite-time disentanglement induced by a
  reservoir.
\newblock {\em Phys. Rev. A}, 73:040305, 2006.

\bibitem{liu06}
A.~Abliz, H.~J. Gao, X.~C. Xie, Y.~S. Wu, and W.~M. Liu.
\newblock Entanglement control in an anisotropic two-qubit heisenberg $xyz$
  model with external magnetic fields.
\newblock {\em Phys. Rev. A}, 74(5):052105, Nov 2006.

\bibitem{benatti06}
F.~Benatti and R.~Floreanini.
\newblock Entangling oscillators through environment noise.
\newblock {\em J. Phys. A: Math. Gen.}, 39:2689, 2006.

\bibitem{yonac061}
M~Y{\"{o}}na\c, T.~Yu, and J.~H. Eberly.
\newblock Sudden death of entanglement of two jaynes{--}cummings atoms.
\newblock {\em J. Phys. B: Atom. Mol. Opt. Phys.}, 39:--621, 2006.

\bibitem{ficek2}
Z.~Ficek and R.~Tana\'{s}.
\newblock Dark periods and revivals of entanglement in a two-qubit system.
\newblock {\em Phys. Rev. A}, 74:024304, 2006.

\bibitem{eberly07}
J.~H. Eberly and T.~Yu.
\newblock The end of an entanglement.
\newblock {\em Science}, 316:555--557, 2007.

\bibitem{yu:459}
T.~Yu and J.~H. Eberly.
\newblock Evolution from entanglement to decoherence of bipartite mixed "x"
  states.
\newblock {\em Quant. Inf. Comp.}, 7:459, 2007.

\bibitem{yonac:s45}
M.~Y{\"{o}}na{\c{c}}, T.~Yu, and J.~H. Eberly.
\newblock Pairwise concurrence dynamics: a four-qubit model.
\newblock {\em J. Phys. B}, 40(9):S45, 2007.

\bibitem{zubairy07}
M.~Ikram, F-L. Li, and M.~S. Zubairy.
\newblock Disentanglement in a two-qubit system subjected to dissipation
  environments.
\newblock {\em Phys. Rev. A}, 75(6):062336, Jun 2007.

\bibitem{liu07}
K-L. Liu and Hsi-Sheng Goan.
\newblock Non-markovian entanglement dynamics of quantum continuous variable
  systems in thermal environments.
\newblock {\em Phys. Rev. A}, 76(2):022312, Aug 2007.

\bibitem{seligman07}
T.~Gorin, C.~Pineda, and T.~H. Seligman.
\newblock Decoherence of an $n$-qubit quantum memory.
\newblock {\em Phys. Rev. Lett.}, 99(24):240405, Dec 2007.

\bibitem{seligman2007}
C.~Pineda, T.~Gorin, and T.~H. Seligman.
\newblock Decoherence of two-qubit systems: a random matrix description.
\newblock {\em New J. Phys.}, 9:106, 2001.

\bibitem{sabrina07}
S.~Maniscalco, S.~Olivares, and M.~G.~A. Paris.
\newblock Entanglement oscillations in non-markovian quantum channels.
\newblock {\em Phys. Rev. A}, 75(6):062119, Jun 2007.

\bibitem{terra01}
M.~O.~Terra Cunha.
\newblock The geometry of entanglement sudden death.
\newblock {\em New J. Phys.}, 9:237, 2007.

\bibitem{ficek08}
Z.~Ficek and R.~Tana\ifmmode~\acute{s}\else \'{s}\fi{}.
\newblock Delayed sudden birth of entanglement.
\newblock {\em Phys. Rev. A}, 77(5):054301, May 2008.

\bibitem{lopez-2008}
C.~E. Lopez, G.~Romero, F.~Lastra, E.~Solano, and J.~C. Retamal.
\newblock Sudden birth versus sudden death of entanglement in multipartite
  systems.
\newblock {\em Phys. Rev. Lett.}, 101:080503, 2008.
\newblock arXiv.org:0802.1825.

\bibitem{marek08}
P.~Marek, J.~Lee, and M.~S. Kim.
\newblock Vacuum as a less hostile environment to entanglement.
\newblock {\em Phys. Rev. A}, 77(3):032302, Mar 2008.

\bibitem{guo08}
Y-X. Gong, Y-S. Zhang, Y-L. Dong, X-L. Niu, Y-F. Huang, and G-C. Guo.
\newblock Dependence of the decoherence of polarization states in phase-damping
  channels on the frequency spectrum envelope of photons.
\newblock {\em Phys. Rev. A}, 78(4):042103, Oct 2008.

\bibitem{james08}
A.~Al-Qasimi and D.~F.~V. James.
\newblock Sudden death of entanglement at finite temperature.
\newblock {\em Phys. Rev. A}, 77(1):012117, Jan 2008.

\bibitem{concentration}
M.~Tiersch, F.~de~Melo, and A.~Buchleitner.
\newblock Universality in open system entanglement dynamics.
\newblock {\em J. Phys. A: Math. Theor.}, 46(8):085301, 2013.

\bibitem{hu08}
C-H. Chou, T.~Yu, and B.~L. Hu.
\newblock Exact master equation and quantum decoherence of two coupled harmonic
  oscillators in a general environment.
\newblock {\em Phys. Rev. E}, 77(1):011112, Jan 2008.

\bibitem{lai08}
C-Y. Lai, J-T. Hung, C-Y. Mou, and Pochung C.
\newblock Induced decoherence and entanglement by interacting quantum spin
  baths.
\newblock {\em Phys. Rev. B}, 77(20):205419, May 2008.

\bibitem{Ferraro08}
A.~Ferraro, D.~Cavalcanti, A.~Garcia-Saez, and A.~Ac\'in.
\newblock Thermal bound entanglement in macroscopic systems and area law.
\newblock {\em Phys. Rev. Lett.}, 100:080502, 2008.

\bibitem{Paz:220401}
J.~P. Paz and A.~J. Roncaglia.
\newblock Dynamics of the entanglement between two oscillators in the same
  environment.
\newblock {\em Phys. Rev. Lett.}, 100(22):220401, Jun 2008.

\bibitem{paz08}
C.~Cormick and J.~P. Paz.
\newblock Decoherence of bell states by local interactions with a dynamic spin
  environment.
\newblock {\em Phys. Rev. A}, 78(1):012357, Jul 2008.

\bibitem{yu08b}
M.~Abdel-Aty and T.~Yu.
\newblock Entanglement sudden birth of two trapped ions interacting with a
  time-dependent laser field.
\newblock {\em J. Phys. B: Atom. Mol. Opt. Phys.}, 41:235503, 2008.

\bibitem{aolita08}
L.~Aolita, R.~Chaves, D.~Cavalcanti, A.~Ac\'{i}n, and L.~Davidovich.
\newblock Scaling laws for the decay of multiqubit entanglement.
\newblock {\em Phys. Rev. Lett.}, 100:080501, 2008.

\bibitem{xu09}
J-S. Xu, C-F. Li, X-Y. Xu, C-H. Shi, X-B. Zou, and G-C. Guo.
\newblock Experimental characterization of entanglement dynamics in noisy
  channels.
\newblock {\em Phys. Rev. Lett.}, 103(24):240502, Dec 2009.

\bibitem{Paz:032102}
J.~P. Paz and A.~J. Roncaglia.
\newblock Dynamical phases for the evolution of the entanglement between two
  oscillators coupled to the same environment.
\newblock {\em Phys. Rev. A}, 79(3):032102, Mar 2009.

\bibitem{cavalcanti09}
D.~Cavalcanti, R.~Chaves, L.~Aolita, L.~Davidovich, and A.~Ac{\'{i}}n.
\newblock Open-system dynamics of graph-state entanglement.
\newblock {\em Phys. Rev. Let.}, 103:030502, 2009.

\bibitem{hor09}
M.~Hor-Meyll, A.~Auyuanet, C.~V.~S. Borges, A.~Arag{\~a}o, J.~A.~O. Huguenin,
  A.~Z. Khoury, and L.~Davidovich.
\newblock Environment-induced entanglement with a single photon.
\newblock {\em Phys. Rev. A}, 80:042327, 2009.

\bibitem{terra02}
R.~C. Drumond and M.~O.~T. Cunha.
\newblock Asymptotic {E}ntanglement {D}ynamics and {G}eometry of {Q}uantum
  {S}tates.
\newblock {\em J. Phys. A}, 42:285308, 2009.

\bibitem{yu09}
T.~Yu and J.~H. Eberly.
\newblock Sudden death of entanglement.
\newblock {\em Science}, 323:598, 2009.

\bibitem{mazzola09}
L.~Mazzola, S.~Maniscalco, J.~Piilo, K.-A. Suominen, and B.~M. Garraway.
\newblock Sudden death and sudden birth of entanglement in common structured
  reservoirs.
\newblock {\em Phys. Rev. A}, 79(4):042302, Apr 2009.

\bibitem{zell09}
T.~Zell, F.~Queisser, and R.~Klesse.
\newblock Distance dependence of entanglement generation via a bosonic heat
  bath.
\newblock {\em Phys. Rev. Lett.}, 102(16):160501, Apr 2009.

\bibitem{sumanta}
S.~Das and G.~S. Agarwal.
\newblock Decoherence effects in interacting qubits under the influence of
  various environments.
\newblock {\em J. Phys. B.}, 42:205502, 2009.

\bibitem{papp09}
S.~B. Papp, K.~S. Choi, H.~Deng, P.~Lougovski, S.~J. van Enk, and H.~J. Kimble.
\newblock Characterization of multipartite entanglement for one photon shared
  among four optical modes.
\newblock {\em Science}, 324:764, 2009.

\bibitem{viviescas10}
C.~Viviescas, I.~Guevara, A.~R. Carvalho, M.~Busse, and A.~Buchleitner.
\newblock Entanglement dynamics in open two-qubit systems via diffusive quantum
  trajectories.
\newblock {\em Phys. Rev. Lett.}, 21:210502, 2010.

\bibitem{cavalcanti10}
D.~Cavalcanti, L.~Aolita, A.~Ferraro, A.~Garcia-Saez, and A.~Ac\'in.
\newblock Macroscopic bound entanglement in thermal graph states.
\newblock {\em New J. Phys.}, 12:025011, 2010.

\bibitem{dur11}
F.~Fro{\"{o}}wis and W.~D{\"{u}}r.
\newblock Stable macroscopic quantum superpositions.
\newblock {\em Phys. Rev. Lett.}, 106:110402, 2011.

\bibitem{aolita2011}
L.~Aolita, A.~J. Roncaglia, A.~Ferraro, and A.~Ac{\'{i}}n.
\newblock Gapped two-body hamiltonian for continuous-variable quantum
  computation.
\newblock {\em Phys. Rev. Lett.}, 106:090501, Feb 2011.

\bibitem{almeida07}
M.~P. Almeida, F.~de Melo, M.~Hor-Meyll, A.~Salles, S.~P. Walborn, P.~H.~Souto
  Ribeiro, and L.~Davidovich.
\newblock Environment-induced sudden death of entanglement.
\newblock {\em Science}, 316:579, 2007.

\bibitem{Laurat}
J.~Laurat, K.~S. Choi, H.~Deng, C.~W. Chou, and H.~J. Kimble.
\newblock Heralded entanglement between atomic ensembles: Preparation,
  decoherence, and scaling.
\newblock {\em Phys. Rev. Lett.}, 99:180504, 2007.

\bibitem{alejo}
A.~Salles, F.~de Melo, M.~P. Almeida, M.~Hor-Meyll, S.~P. Walborn, P.~H.~Souto
  Ribeiro, and L.~Davidovich.
\newblock Experimental investigation of the dynamics of entanglement: Sudden
  death, complementarity, and continuous monitoring of the environment.
\newblock {\em Phys. Rev. A}, 78:022322, 2008.

\bibitem{nussenzveig09}
A.~S. Coelho, F.~A.~S. Barvosa, K.~N. Cassemiro, A.~S. Villar, M.~Martinelli,
  and P.~Nussenzveig.
\newblock Three-color entanglement.
\newblock {\em Science}, 326:823, 2009.

\bibitem{barbosa10}
F.~A.~S. Barbosa, A.~S. Coelho, A.~J. de~Faria, K.~N. Cassemiro, A.~S. Villar,
  P.~Nussenzveig, and M.~Martinelli.
\newblock Robustness of bipartite gaussian entangled beams propagating in lossy
  channels.
\newblock {\em Nature Photon.}, 4:858--861, 2010.

\bibitem{Barreiro10}
J.~T. Barreiro, P.~Schindler, O.~G{\"{u}}hne, T.~Monz, M.~Chwalla, C.~F. Roos,
  M.~Hennrich, and R.~Blatt.
\newblock Experimental multiparticle entanglement dynamics induced by
  decoherence.
\newblock {\em Nat. Phys.}, 6:943, 2010.

\bibitem{farias2012}
O.~J. Far{\'{i}}as, A.~Vald{\'{e}}s-Hern{\'{a}}ndez, G.~H. Aguilar, P.~H.~S.
  Ribeiro, S.~P. Walborn, L.~Davidovich, Xiao-Feng Qian, and J.~H. Eberly.
\newblock Experimental investigation of dynamical invariants in bipartite
  entanglement.
\newblock {\em Phys. Rev. A}, 85:012314, Jan 2012.

\bibitem{osvaldo2012}
O.~J. Far{\'{i}}as, G.~H. Aguilar, A.~Vald{\'{e}}s-Hern{\'{a}}ndez, P.~H.~S.
  Ribeiro, L.~Davidovich, and S.~P. Walborn.
\newblock Observation of the emergence of multipartite entanglement between a
  bipartite system and its environment.
\newblock {\em Phys. Rev. Lett.}, 109:150403, 2012.

\bibitem{adriana10}
A.~Auyuanet and L.~Davidovich.
\newblock Quantum correlations as precursors of entanglement.
\newblock {\em Phys. Rev. A}, 82:032112, 2010.

\bibitem{aolitapra09}
L.~Aolita, D.~Cavalcanti, A.~Ac{\'{i}}n, A.~Salles, Markus Tiersch,
  A.~Buchleitner, and F.~de Melo.
\newblock Scalability of {G}{H}{Z} and random-state entanglement in the
  presence of decoherence.
\newblock {\em Phys Rev A}, 79:032322 -- 032330, 2009.

\bibitem{kwiat00}
P.~G. Kwiat, A.~J. Berglund, J.~B. Altepeter, and A.~G. White.
\newblock Experimental verification of decoherence-free subspaces.
\newblock {\em Science}, 290:498--501, 2000.

\bibitem{haeffner05b}
H.~H{\"{a}}ffner, F.~Schmidt-Kaler, W.~H{\"{a}}nsel, C.~F. Roos,
  T.~K{\"{o}}rber, M.~Chwalla, M.~Riebe, J.~Benhelm, U.~D. Rapol, C.~Becher,
  and R.~Blatt.
\newblock Robust entanglement.
\newblock {\em Appl. Phys. B}, 81:151, 2005.

\bibitem{poyatos}
J.~F. Poyatos, J.~I. Cirac, and P.~Zoller.
\newblock Quantum reservoir engineering with laser cooled trapped ions.
\newblock {\em Phys. Rev. Lett.}, 77:4728--4731, Dec 1996.

\bibitem{andre01}
A.~R.~R. Carvalho, P.~Milman, R.~L. de~Matos~Filho, and L.~Davidovich.
\newblock Decoherence, pointer engineering, and quantum state protection.
\newblock {\em Phys. Rev. Lett.}, 86:4988--4991, 2001.

\bibitem{pielawa}
S.~Pielawa, G.~Morigi, D.~Vitali, and L.~Davidovich.
\newblock Generation of {E}instein-{P}odolsky-{R}osen-entangled radiation
  through an atomic reservoir.
\newblock {\em Phys. Rev. Lett.}, 98:240401, 2007.

\bibitem{pielawa2}
S.~Pielawa, L.~Davidovich, D.~Vitali, and G.~Morigi.
\newblock Engineering atomic quantum reservoirs for photons.
\newblock {\em Phys. Rev. A}, 81:043802, 2010.

\bibitem{diehl08}
S.~Diehl, A.~Micheli, A.~Kantian, B.~Kraus, H.~P. Buchler, and P.~Zoller.
\newblock Quantum states and phases in driven open quantum systems with cold
  atoms.
\newblock {\em Nat Phys}, 4(11):878--883, 11 2008.

\bibitem{kraus08}
B.~Kraus, H.~P. B\"uchler, S.~Diehl, A.~Kantian, A.~Micheli, and P.~Zoller.
\newblock Preparation of entangled states by quantum {M}arkov processes.
\newblock {\em Phys. Rev. A}, 78:042307, 2008.

\bibitem{verstraete09}
F.~Verstraete, M.~M. Wolf, and J.~I. Cirac.
\newblock Quantum computation and quantum-state engineering driven by
  dissipation.
\newblock {\em Nat. Phys.}, 5:633--636, 2009.

\bibitem{andre07}
A.~R.~R. Carvalho and J.~J. Hope.
\newblock Stabilizing entanglement by quantum-jump-based feedback.
\newblock {\em Phys. Rev. A}, 76(1):010301, Jul 2007.

\bibitem{andre08}
A.~R.~R. Carvalho, A.~J.~S. Reid, and J.~J. Hope.
\newblock Controlling entanglement by direct quantum feedback.
\newblock {\em Phys. Rev. A}, 78(1):012334, Jul 2008.

\bibitem{devoret}
S.~Shankar, M.~Hatridge, Z.~Leghtas, K.~M. Sliwa, A.~Narla, U.~Vool, S.~M.
  Girvin, L.~Frunzio, M.~Mirrahimi, and M.~H. Devoret.
\newblock Autonomously stabilized entanglement between two superconducting
  quantum bits.
\newblock {\em Nature}, 504:419, 2013.

\bibitem{lin}
Y.~Lin, J.~P. Gaebler, F.~Reiter, T.~R. Tan, R.~Bowler, A.~S. Sorensen,
  D.~Leibfried, and D.~J. Wineland.
\newblock Dissipative production of a maximally entangled steady state of two
  quantum bits.
\newblock {\em Nature}, 504:415, 2013.

\bibitem{helstrom}
A.~S. Helstrom.
\newblock {\em Quantum detection and estimation theory}.
\newblock Academic Press, New York, 1976.

\bibitem{lloyd04}
V.~Giovannetti, S.~Lloyd, and L.~Maccone.
\newblock Quantum-enhanced measurements: beating the standard quantum limit.
\newblock {\em Science}, 306:1330--1336, 2004.

\bibitem{escherbjp}
B.~M. Escher, R.~L. de~Matos~Filho, and L.~Davidovich.
\newblock Quantum metrology for noisy systems.
\newblock {\em Braz. J. Phys.}, 41:229--247, 2011.

\bibitem{escher12}
B.~M. Escher, L.~Davidovich, N.~Zagury, and R.~L. de~Matos~Filho.
\newblock Quantum metrological limits via a variational approach.
\newblock {\em Phys. Rev. Lett.}, 109:190404, Nov 2012.

\bibitem{bollinger}
J.~J. Bollinger, W.~M. Itano, D.~J. Wineland, and D.~J. Heinzen.
\newblock Optimal frequency measurements with maximally correlated states.
\newblock {\em Phys. Rev. A}, 54:R4649--R4652, 1996.

\bibitem{blatt08}
R.~Blatt and D.~Wineland.
\newblock Entangled states of trapped atomic ions.
\newblock {\em Nature}, 453:1008--1015, 2008.

\bibitem{Huelga1997}
S.~F. Huelga, C.~Macchiavello, T.~Pellizzari, A.~K. Ekert, M.~B. Plenio, and
  J.~I. Cirac.
\newblock Improvement of frequency standards with quantum entanglement.
\newblock {\em Phys. Rev. Lett.}, 79:3865--3868, 1997.

\bibitem{dowling02}
H.~Lee, P.~Kok, and J.~P. Dowling.
\newblock A quantum rosetta stone for interferometry.
\newblock {\em J. Mod. Opt.}, 49:2325--2338, 2002.

\bibitem{Kacprowicz2010}
M.~Kacprowicz, R.~Demkowicz-Dobrzanski, W.~Wasilewski, K.~Banaszek, and I.~A.
  Walmsley.
\newblock Experimental quantum-enhanced estimation of a lossy phase shift.
\newblock {\em Nat. Photon.}, 4(6):357--360, 2010.

\bibitem{bryn}
B.~Bell, S.~Kannan, A.~McMillan, A.~S. Clark, W.~J. Wadsworth, and J.~G.
  Rarity.
\newblock Multicolor quantum metrology with entangled photons.
\newblock {\em Phys. Rev. Lett.}, 111:093603, Aug 2013.

\bibitem{KarolBook}
I.~Bengtsson and K.~{\.{Z}}yczkowski.
\newblock {\em Geometry of quantum states. An introduction to quantum
  Entanglement}.
\newblock Cambridge University Press, 2006.

\bibitem{amico}
L.~Amico, R.~Fazio, A.~Osterloh, and V.~Vedral.
\newblock Entanglement in many-body systems.
\newblock {\em Rev. Mod. Phys.}, 80:517--576, 2008.

\bibitem{plenio07}
M.~B. Plenio and S.~Virmani.
\newblock An introduction to entanglement measures.
\newblock {\em Quant. Inf. Comp.}, 7:1, 2007.

\bibitem{horodecki09}
R.~Horodecki, P.~Horodecki, M.~Horodecki, and K.~Horodecki.
\newblock Quantum entanglement.
\newblock {\em Rev. Mod. Phys.}, 81:865--942, 2009.

\bibitem{guehne09}
O.~G{\"{u}}hne and G.~T{\'{o}}th.
\newblock Entanglement detection.
\newblock {\em Phys. Rep.}, 474:1, 2009.

\bibitem{arealaw}
J.~Eisert, M.~Cramer, and M.~B. Plenio.
\newblock Colloquium: Area laws for the entanglement entropy.
\newblock {\em Rev. Mod. Phys.}, 82(1):277--306, Feb 2010.

\bibitem{breuer}
H.~P. Breuer and F.~Petruccione.
\newblock {\em The Theory of Open Quantum Systems}.
\newblock Oxford University Press, 2002.

\bibitem{schmidt07}
E.~Schmidt.
\newblock A four-party unlockable bound-entangled state.
\newblock {\em Math. Ann}, 63:433, 1907.

\bibitem{werner89}
R.~F. Werner.
\newblock Quantum {S}tates {W}ith {E}instein-{P}odolsky-{R}osen {C}orrelations
  {A}dmitting {A} {H}idden-{V}ariable {M}odel.
\newblock {\em Phys. Rev. A}, 40:4277--4281, 1989.

\bibitem{ghz89}
D.~M. Greenberger, M.~Horne, and A.~Zeilinger.
\newblock Going beyond bell's theorem.
\newblock In M.~Kafatos, editor, {\em Bell{'}s Theorem, Quantum Theory and
  Conceptions of the Universe}, pages 69--72, Dordrecht, 1989. Kluwer.

\bibitem{ghz90}
D.~M. Greenberger, M.~A. Horne, A.~Shimony, and A.~Zeilinger.
\newblock A {B}ell's theorem without inequalities.
\newblock {\em Am. J. Phys.}, 58:1131, 1990.

\bibitem{ghz93}
D.~M. Greenberger, M.~Horne, and A.~Zeilinger.
\newblock Multiparticle interferometry and the superposition principle.
\newblock {\em Phys. Today}, 46(8):22--29, August 1993.

\bibitem{duer00b}
W.~Dur, G.~Vidal, and J.~I. Cirac.
\newblock Three qubits can be entangled in two inequivalent ways.
\newblock {\em Phys. Rev. A}, 62:062314, 2000.

\bibitem{Gurvits03}
L.~Gurvits.
\newblock Classical deterministic complexity of edmonds' problem and quantum
  entanglement.
\newblock In {\em Proceedings of the 35th {ACM} {S}ymposium on the {T}heory of
  {C}omputing}. ACM Press, New York, 2003.

\bibitem{peres96}
A.~Peres.
\newblock Separability criterion for density matrices.
\newblock {\em Phys. Rev. Lett.}, 77:1413--1415, 1996.

\bibitem{horodecki96}
M.~Horodecki, P.~Horodecki, and R.~Horodecki.
\newblock Separability of n-particle mixed states: necessary and sufficient
  conditions in terms of linear maps.
\newblock {\em Phys. Lett. A}, 283:1--7, 2001.

\bibitem{Simon00}
R.~Simon.
\newblock Peres-{H}orodecki separability criterion for continuous variable
  systems.
\newblock {\em Phys. Rev. Lett.}, 84:2726--2729, 2000.

\bibitem{Horodecki97b}
P.~Horodecki.
\newblock Separability {C}riterion and {I}nseparable {M}ixed {S}tates with
  {P}ositive {P}artial {T}ransposition.
\newblock {\em Phys. Lett. A}, 232:333, 1997.

\bibitem{Chen03b}
K.~Chen and L.-A. Wu.
\newblock A matrix realignment method for recognizing entanglement.
\newblock {\em Quant. Inf. Comp.}, 3:193, 2003.

\bibitem{Rudolph05}
O.~Rudolph.
\newblock On the cross norm criterion for separability.
\newblock {\em Quantum Inf. Proc.}, 4:219, 2005.

\bibitem{Ferraro05}
M.~G. A.~Paris A.~Ferraro, S.~Olivares.
\newblock Gaussian states in continuous variable quantum information.
\newblock {\em Bibliopolis, Napoli}, pages ISBN 88--7088--483--X, 2005.

\bibitem{adesso2007}
G.~Adesso and F.~Illuminati.
\newblock Entanglement in continuous-variable systems: recent advances and
  current perspectives.
\newblock {\em J. Phys. A: Math. Theor.}, 40:7821--7880, 2007.

\bibitem{scully97}
M.~O. Scully and M.~S. Zubairy.
\newblock {\em Quantum Optics}.
\newblock Cambridge University Press, Cambridge, 1997.

\bibitem{gardiner}
C.~W. Gardiner and P.~Zoller.
\newblock {\em Quantum Noise}.
\newblock Springer-Verlag, Berlin, 2nd edition, 1999.

\bibitem{werner01}
R.~F. Werner and M.~M. Wolf.
\newblock Bound entangled {G}aussian states.
\newblock {\em Phys. Rev. Lett.}, 86:3658, 2001.

\bibitem{serafini05}
A.~Serafini, G.~Adesso, and F.~Illuninati.
\newblock Unitarily localizable entanglement of {G}aussian states.
\newblock {\em Phys. Rev. A}, 71(032349), 2005.

\bibitem{shchukin}
E.~Shchukin and W.~Vogel.
\newblock Inseparability criterion for continuous bipartite systems.
\newblock {\em Phys. Rev. Lett.}, 95:230502, 2005.

\bibitem{gomes}
R.~M. Gomes, A.~Salles, F.~Toscano, P.~H. Souto~Ribeiro, and S.~P. Walborn.
\newblock Quantum entanglement beyond gaussian criteria.
\newblock {\em Proc. Natl. Acad. Sci. USA}, 106:21517--21520, 2009.

\bibitem{terhal00}
B.~M. Terhal.
\newblock Bell inequalities and the separability criterion.
\newblock {\em Phys. Lett. A}, 271:319, 2000.

\bibitem{lewenstein00}
M.~Lewenstein, B.~Kraus, J.~I. Cirac, and P.~Horodecki.
\newblock Optimization of entanglement witnesses.
\newblock {\em Phys. Rev. A}, 62:230504, 2000.

\bibitem{bruss02b}
D.~Bru{\ss{}}, J.~I. Cirac, P.~Horodecki, F.~Hulpke, B.~Kraus, M.~Lewenstein,
  and A.~Sanpera.
\newblock Reflections upon separability and distillability.
\newblock {\em J. Mod. Opt.}, 49:1399--1418, 2002.

\bibitem{Acin01}
A.~Ac\'in, D.~Bru{\ss{}}, M.~Lewenstein, and A.~Sanpera.
\newblock Classification of mixed three-qubit states.
\newblock {\em Phys. Rev. Lett.}, 87:040401, 2001.

\bibitem{Bourennane04}
M.~Bourennane, M.~Eibl, C.~Kurtsiefer, S.~Gaertner, H.~Weinfurter,
  O.~G{\"{u}}hne, P.~Hyllus, D.~Bruss, M.~Lewenstein, and A.~Sanpera.
\newblock Experimental detection of multipartite entanglement using witness
  operators.
\newblock {\em Phys. Rev. Lett.}, 92:087902, 2004.

\bibitem{GuehneSeevinck}
O.~G{\"{u}}hne and M.~Seevinck.
\newblock Separability criteria for genuine multiparticle entanglement.
\newblock {\em New J. Phys.}, 12:053002, 2010.

\bibitem{Huber10}
M.~Huber, F.~Mintert, A.~Gabriel, and B.~C. Hiesmayr.
\newblock Detection of high-dimensional genuine multi-partite entanglement of
  mixed states.
\newblock {\em Phys. Rev. Lett.}, 104:210501, 2010.

\bibitem{guehne11b}
O.~G\"{u}hne, B.~Jungnitsch, T.~Moroder, and Y.~S. Weinstein.
\newblock Multiparticle entanglement in graph-diagonal states: {N}ecessary and
  sufficient conditions for four qubits.
\newblock {\em Phys. Rev. A}, 84:052319, 2011.

\bibitem{guehne2011}
B.~Jungnitsch, T.~Moroder, and O.~G{\"{u}}hne.
\newblock Taming multiparticle entanglement.
\newblock {\em Phys. Rev. Lett.}, 106:190502, 2011.

\bibitem{bennett96b}
C.~H. Bennett, D.~P. DiVincenzo, J.~A. Smolin, and W.~K. Wootters.
\newblock Mixed state entanglement and quantum errror correction.
\newblock {\em Phys. Rev. A}, 54:3824, 1996.

\bibitem{donald02}
M.~J. Donald, M.~Horodecki, and O.~Rudolph.
\newblock The uniqueness theorem for entanglement measures.
\newblock {\em Journal Of Mathematical Physics}, 43:4252--4272, 2002.

\bibitem{Gheorghiu}
V.~Gheorghiu and R.~B. Griffiths.
\newblock Separable operations on pure states.
\newblock {\em Phys. Rev. A}, 78:020304(R), 2008.

\bibitem{bennett00}
C.~H. Bennett, S.~Popescu, D.~Rohrlich, J.~Smolin, and A.~V. Thapliyal.
\newblock Exact and asymptotic measures of multipartite pure state
  entanglement.
\newblock {\em Phys. Rev. A}, 63:012307, 2000.

\bibitem{bennett96c}
C.~H. Bennett, G.~Brassard, S.~Popescu, B.~Schumacher, J.~Smolin, and W.~K.
  Wooters.
\newblock Purification of noisy entanglement and faithful teleportation via
  noisy channels.
\newblock {\em Phys. Rev. Lett.}, 78:2031, 1996.

\bibitem{horodecki98}
M.~Horodecki, P.~Horodecki, and R.~Horodecki.
\newblock Mixed-state entanglement and distillation: Is there a {``}bound"
  entanglement in nature?
\newblock {\em Phys. Rev. Lett.}, 80:5239, 1998.

\bibitem{DiVincenzo00}
D.~P. DiVincenzo, P.~W. Shor, J.~A. Smolin, B.~M. Terhal, and A.~V. Thapliyal.
\newblock Evidence for bound entangled states with negative partial transpose.
\newblock {\em Phys. Rev. A}, 61:062312, 2000.

\bibitem{Duer00d}
W.~D\"{u}r, J.~I. Cirac, M.~Lewenstein, and D.~Bruss.
\newblock Distillability and partial transposition in bipartite systems.
\newblock {\em Phys. Rev. A}, 61:062313, 2000.

\bibitem{clarisse06}
L.~Clarisse.
\newblock {\em Entanglement Distillation; a Dicourse on Bound Entanglement in
  Quantum}.
\newblock PhD thesis, University of York, 2006.
\newblock Eprint: quant-ph/0612072.

\bibitem{Yu13}
N.~Yu, C.~Guo, and R.~Duan.
\newblock Obtain {W}-state from three-qubit {G}{H}{Z}-state on rate 1.
\newblock {\em ar{X}iv:1309.4833}, 2013.

\bibitem{Dur00}
W.~D\"{u}r and J.~I. Cirac.
\newblock Classification of multiqubit mixed states: Separability and
  distillability properties.
\newblock {\em Phys. Rev. A}, 61:042314 -- 042325, 2000.

\bibitem{Duer&Briegel07}
W.~D\"{u}r and H.~Briegel.
\newblock Entanglement purification and quantum error correction.
\newblock {\em Rep. Prog. Phys.}, 70:1381, 2007.

\bibitem{duer00c}
W.~D{\"{u}}r and J.~I. Cirac.
\newblock Activating bound entanglement in multiparticle systems.
\newblock {\em Phys. Rev. A}, 62:022302, 2000.

\bibitem{Piani&Mora07}
M.~Piani and C.~Mora.
\newblock Class of positive-partial-transpose bound entangled states associated
  with almost any set of pure entangled states.
\newblock {\em Phys. Rev. A}, 75:012305, 2007.

\bibitem{bennett99}
C.~H. Bennett, D.~P. DiVincenzo, Tal Mor, P.~W. Shorand J.~A. Smolin, and B.~M.
  Terhal.
\newblock Unextendible product states and bound entanglement.
\newblock {\em Phys. Rev. Lett.}, 82:5385, 1999.

\bibitem{smolin01}
J.~A. Smolin.
\newblock A four-party unlockable bound-entangled state.
\newblock {\em Phys. Rev. A}, 63:032306, 2001.

\bibitem{Amselem09}
E.~Amselem and M.~Bourennane.
\newblock Experimental four-qubit bound entanglement.
\newblock {\em Nat. Phys.}, 5:748, 2009.

\bibitem{Lavoie10}
J.~Lavoie, R.~Kaltenbaek, M.~Piani, and K.~J. Resch.
\newblock Experimental bound entanglement in a four-photon state.
\newblock {\em Phys. Rev. Lett.}, 105:130501, 2010.

\bibitem{Bruss99}
D.~Bruss and C.~Macchiavello.
\newblock Optimal state estimation for d-dimensional quantum systems.
\newblock {\em Phys. Lett. A}, 253:249, 1999.

\bibitem{vedral97}
V.~Vedral, M.~B. Plenio, K.~Jacobs, and P.~L. Knight.
\newblock Statistical inference, distinguishability of quantum states, and
  quantum entanglement.
\newblock {\em Physical Review A}, 56:4452--4455, 1997.

\bibitem{vidal00}
G.~Vidal.
\newblock Entanglement monotones.
\newblock {\em J. Mod. Opt.}, 47:355, 2000.

\bibitem{Chitambar09}
E.~Chitambar and R.~Duan.
\newblock Nonlocal {E}ntanglement {T}ransformations {A}chievable by {S}eparable
  {O}perations.
\newblock {\em Phys. Rev. Lett.}, 103:110502, 2009.

\bibitem{bruss02}
D.~Bru{\ss{}}.
\newblock Characterizing entanglement.
\newblock {\em J. Math. Phys.}, 43:4237, 2002.

\bibitem{shor01}
P.~W. Shor, J.~Smolin, and B.~M. Terhal.
\newblock Nonadditivity of bipartite distillable entanglement follows from a
  conjecture on bound entangled werner states.
\newblock {\em Phys. Rev. Lett.}, 86:2681, 2001.

\bibitem{shor03}
P.~W. Shor, J.~A. Smolin, and A.~V. Thapliyal.
\newblock Superactivation of bound entanglement.
\newblock {\em Phys. Rev. Lett.}, 90:107901, 2003.

\bibitem{uhlmann00}
A.~Uhlmann.
\newblock Fidelity and concurrence of conjugated states.
\newblock {\em Phys. Rev. A}, 62:032307, 2000.

\bibitem{hayden01}
P.~M. Hayden and M.~Horodecki andB.~M. Terhal.
\newblock The assymptotic entenglement cost of preparing a quantum state.
\newblock {\em J. Phys. A: Math. Gen}, 34:6891, 2001.

\bibitem{Shor04}
P.~W. Shor.
\newblock Equivalence of additivity questions in quantum information theory.
\newblock {\em Commun. Math. Phys.}, 246:453 -- 472, 2004.

\bibitem{Matsumoto}
K.~Matsumoto.
\newblock On additivity questions.
\newblock In H.~Imai and M.~Hayashi, editors, {\em Quantum Computation and
  Information: From Theory to Experiments}, Topics in Applied Physics, pages
  133--166. Springer, Berlin, 2006.

\bibitem{Hastings09}
M.~B. Hastings.
\newblock Super additivity of communication channels using entangled inputs.
\newblock {\em Nat. Phys.}, 5:255, 2009.

\bibitem{hills97}
S.~Hills and W.~K. Wootters.
\newblock Entanglement of a pair of quantum bits.
\newblock {\em Phys. Rev. Lett.}, 78:5022, 1997.

\bibitem{wooters98}
W.~K. Wootters.
\newblock Entanglement of formation of an arbitrary state of two qubits.
\newblock {\em Phys. Rev. Lett.}, 80:2245, 1998.

\bibitem{rungta01}
P.~Rungta, V.~Buzek, C.~M. Caves, M.~Hillery, and G.~J. Milburn.
\newblock Universal state inversion and concurrence in arbitrary dimensions.
\newblock {\em Phys. Rev. A}, 64:042315, 2001.

\bibitem{meyer02}
D.~A. Meyer and N.~R. Wallach.
\newblock Global entanglement in multiparticle systems.
\newblock {\em J. Math. Phys. (N. Y.)}, 43:4273, 2002.

\bibitem{brennen03}
G.~K. Brennen.
\newblock Experimental realization of teleporting an unknown pure quantum state
  via dual classical and {E}instein-{P}odolsky-{R}osen channels.
\newblock {\em Quantum Inf. Comput.}, 3:619, 2003.

\bibitem{plenio97}
V.~Vedral, M.~B. Plenio, M.~A. Rippin, and P.~L. Knight.
\newblock Quantifying entanglement.
\newblock {\em Phys. Rev. Lett.}, 78(12):2275--2279, Mar 1997.

\bibitem{plenio}
V.~Vedral and M.~B. Plenio.
\newblock Entanglement measures and purification procedures.
\newblock {\em Phys. Rev. A}, 57(3):1619--1633, Mar 1998.

\bibitem{Wei03}
T.~C. Wei and P.~M. Goldbart.
\newblock Geometric measure of entanglement and applications to bipartite and
  multipartite quantum states.
\newblock {\em Phys. Rev. A}, 68:042307 -- 042319, 2003.

\bibitem{vidal02}
G.~Vidal and R.~F. Werner.
\newblock Computable measure of entanglement.
\newblock {\em Phys. Rev. A}, 65:032314, 2002.

\bibitem{ziczkowski98}
K.~Zyczkowski, P.~Horodecki, A.~Sanpera, and M.~Lewenstein.
\newblock Volume of the set of separable states.
\newblock {\em Physical Review A}, 58:883--892, 1998.

\bibitem{Eisert99}
J.~Eisert and M.~Plenio.
\newblock A comparison of entanglement measures.
\newblock {\em J. Mod. Opt.}, 46:145 -- 154, 1999.

\bibitem{Virmani00}
S.~Virmani and M.~Plenio.
\newblock Ordering states with entanglement measures.
\newblock {\em Phys. Lett. A}, 268:31 -- 34, 2000.

\bibitem{Verstraete01}
F.~Verstraete, K.~Audenaert, J.~Dehanene, and B.~De Moor.
\newblock A comparison of the entanglement measures negativity and concurrence.
\newblock {\em J. Phys. A: Math. Gen}, 34:10327, 2001.

\bibitem{Zykowski02}
K.~\. Zyczkowski and I.~Bengtsson.
\newblock Relativity of pure states entanglement.
\newblock {\em Ann. Phys. (N. Y.)}, 299:115 -- 135, 2002.

\bibitem{Miranowicz04}
A.~Miranowicz and A.~Grudka.
\newblock Ordering two-qubit states with concurrence and negativity.
\newblock {\em Phys. Rev. A}, 70:032326 -- 032330, 2004.

\bibitem{vidal99}
G.~Vidal and R.~Tarrach.
\newblock Robustness of entanglement.
\newblock {\em Phys. Rev. A}, 59:141, 1999.

\bibitem{audenaert}
K.~Audenaert, F.~Verstraete, T.~De Bie, and B.~De Moor.
\newblock Negativity and concurrence of mixed 2x2 states.

\bibitem{CHSH69}
J.~F. Clauser, M.~A. Horne, A.~Shimony, and R.~A. Holt.
\newblock Proposed experiment to test local hidden-variable theories.
\newblock {\em Phys. Rev. Lett.}, 23:880 -- 884, 1969.

\bibitem{CHSH70}
J.~F. Clauser, M.~A. Horne, A.~Shimony, and R.~A. Holt.
\newblock Proposed experiment to test local hidden variable theories.
\newblock {\em Phys. Rev. Lett.}, 24:549, 1970.

\bibitem{Cirelson80}
B.~S. Cirel{'}son.
\newblock Quantum generalizations of {B}ell's inequality.
\newblock {\em Lett. Math. Phys.}, 4:93 -- 100, 1980.

\bibitem{Gisin91}
N.~Gisin.
\newblock Bell's inequality holds for all non-product states.
\newblock {\em Phys. Lett. A}, 154:201 -- 202, 1991.

\bibitem{Gisin92}
N.~Gisin and A.~Peres.
\newblock Maximal violation of bell's inequality for arbitrarily large spin.
\newblock {\em Phys. Lett. A}, 162:15 -- 17, 1992.

\bibitem{Popescu92}
S.~Popescu and D.~Rohrlich.
\newblock Generic quantum nonlocality.
\newblock {\em Phys. Lett. A}, 166:293, 1992.

\bibitem{Carpasso73}
V.~Carpasso, D.~Fortunato, and F.~Selleri.
\newblock Sensitive observables of quantum mechanics.
\newblock {\em Int. J. Theor. Phys.}, 7:319 -- 326, 1973.

\bibitem{barret02}
J.~Barrett.
\newblock Nonsequential positive-operator-valued measurements on entangled
  mixed states do not always violate a {B}ell inequality.
\newblock {\em Phys. Rev. A}, 65:042302 -- 042306, 2002.

\bibitem{acin06}
A.~Ac{\'{i}}n, N.~Gisin, and B.~Toner.
\newblock Grothendieck's constant and local models for noisy entangled quantum
  states.
\newblock {\em Phys. Rev. A}, 73:062105 -- 062110, 2006.

\bibitem{Cavalcanti11}
D.~Cavalcanti, M.~L. Almeida, V.~Scarani, and A.~Ac\'in.
\newblock Quantum networks reveal quantum nonlocality.
\newblock {\em Nat. Commun.}, 2:184, 2011.

\bibitem{Horodecki95}
R.~Horodecki, P.~Horodecki, and M.~Horodecki.
\newblock Violating bell inequality by mixed spin 1/2 states: necessary and
  sufficient condition.
\newblock {\em Phys. Lett. A}, 200:340 -- 344, 1995.

\bibitem{Vertesi08}
T.~V{\'{e}}rtesi.
\newblock More efficient {B}ell inequalities for {W}erner states.
\newblock {\em Phys. Rev. A}, 78:032112 -- 032118, 2008.

\bibitem{Freedman72}
S.~J. Freedman and J.~F. Clauser.
\newblock Experimental test of local hidden-variable theories.
\newblock {\em Phys. Rev. Lett.}, 28:938, 1972.

\bibitem{aspect81}
A.~Aspect, P.~Grangier, and G.~Roger.
\newblock Experimental realization of {E}instein-{P}odolsky-{R}osen-bohm: a new
  violation of {B}ell{'}s inequalities.
\newblock {\em Phys. Rev. Lett.}, 49:91, 1982.

\bibitem{aspect82}
A.~Aspect, J.~Dalibard, and G.~Roger.
\newblock Experimental test of {B}ell{'}s inequalities using time-varying
  analyzers.
\newblock {\em Phys. Rev. Lett.}, 49:1804, 1982.

\bibitem{Tapster94}
P.~R. Tapster, J.~G. Rarity, and P.~C.~M. Owens.
\newblock Violation of {B}ell{'}s inequality over 4 km of optical fiber.
\newblock {\em Phys. Rev. Lett.}, 73:1923, 1994.

\bibitem{Weihs98}
G.~Weihs, T.~Jennewein, C.~Simon, H.~Weinfurter, and A.~Zeilinger.
\newblock Violation of {B}ell{'}s inequality under strict {E}instein locality
  condition.
\newblock {\em Phys. Rev. Lett.}, 81:5039, 1998.

\bibitem{tittel99}
W.~Tittel, J.~Brendel, H.~Zbinden, and N.~Gisin.
\newblock Violation of {B}ell inequalities by photons more than 10 km apart.
\newblock {\em Phys. Rev. Lett.}, 81:3563--3566, 1998.

\bibitem{Rowe01}
M.~A. Rowe, D.~Kielpinski, V.~Meyer, C.~A. Sackett, W.~M. Itano, C.~Monroe, and
  D.~J. Wineland.
\newblock Experimental violation of a {B}ell's inequality with efficient
  detections inequality with efficient detection.
\newblock {\em Nature}, 409:791, 2001.

\bibitem{Baas08}
D.~Salart, A.~Baas, C.~Branciard, N.~Gisin, and H.~Zbinden.
\newblock Testing the speed of 'spooky action at a distance'.
\newblock {\em Nature}, 454:861 -- 864, 2008.

\bibitem{Matsukevich08}
D.~Matsukevich, P.~Maunz, D.~Moehring, S.~Olmschenk, and C.~Monroe.
\newblock Bell {I}nequality {V}iolation with {T}wo {R}emote {A}tomic qubits.
\newblock {\em Phys. Rev. Lett.}, 100:150404 -- 150408, 2008.

\bibitem{Cinelli05}
C.~Cinelli, M.~Barbieri, R.~Perris, P.~Mataloni, and F.~De~Martini.
\newblock All-versus-nothing nonlocality test of quantum mechanics by
  two-photon hyperentanglement.
\newblock {\em Phys. Rev. Lett.}, 95:240405, 2005.

\bibitem{Yang05}
T.~Yang, Q.~Zhang, J.~Zhang, J.~Yin, Z.~Zhao, M.~\.{Z}ukowski, Z.-B. Chen, and
  J.-W. Pan.
\newblock All-versus-nothing violation of local realism by two-photon,
  four-dimensional entanglement.
\newblock {\em Phys. Rev. Lett.}, 95:240406, 2005.

\bibitem{Pomarico11}
E.~Pomarico, J.-D. Bancal, B.~Sanguinetti, A.~Rochdi, and N.~Gisin.
\newblock Various quantum nonlocality tests with a simple 2-photon entanglement
  source.
\newblock {\em Phys. Rev. A}, 83:052104, 2011.

\bibitem{Aolita11}
L.~Aolita, R.~Gallego, A.~Cabello, and A.~Ac\'{\i}n.
\newblock Fully nonlocal, monogamous, and random genuinely multipartite quantum
  correlations.
\newblock {\em Phys. Rev. Lett.}, 108:100401, Mar 2012.

\bibitem{raschenbeutel00}
A.~Raschenbeutel, G.~Nogues, S.~Osnaghi, P.~Bertet, M.~Brune, and S.~Haroche
  J.~M.~Raimond.
\newblock Step-by-step engeneered multiparticle entanglement.
\newblock {\em Science}, 288:2024, 2000.

\bibitem{Bowmeester99}
D.~Bouwmeester, J.~W. Pan, M.~Daniell, H.~Weinfurter, and A.~Zeilinger.
\newblock Observation of three-photon greenberger-horn-zeilinger entanglement.
\newblock {\em Phys. Rev. Lett.}, 82:1345, 1999.

\bibitem{zhao04}
Z.~Zhao, Y.~A. Chen, A.~N. Zhang, T.~Yang, H.~J. Briegel, and J.~W. Pan.
\newblock Experimental demonstration of five-photon entanglement and
  open-destination teleportation.
\newblock {\em Nature}, 430:54, 2004.

\bibitem{Chen06b}
Yu-Ao Chen, Tao Yang, An-Ning Zhang, Zhi Zhao, Ad\'an Cabello, and Jian-Wei
  Pan.
\newblock Experimental violation of bell's inequality beyond tsirelson's bound.
\newblock {\em Phys. Rev. Lett.}, 97:170408, Oct 2006.

\bibitem{Lavoie09}
J.~Lavoie, R.~Kaltenbaek, and K.~J. Resch.
\newblock Experimental violation of {S}vetlichny's inequality.
\newblock {\em New J. Phys.}, 11:073051 -- 073062, 2009.

\bibitem{Mermin90}
N.~D. Mermin.
\newblock Extreme quantum entanglement in a superposition of macroscopically
  distinct states.
\newblock {\em Phys. Rev. Lett.}, 65:1838 -- 1840, 1990.

\bibitem{Svetlichny87}
G.~Svetlichny.
\newblock Distinguishing three-body from two-body nonseparability by a
  {B}ell-type inequality.
\newblock {\em Phys. Rev. D}, 35:3066 -- 3069, 1987.

\bibitem{Ansmann09}
M.~Ansmann, H.~Wang, R.~C. Bialczak, M.~Hofheinz, E.~Lucero, M.~Neeley, A.~D.
  O'Connell, D.~Sank, M.~Weides, J.~Wenner, A.~N. Cleland, and J.~M. Martinis.
\newblock Violation of {B}ell's inequality in {J}osephson phase qubits.
\newblock {\em Nature}, 461:504, 2009.

\bibitem{Hofmann12}
J.~Hofmann, M.~Krug, N.~Ortegel, L.~G{\'e}rard, M.~Weber, W.~Rosenfeld, and
  H.~Weinfurter.
\newblock Heralded entanglement between widely separated atoms.
\newblock {\em Science}, 337:72, 2012.

\bibitem{eberhard93}
P.~H. Eberhard.
\newblock Background level and counter efficiencies requires for a
  loophole-free {E}instein-{P}odolsky-{R}osen experiment.
\newblock {\em Phys. Rev. A}, 47:747--750, 1993.

\bibitem{zeilinger2013}
M.~Giustina, A.~Mech, S.~Ramelow, B.~Wittmann, J.~Kofler, J.~Beyer, A.~Lita,
  B.~Calkins, T.~Gerrits, S.~W. Nam, R.~Ursin, and A.~Zeilinger.
\newblock Bell violation using entangled photons without the fair-sampling
  assumption.
\newblock {\em Nature}, 497:227--230, 2013.

\bibitem{kwiat2013}
B.~G. Christensen, K.~T. McCusker, J.~B. Altepeter, B.~Calkins, T.~Gerrits,
  A.~E. Lita, A.~Miller, L.~K. Shalm, Y.~Zhang, S.~W. Nam, N.~Brunner, C.~C.~W.
  Lim, N.~Gisin, and P.~G. Kwiat.
\newblock Detection-loophole-free test of quantum nonlocality, and
  applications.
\newblock {\em Phys. Rev. Lett.}, 111:130406, Sep 2013.

\bibitem{Brunner13}
N.~Brunner, D.~Cavalcanti, S.~Pironio, V.~Scarani, and S.~Wehner.
\newblock Bell non-locality.
\newblock {\em arXiv:1303.2849}, 2013.

\bibitem{Guehne02}
O.~G{\"{u}}hne, P.~Hyllus, D.~Bru{\ss{}}, A.~Ekert, M.~Lewenstein,
  C.~Macchiavello, and A.~Sanpera.
\newblock Detection of entanglement with few local measurements.
\newblock {\em Phys. Rev. A}, 66:062305, 2002.

\bibitem{Guehne03}
O.~G{\"{u}}hne, P.~Hyllus, D.~Bru{\ss{}}, A.~Ekert, M.~Lewenstein,
  C.~Macchiavello, and A.~Sanpera.
\newblock Experimental detection of entanglement via witness operators and
  local measurements.
\newblock {\em J. Mod. Opt.}, 50 (6-7):1079--1102, 2003.

\bibitem{Barbieri03}
M.~Barbieri, F.~De Martini, G.~Di Nepi, P.~Mataloni, G.~M D'Ariano, and
  C.~Macchiavello.
\newblock Detection of {E}ntanglement with {P}olarized {P}hotons:
  {E}xperimental {R}ealization of an {E}ntanglement {W}itness.
\newblock {\em Phys. Rev. Lett.}, 91:227901, 2003.

\bibitem{kiesel07}
N.~Kiesel, C.~Schmid, G.~T{\'{o}}th, E.~Solano, and H.~Weinfurter.
\newblock Experimental observation of four-photon entangled {D}icke state with
  high fidelity.
\newblock {\em Phys. Rev. Lett.}, 98:063604 -- 063608, 2007.

\bibitem{Wieczorek09}
W.~Wieczorek, R.~Krischek, N.~Kiesel, P.~Michelberger, G.~Toth, and
  H.~Weinfurter.
\newblock Experimental entanglement of a six-photon symmetric {D}icke state.
\newblock {\em Phys. Rev. Lett.}, 103:020504, 2009.

\bibitem{Radmark09}
M.~R{\aa}dmark, M.~\.Zukowski, and M.~Bourennane.
\newblock Experimental high fidelity six-photon entangled state for telecloning
  protocols.
\newblock {\em New J. Phys.}, 11:103016, 2009.

\bibitem{bovino05}
F.~A. Bovino, G.~Castagnol, A.~Ekert, P.~Horodecki, C.~M. Alves, and A.~V.
  Sergienko.
\newblock Direct measurement of nonlinear properties of bipartite quantum
  states.
\newblock {\em Phys. Rev. Lett.}, 95:240407 -- 240411, 2005.

\bibitem{roos04a}
C.~F. Roos, M.~Riebe, H.~H{\"{a}}ffner, W.~H{\"{a}}nsel, J.~B. Elm, G.~P.~T.
  Lancaster, C.~Becher, F.~Schmidt-Kaler, and R.~Blatt.
\newblock Control and measurement of three-qubit entangled states.
\newblock {\em Science}, 304:1478, 2004.

\bibitem{Leonhardt95}
U.~Leonhardt.
\newblock Quantum-{S}tate {T}omography and {D}iscrete {W}igner {F}unction.
\newblock {\em Phys. Rev. Lett.}, 74:4101, 1995.

\bibitem{white99}
A.~G. White, D.~F.~V. James, P.~H. Eberhard, and P.~G. Kwiat.
\newblock Nonmaximally entangled states: Production, characterization, and
  utilization.
\newblock {\em Phys. Rev. Lett.}, 83:3103, 1999.

\bibitem{roos04b}
C.~F. Roos, G.~P.~T. Lancaster, M.~Riebe, H.~H{\"{a}}ffner, W.~H{\"{a}}nsel,
  S.~Gulde, C.~Becher, J.~Eschner, F.~Schmidt-Kaler, and R.~Blatt.
\newblock Bell states of atoms with ultralong lifetimes and their tomographic
  state analysis.
\newblock {\em Phys. Rev. Lett.}, 92:220402, 2004.

\bibitem{Smithey93}
D.~T. Smithey, M.~Beck, M.~G. Raymer, and A.~Faridani.
\newblock Measurement of the {W}igner distribution and the density matrix of a
  light mode using optical homodyne tomography: {A}pplication to squeezed
  states and the vacuum.
\newblock {\em Phys. Rev. Lett.}, 70:1244, 1993.

\bibitem{Leibfried96}
D.~Leibfried, D.~M. Meekhof, B.~E. King, C.~Monroe, W.~M. Itano, and D.~J.
  Wineland.
\newblock Experimental {D}etermination of the {M}otional {Q}uantum {S}tate of a
  {T}rapped {A}tom.
\newblock {\em Phys. Rev. Lett.}, 77:4281, 1996.

\bibitem{Lutterbach97}
L.~G. Lutterbach and L.~Davidovich.
\newblock Method for {D}irect {M}easurement of the {W}igner {F}unction in
  {C}avity {QED} and {I}on {T}raps.
\newblock {\em Phys. Rev. Lett.}, 78:2547, 1997.

\bibitem{deMelo06}
F.~de~Melo, L.~Aolita, F.~Toscano, and L.~Davidovich.
\newblock Direct measurement of the quantum state of the electromagnetic field
  in a superconducting transmission line.
\newblock {\em Phys. Rev. A}, 73:030303, 2006.

\bibitem{Bendersky08}
A.~Bendersky, F.~Pastawski, and J.~P. Paz.
\newblock Selective and {E}fficient {E}stimation of {P}arameters for {Q}uantum
  {P}rocess {T}omography.
\newblock {\em Phys. Rev. Lett.}, 190403:190403, 2008.

\bibitem{Schmiegelow10}
C.~T. Schmiegelow, M.~A. Larotonda, , and J.~P. Paz.
\newblock Selective and {E}fficient {Q}uantum {P}rocess {T}omography with
  {S}ingle {P}hotons.
\newblock {\em Phys. Rev. Lett.}, 104:123601, 2010.

\bibitem{Gross10}
D.~Gross, Y.-K. Liu, S.~T. Flammia, S.~Becker, and J.~Eisert.
\newblock Quantum {S}tate {T}omography via {C}ompressed {S}ensing.
\newblock {\em Phys. Rev. Lett.}, 105:150401, 2010.

\bibitem{brun04}
T.~A. Brun.
\newblock Measuring polynomial functions of states.
\newblock {\em Quantum Information and Computation}, 4:401, 2004.

\bibitem{mintert05a}
F.~Mintert, M.~Ku{\'{s}}, and A.~Buchleitner.
\newblock Concurrence of mixed multipartite quantum states.
\newblock {\em Phys. Rev. Lett.}, 95:260502, 2005.

\bibitem{aolita06concu}
L.~Aolita and F.~Mintert.
\newblock Measuring multipartite concurrence with a single factorizable
  observable.
\newblock {\em Phys. Rev. Lett.}, 97:050501, 2006.

\bibitem{wootters82}
W.~K. Wootters and W.~H. Zurek.
\newblock A single quantum cannot be cloned.
\newblock {\em Nature}, 299:802--803, 1982.

\bibitem{Dieks82}
D.~Dieks.
\newblock Communication by {E}{P}{R} devices.
\newblock {\em Phys. Lett. A}, 92:6, 1982.

\bibitem{walborn06}
S.~P. Walborn, P.~H. Souto, L.~Davidovich, F.~Mintert, and A.~Buchleitner.
\newblock Experimental determination of entanglement with a single measurement.
\newblock {\em Nature}, 440:1022, 2006.

\bibitem{mintert07b}
F.~Mintert and A.~Buchleitner.
\newblock Observable entanglement measure for mixed quantum states.
\newblock {\em Phys. Rev. Lett.}, 98:140505, 2007.

\bibitem{aolita07concu}
L.~Aolita, F.~Mintert, and A.~Buchleitner.
\newblock Scalable method to estimate experimentally the entanglement of
  multipartite systems.
\newblock {\em Phys. Rev. A}, 78:022308, 2008.

\bibitem{Sun07}
F.~W. Sun, J.~M. Cai, J.~S. Xu, G.~Chen, B.~H. Liu, C.~F. Li, Z.~W. Zhou, and
  G.~C. Guo.
\newblock Experimental measurement of multidimensional entanglement via
  equivalent symmetric projection.
\newblock {\em Phys. Rev. A}, 76:052303, 2007.

\bibitem{Zhang08}
C.-J. Zhang, Y.-Xiao G., Y.-S. Zhang, and G.-C. Guo.
\newblock Observable estimation of entanglement for arbitrary
  finite-dimensional mixed states.
\newblock {\em Phys. Rev. A}, 78:042308, 2008.

\bibitem{Bendersky09}
A.~Bendersky, J.~P. Paz, and M.~T. Cunha.
\newblock General theory of measurement with two copies of a quantum state.
\newblock {\em Phys. Rev. Lett.}, 103:040404, 2009.

\bibitem{horodecki03}
P.~Horodecki.
\newblock Measuring quantum entanglement without prior state reconstruction.
\newblock {\em Phys. Rev. Lett.}, 90:167901, 2003.

\bibitem{schlosshauer:1267}
M.~Schlosshauer.
\newblock Decoherence, the measurement problem, and interpretations of quantum
  mechanics.
\newblock {\em Rev. Mod. Phys.}, 76(4):1267, 2004.

\bibitem{kraus83}
K.~Kraus.
\newblock {\em States, Effects, And Operations - Fundamental Notions Of
  Quantum-Theory}, volume 190.
\newblock Springer-Verlag, 1983.

\bibitem{preskill98}
J.~Preskill.
\newblock Lecture notes: Caltech physics 219/computer science 219 quantum
  computation, 1998.
\newblock http://www.theory.caltech.edu/people/preskill/ph229.

\bibitem{jamiol}
A.~Jamio{\l}kowsky.
\newblock Linear transformations which preserve trace and positive
  semidefiniteness of operators.
\newblock {\em Rep. Math. Phys.}, 3(4):275, 1972.

\bibitem{yao}
W.~Yao, R.-B. Liu, and L.~J. Sham.
\newblock Restoring coherence lost to a slow interacting mesoscopic spin bath.
\newblock {\em Phys. Rev. Lett.}, 98:077602, 2007.

\bibitem{ebc}
M.~Horodecki, P.~W. Shor, and M.~B. Ruskai.
\newblock General entanglement breaking channels.
\newblock {\em Rev. Math. Phys.}, 15:629, 2003.

\bibitem{Holevo99}
A.~S. Holevo.
\newblock {C}oding {T}heorems for {Q}uantum {C}hannels.
\newblock {\em Russian Math. Surveys}, 53:1295, 1999.

\bibitem{Davies76}
E.~B. Davies.
\newblock {\em {Q}uantum {T}heory of {O}pen {S}ystems}.
\newblock Academic Press. London, 1976.

\bibitem{cohenAP}
C.~Cohen-Tannoudji, J.~Dupont-Roc, and G.~Grynberg.
\newblock {\em Atom{--}-Photon Interactions: Basic Processes and Applications}.
\newblock Wiley-Interscience, 1992.

\bibitem{carmichael}
H.~Carmichael.
\newblock {\em An Open Systems Approach to Quantum Optics}.
\newblock Springer, 1993.

\bibitem{Gorini76}
V.~Gorini, A.~Kossakowski, and E.~C.~G. Sudarshan.
\newblock Completely positive dynamical semigroups of $n$-level systems.
\newblock {\em J. Math. Phys.}, 17:821, 1976.

\bibitem{Lindblad76}
G.~Lindblad.
\newblock On the generator of quantum dynamical semigroups.
\newblock {\em Commun. Math. Phys.}, 48:119, 1976.

\bibitem{jcm1}
E.~T. Jaynes and F.~W. Cummings.
\newblock Comparison of quantum and semiclassical radiation theories with
  application to the beam maser.
\newblock {\em Proc. IEEE}, 51:89, 1963.

\bibitem{jcm2}
F.~W. Cummings.
\newblock Stimulated emission of radiation in a single mode.
\newblock {\em Phys. Rev.}, 140:1051, 1965.

\bibitem{weisskopf}
V.~Weisskopf and E.~Wigner.
\newblock Berechnung der nat{\"{u}}rlichen linienbreite auf grund der
  diracschen lichttheorie.
\newblock {\em Z. f{\"{u}}r Phys.}, 63:54--73, 1930.

\bibitem{xxy}
X.X. Yi and C.P. Sun.
\newblock Factoring the unitary evolution operator and quantifying
  entanglement.
\newblock {\em Phys. Lett. A}, 262:287, 1999.

\bibitem{braun02}
D.~Braun.
\newblock Creation of entanglement by interaction with a common heat bath.
\newblock {\em Phys. Rev. Lett.}, 89:277901, Dec 2002.

\bibitem{benatti03}
F.~Benatti, R.~Floreanini, and M.~Piani.
\newblock Environment induced entanglement in markovian dissipative dynamics.
\newblock {\em Phys. Rev. Lett.}, 91:070402, Aug 2003.

\bibitem{oh06}
S.~Oh and J.~Kim.
\newblock Entanglement between qubits induced by a common environment with a
  gap.
\newblock {\em Phys. Rev. A}, 73:062306, Jun 2006.

\bibitem{das09}
S.~Das and G.~S. Agarwal.
\newblock Bright and dark periods in entanglement dynamics of interacting
  qubits in contact with the environment.
\newblock {\em J. Phys. B: Atom. Mol. Opt. Phys.}, 42:141003, 2009.

\bibitem{haroche}
M.~Gross and S.~Haroche.
\newblock Superradiance: An essay on the theory of collective spontaneous
  emission.
\newblock {\em Phys. Rep.}, 93:301--396, 1982.

\bibitem{dicke}
R.~H. Dicke.
\newblock Coherence in spontaneous radiation processes.
\newblock {\em Phys. Rev.}, 93:99--110, 1954.

\bibitem{isotropic}
M.~Horodecki and P.~Horodecki.
\newblock Reduction criterion of separability and limits for a class of
  distillation protocols.
\newblock {\em Phys. Rev. A}, 59(6):4206--4216, Jun 1999.

\bibitem{EoFIsoTerhal}
B.~M. Terhal and K.~G.~H. Vollbrecht.
\newblock Entanglement of formation for isotropic states.
\newblock {\em Phys. Rev. Lett.}, 85(12):2625--2628, Sep 2000.

\bibitem{chen06c}
K.~Chen, S.~Albeverio, and S-M. Fei.
\newblock Concurrence-based entanglement measure for {W}erner states.
\newblock {\em Rep. Math. Phys.}, 58:325, 2006.

\bibitem{Rungta03}
P.~Rungta and C.~M. Caves.
\newblock Concurrence-based entanglement measures for isotropic states.
\newblock {\em Phys. Rev. A}, 67(1):012307, Jan 2003.

\bibitem{Ann07-1}
K.~Ann and G.~Jaeger.
\newblock Local-dephasing-induced entanglement sudden death in two-component
  finite-dimensional systems.
\newblock {\em Phys. Rev. A}, 76(4):044101, Oct 2007.

\bibitem{paris02}
M.~G.~A. Paris.
\newblock Evolution of a twin beam in active optical media.
\newblock {\em J. Opt. B}, 4(6):442, 2002.

\bibitem{jakub04}
J.~S. Prauzner-Bechcicki.
\newblock Two-mode squeezed vacuum state coupled to the common thermal
  reservoir.
\newblock {\em J. Phys. A: Math. Gen.}, 37:L173, 2004.

\bibitem{Ban06}
M.~Ban.
\newblock Decoherence of continuous variable quantum information in
  non-markovian channels.
\newblock {\em J. Phys. A}, 39(8):1927, 2006.

\bibitem{An007}
J-H. An and W-M. Zhang.
\newblock Non-markovian entanglement dynamics of noisy continuous-variable
  quantum channels.
\newblock {\em Phys. Rev. A}, 76:042127, Oct 2007.

\bibitem{hammer08}
C.~H\"orhammer and H.~B\"uttner.
\newblock Environment-induced two-mode entanglement in quantum brownian motion.
\newblock {\em Phys. Rev. A}, 77:042305, Apr 2008.

\bibitem{Hu92}
B.~L. Hu, J.~P. Paz, and Y.~Zhang.
\newblock Quantum brownian motion in a general environment: {E}xact master
  equation with nonlocal dissipation and colored noise.
\newblock {\em Phys. Rev. D}, 45:2843--2861, Apr 1992.

\bibitem{Martinez13}
E.~A. Martinez and J.~P. Paz.
\newblock Dynamics and {T}hermodynamics of {L}inear {Q}uantum {O}pen {S}ystems.
\newblock {\em Phys. Rev. Lett.}, 110:130406, 2013.

\bibitem{florian09}
F.~Mintert.
\newblock Robust entangled states.
\newblock {\em J. Phys. A: Math. Theor.}, 43(24):245303, 2010.

\bibitem{Froewis11}
F.~Fr\"{o}wis and W.~D\"{u}r.
\newblock Stable macroscopic quantum superpositions.
\newblock {\em Phys. Rev.. Lett.}, 106:110402, 2011.

\bibitem{Froewis12}
F.~Fr\"{o}wis and W.~D\"{u}r.
\newblock Stability of encoded macroscopic quantum superpositions.
\newblock {\em Phys. Rev. A}, 85:052329, 2012.

\bibitem{chaves12b}
R.~Chaves, L.~Aolita, and A.~Ac\'in.
\newblock Robust macroscopic quantum correlations without complex encodings.
\newblock {\em Phys. Rev. A}, 86:020301, 2012.

\bibitem{Ali13}
M.~Ali and O.~G{\"{u}}hne.
\newblock Robustness of multiparticle entanglement: specific entanglement
  classes and random states.
\newblock {\em arXiv:1310.7336}, 2013.

\bibitem{wiseman:548}
H.~M. Wiseman and G.~J. Milburn.
\newblock Quantum theory of optical feedback via homodyne detection.
\newblock {\em Phys. Rev. Lett.}, 70(5):548--551, Feb 1993.

\bibitem{doherty:2700}
A.~C. Doherty and K.~Jacobs.
\newblock Feedback control of quantum systems using continuous state
  estimation.
\newblock {\em Phys. Rev. A}, 60(4):2700--2711, Oct 1999.

\bibitem{mancini:010304}
Stefano Mancini.
\newblock Markovian feedback to control continuous-variable entanglement.
\newblock {\em Phys. Rev. A}, 73(1):010304, Jan 2006.

\bibitem{kwiat01}
P.~G. Kwiat, S.~Barraza-Lopez, A.~Stefanov, and N.~Gisin.
\newblock Experimental entanglement distillation and `hidden' non-locality.
\newblock {\em Nature}, 409:1014--1017, 2001.

\bibitem{felix09}
F.~Platzer, F.~Mintert, and A.~Buchleitner.
\newblock Optimal dynamical control of many-body entanglement.
\newblock {\em Phys. Rev. Lett.}, 105:020501, Jul 2010.

\bibitem{sun10}
Q.~Sun, M.~Al-Amri, L.~Davidovich, and M.~S. Zubairy.
\newblock Reversing entanglement change by a weak measurement.
\newblock {\em Phys. Rev. A}, 82:052323, 2010.

\bibitem{kim11}
Y.-S. Kim, J.-C. Lee, O.~Kwon, and Y.-H. Kim.
\newblock Protecting entanglement from decoherence using weak measurement and
  quantum measurement reversal.
\newblock {\em Nat. Phys.}, 8:117, 2011.

\bibitem{viola98}
L.~Viola and S.~Lloyd.
\newblock Dynamical suppression of decoherence in two-state quantum systems.
\newblock {\em Phys. Rev. A}, 58:2733--2744, Oct 1998.

\bibitem{viola99}
L.~Viola, E.~Knill, and S.~Lloyd.
\newblock Dynamical decoupling of open quantum systems.
\newblock {\em Phys. Rev. Lett.}, 82:2417--2421, Mar 1999.

\bibitem{rabitz00}
H.~Rabitz, R.~de~Vivie-Riedle, M.~Motzkus, and K.~Kompa.
\newblock Whither the future of controlling quantum phenomena?
\newblock {\em Science}, 288(5467):824--828, 2000.

\bibitem{dong10}
D.~Daoyi and I.~R. Petersen.
\newblock Quantum control theory and applications: A survey.
\newblock {\em IET Control Theory Appl.}, 4(12):2651 -- 2671, 2010.

\bibitem{trail10}
C.~M. Trail, P.~S. Jessen, and I.~H. Deutsch.
\newblock Strongly enhanced spin squeezing via quantum control.
\newblock {\em Phys. Rev. Lett.}, 105:193602, Nov 2010.

\bibitem{lee13}
J.~H. Lee, E.~Montano, I.~H. Deutsch, and P.~S. Jessen.
\newblock Robust site-resolvable quantum gates in an optical lattice via
  inhomogeneous control.
\newblock {\em Nat. Commun.}, 4(2027), 2013.

\bibitem{konrad}
T.~Konrad, F.~{de Melo}, M.~Tiersch, C.~Kasztelan, A.~Aragao, and
  A.~Buchleitner.
\newblock A factorization law for entanglement decay.
\newblock {\em Nat. Phys.}, 4:99, 2008.

\bibitem{tiersch09}
M.~Tiersch, F.~{de Melo}, T.~Konrad, and A.~Buchleitner.
\newblock Equation of motion for entanglement.
\newblock {\em Quant. Inf. Proc.}, 8:523, 2009.

\bibitem{oswaldo09}
O.~J. Far{\'{i}}as, C.~L. Latune, S.~P. Walborn, Davidovich, and P.~H.~S.
  Ribeiro.
\newblock Determining the dynamics of entanglement.
\newblock {\em Science}, 324:1414, 2009.

\bibitem{gour05}
G.~Gour.
\newblock Family of concurrence monotones and its applications.
\newblock {\em Phys. Rev. A}, 71(1):012318, Jan 2005.

\bibitem{gour05b}
G.~Gour.
\newblock Mixed-state entanglement of assistance and the generalized
  concurrence.
\newblock {\em Phys. Rev. A}, 72(4):042318, Oct 2005.

\bibitem{gour10}
G.~Gour.
\newblock Evolution and symmetry of multipartite entanglement.
\newblock {\em Phys. Rev. Lett.}, 105:190504, Nov 2010.

\bibitem{gour12}
V.~Gheorghiu and G.~Gour.
\newblock Multipartite entanglement evolution under separable operations.
\newblock {\em Phys. Rev. A}, 86:050302, Nov 2012.

\bibitem{tiersch08}
M.~Tiersch, F.~de~Melo, and A.~Buchleitner.
\newblock Entanglement evolution in finite dimensions.
\newblock {\em Phys. Rev. Lett.}, 101(17):170502, Oct 2008.

\bibitem{VolumeSepState2}
K.~\ifmmode~\dot{Z}\else \.{Z}\fi{}yczkowski.
\newblock Volume of the set of separable states. {II}.
\newblock {\em Phys. Rev. A}, 60(5):3496--3507, Nov 1999.

\bibitem{Yu07}
T.~Yu and J.~H. Eberly.
\newblock Many-body separability of warm qubits.
\newblock Eprint: arXiv: 0707.3215.

\bibitem{hayden}
P.~Hayden, D.~W. Leung, and A.~Winter.
\newblock Aspects of generic entanglement.
\newblock {\em Comm. Math. Phys.}, 265:95, 2006.

\bibitem{MarkusThesis}
M.~Tiersch.
\newblock {\em Benchmarks and Statistics of Entanglement Dynamics}.
\newblock PhD thesis, Freiburg University, 2009.

\bibitem{ledoux}
M.~Ledoux.
\newblock {\em The Concentration of Measure Phenomenon}, volume~89 of {\em
  Mathematical Surveys and Monographs}.
\newblock AMS, 2001.

\bibitem{LiLi2012}
X.~Li and D.~Li.
\newblock Classification of general n-qubit states under stochastic local
  operations and classical communication in terms of the rank of coefficient
  matrix.
\newblock {\em Phys. Rev. Lett.}, 108:180502, 2012.

\bibitem{bose98}
S.~Bose, B.~Vedral, and P.~L. Knight.
\newblock Multiparticle generalization of entanglement swapping.
\newblock {\em Phys. Rev. A}, 57:822, 1998.

\bibitem{hillery99}
M.~Hillery, V.~Bu{\v{z}}ek, and A.~Berthiaume.
\newblock Quantum secret sharing.
\newblock {\em Phys. Rev. A}, 59:1829, 1999.

\bibitem{dhondt05}
E.~D'Hondt and P.~Panangaden.
\newblock The {C}omputational {P}ower of the {W} and {GHZ} states.
\newblock {\em Quantum Information \& Computation}, 6:173--183, 2006.

\bibitem{duer99}
W.~D\"{u}r, J.~I. Cirac, and R.~Tarrach.
\newblock Separability and distillability of multiparticle quantum systems.
\newblock {\em Physical Review Letters}, 83:3562--3565, 1999.

\bibitem{Dur01}
J.~I.~Cirac W.~D\"{u}r.
\newblock Multiparticle entanglement and its experimental detection.
\newblock {\em J. Phys. A}, 34:6837, 2001.

\bibitem{Hein04}
M.~Hein, J.~Eisert, and H.~J. Briegel.
\newblock Multiparty entanglement in graph states.
\newblock {\em Phys. Rev. A}, 69:062311, 2004.

\bibitem{Hein_Review}
M.~Hein, W.~D{\"{u}}r, J.~Eisert, R.~Raussendorf, M.~Van~den Nest, and H.~J.
  Briegel.
\newblock {E}ntanglement in {G}raph {S}tates and its {A}pplications.
\newblock In {\em {P}roceedings of the {I}nternational {S}chool of {P}hysics
  ``{E}nrico {F}ermi'' on {Q}uantum {C}omputers, {A}lgorithms and {C}haos},
  2006.

\bibitem{Raussendorf01}
R.~Raussendorf and H.~J. Briegel.
\newblock A one-way quantum computer.
\newblock {\em Phys. Rev. Lett.}, 86:5188 -- 5191, 2001.

\bibitem{Briegel09}
H.~J. Briegel, D.~E. Browne, W.~D\"{u}r, R.~Raussendorf, and M.~Van den Nest.
\newblock Measurement-based quantum computation.
\newblock {\em Nat. Phys.}, 5:19, 2009.

\bibitem{Schlingemann01}
D.~Schlingemann and R.~F. Werner.
\newblock Quantum error-correcting codes associated with graphs.
\newblock {\em Phys. Rev. A}, 65:012308 -- 012316, 2001.

\bibitem{dur05}
W.~D{\"{u}}r, J.~Calsamiglia, and H.~J. Briegel.
\newblock Multipartite secure state distribution.
\newblock {\em Phys. Rev. A}, 71:042336 -- 042344, 2005.

\bibitem{chen07}
K.~Chen and H.~K. Lo.
\newblock Multi-partite quantum cryptographic protocols with noisy {GHZ}
  states.
\newblock {\em Quant. Inf. and Comp.}, 7(8):689 -- 715, 2007.

\bibitem{Dahlsten06}
O.~C. Dahlsten and M.~B. Plenio.
\newblock Entanglement probability distribution of bi-partite randomised
  stabilizer states.
\newblock {\em Quant. Inf. Comp.}, 6:527 -- 538, 2006.

\bibitem{Kiesel05}
N.~Kiesel, C.~Schmid, U.~Weber, G.~T{\'{o}}th, O.~G{\"{u}}hne, R.~Ursin, and
  H.~Weinfurter.
\newblock Experimental analysis of a four-qubit photon cluster state.
\newblock {\em Phys. Rev. Lett.}, 95:210502--210506, 2005.

\bibitem{Chen07b}
K.~Chen, C-M. Li, Q.~Zhang, Y-A. Chen, A.~Goebel, S.~Chen, A.~Mair, and J-W.
  Pan.
\newblock Experimental realization of one-way quantum computing with two-photon
  four-qubit cluster states.
\newblock {\em Phys. Rev. Lett.}, 99:120503 -- 120507, 2007.

\bibitem{Vallone08}
G.~Vallone, E.~Pomarico, F.~De Martini, and P.~Mataloni.
\newblock Active one-way quantum computation with two-photon four-qubit cluster
  states.
\newblock {\em Phys. Rev. Lett.}, 100:160502 -- 160506, 2008.

\bibitem{Duer03}
W.~D{\"{u}}r, H.~Aschauer, and Briegel H.~J.
\newblock Multiparticle entanglement purification for graph states.
\newblock {\em Phys. Rev. Lett.}, 91:107903, 2003.

\bibitem{Aschauer05}
H.~Aschauer, W.~D{\"{u}}r, and H.~J. Briegel.
\newblock Multiparticle entanglement purification for two-colorable graph
  states.
\newblock {\em Phys. Rev. A}, 71:012319, 2005.

\bibitem{chirag10}
L.~Aolita, D.~Cavalcanti, R.~Chaves, C.~Dhara, L.~Davidovich, and A.~Ac\'{\i}n.
\newblock Noisy evolution of graph-state entanglement.
\newblock {\em Phys. Rev. A}, 82:032317, 2010.

\bibitem{Raussendorf05}
R.~Raussendorf, S.~Bravyi, and J.~Harrington.
\newblock Long-range quantum entanglement in noisy cluster states.
\newblock {\em Phys. Rev. A 71}, 71:062313, 2005.

\bibitem{kay06}
A.~Kay, J.~Pachos, W.~D\"ur, and H.~Briegel.
\newblock Optimal purification of thermal graph states.
\newblock {\em New J. Phys.}, 8:147, 2006.

\bibitem{SenDe03}
A.~Sen(De), U.~Sen, M.~Wie{\'{}}sniak{\'{}}sand, D.~Kaszlikowski, and M.~\.
  Zukowski.
\newblock Multiqubit {W} states lead to stronger nonclassicality than
  greenberger-horne-zeilinger states.
\newblock {\em Phys. Rev. A}, 68:062306, 2003.

\bibitem{Dicke54}
R.~H. Dicke.
\newblock Coherence in spontaneous radiation processes.
\newblock {\em Phys. Rev.}, 93:99 -- 110, 1954.

\bibitem{Kitagawa93}
M.~Kitagawa and M.~Ueda.
\newblock Squeezed spin states.
\newblock {\em Phys. Rev. A}, 47:5138 -- 5143, 1993.

\bibitem{Sorensen01}
A.~S{\o}rensen, L.~M. Duan, J.~I. Cirac, and P.~Zoller.
\newblock Many-particle entanglement with bose--einstein condensates.
\newblock {\em Nature}, 409:63 -- 66, 2001.

\bibitem{ba_an09}
N.~Ba~An and J.~Kim.
\newblock Finite-time and infinite-time disentanglement of multipartite
  {G}reenberger-{H}orne-{Z}eilinger-type states under the collective action of
  different types of noise.
\newblock {\em Phys. Rev. A}, 79:022303, 2009.

\bibitem{Schack00}
R.~Schack and C.~M. Caves.
\newblock Explicit product ensembles for separable quantum states.
\newblock {\em J. Mod. Optics}, 47:387, 2000.

\bibitem{Guehne08}
O.~G{\"{u}}hne, F.~Bodoky, and M.~Blaauboer.
\newblock Multiparticle entanglement under the influence of decoherence.
\newblock {\em Phys. Rev. A}, 78:060301, 2008.

\bibitem{Bandyopadhyay04}
S.~Bandyopadhyay and D.~A. Lidar.
\newblock Robustness of multi-qubit entanglement in the independent decoherence
  model.
\newblock {\em Phys. Rev. A}, 72:042339 -- 042345, 2005.

\bibitem{chaves12}
R.~Chaves, D.~Cavalcanti, L.~Aolita, and A.~Ac\'in.
\newblock Multipartite quantum non-locality under local decoherence.
\newblock {\em Phys. Rev. A}, 86:012108, 2012.

\bibitem{borras09}
A.~Borras, A.~P. Majtey, A.~R. Plastino, M.~Casas, and A.~Plastino.
\newblock Robustness of {H}ighly {E}ntangled {M}ulti-{Q}ubit {S}tates {U}nder
  {D}ecoherence.
\newblock {\em Phys. Rev. A}, 79:022108, 2009.

\bibitem{chaves12c}
R.~Chaves, J.~B. Brask, M.~Markiewicz, J.~Kolodynski, and A.~Ac\'{i}n.
\newblock Noisy metrology beyond the standard quantum limit.
\newblock {\em Phys. Rev. Lett.}, 111:120401, 2013.

\bibitem{nielsen00}
M.~A. Nielsen and I.~L. Chuang.
\newblock {\em Quantum Computation and Quantum Information}.
\newblock Cambridge University Press, 2000.

\bibitem{Kessler13}
E.~M. Kessler, I.~Lovchinsky, A.~O. Sushkov, and M.~D. Lukin.
\newblock Quantum {E}rror {C}orrection for {M}etrology.
\newblock {\em arXiv:1310.3260}, 2013.

\bibitem{Arrad13}
G.~Arrad, Y.~Vinkler, D.~Aharonov, and A.~Retzker.
\newblock Increasing sensing resolution with error correction.
\newblock {\em arXiv:1310.3016}, 2013.

\bibitem{Duer13}
W.~D\"{u}r, M.~Skotiniotis, F.~Fr\"{o}wis, and B.~Kraus.
\newblock Improved quantum metrology using quantum error-correction.
\newblock {\em arXiv:1310.3750}, 2013.

\bibitem{Ozeri13}
R.~Ozeri.
\newblock Heisenberg limited metrology using {Q}uantum {E}rror-{C}orrection
  {C}odes.
\newblock {\em arXiv:1310.3432}, 2013.

\bibitem{Kesting13}
F.~Kesting, F.~Fr\"owis, and W.~D\"ur.
\newblock Effective noise channels for encoded quantum systems.
\newblock {\em Phys. Rev. A}, 88:042305, Oct 2013.

\bibitem{laflamme96}
R.~Laflamme, C.~Miquel, J.~P. Paz, and W.~H. Zurek.
\newblock Perfect quantum error correcting code.
\newblock {\em Phys. Rev. Lett.}, 77:198, 1996.

\bibitem{Montakhab08}
A.~Montakhab and A.~Asadian.
\newblock Dynamics of global entanglement under decoherence.
\newblock {\em Phys. Rev. A}, 77:062322, 2008.

\bibitem{Chaves10}
R.~Chaves and L.~Davidovich.
\newblock Robustness of entanglement as a resource.
\newblock {\em Phys. Rev. A.}, 82:052308, 2010.

\bibitem{Campbell09}
S.~Campbell, M.~S. Tame, and M.~Paternostro.
\newblock Characterizing multipartite symmetric dicke states under the effects
  of noise.
\newblock {\em New J. Phys.}, 11:073039, 2009.

\bibitem{Schmid08}
C.~Schmid, N.~Kiesel, W.~Laskowski, W.~Wieczorek, M.~{\.{Z}}ukowski, and
  H.~Weinfurter.
\newblock Discriminating multipartite entangled states.
\newblock {\em Phys. Rev. Lett.}, 100:200407 -- 200411, 2008.

\bibitem{Toth07b}
G.~T{\'{o}}th.
\newblock Detection of multipartite entanglement in the vicinity of symmetric
  {D}icke states.
\newblock {\em J. Opt. Soc. Am. B}, 24:275 -- 282, 2007.

\bibitem{toth07}
G.~T{\'{o}}th, C.~Knapp, O.~G{\"{u}}hne, and H.~J. Briegel.
\newblock Optimal spin squeezing inequalities detect bound entanglement in spin
  models.
\newblock {\em Phys. Rev. Lett.}, 99:250405, 2007.

\bibitem{patane07}
D.~Patan{\`{e}}, R.~Fazio, and L.~Amico.
\newblock Bound entanglement in the xy model.
\newblock {\em New J. Phys.}, 9:322, 2007.

\bibitem{Jack07}
D.~Cavalcanti, A.~Ferraro, A.~Garc{\'{i}}a-Saez, and A.~Ac{\'{i}}n.
\newblock Thermal bound entanglement and area laws.
\newblock {\em Phys. Rev. A.}, 78:012335, 2008.

\bibitem{kay07}
A.~Kay and J.~K. Pachos.
\newblock Multipartite purification protocols: Upper and optimal bounds.
\newblock {\em Phys. Rev. A}, 75:062307, 2007.

\bibitem{Kaszlikowski08}
D.~Kaszlikowski and A.~Kay.
\newblock A witness of multipartite entanglement strata.
\newblock {\em New J. Phys.}, 10:053026, 2008.

\bibitem{kay10}
A.~Kay.
\newblock Arboreal bound entanglement.
\newblock {\em J. Phys. A: Math. Theor.}, 43:495301, 2010.

\bibitem{DLCZ}
L.~M. Duan, M.~D. Lukin, J.~I. Cirac, and P.~Zoller.
\newblock Long-distance quantum communication with atomic ensembles and linear
  optics.
\newblock {\em Nature}, 414:413, 2001.

\bibitem{kimbleQN}
H.~J. Kimble.
\newblock The quantum network.
\newblock {\em Nature}, 453:1023 -- 1030, 2008.

\bibitem{farias}
O.~J. Far{\'{i}}as, C.~L. Latune, S.~P. Walborn, L.~Davidovich, and P.~H.~Souto
  Ribeiro.
\newblock Determining the dynamics of entanglement.
\newblock {\em Science}, 324:1414--1417, 2009.

\bibitem{jinshi02}
J.-S. Xu, C.-F. Li, M.~Gong, X.-B. Zou, C.-H. Shi, G.~Chen, and G.-C. Guo.
\newblock Experimental demonstration of photonic entanglement collapse and
  revival.
\newblock {\em Phys. Rev. Lett.}, 104(10):100502, Mar 2010.

\bibitem{kielpinski01}
D.~Kielpinski, V.~Meyer, M.~A. Rowe, C.~A. Sackett, W.~M. Itano, C.~Monroe, and
  D.~J. Wineland.
\newblock A decoherence-free quantum memory using trapped ions.
\newblock {\em Science}, 291:1013--1015, 2001.

\bibitem{roos06}
C.~F. Roos, M.~Chwalla, K.~Kim, M.~Riebe, and R.~Blatt.
\newblock {`}designer atoms{'} for quantum metrology.
\newblock {\em Nature}, 443:316--319, 2006.

\bibitem{yamamoto}
T.~Yamamoto, K.~Hayashi, S.~K. Ozdemir, M.~Koashi, and N.~Imoto.
\newblock Robust photonic entanglement distribution by state-independent
  encoding onto decoherence-free subspace.
\newblock {\em Nature Photon.}, 2:488--491, 2008.

\bibitem{sorensen00}
A.~S{{\o}}rensen and K.~M{{\o}}lmer.
\newblock Entanglement and quantum computation with ions in thermal motion.
\newblock {\em Phys. Rev. A}, 62:022311, 2000.

\bibitem{benhelm08}
J.~Benhelm, G.~Kirchmair, C.~F. Roos, and R.~Blatt.
\newblock Towards fault-tolerant quantum computing with trapped ions.
\newblock {\em Nat. Phys.}, 4:463--466, 2008.

\bibitem{marshall03}
W.~Marshall, C.~Simon, R.~Penrose, and D.~Bouwmeester.
\newblock Towards quantum superpositions of a mirror.
\newblock {\em Phys. Rev. Lett.}, 91:130401, 2003.

\bibitem{groblacher}
S.~Groblacher, K.~Hammerer, M.~R. Vanner, and M.~Aspelmeyer.
\newblock Observation of strong coupling between a micromechanical resonator
  and an optical cavity field.
\newblock {\em Nature}, 460(7256):724--727, 08 2009.

\bibitem{kleckner08}
D.~Kleckner, I.~Pikovski, E.~Jeffrey, L.~Ament, E.~Eliel, J.~van~den Brink, and
  2~D.~Bouwmeester1.
\newblock Creating and verifying a quantum superposition in a
  micro-optomechanical system.
\newblock {\em New J. Phys.}, 10:095020, 2008.

\bibitem{aolita07ions}
L.~Aolita, L.~Davidovich, K.~Kim, and H.~H\"affner.
\newblock Universal quantum computation in decoherence-free subspaces with hot
  trapped ions.
\newblock {\em Physical Review A}, 75:052337, 2007.

\bibitem{Monz09}
T.~Monz, K.~Kim, A.~S. Villar, P.~Schindler, M.~Chwalla, M.~Riebe, C.~F. Roos,
  H.~H\"affner, W.~H\"ansel~M. Hennrich, and R.~Blatt.
\newblock Realization of {U}niversal {I}on-{T}rap {Q}uantum {C}omputation with
  {D}ecoherence-{F}ree {Q}ubits.
\newblock {\em Phys. Rev. Lett.}, 103:200503, 2009.

\bibitem{Wu08}
C.~Wu, X.-L. Feng, X.~X. Yi, I.~M. Chen, and C.~H. Oh.
\newblock Quantum gate operations in the decoherence-free subspace of
  superconducting quantum-interference devices.
\newblock {\em Phys. Rev. A}, 78:062321, 2008.

\bibitem{Xue09}
Z.-Y. Xue, S.-L. Zhu, and Z.~D. Wang.
\newblock Quantum computation in decoherence-free subspace with superconducting
  devices.
\newblock {\em Eur. Phys. J. D}, 55:223, 2009.

\bibitem{aolita07quantcom}
L.~Aolita and S.~P. Walborn.
\newblock Quantum communication without alignment using multiple-qubit
  single-photon states.
\newblock {\em Phys. Rev. Lett.}, 98:100501, 2007.

\bibitem{souza08}
C.~E.~R. Souza, C.~V.~S. Borges, J.~A.~O. Huguenin, L.~Aolita, S.~P. Walborn,
  and A.~Z. Khoury.
\newblock Quantum key distribution without a shared reference frame.
\newblock {\em Phys. Rev. A}, 77:032345, 2008.

\bibitem{Dambrosio12}
V.~D'Ambrosio, E.~Nagali, S.~P. Walborn, L.~Aolita, S.~Slussarenko,
  L.~Marrucci, and F.~Sciarrino.
\newblock Complete experimental toolbox for alignment-free quantum
  communication.
\newblock {\em Nat. Commun.}, 3:961, 2012.

\bibitem{Vallone14}
G.~Vallone, V.~D'Ambrosio, A.~Sponselli, S.~Slussarenko, L.~Marrucci,
  F.~Sciarrino, and P.~Villoresi.
\newblock Free-space quantum key distribution by rotation-invariant twisted
  photons.
\newblock {\em ArXiv: 1402.2932}.

\bibitem{Fickler12}
R.~Fickler, R.~Lapkiewicz, W.~N. Plick, M.~Krenn, C.~Schaeff, S.~Ramelow, and
  A.~Zeilinger.
\newblock Quantum {E}ntanglement of {V}ery {H}igh {A}ngular {M}omenta.
\newblock {\em Science}, 338:640, 2012.

\bibitem{DAmbrosio13}
V.~D'Ambrosio, N.~Spagnolo, L.~Del Re, S.~Slussarenko, Y.~Li, L.~C. Kwek,
  L.~Marrucci, S.~P. Walborn, L.~Aolita, and F.~Sciarrino.
\newblock Photonic polarization gears for ultra-sensitive angular measurements.
\newblock {\em Nat. Commun.}, 4:2432, 2013.

\bibitem{Marrucci11}
L.~Marrucci, E.~Karimi, S.~Slussarenko, B.~Piccirillo, E.~Santamato, E.~Nagali,
  and F.~Sciarrino.
\newblock Spin-to-orbital conversion of the angular momentum of light and its
  classical and quantum applications.
\newblock {\em J. Opt.}, 13:064001, 2011.

\bibitem{Flicker14}
R.~Fickler, R.~Lapkiewicz, M.~Huber, M.~P.~J. Lavery, M.~J. Padgett, and
  A.~Zeilinger.
\newblock Interface between path and {O}{A}{M} entanglement for
  high-dimensional photonic quantum information.
\newblock {\em ArXiv: 1402.2423}.

\bibitem{Krenn14}
M.~Krenn, R.~Fickler, M.~Fink, J.~Handsteiner, M.~Malik, T.~Scheidl, R.~Ursin,
  and A.~Zeilinger.
\newblock Twisted light communication through turbulent air across {V}ienna.
\newblock {\em Arxiv: 1402.2602}.

\bibitem{Ozzy14}
O.~J. Far{\'{i}}as, V.~D'Ambrosio, C.~Taballione, F.~Bisesto, S.~Slussarenko,
  L.~Aolita, L.~Marrucci, S.~P. Walborn, and F.~Sciarrino.
\newblock Resilience of hybrid optical angular momentum qubits to turbulence.
\newblock {\em In preparation}, (2014).

\bibitem{Schliemann01}
J.~Schliemann, J.~I. Cirac, M.~Ku\ifmmode~\acute{s}\else \'{s}\fi{},
  M.~Lewenstein, and D.~Loss.
\newblock Quantum correlations in two-fermion systems.
\newblock {\em Phys. Rev. A}, 64:022303, 2001.

\bibitem{eckert02}
K.~Eckert, J.~Schliemann, D.~Bru{\ss{}}, and M.~Lewenstein.
\newblock Quantum correlations in systems of indistinguishable particles.
\newblock {\em Ann. Phys.}, 299:88--127, 2002.

\bibitem{Ghirardi04}
G.~Ghirardi and L.~Marinatto.
\newblock General criterion for the entanglement of two indistinguishable
  particles.
\newblock {\em Phys. Rev. A}, 70:012109, 2004.

\bibitem{Benatti20121304}
F.~Benatti, R.~Floreanini, and U.~Marzolino.
\newblock Bipartite entanglement in systems of identical particles: The partial
  transposition criterion.
\newblock {\em Ann. Phys.}, 327:1304 -- 1319, 2012.

\bibitem{tichy2013}
M.~C. Tichy, F.~de~Melo, M.~Ku{\'s}, F.~Mintert, and A.~Buchleitner.
\newblock Entanglement of identical particles and the detection process.
\newblock {\em Fortschr. Phys.}, 61:225, 2013.

\bibitem{Eisert10}
J.~Eisert and T.~Prosen.
\newblock Noise-driven quantum criticality.
\newblock {\em ArXiv: 1012.5013}, 2010.

\bibitem{barreiro11}
J.~T. Barreiro, M.~Muller, P.~Schindler, D.~Nigg, T.~Monz, M.~Chwalla,
  M.~Hennrich, Christian~F. Roos, P.~Zoller, and R.~Blatt.
\newblock An open-system quantum simulator with trapped ions.
\newblock {\em Nature}, 470(7335):486--491, 02 2011.

\bibitem{lieb72}
E.~H. Lieb and D.~W. Robinson.
\newblock The finite group velocity of quantum spin systems.
\newblock {\em Commun. Math. Phys.}, 28:251, 1972.

\bibitem{hastings04}
M.~B. Hastings.
\newblock Locality in quantum and markov dynamics on lattice and networks.
\newblock {\em Phys. Rev. Lett.}, 93:140402, 2004.

\bibitem{sims}
B.~Nachtergaele, Y.~Ogata, and R.~Sims.
\newblock Propagation of correlations in quantum lattice systems.
\newblock {\em J. of Stat. Phys.}, 124:1--13, 2006.

\bibitem{braviy06}
S.~Bravyi, M.~B. Hastings, and F.~Verstraet.
\newblock Lieb-{R}obinson bounds and the generation of correlations and
  topological quantum order.
\newblock {\em Phys. Rev. Lett.}, 97:050401, 2006.

\bibitem{eisert13}
M.~Kliesch, C.~Gogolin, and J.~Eisert.
\newblock Lieb-{R}obinson bounds and the simulation of time evolution of local
  observables in lattice systems.
\newblock {\em ArXiv:1306.0716v1}, 2013.

\bibitem{Poulin}
D.~Poulin.
\newblock Lieb-{R}obinson bound and locality for general {M}arkovian quantum
  dynamics.
\newblock {\em Phys. Rev. Lett.}, 104:190401, 2010.

\bibitem{hauke13}
P.~Hauke and L.~Tagliacozzo.
\newblock Spread of correlations in long-range interacting quantum systems.
\newblock {\em Phys. Rev. Lett.}, 111:207202, 2013.

\bibitem{roos13}
J.~Schachenmayer, B.~P. Lanyon, C.~F. Roos, and A.~J. Daley.
\newblock Entanglement growth in quench dynamics with variable interactions.
\newblock {\em Phys. Rev. X}, 3:031015, 2013.

\bibitem{jurcevic14}
P.~Jurcevic, B.~P. Lanyon, P.~Hauke, C.~Hempel, P.~Zoller, R.~Blatt, and C.~F.
  Roos.
\newblock Observation of entanglement propagation in a quantum many-body
  system.
\newblock {\em arXiv:1401.5387v3}, 2014.

\end{thebibliography}
%
\end{document}